%% file: phd_thesis.tex
\documentclass[a4paper,twoside,openright]{report}
\usepackage{latexsym}
\usepackage{amsmath}
\usepackage{amsfonts}
\usepackage{amssymb}
\usepackage{multirow}
\usepackage[dvips]{graphicx}
\usepackage[T1]{fontenc}
\usepackage{verbatim}
\usepackage{mathrsfs}
\usepackage{fancyhdr}
\usepackage{pstricks,pst-node,pst-text,pst-3d}
\usepackage{mathrsfs}
\usepackage{fancyhdr}
\usepackage{lastpage}
\usepackage{longtable}
\title{Simplicial and Modular aspects of String Dualities}
\author{Valeria Gili}
\newcommand{\cyl}[1]{$\Delta^*_{\varepsilon (#1)}$}
\newcommand{\boup}{S^{(+)}_{\varepsilon (k)}}
\newcommand{\boum}{S^{(-)}_{\varepsilon (k)}}
\newcommand{\z}{\tilde{\zeta}}
\newcommand{\fa}{\mathfrak{a}}
\newcommand{\ofa}{\overline{\mathfrak{a}}}
\newcommand{\vacuum}[1]{\vert \left( \mu^{#1}, \nu^{#1}\right)\rangle}
\newcommand{\ie}{\emph{i.e.}}
\newcommand{\q}{\frac{2\pi}{2\pi -\varepsilon(k)}}
\newcommand{\Rangle}{\rangle\negthinspace\rangle}
\newcommand{\ordprod}[1]{:\negmedspace{#1}\negmedspace:}
\newcommand{\bz}{\bar{z}}
\newcommand{\bgz}{\bar{\zeta}}
\newcommand{\e}{\mbox{exp}}                       % esponenziale
\newcommand{\vac}{\vert0\rangle}
\def\IL{\relax{\rm I\kern-.18em L}}

\def\a{\alpha}  
\def\e{\epsilon}

\newcommand{\ft}[2]{{\textstyle\frac{#1}{#2}}}
\newtheorem{lemma}{Lemma}
\newtheorem{teo}{Theorem}
\newtheorem{rem}{Remark}
\newtheorem{prop}{Proposition}
\newtheorem{defin}{Definition}
\begin{document}
\begin{titlepage}
\begin{center}
\begin{figure}[t!]
  \centering
  \includegraphics[width=3cm]{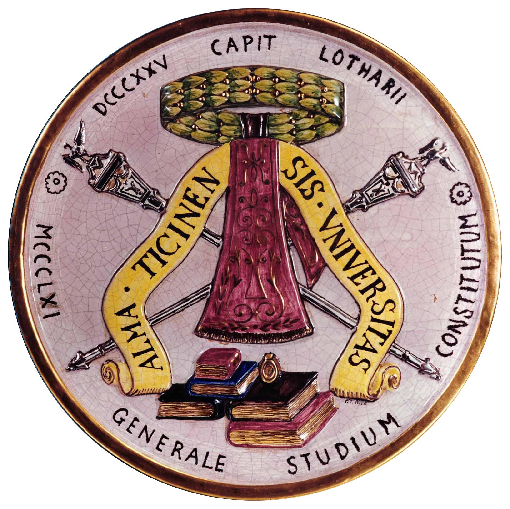}
\end{figure}
{\Large \textsc{University of Pavia}} 
{\large\mbox{\textsc{Department of Theoretical and Nuclear Physics}}}

\vspace*{5cm}

{\Huge \centering \textbf{Simplicial and modular\\ 
\vspace*{3mm}
aspects of string dualities}}
\end{center}
\vspace{2,5cm}
Advisor:\\
\textbf{Prof. Mauro Carfora}
\vspace{1,5cm}
\begin{flushright}
Doctoral thesis of:\\
\textbf{Valeria L. Gili}
\vfill
\end{flushright}
\begin{center}
\textbf{Dottorato di Ricerca XVIII Ciclo}
\end{center}
\end{titlepage}

\tableofcontents

\include{introduction}

\part[Simplicial aspects of string dualities]
{\vspace{1cm}\psshadowbox[shadowsize=7pt,framesep=11pt]{\parbox{11cm}{\sc
      \centering Simplicial aspects of string dualities}}} 
\label{part:ssd}

\include{uniformizations}

\include{bcft_langlands}

\include{bio}

\include{osgauge}

\part[Cosmological backgrounds in String Theory]
{\vspace{1cm}\psshadowbox[shadowsize=7pt,framesep=11pt]{\parbox{13cm}{\sc
    \centering Cosmological  backgrounds of superstring theory:\\
    oxidation and branes}}}
\label{part:cb}

\include{cosmo}

\appendix

\part[Appendices]
{\vspace{1cm}\psshadowbox[shadowsize=7pt,framesep=11pt]{\parbox{7cm}{\sc
    \centering Appendices}}}

\include{lang_proj}

\include{biobcft}

\include{bcft_deform}

\include{formulae}

\include{roots}

\include{ackn}

\addcontentsline{toc}{chapter}{References}
\bibliographystyle{unsrt}
\bibliography{dualitybib_db.bib,AdS_CFT-bibliodb.bib,cosmodb.bib}

\end{document}

%% file: introduction.tex
\chapter*{Introduction}
\addcontentsline{toc}{chapter}{Introduction}

My Ph.D. work has been dedicated to the study of different aspects of
modern unifying theories. 

Part of the time has been devoted to search for cosmological
backgrounds in superstring theories, thus continuing the research line
started in my Laurea thesis.

This work, which has been done in collaboration with Prof. Pietro
Fr\'e's group, fits into a wide set of research lines merging string
theory, which try to express all fundamental constituents of the
universe and their interaction into a unique framework, with
the undertaken of cosmological models, whose aim is to describe the
origin, the evolution and the structure of the Universe as a whole.
The great number of success of both models has been suggesting since
middle of 90's a possible unification of their investigation areas.  In
view of the new observational data which seems to provide evidence for
a small but positive cosmological constant
\cite{experiment1,experiment2,experiment3,linde90}, there has been
wide interest in the context of M--theory/string theory and extended
supergravities for the search of de Sitter like vacua (see for
instance
\cite{Kachru:2003sx,Kachru:2003aw,Burgess:2003ic,Fre:2002pd,deRoo:2003rm}
and references therein).  More generally, the analysis of
time--dependent backgrounds
\cite{Gutperle:2002ai,Ivashchuk:1997pk,Cornalba:2002nv,Cornalba:2003ze,Leblond:2003ac,Kruczenski:2002ap,Ohta:2003pu,Emparan:2003gg,Buchel:2003xa,Papadopoulos:2002bg,que,GV,craps,ban,setu,cope,mart}
has been done in various approaches and at different levels, namely
both from the microscopic viewpoint, considering time--dependent
boundary states and boundary CFTs (see for instance
\cite{Sen:2002vv,Sen:2002nu} and references therein) and from the
macroscopic viewpoint studying cosmological solutions in both gauged
and ungauged supergravities.

Successively, in collaboration with Prof. Mauro Carfora's group, I
started investigating modular and simplicial properties of string
dualities.  Both collaborations have been characterized by a deep
analysis of dynamic symmetries associated with Lie algebras of both
finite and affine kind.

The first hint of role {Ka\v c-Moody} algebras plays in describing
dynamical symmetries of gravitational models goes back to early '70,
when Geroch showed that the automorphism group of solutions of
Einstein Gravity reduced to $D=2$ is infinite dimensional and it has
as underlying Lie algebra $A_{\hat{1}}$, the untwisted affine {Ka\v
  c-Moody} extension of $A_1$ \cite{Geroch:1970nt,Geroch:1972yt}, this
being the three dimensional Lie algebra of $SL(2,\mathbb{R})$.  This
discovery triggered a lot of interest in the analysis of these
symmetries, but it was with the advent of string theory and
supergravity that the interest in infinite symmetries received a
boost.  As a matter of fact, by dimensionally reducing $D=10$ or
$D=11$ supergravity to lower dimensions, new dynamical symmetries
arise. The so called hidden symmetries act as isometries of the metric
on the scalar manifold and as generalized electric/magnetic duality
rotations on the various $p$-forms. Formalization of these concepts
goes back to the work of Cremmer and Julia, who clarified that $E_{11
  - D (11 - D)}$ is the duality symmetry group of maximal supergravity
in $D \geq 3$ dimensions, obtained by a Kaluza-Klein compactification
of $M$-theory on a $T^{11 - D}$ torus.  In particular, they showed
that the massless scalars which emerge from the Kaluza-Klein mechanism
in $D$ dimensions, just parametrize the maximally non--compact coset
manifold
\begin{equation*}
  \mathcal{M}_{scalar} \, = \, \frac{E_{11 - D (11 -
      D)}}{H_{11 - D }}
\label{eseries}
\end{equation*}
where $H_{(11-D)}$ is the maximally compact subgroup of the simple Lie
group $E_{11 - D (11 - D)}$ \cite{cremmer80,julia81}. 

Julia noted that the extension of this process to $D < 3$ calls into
play algebras which are no longer finite dimensional, but rather they
are infinite dimensional Ka\v c-Moody ones\cite{Julia:1981wc}. As a
matter of fact, $E_{9(9)}$ is the affine extension of $E_{8(8)}$,
while $E_{10(10)}$ is its double hyperbolic extension. This last case
is particularly intriguing. Since compactification and truncation to the
massless modes is an alternative way of saying that we just focus on
field configurations \textsc{that depend only the remaining $D$
  coordinates}, the compactification on a $T^{10}$ torus leads to non
trivial dependence of supergravity fields just on \textit{one
  coordinate}, namely \textit{time}, thus linking the arising of
cosmological backgrounds to the $E_{10}$ algebra.

In this connection, a very much appealing and intriguing scenario has
been proposed in a series of papers by Damour et al.
\cite{bill99,dualiza2,Henneaux:2003kk,deBuyl:2003za,Damour:2002tc,Damour:2000pq,Damour:2001sa,Damour:2000hv,Damour:2000th,Damour:2000wm,Demaret:sg}:
that of cosmological billiards.  Studying the asymptotic behavior of
ten (eleven) dimensional supergravity field equations near time
(space--like) singularities, these authors envisaged the
possibility that the nine (ten) cosmological scale factors relative to
the different space dimensions of string theory plus the dilaton could
be assimilated to the Lagrangian coordinates of a fictitious ball
moving in a ten(resp. eleven)--dimensional space. 

\begin{figure}[t]
  \centering
\includegraphics[width=12cm]{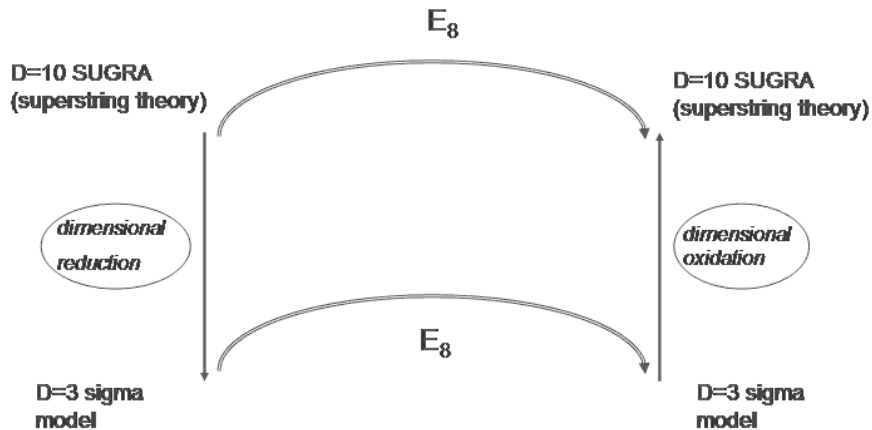}
  \caption{Time--dependent homogeneous supergravity backgrounds in
$D=10$ can be obtained by first dimensionally reducing to $D=3$,
solving the differential equations of the sigma model and then
oxiding back the result to $D=10$. This procedure defines also the
action of the hidden symmetry group $E_{8(8)}$ on the ten
dimensional configurations.}
  \label{disegnino}
\end{figure}

In their approach, Damour and collaborators analyzed the cosmic
billiard phenomenon as an asymptotic regime in the neighborhood of
space-like singularities. The billiard walls were seen as the various
$p$-forms of supergravity acting as exponential potential walls which
become sharper and sharper in the evolution toward the singularity,
ending in infinite potential walls in the $t \rightarrow 0^+$ limit.
The main focus attention was centered on establishing whether and under
which conditions there may be a chaotic behavior in the evolutions of
scale factors of the universe. Eventually, the authors established
that the billiard table to be hyperbolic was a necessary condition,
where the billiard table was identified with the Weyl chamber of the
$E_{(10)}$ algebra.

There is a clear relation between this picture and the duality groups
of superstring theories. As a matter of fact, the Cartan generators of
the $\mathrm{E_{r(r)}}$ algebra are dual to the \textit{radii} of the
$T^{r-1}$ torus plus the dilaton. So it is no surprising that the
evolution of the cosmological scale factor should indeed represent
some kind of motion in the dual of the Cartan subalgebra of $E_{10}$.
Although naturally motivated, the $E_{10}$ billiard picture was so far
considered only in the framework of an approximated asymptotic
analysis and no exact solution with such a behavior was actually
constructed.

In our work, we investigated the problem from a different viewpoint.
We focused on three--dimensional maximal supergravity
\cite{Marcus:1983hb}--\cite{Nicolai:2001sv}, namely on the dimensional
reduction of type II theories on a $\mathrm{T^7}$ torus, instead of
going all the way down to reduction to one--dimension. The advantage
of this choice is that all the bosonic fields are already scalar
fields, described by a non--linear sigma model without, however, the
need of considering Ka\v c--Moody algebras which arise as isometry
algebras of scalar manifolds in $D<3$ space--times. In this way we had
been able to utilize the \textit{solvable Lie algebra approach} to the
description of the whole bosonic sector which enabled us to give a
completely algebraic characterization of the microscopic origin of the
various degrees of freedom
\cite{Andrianopoli:1996bq,Andrianopoli:1996zg}.  Within this framework
the supergravity field equations for bosonic fields restricted to only
time dependence reduce simply to the geodesic equations in the target
manifold $\mathrm{E_{8(8)}/SO(16)}$.  These latter can be further
simplified to a set of differential equations whose structure is
completely determined in Lie algebra terms. This was done through the
use of the so called \textit{Nomizu operator}.  The concept of Nomizu
operator coincides with the concept of covariant derivative for
solvable group manifolds. The possibility of writing covariant
derivatives in this algebraic way as linear operators on solvable
algebras relies on the theorem that states that a non--compact coset
manifold with a transitive solvable group of isometries is isometrical
to the solvable group itself.

The underlying idea for our approach was rooted in the concept of
hidden symmetries. Cosmological backgrounds of superstring theory,
being effectively one--dimensional fill orbits under the action of a
very large symmetry group, possibly $\mathrm{E_{10}}$, that necessarily
contains $\mathrm{E_{8(8)}}$, as the manifest subgroup in three
dimensions.  Neither $\mathrm{E_{10}}$ nor $\mathrm{E_{8(8)}}$ are
manifest in $10$--dimensions but become manifest in lower dimension.
So an efficient approach to finding spatially homogeneous solutions in
ten dimensions consists of the process schematically described in
fig.\ref{disegnino}. First one reduces to $D=3$, then solves the
geodesic equations in the algebraic setup provided by the
Nomizu--operator--formalism and then oxides back the result to a full
fledged $D=10$ configuration. Each possible $D=3$ solution is
characterized by a non--compact subalgebra
\begin{equation*}
  \mathbf{G} \, \subset \,\mathrm{ {E}_{8(8)}}
\label{Gsubalg}
\end{equation*}
which defines the smallest consistent truncation of the full
supergravity theory within which the considered solution can be
described. The inverse process of oxidation is not unique but leads to
as many physically different ten dimensional solutions as there are
algebraically inequivalent ways of embedding $\mathbf{G}$ into
$\mathrm{E_{8(8)}}$. In \cite{Fre:2003ep} we illustrated this
procedure by choosing for $\mathbf{G}$ the smallest non abelian rank
two algebra, namely $\mathbf{G}=\mathrm{A_2}$. The solvable Lie
algebra formalism allowed us to control, through the choice of the
$\mathbf{G}$--embedding, the physical ten--dimensional interpretation
of any given $\sigma$--model solution. Focusing on a particular
embedding for the subalgebra $\mathrm{A_2}$, we got a type IIB time
dependent background generated by a system of two euclidean D-branes
or S-branes
\cite{Gutperle:2002ai,Ivashchuk:1997pk,Cornalba:2002nv,Cornalba:2003ze,
  Leblond:2003ac,Kruczenski:2002ap,Ohta:2003pu,Emparan:2003gg,Buchel:2003xa}:
a D3 and a D1, whose world volumes are respectively four and two
dimensional. This physical system contains also an essential non
trivial B--field reflecting the three positive root structure of the
$\mathrm{A_2}$ Lie algebra, one root being associated with the RR
$2$--form $C^{[2]}$, a second with the RR $4$--form $C^{[4]}$ and the
last with the NS $2$--form $B^{[2]}$. In the time evolution of this
exact solution of type II B supergravity we retrieved a smooth
realization of the bouncing phenomenon envisaged by the \textit{cosmic
  billiards} of \cite{dualiza2}--\cite{Demaret:sg}.  Indeed, in the
oxided picture, the scale factors corresponding to the dimensions
parallel to the S--branes first expand and then, after reaching a
maximum, contract. The reverse happens to the dimensions transverse to
the $S$--branes. They display a minimum approximately at the same time
when the parallel ones are maximal.  Transformed to the dual CSA space
this is the bouncing of the cosmic ball on a Weyl chamber wall. This
is not the full cosmic billiard, but it illustrates the essential
physical phenomena underlying its implementation.

In the meanwhile I started, in collaboration with Prof. Mauro
Carfora and Dr. Claudio Dappiaggi, to investigate simplicial and
modular aspects of string dualities.

\begin{figure}[t]
  \centering
  \includegraphics[width=\textwidth]{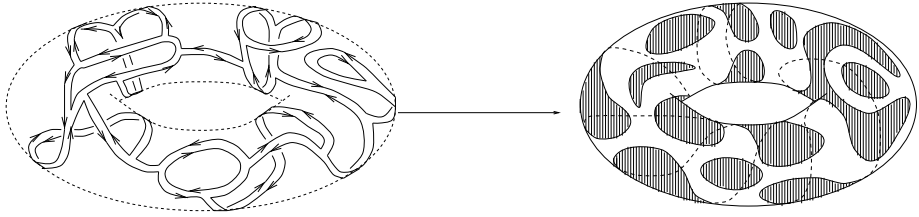}
  \caption{A ribbon graph with $g=1$ and $h=9$ and the associated
    Riemann surface, in wich holes are filled with regions having the
    topology of a disk}
  \label{fig:filling}
\end{figure}

Since 't Hooft seminal paper\cite{tHooft}, the idea that a large N
gauge theory has a dual description in term of a closed string theory
has been drawing the attention of large part of theoretical physicists
community.  According to 't Hooft prescription, if we consider a
$U(N)$ gauge theory whose action is written as $\frac{1}{g_{YM}^2}
\int \mathcal{L}(A)$ ($A$ is the gauge connection), the group
theoretical part of Feynman diagrams arises as a ribbon graph, whose
edges acquire a gauge coloring in the first fundamental times first
anti-fundamental representation of the gauge group. In this connection,
amplitudes and Green functions are obtained summing over all possible
(planar and non-planar) diagrams, with the appropriate combinatorial
factor. Ribbon graphs associated to vacuum amplitudes can be
classified according to their dependence from $N$ and $g_{YM}$. As a
matter of fact, if we consider a ribbon graph with  $h$ boundary
components, it comes with the factor $(g_{YM}^2)^{-V + E} N^h$, where $V$
and $E$ are respectively the number of vertexes and edges of the
graph. Since each ribbon graph can be viewed as a closed Riemann
surface with $h$ holes, the dependence of the amplitude from  $N$ and
$g_{YM}$ can be captured by the topology of the underlying
surface. If we consider an associated Riemann surface of genus $g$, we
can write:
\begin{equation}
  (g_{YM}^2)^{-V + E} N^h \,=\, (g_{YM}^{2g -2})\,(g_{YM}^2 N)^h
\,=\,
 (g_{YM}^{2g -2})\,\lambda^h
\end{equation}
where we have exploited the topological relation $V - E + h = 2 - 2g$,
and where we have introduced the 't Hooft coupling $\lambda = g_{YM}^2
N$. Thus, the sum over all ribbon graphs remarkably translates into a
sum over all topologies. The associated amplitude is then given by:
\begin{equation}
  \label{eq:amc}
  \mathcal{F} \,=\, 
  \sum_{g=0}^\infty
  \sum_{h=1}^\infty  
  (g_{YM}^{2g -2})\,\lambda^h\,
  \mathcal{F}_{g,h} 
  \,=\,
  \mathcal{F} \,=\, 
  \sum_{g=0}^\infty
  (g_{YM}^{2g -2})\,
  \mathcal{F}_{g} ,
\end{equation}
where we have defined 
\begin{equation}
  \label{eq:am}
  \mathcal{F}_{g} ,
  \,=\,  
  \sum_{h=1}^\infty 
  \lambda^h\,
  \mathcal{F}_{g,h}  
\end{equation}
and the coefficients $\mathcal{F}_{g,h}$ being functions of other
parameters of the theory. 't~Hooft's conjecture states that, since at
fixed $\lambda$ formula \eqref{eq:am} can be seen both as an $g_{YM}
\rightarrow 0$ or as a $N \rightarrow \infty$ expansion, in the large
$N$ (and fixed $\lambda$) limit an equivalent description arises,
involving closed Riemann surfaces which are obtained ``filling holes
of the ribbon graphs with disks''. This naturally promotes the
expansion \eqref{eq:amc} to a sum over genus, \ie{} a closed string
expansion.

Furthermore, at the beginning of 90s', Susskind\cite{Susskind:1994vu}
and 't~Hooft\cite{'tHooft:1993gx} himself introduced the
\textsc{holographic principle}, according to which classical spacetime
geometry and its matter contents arise from an underling (non a priori
gravitational) quantum theory in such a way that the covariant entropy
bound should be satisfied.

In this connection, Maldacena's conjecture about $AdS/CFT$
correspondence~\cite{Maldacena:1997re} had merged elegantly these two
aspects. It is perhaps the best example of such a kind of holographic
correspondence, in which the supergravity limit of a string theory
in $d$ dimensions admits an holographically dual description in term
of a gauge theory living on a codimension one manifold, and where the
connection between the gauge theory in the open string sector and the
closed string theory is guided by a large N transition. Not less
important, it provides a powerful tool in studying non-perturbative
effects of a quantum field theory by means of perturbative techniques
proper of string theory.

However, dealing with $AdS/CFT$ correspondence, a strong working
assumption is usually performed. The large N transition is, a priori,
a worldsheet effect: as remarked above, the leading (gauge theory)
large $N$ correlator functions arise from planar open string diagrams
with some insertion on their boundaries. In the expected worldsheet
scenario, the loops of the original conformal field theory glue up to
form a closed Riemann surface with $n$ closed string insertions on a
different background.  Thus, the open-closed string equivalence works
at the level of the worldsheet and it has to be implemented at a
worldsheet moduli space level, though the very process is unknown.

On the contrary, in a topological setup, even if the duality is guided
by a geometrical transition at target space level (for a review see
\cite{Marino:2004eq,Grassi:2002tz}), the duality process acting at a
worldsheet level was identified.  As a matter of fact, in their
original definition, topological large N open/closed string duality
are based on the gauge theory description of topological string
theory. In particular, when the $A$ topological string model is
defined on a Calabi-Yau $X=T^*M$, $M$ being a three-manifold, and
there are $N$ topological $D$-branes wrapping on $M$, the full Open
String Field Theory is equivalent to a $U(N)$ Chern-Simons on $M$.
The open string amplitudes are then just numbers computed by the
fatgraphs of the corresponding gauge theories.  In this background,
the large N transition act at a geometric level as a conifold
transition \cite{Grassi:2002tz} relating the open string Calabi-Yau
background X underling the gauge theory and the closed string
Calaby-Yau background Y. As usual, at worldsheet level, the scenario
one expects is that the boundary of the open Riemann surface get glued
up to form a closed surface with $n$ close string insertion.  In this
connection, at worldsheet level, when the 't Hooft coupling is small,
a new branch of the conformal theory arises and the compact domains of
this new phase can be viewed as holes for an open string theory living
in a different regime. In such a context, assumptions are no longer
necessary.  However, the theory is topological and it excludes a
dynamical coupling with gravity in the target space. As pointed out,
open string amplitudes are just numbers which the duality process
rereads in term of the topological invariants of the closed surface.

In this connection, a paradigmatical result has been recently
established by Gopakumar \cite{Gopakumar:2003ns,Gopakumar:2004qb} in
the setup of AdS/CFT correspondence. His intuition has as starting
point $\mathcal{N} = 4$ SYM theory, namely the $\alpha' \rightarrow 0$
limit of open string theory.  Starting from a Schwinger
parametrization of free fields correlators, the author was able to
reorganize $n$ points amplitudes (at genus $g$ in a 't Hooft sense) in
terms of skeleton diagrams. These are simply the graphs obtained by
merging together all the homotopically equivalent contractions between
any two pairs of vertexes. In this connection, the generic skeleton
diagram associated to a particular merging of contractions is simply a
triangulation of the underlying genus $g$ surface, with as many
vertexes there are internal and external ones.  As a consequence, the
entire expansion for the $n$-points function can be completely
expressed as a sum over all the inequivalently connected skeleton
graphs contributing to the amplitude.

The counting over degrees of freedoms allowed him to
interpret the integral over Schwinger times plus the sum over
inequivalent skeleton graphs as sum over the decorated moduli space
$\mathcal{M}_{g,n} \times \mathbb{R}_+^n$. Thus, the arising of a
closed string moduli space, was claimed as the manifestation that a
glueing process might be seen as a change of variables in the
integrand of amplitudes. 

This transition was actually evident in a target space connection: the
author introduced an algorithm which, exploiting the
\emph{bulk-to-boundary} propagator introduced by Witten, Gubser,
Klebanov and Polyakov in \cite{Witten:1998qj,Gubser:1998bc}, allows to
recast gauge amplitudes as amplitudes in $AdS$\cite{Gopakumar:2004qb}.

In view of these results, in order to analyze string dualities in a general
framework, it seems that a discretized approach can play a fundamental
role, in particular dealing with triangulations with varying
connectivity.  Thus, motivated by the ubiquitous role that simplicial
methods play in the above results, in collaboration with Prof. Mauro
Carfora and Dr. Claudio Dappiaggi, we have tried to implement examples
of open/closed string duality in a recently introduced new geometrical
framework\cite{Carfora1}.

Our aim is to extend the large $N$ process
to the appealing context of dynamical background fields, while
retaining the \emph{open-to-closed} transition to be implemented at a
worldsheet level.

This approach is based on a careful use of uniformization theory for
triangulated surfaces carrying curvature degrees of freedom.  Starting
from a closed string theory point of view i.e.  an n-marked Riemann
surface with conical singularities, we can switch through geometrical
arguments to a open string theory point of view where the ''cones''
are traded with finite cylindrical ends.  This
transition between the open and the closed Riemann surface exploits a
particular connection between the $N_0$ localized curvature degrees of
freedom in the closed sector and the modular data associated to finite
cylindrical ends in the open sector\cite{Carfora1}.

In order to show how this uniformization arises, let us consider the
dual polytope associated with a Random Regge Triangulation
Triangulation \cite{Carfora3} $|T_l| \rightarrow M$ of a Riemannian
manifold $M$.  The singular Euclidean structure around each puncture
can be uniformized by a conformal class of conical metrics (see
formula \eqref{eq:metric}), while the edge-refinement of the Regge
dual polytope 1-skeleton is in one-to-one correspondence with
trivalent Ribbon graphs.  

Using properties of Jenkins-Strebel quadratic differentials
\cite{Carfora1}, it is possible to fix a point in the conformal class
decorating the neighborhood of each curvature supporting vertex,
uniformizing it with a punctured disk endowed with a conical metric
\begin{equation*}
  ds_{(k)}^{2}\doteq
  \frac{\lbrack L(k)]^{2}}{4\pi ^{2}}|\zeta (k)|^{-2(
    \frac{\varepsilon (k)}{2\pi })}|d\zeta (k)|^{2}.
\end{equation*}
Moreover, exploiting the Stebel theorem\cite{mulase}, it is possible
to introduce an uniformizing coordinate both along the edges of the
Ribbon Graph and on each trivalent vertex.

Alternatively, we can blow up every cone into a corresponding finite
cylindrical end, by introducing a finite annulus
\begin{equation*}
  \Delta_{\varepsilon (k)}^\ast\doteq 
  \left\{ 
    \zeta(k) \in \mathbb{C} | \quad 
    \exp{-\frac{2\pi }{2\pi -\varepsilon(k)}}
    \leq |\zeta (k)|\leq 1
  \right\}
\end{equation*}
endowed with the cylindrical metric:
\begin{equation*}
  |\phi (k)|\doteq \frac{\lbrack L(k)]^{2}}{4\pi ^{2}}|
  \zeta (k)|^{-2}|d\zeta(k)|^{2} 
\end{equation*}

It is important to stress the different role that the deficit angle
plays in such two uniformization.  In the ``closed'' uniformization
the deficit angles $\varepsilon(k)$ plays the usual role of localized
curvature degrees of freedom and, together with the perimeter of the
polytopal cells, provide the geometrical information of the underlying
triangulation.  Conversely, in the ``open'' uniformization, the
deficit angle associated with the $k$-th polytope cell defines the
geometric moduli of the $k$-th cylindrical end. As a matter of fact
each annulus can be mapped into a cylinder of circumpherence $L(k)$
and height $\frac{L(k)}{2 \pi - \varepsilon(k)}$, thus $\frac{1}{2 \pi
  - \varepsilon(k)}$ is the geometrical moduli of the cylinder.  This
shows how the uniformization process works quite differently from the
one used in Kontsevich-Witten models, in which the whole punctured
disk is uniformized with a cylindrical metric. In this case the disk
can be mapped into a semi-infinite cylinder, no role is played by the
deficit angle and the model is topological; conversely, in our case,
we are able to deal with a non topological theory.  

Very recently\cite{Gopakumar:2005fx} Gopakumar pursued further the
identification of the field theory expression for the integrand over
the Schwinger parameters and with a correlator of closed string vertex
operator. In his construction, he exploited the isomorphism between
the space of metric Ribbon graphs and $\mathcal{M}_{g,n} \times
\mathbb{R}_+^n$ to associate to each skeleton diagram naturally
coupled to the Schwinger parametrization of gauge amplitudes the
unique dual Ribbon graph. To this end, he proposed a precise
dictionary between Strebel lengths $l_r$ (\ie{} lengths of ribbon graph
edges) and Schwinger times $\tau_r$, namely:
\begin{equation}
  l_r \,=\, \frac{1}{\tau_r}
\end{equation}
With this dictionary in mind, exploiting the Strebel theorem, which
uniquely associate a closed Riemann surface uniformization to a given
ribbon graph, he gave a concrete proposal to reconstruct a
particular closed Riemann worldsheet associated to a given gauge
amplitude.

We would like to remark differences between this approach and our
construction. Gopakumar starts with gauge theory correlator, and,
identifying Schwinger times with the inverse of Strebel lengths, he is
able to associate to each correlator a dual closed surface. On the
contrary, in our construction, Strebel theorem allows to introduce a
suitable uniformization of both the open and the closed surfaces,
while the transition between the two geometries is deeply rooted into
the discrete structure of the surface. It is obtained exploiting
conformal properties of the singular Euclidean metric around each
vertex in order to trade curvature assignment of the closed surface into moduli
of the open surface $M_\partial$, which thus inherits a precise discrete
structure.  As a final result, in the open sector the overall picture
sees the decomposition of the Riemann surface into its fundamental
cylindrical components. 

In our picture, each cylindrical end can be interpreted as an open
string connected at one boundary to the ribbon graph associated to the
discretized worldsheet and, at the other boundary, to a D-brane which
acts naturally as a source for gauge fields. This provides a new
kinematical set-up for discussing gauge/gravity correspondence.

While, at fixed genus $g$ and number of vertexes $N_0$, in the closed
sector both the coupling of the geometry of the triangulation with $D$
bosonic fields and the quantization of the theory can be performed
under the paradigm of critical field theory, the coupling of the
two-dimensional open Riemann surface $M_\partial$ in formula
\eqref{eq:bndry_surf} with a quantum bosonic open string has called
into play Boundary Conformal Field Theory (BCFT) techniques. We have
discussed the quantization of bosonic fields on each finite
cylindrical end, then we have glued together the resulting BCFTs along
the intersection pattern defined by a the ribbon graph naturally
associated to the Regge polytope dual to the original
triangulation\cite{Carfora:2004fd,Noi1}.

In particular, the
unwrapping of the cones into finite cylinders has suggested to compactify
each field defined on the $k$-th cylindrical end along a circle of
radius $\Omega(k) = \frac{R(k)}{l(k)}$: 
\begin{equation*}
  X^{\alpha }(k)
  \xrightarrow{\vartheta (k)\rightarrow \vartheta (k)+2\pi}
  X^{\alpha }(k) + 2\pi \nu^{\alpha} (k)\frac{R^{\alpha}(k)}{l(k)}
  \qquad
  \nu (k)\in \mathbb{Z}
\end{equation*}  
where $l(k)$ is an unspecified length defined as a function of all
characteristic scales defining the geometry of the underlying
triangulation.

Under these assumptions, it is possible to quantize the theory and to
compute the quantum amplitude over each cylindrical end: writing it as
an amplitude between an initial and final state, we can extract
suitable boundary states which arise as a generalization of the states
introduced by Langlands in \cite{Langlands}.  As they stand, these
boundary state did not preserve neither the conformal symmetry nor the
$U(1)_L \times U(1)_R$ symmetry generated by the cylindrical geometry.
It has then been necessary to impose on them suitable gluing conditions
relating the holomorphic and anti-holomorphic generators on the
boundary. These restrictions generated the usual families of Neumann
and Dirichlet boundary states.

Within this framework, the next step in the quantization of the theory
has been to define the correct interaction of the $N_0$ copies of the
cylindrical CFT on the ribbon graph associated with the underlying
Regge Polytope.  This has been achieved via the introduction over each
strip of the graph of Boundary Insertion Operators (BIO)
$\psi^{\lambda(p) \lambda(q)}_{\lambda(p,q)}$ which act as a
coordinate dependent homomorphism from $V_{\lambda(p)} \star
V_{\lambda(p,q)}$ and $V_{\lambda(q)}$, so mediating the changing in
boundary conditions.  Here $V_{\lambda(\bullet)}$ denotes the Verma
module generated by the action of the Virasoro generators over the
$\lambda(\bullet)$ highest weight and $\star$ denotes the fusion of
the two representations.

In the limit in which the theory is rational (\ie when the
compactification radius is an integer multiple of the self dual radius
$\Omega_{s.d.} \,=\, \sqrt{2}$) the Hilbert space of the BCFT can be
rewritten as an $SU(2)_{k=1}$ WZW model Hilbert space. It is then
possible to identify BIO as marginal deformations of boundary
conditions changing operators. Thus they are primary operators with
well defined conformal dimension and correlators.  Moreover,
considering the coordinates of three points in the neighborhood of a
generic vertex of the ribbon graph, we can write the OPEs describing
the insertion of such operators in each vertex.  Considering four
adjacent boundary components, it has then been possible to show that the OPE
coefficients $C_{j_{(r,p)}j_{(q,r)}j_{(p,q)}}^{j_{p}j_{r}j_{q}}$ are
provided by the fusion matrices
$F_{j_{r}j_{(p,q)}}\left[\begin{smallmatrix}
    j_p & j_q\\
    j_{(r,p)} & j_{(q,r)}\end{smallmatrix}\right]$, which in WZW
models coincide with the $6j$-symbols of the quantum group
$SU(2)_{e^{\frac{\pi}{3} i}}$:
\begin{equation*}
C_{j_{(r,p)}j_{(q,r)}j_{(p,q)}}^{j_{p}j_{r}j_{q}}=\left\{
\begin{smallmatrix}
j_{(r,p)} & j_{p} & j_{r} \\
j_{q} & j_{(q,r)} & j_{(p,q)}
\end{smallmatrix}
\right\} _{Q=e^{\frac{\pi }{3} i}}
\end{equation*}

From these data, through edge-vertex factorization we have characterized
the general structure of the partition function for this model
\cite{Carfora2} as a sum over all possible $SU(2)$ primary quantum
numbers describing the propagation of the Virasoro modes along the
$N_0$ cylinders $\{\Delta^*_{\varepsilon(k)}\}$.

The emerging overall picture is that of $N_0$ cylindrical ends
glued through their inner boundaries to the ribbon graph, while their
outer boundaries lay on D-branes. Each D-brane acts naturally as a
source for gauge fields: it has allowed us to introduce open string degrees
of freedom whose information is traded through the cylinder to the
ribbon graph, whose edges thus acquire naturally a gauge coloring.
This provides a new kinematical set-up for discussing gauge/gravity
correspondence.

\subsubsection{Outline}
In writing this thesis, I have separated the two different topics in
two distinct parts.

The first part has been dedicated to the investigation of simplicial
and modular aspects of string dualities.  

Chapter \ref{ch:uniformizations} contains an introduction to the
peculiar geometry arising when we uniformize Riemann surfaces carrying
curvature degrees of freedom. After a short presentation of
fundamental concepts of simplicial geometry, I have reviewed the two
dual uniformizations of a RRT cited above, thus summarizing results in
\cite{Carfora1,Carfora3}. 

With chapter \ref{ch:bcft_langlands} it starts the description of the
coupling between non-critical Polyakov string with the geometry
defined by the dual uniformized open Riemann surface $\partial M$. In
particular, this chapter is devoted to the quantization of the BCFT
arising over each cylindrical end \cyl{k}.

Chapter \ref{ch:bio} introduces Boundary Insertion Operators and their
conformal properties, laying the basis of the description of
interaction of the $N_T(0)$ distinct copies of the BCFT along the
pattern defined by the ribbon graph $\Gamma$. This project is
completed in chapter \ref{ch:bcft_deform}, where we will show that, at
enhanced symmetry point, we are able to coherentely describe the
dynamical glueing calling into play truly marginal deformation of the
original boundary conformal field theory.

Eventually, chapter \ref{ch:os_gauge} is devoted to the analysis of
the coupling of the model with background gauge fields.

Part \ref{part:cb} reports results obtained in \cite{Fre:2003ep} in
searching for cosmological background of superstring theories.
Chapter \ref{geodesinomi}, after summarizing the Kaluza-Klein
reduction of th bosonic sector of type $IIA/B$ superstring theory on
$T^7$, introduces a particular ansatz for the background metric in
$D=3$, allowing the decoupling on the scalar sigma model from gravity.
In particular, it shows how to rephrase the supergravity scalar fields
equations of motions as geodesic equations on $\mathcal{M}_{128}$.
Moreover, it presents the compensator method which allows, once we
have found a solution of the geodesic problem via the solvable Lie
algebra parametrization of the above coset manifold, to generate new
solutions exploiting isometries of the model.

In chapter \ref{exampsolv} we have applied the compensator method to
the simplest (non-trivial) coset manifold $\mathcal{M}_5\,=\,\exp{[Solv
  A_2]}$, while in chapter \ref{genoxide} shows that the hierarchical
dimensional reduction/oxidation of supergravity backgrounds is
algebraically encoded in the hierarchical embeddings of subalgebras into
$E_8$ algebra.  Thus, in chapter \ref{occidoa2}, after analyzing
possible regular embeddings $A_2 \hookrightarrow E_{8(8)}$,
promoting the solution found with the compensator method to solutions
of type $IIA/B$ supergravities, we have applied the oxidation process
to derive the $D=10$ supergravity cosmological backgrounds associated
to the previous solutions.

%% file: uniformizations.tex
\chapter{Dual uniformizations of triangulated surfaces}
\label{ch:uniformizations}

In this chapter, after a short introduction to the concepts of
triangulation of a Riemann surface and of random Regge triangulations
(an exhaustive introduction to simplicial geometry can be found in
\cite{nakahara}, while a statistical field theory approach to
discrete geometry can be found in \cite{ambjorn}), we will review
explicitly the geometrical framework arising when we uniformize
triangulated surfaces carrying curvature degrees of freedom. This
framework has been developed in \cite{Carfora1,Carfora3} and we refer
to these papers and to references therein for a deeper analysis.

\section{Random Regge triangulated surfaces as singular Euclidean structures}
\label{sec:RRT}
One way to define the topological properties of a surface is to
construct a polyhedron homeomorphic to it. Then we
will be able to define the Euler characteristic and the homological
properties of the given surface via the properties of the associated
polyhedron. The main procedure, then, is to associate to each surface a
collection of standard object (triangles in two dimensions, simplexes
in higher ones)in such a way that it will become possible to associate
to each surface a standard abelian structure.

Simplexes are defined as the building blocks of polyhedron. An
$r$-simplex $\langle p_0\,p_1\,\ldots\,p_r \rangle$ is an $r$-dimensional
object whose vertexes $p_i$ are geometrically independent, that is, no
$(r-1)$-dimensional hyperplane contains all the $r + 1$ points. 
Let $p_0,\ldots,\,p_r$ be geometrically independent points in
$\mathbb{R}^m$, with $m\,\geq\,r$. The $r$-simplex $\sigma_r = \langle
p_0\,p_1\,\ldots\,p_r \rangle$is expressed as:
\begin{equation*}
  \sigma_r \,=\, 
  \left\{
    x \,\in\, \mathbb{R}^m 
    \quad \vert \quad
    x \,=\, \sum_{i=0}^r c_i\,p_i \,\geq\, 0,\,
    \sum_{i=0}^r c_i \,=\, 1
  \right\}
\end{equation*}
where the $(c_1,\,\ldots,\,c_r)$ are the barycentric coordinates of
$x$. Since $\sigma_r$ is a bounded closed subset of $\mathbb{R}^m$, it
is compact.

Let $q$ be an integer with $0\,\geq\,q\,\geq\,r$. If we choose
$q\,+\,1$ points $p_{i_0},\ldots,p_{i_q}$ out of $p_0\ldots,p_r$,
these $q\,+\,1$ points define a $q$-simplex $\sigma_q \,=\, \langle
p_{i_0},\ldots,p_{i_q} \rangle$, which is called a $q$-face of
$\sigma_r$ and is denoted by $\sigma_q\,\leq\,\sigma_r$. If
$\sigma_q\,<\,\sigma_r$, then $\sigma_q$ is called a proper face of
$\sigma_r$.
We can further define the star of a face $\sigma_p$, denoted with
$st(\sigma_p)$ as the union of all simplices of which $\sigma_p$ is a
face, and the link of a face $\sigma_p$, $lk(\sigma_p)$, as the union
of all faces $\sigma_f$, $f \leq p$ in $st(\sigma_p)$ such that
$\sigma_f \cap \sigma_p = \emptyset$  

Let $T$ be a finite set of simplices . We define a
\textsc{simplicial complex} in $\mathbb{R}^m$ a finite set of
simplexes $T$ nicely fitted together, \ie: 
\begin{enumerate}
\item If $\sigma\,\in\,T$ and $\sigma'\,\leq\,\sigma$, then
  $\sigma'\,\in\,T$.
\item If $\sigma,\,\sigma'\,\in\,T$, then either
  $\sigma\,\cap\,\sigma'\,=\,\emptyset$ or
  $\sigma\,\cap\,\sigma'\,\leq\,\sigma$ and
  $\sigma\,\cap\,\sigma'\,\leq\,\sigma'$.  
\end{enumerate}
The dimension of a simplicial complex $T$ is defined to be the maximum
dimension of  simplexes in $T$.

Thus, a simplicial complex is a collection of simplexes. If each
simplex is regarded as a subset of $\mathbb{R}^m$, $m\geq\text{dim}T$,
the union of all the elements in $T$ is a subset of $\mathbb{R}^m$
too. It is called the polyhedron $\vert T\vert$ associated to a
simplicial complex $T$ and it holds $\text{dim}\vert T \vert =
\text{dim}T$. 

Let $X$ be a topological space. If there exist a simplicial complex
$T$ and an homeomorphism $f: \vert{}T\vert \rightarrow X$, then $X$ is
said to be triangulable and the pair $(T,f)$ is called a
\textsc{triangulation of $X$}.

To be more precise let us consider a two dimensional simplicial
complex $T$ with underling polyhedron $\vert T \vert$ and a vector
$\left(N_0(T),\,N_1(T),\,N_2(T)\right)$, where
$N_i(T)\,\in\,\mathbb{N}$ is the number of $i$-dimensional simplexes
in $T$.  A Regge triangulation of a two dimensional Piecewise Linear
(PL from now on) manifold $M$ is an homeomorphism $\vert T_l \vert
\rightarrow M$, where each face of $T$ is realized by a rectilinear
simplex of variable edge length $l(\sigma_1(i)),\,i =
1,\,\ldots,\,N_1(T)$ of the appropriate dimension.

A dynamical triangulation is a particular case $\vert T_{l=a} \vert
\rightarrow M$ of a Regge triangulation in which a PL manifold is
realized by rectilinear and equilateral simplexes of constant edge
length: $l(\sigma_1(i)) = a ,\,\forall i = 1,\,\ldots,\,N_1(T)$. 

A triangulation of a two dimensional PL surface can be further
described by its connectivity. This is specified via an adjacency
matrix, \ie{} the $N_0\times{}N_0$ matrix:
\begin{equation}
  \label{eq:adj}
  B_{ij} \,=\,
  \begin{cases}
    1 & \text{if $\langle p_i p_j \rangle$ exists}\\
    0 & \text{otherwise}
  \end{cases}
\end{equation}

A Regge Triangulation has connectivity fixed \emph{a priori}. However, in the
following we will deal with triangulations with both variable edge length
(thus not dynamical) and variable connectivity. These structure have
been first introduced in \cite{Carfora1}, where they have been
baptized Random Regge Triangulations (RRT).

The metric structure of a RRT is locally Euclidean everywhere, with
the exception of the vertexes $\sigma_0(i)$ (also called \emph{the
  bones}), where the sum of the dihedral angles $\theta(\sigma_2)$ of
the incident triangles $\sigma$ can be in excess or in defect respect
to the flatness condition
$\theta(i)=\sum_{{\sigma_2}_{inc}}\theta(\sigma_2) = 2 \pi$, where the
sum extends to all the triangles incident in the $i$-th bone. The
first case is the negative curvature one, while in the second case, we
speak of positive curvature. The correspondent deficit angle is
defined as:
\begin{equation}
  \label{eq:defangle}
  \varepsilon(i) \,=\, 2\pi - \sum_{{\sigma_2}_{inc}}\theta(\sigma_2),
  \qquad 
  i=1,\ldots,\,N_0(T)
\end{equation}

If we denote with $K_0(T) = \{\sigma_0(i)\}, i=1,\,N_0(T)$ the
0-skeleton of the triangulation, then $M\backslash{}K_0(T)$ is a flat
Riemannian manifold, and any point in the interior of an $r$-simplex
$\sigma_r$ has a neighborhood homeomorphic to $B^r \times
C(lk(\sigma_r))$, where $B^r$ is the $r$-dimensional ball in
$\mathbb{R}^m$ and $C(lk(\sigma_r))$ is the cone over the link of
$\sigma_r$ (\ie{} the product $lk(\sigma_r)\times{}[0,1]$, with
$lk(\sigma_r)\times{}\{1\}$ shrinked to a point). 

For the simpler case of dynamical triangulations, the deficit angles
are generated by the \textsc{curvature assignment}, \ie{} the string of
integers $\{q(i)\}_{i=1}^{N_0(T)}$ which specifies the number of equilater
triangles incident in the $i$-th bone, via the relation:
\begin{equation}
  \label{eq:defangle2}
  \varepsilon(i)
  \,=\,
  2 \pi
  \,-\,
  q(i)\,
  \arccos{\frac{1}{2}} 
  \qquad
  i\,=\,1,\,\ldots,\,N_0(T).
\end{equation}

For a regular (not necessary dynamical) triangulation we have $q(k)
\geq 3\,\forall k$ and, since each triangle has three internal angles,
the curvature assignments obey the following
constraint\cite{Carfora1}:
\begin{equation}
  \label{eq:curvass_constr}
  \sum_{k=1}^{N_0(T)} q(k) 
  \,=\, 
  3 N_2(T) 
  \,=\, 
  6 \left[
    1 - \frac{\chi(M)}{N_0(T)} 
  \right]\,N_0(t)
\end{equation}
where $\chi(M)$ is the Euler-Poincar\'e characteristic of the surface
and $6 \left[ 1 - \frac{\chi(M)}{N_0(T)} \right]$ is the average value
of the curvature assignments.  In particular, if we remove the
constraint $q(k) \geq 3$ we are dealing with generalized
RRT, namely configurations in which the star of a
vertex contains only two or one bidimensional simplexes.

In what follow, we will summarize the main results of \cite{Carfora3}
showing that it is possible to geometrically characterize the metrical
structure of a generalized RRT described above as a particular case of
the theory of singular Riemann surfaces endowed with a
\textsc{Singular Euclidean Structure}\cite{troyanov,thurston}. In the
next section, we will describe results, of \cite{Carfora1}, showing
that this structure can be described in terms of complex functions
theory.

Let us consider the first barycentric subdivision of $\vert T_l \vert
\rightarrow M$ and, within this subdivision, the closed stars of the
vertexes of the original triangulation. They form a collection of
two-cells $\{\rho^2(i)\}_{i=1}^{N_0(T)}$ characterizing a polytope $P$
baricentrically dual to $T$. As remarked in \cite{Carfora3}, it is
important to stress that we are not dealing with a rectilinear
presentation of $P$, where the cells are realized by rigid polytopes,
but with a geometrical presentation of $P$, where the two-cells retain
the conical geometry induced by the deficitary sum of dihedral angles
of the original triangulation. To this end, we endow each cell
$\{\rho^2(i)\}_{i=1}^{N_0(T)}$ with a polar reference frame, centered
on the vertex $\sigma_0(i)$. Denoting with $(\lambda(i), \chi(i))$ the
polar coordinate of a point $p \in \rho^2(i)$, then $\rho^2(i)$ is
geometrically realized as the space\cite{Carfora3}
\begin{equation}
  \label{eq:dual_cell}
  \rho^2(i) \,=\,
  \frac{\left\{
      (\lambda(i), \chi(i))
      \quad \vert \quad
      \lambda(i) \geq 0; \,\, 
      \chi(i) \in \mathbb{R}/(2\pi - \varepsilon)\mathbb{Z}
    \right\}}
  {
    (0,\,\chi(i)) \sim (0,\,\chi'(i))
  }
\end{equation}
endowed with the metric
\begin{equation}
  \label{eq:dual_cell_metric}
  d^2 \lambda(i) + \lambda^2(i)\, d^2\chi(i)  
\end{equation}
This definition characterize the conical Regge polytope $\vert P_{T_l}
\vert \rightarrow M$ barycentrically dual to $\vert T_l\vert \rightarrow
M$.

\begin{figure}[!t]
  \centering
  \includegraphics[width=.5\textwidth]{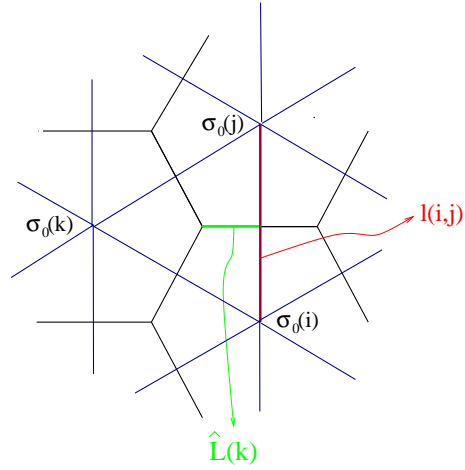}
  \caption{Relations between the edge-lengths of the conical polytope
  and the edge-lengths of the triangulation}
  \label{fig:geom}
\end{figure}

The relation between the conical geometry of the barycentrically dual
polytope $\vert P_{T_l} \vert \rightarrow M$ and the original
triangulation data are fixed once they are given the lengths of the
1-simplexes of $\vert T_l \vert \rightarrow M$. A direct calculation
provides the lengths of the (third part of the) medians connecting
the barycenter $\rho_0(i,j,k)$ of the triangle defined by the bones
$\sigma_0(i),\,\sigma_0(j) \text{and} \sigma_0(k)$, $i,\,j,\,k =
1,\,\ldots,\,N_2(T)$, to the middle points of its edges. Following
notation in \cite{Carfora1}, let  us denote
with $l(i,j)$ the length of the edge of the triangle connecting the
bones $\sigma_0(i)$ and $\sigma_0(j)$, while $\hat{L}(k)$ is the
length of the segment connecting its middle point with $\rho_0(i,j,k)$
(\ie{} the third part of the median drawn from $\sigma_0(k)$, see fig.
\ref{fig:geom}). We can write:
\begin{gather*}
  \hat{L}^2(i) 
  \,=\,
  \frac{1}{18} l^2(i,j) 
  \,+\, 
  \frac{1}{18} l^2(k,i) 
  \,-\,
  \frac{1}{36} l^2(j,k) \\
  \hat{L}^2(j) 
  \,=\,
  \frac{1}{18} l^2(j,k) 
  \,+\, 
  \frac{1}{18} l^2(i,j) 
  \,-\,
  \frac{1}{36} l^2(k,i) \\
  \hat{L}^2(k) 
  \,=\,
  \frac{1}{18} l^2(j,k) 
  \,+\, 
  \frac{1}{18} l^2(k,i) 
  \,-\,
  \frac{1}{36} l^2(i,j) \\
  \intertext{and}
  l^2(i,j)
  \,=\, 
  8 \hat{L}^2(i) 
  \,+\,
  8 \hat{L}^2(j)
  \,-\,
  4 \hat{L}^2(k) \\
  l^2(j,k)
  \,=\, 
  8 \hat{L}^2(j) 
  \,+\,
  8 \hat{L}^2(k)
  \,-\,
  4 \hat{L}^2(i)  \\
  l^2(k,i)
  \,=\, 
  8 \hat{L}^2(k) 
  \,+\,
  8 \hat{L}^2(i)
  \,-\,
  4 \hat{L}^2(j) 
\end{gather*}

In such a geometrical framework, let $\rho^2(k)$ be the generic 2-cell 
barycentrically dual to the bone
$\sigma_0(k),\,k=1,\,\ldots,\,N_0(T)$, and let us denote with:
\begin{equation}
  \label{eq:boundary}
  L(k)\,\doteq\,
\sum_{i=1}^{q(k)} L(\rho^1(i))
\end{equation}
the length of its boundary, where $L(\rho^1(i))$ is the length of its
$q(k)$ ordered edges $\rho^1(i)\,\in\,\vert P_{T_l} \vert \rightarrow
M$. If $\varepsilon(k)$ denotes the deficit angle associated to the
bone $\sigma_0(k)\,\in\,\vert T_l \vert \rightarrow M$, the slant
radius associated with the cell $\rho^2(k)\,\in\,\vert P_{T_l} \vert
\rightarrow M$ is defined as:
\begin{equation}
  \label{eq:slantrad}
  r(k)
\,\doteq\,
\frac{L(k)}{2\pi\,-\,\varepsilon(k)},
\qquad k=1,\,\ldots,\,N_0(T)
\end{equation}

We can associate to each two-cell $\rho^2(k)\,\in\,\vert P_{T_l}
\vert \rightarrow M$ an open ball $B^2(k) \,=\, \left\{p \in \rho^2(k)
/ \partial(\rho^2(k)) \right\}$; it is contained in
$st(\sigma_0(k))$.  To any vertex we can associate a complex
uniformizing coordinate $\zeta(k)$ defined on an open disk of radius
$r(k)$:
\begin{equation}
\label{eq:ball}
  B^2(k)
  \,\longrightarrow\,
  D_k(r(k))
  \,\doteq\,
  \left\{
    \zeta(k) \,\in\,\mathbb{C}
    \quad\vert\quad
    0\,\leq\,\zeta(k)\,\leq\,r(k)
  \right\}
\end{equation}

In terms of $\zeta(k)$ we can write explicitly the singular Euclidean
metric characterizing the singular Euclidean structure on each open
ball as\cite{Carfora3}:
\begin{equation}
  \label{eq:metric}
  d s^2_k\,\doteq\,
  e^{2 u}\,
  \vert 
  \zeta(k) - \zeta(k)(\sigma_0(k))
  \vert^{-2 \frac{\varepsilon(k)}{2 \pi}} 
  \vert d \zeta(k) \vert^2, 
\end{equation}
where $u:\,B^2 \rightarrow \mathbb{R}$ is a continuous function such
that
\begin{subequations}
  \begin{gather}
    \left\vert
      \zeta(k) - \zeta(k)(\sigma_0(k))
    \right\vert
    \frac{\partial\,u}{\partial \zeta(k)} 
    \,\xrightarrow[\zeta(k) \rightarrow \zeta(k)(\sigma_0(k))]{}\, 
    0 \\
    \left\vert
      \zeta(k) - \zeta(k)(\sigma_0(k))
    \right\vert
    \frac{\partial\,u}{\partial \zeta(k)} 
    \,\xrightarrow[\zeta(k) \rightarrow \overline{\zeta}(k)(\sigma_0(k))]{}\, 
    0
  \end{gather}
\end{subequations}

Up to the conformal factor $e^{2 u}$, \eqref{eq:metric} is the metric
of a Euclidean cone of total angle $\theta(k) \,=\, 2\pi -
\varepsilon(k)$.  We can glue together the uniformizations
$\{D_k(r(k))\}_{k=1}^{N_0(T)}$ along the pattern defined by the
1-skeleton of the dual Regge polytope and generate on $M$ the quasi
conformal structure:
\begin{equation}
  (M,\,C_{sg}) \,\doteq\,
  \bigcup_{|P_{T_l}| \rightarrow M} 
  \{D_k(r(k));\,d s^2_{(k)} \}_{k=1}^{N_0(T)}
\end{equation}
If $|d t^2|$ is a conformally flat metric on $M$, then the
quasi-conformal structure $(M,\,C_{sg})$ can be locally represented by
the metric\cite{Carfora3}:
\begin{equation}
  \label{eq:metricfin}
  d s^2_T 
\,=\,
e^{2v} |d t^2|
\end{equation}
with conformal factor $v \equiv u - \sum_{k = 1}^{N_0(T)} \left(-
  \frac{\varepsilon(k)}{2 \pi} \right)\,ln|\zeta(k) -
\zeta(k)(\sigma_0(k))|$.  As there are no natural choice for $u$, the
metric is only well defined up to conformal symmetry: this allows to
move within the conformal class of all the metrics possessing the same
singular structure of the triangulated surface $|T_l| \rightarrow M$.

This quasi conformal structure defines a RRT as a particular case of
the theory of singular Riemann surfaces\cite{Carfora3}. As a matter of
fact, the singular structure described
above can be summarized introducing the associated real divisor:
\begin{equation}
  \label{eq:div}
  Div(T)\,\doteq\,
  \sum_{k = 1}^{N_0(T)} 
  \left(- 
    \frac{\varepsilon(k)}{2 \pi}
  \right)\,
  \sigma_0(k) \,=\
  \sum_{k = 1}^{N_0(T)} 
  \left( 
    \frac{\theta(k)}{2 \pi} - 1
  \right)\,
  \sigma_0(k)
\end{equation}
supported on the set of bones $\{\sigma_0(k)\}_{k=1}^{N_0(T)}$.  The
degree of such a divisor is $|Div(T)| = \sum_{k = 1}^{N_0(T)} \left(
  \frac{\theta(k)}{2 \pi} - 1 \right) = -\chi(M)$. The real divisor
characterizes the Euler class of the pair $(|T_l|\rightarrow
M,\,Div(T))$ (or, shortly, $(T,\,Div(T))$): 
Associating to
$(T,\,Div(T))$ the Euler number\cite{troyanov}
\begin{equation}
  \label{eq:Enum}
  e(T,\,Div(T)) 
  \,=\, 
  \chi(M) - |Div(T)|
\end{equation}
it is possible to rewrite the Gauss-Bonnet formula as:
\begin{lemma}
  \textbf{Gauss-Bonnet formula for triangulated surfaces}\cite{Carfora3}\\
  Let $(T,\,Div(T))$ be a triangulated surface with divisor: 
  \begin{equation*}
 Div(T)
  \,=\ \sum_{k = 1}^{N_0(T)} \left( \frac{\theta(k)}{2 \pi} - 1
  \right)\, \sigma_0(k)   
  \end{equation*}
  associated with the vertexes incidences
  $\{\sigma_0(k)\}_{k=1}^{N_0(T)}$. Let $d s^2$ be the conformal
  metric \eqref{eq:metricfin} representing the divisor $Div(T)$. Then
  \begin{equation*}
    \frac{1}{2\pi} \int K\,d A 
\,=\,
e(T,\,Div(T))
  \end{equation*}
where $K$ and $d A $ are respectively the curvature and the area
element corresponding to the metric \eqref{eq:metricfin}.
\end{lemma}
Since for a RRT it holds $e(T,Div(T)) = 0$, then $\frac{1}{2\pi} \int
K\,d A \,=\,0$, \ie{} a triangulation $|T_l| \leftarrow M$ naturally
carries a conformally flat structure. This result admits a non-trivial
converse:
\begin{teo}
 \textbf{Troyanov-Picard}\cite{troyanov}\\
Let $((M,\,C_{sg}),\,Div)$ be a singular Riemann surface with a
divisor such that $e(T,Div(T)) = 0$. Then there exists on $M$ a unique
(up to homothety) conformally flat metric representing the divisor $Div(T)$. 
\end{teo}

With these last statements, authors of \cite{Carfora3} fully
characterized the metric triangulations as a particular case of the
theory of singular Riemann surfaces.  Moreover, thanks to the
introduction of the real divisor \eqref{eq:div} and exploiting the 
Poncar\'e-Klein-Koebe uniformization theorem, they showed that the
singular Riemann surface $((M,\,C_{sg}),\,Div)$ is a particular case
of singular uniformizations of  $(M,\,N_0) \,=\, M - \sum_{i=1}^{N_0(T)}
\sigma_0(i)$, thus including the study of RRT into the theory of punctured
surfaces\cite{Carfora3}.
However, they stressed that in due to the presence of null euler
number, this approach would have involved non standard techniques to
be dealt with.  Thus, they pursued further this analysis relating the
space of inequivalent singular Euclidean structure to the theory of
uniformization of singular Euclidean surfaces via the properties of
Jenkins-Strebel quadratic differentials.

\begin{rem}
  We would like to stress that the construction described above
  includes naturally particular limiting cases of Regge triangulation
  like those situations in which some of the vertexes are
  characterized by deficit angles $\varepsilon(K) \rightarrow 2\pi$,
  \ie{} $\theta(k)=0$. Such a situation correspond to having the cone
  $C|lk(\sigma_0(k))|$ over the link $lk(\sigma_0(k))$ realized as an
  Euclidean cone of total angle 0. These situations, which had always
  been considered as pathological, are no loger so in the connection
  introduced above: the corresponding two-cells $\rho^2(k) \in
  |P_{T_l}|\rightarrow M$ can be naturally endowed with the conformal
  Euclidean structure obtained by setting $\frac{\varepsilon(k)}{2\pi}
  = 1$ in \eqref{eq:metric}. The remaining object is then (up to the
  conformal factor $e^{2u}$) the flat metric on the semi-infinite
  cylinder $\mathbb{S}^1\times\mathbb{R}_+$ (a cylindrical end).
  Alternatively, it is possible to consider $\rho^2(k)$ endowed with
  the geometry of an hyperbolic cusp, \ie{} an half-infinite cylinder
  endowed with the hyperbolic metric $\lambda(k)^{-2}(d \lambda(k)^2 +
  d \chi(k)^2)$. The triangles incident in $\sigma_0(k)$ are then
  realized as hyperbolic triangles with the vertex located at
  $\lambda(k)=\infty$ ad null incident angle. The two points of view
  are strictly related: one can switch from the Euclidean
  representation to the hyperbolic one simply setting the conformal
  factor to $e^{2u} = \left(\ln{\frac{1}{|\zeta(k) -
        \zeta(k)(\sigma_0(k))|}}\right)^{-2}$. The presence of
  cylindrical ends is consistent with a singular Euclidean structure
  as long as the associated divisor satisfies the topological
  constraint $|Div(t)| = - \chi(M)$\cite{Carfora1}.
\end{rem}

\begin{figure}[!t]
  \centering
  \includegraphics[width=\textwidth]{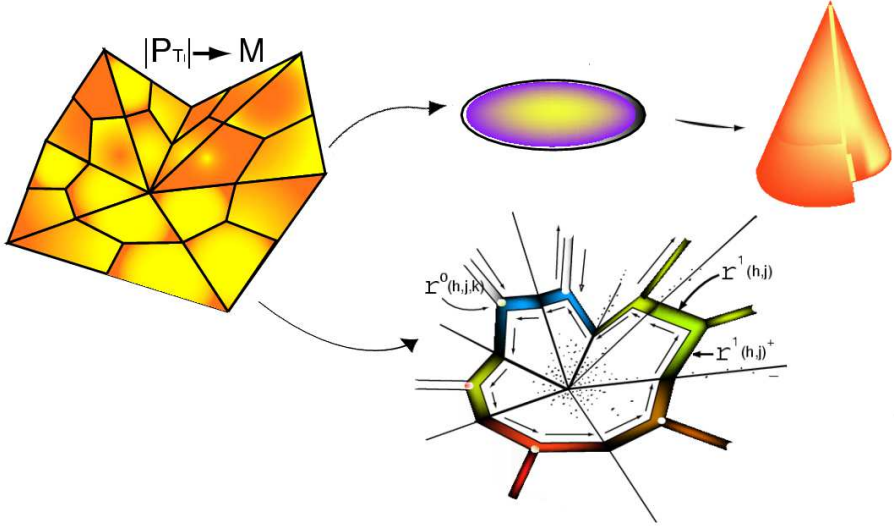}
  \caption{The two-cells $\rho^2(k)$ of the dual polytope $|P_{T_l}
    \rightarrow M|$ can be geometrically represented by a punctured
    disk uniformized by a conical metric, while its 1-skeleton is in
    one-to-one correspondence with a trivalent ribbon graph\cite{Carfora2}}
  \label{fig:ribbon}
\end{figure}

\section[Ribbon graphs on Regge polytopes]{Ribbon graphs on Regge
  polytopes: moduli space parametrization }

In the previous section we have shown that, given a triangulation
$|T_l| \rightarrow M$ of a Riemann surface M, the two-cells of the
dual Regge polytope $|P_{T_l}| \rightarrow M$ are geometrically
realized by punctured disks, uniformized by a conical
metric. Moreover, the geometrical realization of the 1-skeleton of
the dual Regge polytope is a trivalent ribbon graph:
\begin{equation}
  \label{eq:rg}
  \Gamma \,=\, \left( \{\rho^0(k)\} , \{\rho^1(k)\}\right)
\end{equation}
where the $\{\rho^0(k)\}_{k=1}^{N_2(t)}$ are the barycenter of the
triangles $\{\sigma_2(k)\}_{k=1}^{N_2(t)} \in |T_l| \rightarrow M$, and
$\{\rho^1(k)\}$ is the set of edges of $|P_{T_l}| \rightarrow M$. These
are generated by the pairwise joining of the half edges $\rho^1(k)^+$
and $\rho^1(k)^-$ in through the barycenter
$W(k),\,k=1,\ldots,N_1(T)$ of the edges $\{\sigma_1(k)\}_{k=1}^{N_1(t)}
\in |T_l| \rightarrow M$. If we formally introduce a degree-2
ghost-vertex on each  $W(k)$, then the underling graph of $|P_{T_l}|
\rightarrow M$ is the edge refinement of $\Gamma$ (see fig. \ref{fig:ribbon}):
\begin{equation}
  \label{eq:er}
  \Gamma_{\mbox{ref}} \,=\, 
  \left(
    \{\rho^0(k)\}\,\bigsqcup_{h=1}^{N_1(T)} \{W(h)\},\, 
    \{\rho^1(j)^+\}
    \bigsqcup_{j=1}^{N_1(T)}
    \{\rho^1(j)^-\}
  \right) .
\end{equation}

The natural automorphism group $Aut(P_l)$ of $|P_{T_l}| \rightarrow M$
(\ie{} the set of bijective maps $\Gamma \,=\, \left( \{\rho^0(k)\} ,
  \{\rho^1(k)\}\right) \,\rightarrow\, \widetilde{\Gamma} \,=\, \left(
  \{\widetilde{\rho}^0(k)\} , \{\widetilde{\rho}^1(k)\}\right)$
preserving the incidence matrix of the graph), is the automorphism
group of its edge refinement\cite{mulase}: $Aut(P_{T_l}) \doteq
Aut{\Gamma_{\mbox{ref}}}$.

The local uniformizing complex coordinate $\zeta(k)$, in terms of wich
we have explicitly express the conical metric uniformizing the
neighborhood of conical singularities, provides a counterclockwise
orientation of the two cells of $|P_{T_l}| \rightarrow M$. Such an
orientation provides a cyclic ordering of the set of half edges
incidents on each vertex ${\rho^0(k)}_{k=1}^{N_2(T)}$: according to
these remarks, the 1-skeleton of  $|P_{T_l}| \rightarrow M$ is a
ribbon (or fat) graph, \ie{} a graph together with a cyclic ordering
of the set of half edges incident to each vertex of
$\Gamma$. Conversely, any ribbon graph $\Gamma$ characterize a Riemann
surface $M(\Gamma)$ possessing $\Gamma$ as a spine. Thus, the edge
refinement of the 1-skeleton of the Regge dual graph of a RRT is in
one-to-one correspondence with trivalent metric ribbon graphs.

The set of all such trivalent metric Ribbon graphs $\Gamma$ with given
edge set $e(\Gamma)$ can be characterize as a space homeomorphic to
$\mathbb{R}_+^{|e(\Gamma)|}$ (where $|e(\Gamma)|$ is the number of
edges in $e(\Gamma)$). The automorphism group $Aut_{P_{T_l}}$ acts
naturally on $\mathbb{R}_+^{|e(\Gamma)|}$ via the homomorphism
$Aut_{P_{T_l}} \rightarrow \mathfrak{S}_{e(\Gamma)}$
($\mathfrak{S}_{e(\Gamma)})$ being the symmetric group over
$|e(\Gamma)|$ elements). 

In this connection, let $Aut_{\partial }(P_{l})\subset Aut(P_{l})$,
denote the subgroup of ribbon graph automorphisms of the (trivalent)
$1$-skeleton $\Gamma $ of $ |P_{T_{l}}|\rightarrow {M}$ that preserve
the (labeling of the) boundary components of $\Gamma $. Then, the
space $K_{1}RP_{g,N_{0}}^{met}$ of $1$ -skeletons of conical Regge
polytopes $|P_{T_{l}}|\rightarrow {M}$, with $ N_{0}(T)$ labelled
boundary components, on a surface $M$ of genus $g$ can be defined by
\cite{mulase}
\begin{equation}
K_{1}RP_{g,N_{0}}^{met}=\bigsqcup_{\Gamma \in RGB_{g,N_{0}}}\frac{\mathbb{R}
_{+}^{|e(\Gamma )|}}{Aut_{\partial }(P_{l})},  \label{DTorb}
\end{equation}
where the disjoint union is over the subset of all trivalent ribbon
graphs (with labelled boundaries) satisfying the topological
constraint $ 2-2g-N_{0}(T)<0$, and which are dual to generalized
triangulations. It follows, (see \cite{mulase} theorems 3.3, 3.4, and
3.5), that the set $ K_{1}RP_{g,N_{0}}^{met}$ is locally modeled on a
stratified space constructed from the components $\mathbb{R}
_{+}^{|e(\Gamma )|}/Aut_{\partial }(P_{l})$ by means of a (Whitehead)
expansion and collapse procedure for ribbon graphs, which amounts to
collapsing edges and coalescing vertexes, (the Whitehead move in $
|P_{T_{l}}|\rightarrow {M}$ is the dual of the flip move
\cite{ambjorn} for triangulations). Explicitly, if $l(t)=tl$ is the
length of an edge $\rho ^{1}(j)$ of a ribbon graph $\Gamma _{l(t)}\in
$ $K_{1}RP_{g,N_{0}}^{met}$, then, as $t\rightarrow 0$, we get the
metric ribbon graph $\widehat{\Gamma }$ which is obtained from $\Gamma
_{l(t)}$ by collapsing the edge $\rho ^{1}(j)$ . By exploiting such
construction, we can extend the space $ K_{1}RP_{g,N_{0}}^{met}$ to a
suitable closure $\overline{K_{1}RP} _{g,N_{0}}^{met}$
\cite{looijenga}, (this natural topology on $K_{1}RP_{g,N_{0}}^{met}$
shows that, at least in two-dimensional quantum gravity, the set of
Regge triangulations with \emph{fixed connectivity} does not explore
the full configuration space of the theory). The open cells of
$K_{1}RP_{g,N_{0}}^{met}$, being associated with trivalent graphs,
have dimension provided by the number $N_{1}(T)$ of edges of
$|P_{T_{l}}| \rightarrow {M}$, \emph{i.e.}
\begin{equation}
\dim \left[ K_{1}RP_{g,N_{0}}^{met}\right] =N_{1}(T)=3N_{0}(T)+6g-6.
\end{equation}
There is a natural projection\cite{Carfora1} 
\begin{gather}
p:K_{1}RP_{g,N_{0}}^{met}\longrightarrow \mathbb{R}_{+}^{N_{0}(T)} \\
\Gamma \longmapsto p(\Gamma )=(l_{1},...,l_{N_{0}(T)}),  \notag
\end{gather}
where $(l_{1},...,l_{N_{0}(T)})$ denote the perimeters of the polygonal
2-cells $\{\rho ^{2}(j)\}$ of $|P_{T_{l}}|\rightarrow {M}$. With respect to
the topology on the space of metric ribbon graphs, the orbifold 
$K_{1}RP_{g,N_{0}}^{met}$ endowed with such a projection acquires the
structure of a cellular bundle. For a given sequence $\{l(\partial (\rho
^{2}(k)))\}$, the fiber 
\begin{equation}
p^{-1}(\{l(\partial (\rho ^{2}(k)))\})=\left\{ |P_{T_{l}}|\rightarrow {M}\in
K_{1}RP_{g,N_{0}}^{met}:\{l_{k}\}=\{l(\partial (\rho ^{2}(k)))\}\right\}
\end{equation}
is the set of all generalized conical Regge polytopes with the given
set of perimeters\cite{Carfora1}. If we take into account the
$N_{0}(T)$ constraints associated with the perimeters assignments, it
follows that the fibers $p^{-1}(\{l(\partial (\rho ^{2}(k)))\})$ have
dimension provided by
\begin{equation}
\dim \left[ p^{-1}(\{l(\partial (\rho ^{2}(k)))\}\right] =2N_{0}(T)+6g-6,
\end{equation}
which again corresponds to the real dimension of the moduli space
$\mathfrak{M}_{g},_{N_{0}}$ of $N_{0}$-pointed Riemann surfaces of
genus $g$.

The complex analytic geometry of the space of conical Regge polytopes
which we will discuss in the next section generalizes the well-known
bijection (a homeomorphism of orbifolds) between the space of metric
ribbon graphs $ K_{1}RP_{g,N_{0}}^{met}$ (which forgets the conical
geometry) and the moduli space $\mathfrak{M}_{g},_{N_{0}}$ of genus
$g$ Riemann surfaces $((M;N_{0}), \mathcal{C})$ with $N_{0}(T)$
punctures \cite{mulase}, \cite{looijenga}. This bijection results in a
local parametrization of $\mathfrak{M}_{g},_{N_{0}}$ defined by
\begin{gather}
  h:K_{1}RP_{g,N_{0}}^{met}\rightarrow \mathfrak{M}_{g},_{N_{0}}\times
  {R}_{+}^{N}
  \label{bijec} \\
  \Gamma \longmapsto \lbrack ((M;N_{0}),\mathcal{C}),l_{i}] \notag
\end{gather}
where $(l_{1},...,l_{N_{0}})$ is an ordered n-tuple of positive real
numbers and $\Gamma $ is a metric ribbon graphs with $N_{0}(T)$
labelled boundary lengths $\{l_{i}\}$ (figure 8).

If $\overline{K_{1}RP}_{g,N_{0}}^{met}$ is the closure of
$K_{1}RP_{g,N_{0}}^{met}$, then the bijection $h$ extends to
$\overline{ K_{1}RP}_{g,N_{0}}^{met}\rightarrow
\overline{\mathfrak{M}}_{g},_{N_{0}}\times {R }_{+}^{N_{0}}$ in such a
way that a ribbon graph $\Gamma \in \overline{RGP} _{g,N_{0}}^{met}$
is mapped in two (stable) surfaces $M_{1}$ and $M_{2}$ with $N_{0}(T)$
punctures if and only if there exists an homeomorphism between $M_{1}$
and $M_{2}$ preserving the (labeling of the) punctures, and is
holomorphic on each irreducible component containing one of the
punctures.

As remarked in \cite{Carfora1}, it is important to stress that even if
ribbon graphs can be thought of as arising from Regge polytopes (with
variable connectivity), the morphism (\ref {bijec}) only involves the
ribbon graph structure and the theory can be (and actually is)
developed with no reference at all to a particular underlying
triangulation. In such a connection, the role of dynamical
triangulations has been slightly overemphasized, they simply provide a
convenient way of labeling the different combinatorial strata of the
mapping (\ref{bijec}), but, by themselves they do not define a
combinatorial parametrization of $
\overline{\mathfrak{M}}_{g},_{N_{0}}$ for any finite $N_{0}$. However,
it is very useful, at least for the purposes of quantum gravity, to
remember the possible genesis of a ribbon graph from an underlying
triangulation and be able to exploit the further information coming
from the associated conical geometry. Such an information cannot be
recovered from the ribbon graph itself (with the notable exception of
equilateral ribbon graphs, which can be associated with dynamical
triangulations), and must be suitably codified by adding to the
boundary lengths $\{l_{i}\}$ of the graph a further decoration. This
can be easily done by explicitly connecting Regge polytopes to
punctured Riemann surfaces\cite{Carfora1}.

\section{Dual uniformizations of a random Regge triangulated surface}

The genesis of a ribbon graph as underlying structure of a RRT has
been shown very useful because it has allowed to parametrize the
moduli space of a RRT via the morphism between the space of trivalent
metric ribbon graphs and $\mathcal{M}_{g,N} \times \mathbb{R}_+^N$.
Nonetheless, the full set of data we are handling include the conical
geometry defined on the Regge polytopes. This can be codified in the
above context by adding to the ribbon graph data (\ie{} to the
boundary lengths $\{L(k)\}_{k=1}^{N_0(T)}$) a suitable decoration
which arises naturally once we connect the dual Regge polytopes
$|P_{T_l}| \rightarrow M$ to the theory of punctured Riemann surfaces.

In \cite{mulase} it was introduced a ribbon graph uniformization
which, exploiting properties of Jenkins-Strebel quadratic
differential, allowes to associate to a ribbon graph with given edge
lengths $\{L(k)\}_{k=1}^{N_0(T)}$ a complex structure
$((M,N_0),\,\mathcal{C})$, \ie{} a punctured Riemann surface.  This
construction is a priori completely general: given a smooth Riemann
surface with $N$ marked points and $(C,\,(p_1,\,\ldots,\,p_N))$ and an
ordered $N$-tuple $(a_1,\,\ldots,\,a_N) \in \mathbb{R}^N_+$, there is
a unique meromorphic quadratic differential \footnote{Let us remember
  that an \textsc{holomorphic quadratic differential} defined on a
  compact Riemann surface $M$ is an element of
  $H^0(C,\,{K_C}^{\otimes2})$ (being ${K_C}^{\otimes2}$ the second
  symmetric tensor product of the canonical sheaf on $M$, $K_C$). In a
  local coordinate frame on $M$ the quadratic differential is
  represented by $\phi = f(z) (d z)^2$, $f(z)$ being a locally defined
  holomorfic function. Under a coordinate change $w = w(z)$ the local
  expression $\phi = f(z) (d z)^2 = g(w) (d w)^2$ transforms as
  \begin{equation}
    \label{eq:strlaw}
    f(z) = g(w(z))\left(\frac{d w(z)}{d z}\right)^2.
  \end{equation}
  A \textsc{meromorphic} quadratic differential on $M$ is an
  holomorphic quadratic differential with the exception of a finite
  set $(p_1,\,\ldots,\,p_N)$ of points of $M$ such that at each
  singularity $p_j$ of $\phi$ there is a local expression $\phi = f_j(z) (d
  z)^2$ with a meromorphic function $f_j(z)$ having a pole at $z =
  z(p_j)$. If $f_j(z)$ has a pole of order $r$ at $z = z(p_j)$, then
  we say that $\phi$ has an order $r$ pole at $z = z(p_j)$.}
$\phi$ on $M$ satisfying the following conditions:
\begin{enumerate}
\item $\phi$ is holomorphic on $C / (p_1,\,\ldots,\,p_N)$.
\item $\phi$ has a double pole at each $p_j,\,j=1,\,\ldots,\,N$.
\item The union of all non-compact horizontal leaves
  \footnote{Let us recall that given a meromorphic quadratic
    differential $\phi = f(z) (d z)^2$ defined on $M$, a real parametric
    curve 
    \begin{equation*}
      \gamma\,:\; (a,b) \ni t \mapsto \gamma(t) = z \in \mathbb{C}
    \end{equation*}
    parametrized on an open interval $(a,b)$ of the real axis is 
    \begin{itemize}
    \item an \textsc{horizontal leaf} (or \textsc{horizontal trajectory})
      of $\phi$ if 
      \begin{equation*}
        f(\gamma(t)) 
        \left(
          \frac{d \gamma(t)}{d t} 
        \right)^2 > 0 
      \end{equation*}
      for every value of $t \in (a,b)$;
    \item a
      \textsc{vertical leaf} (or \textsc{vertical trajectory}) of $\phi$
      if 
      \begin{equation*}
        f(\gamma(t)) 
        \left(
          \frac{d \gamma(t)}{d t} 
        \right)^2 > 0 
      \end{equation*}
      for every value of $t \in (a,b)$.
    \end{itemize}} 
  form a zero-measure close subset of $M$. 
\item Every compact horizontal leaf $\gamma_j$ is a simple loop
  surrounding one of the poles and satisfying:
  \begin{equation}
\label{eq:per}
    a_j \,=\,\int_{\gamma_j} \sqrt{\phi}
  \end{equation}
where the branch of the square root is chosen so that the integral
has a positive value with respect to the positive orientation of
$\gamma_j$, which is determined by the complex structure on $M$.
\end{enumerate}
This quadratic differential is called \textsc{Strebel differential}.
The collection of all its compact horizontal leaves surrounding a pole
forms a punctured disk centered on the pole itself. The punctured disk
ends on the set non-compact horizontal leaves connecting pairwise the
zeros of the differential. The latters are joined to the pole by a
vertical leaf(see \cite{mulase}). The union of the punctured disks
(with the punctures filled by the points $(p_1,\ldots,\,p_n)$), of the
zeros of $\phi$ and of the non-compact horizontal leaves form the
underlying topological surface of the Riemann surface $M$.

In this sense, a Strebel quadratic differential defined on a
nonsingular Riemann surface $M$ of genus $g$ determines a unique cell
decomposition $\square_\phi$ of $M$ consisting of $N_0$ 2-cells, $N_1$
1-cells and $N_2$ 0-cells, where $N_2$ is the number of zeros of the
Strebel differential and it holds $N_0 - N_1 + N_2 = 2 - 2g$. The
1-skeleton of such a cellular decomposition, \ie{} the union of the
0-cells and 1-cells of the the cell decomposition which is defined by
the Strebel quadratic differential is a metric ribbon graph.

Moreover, the set of all the vertical leaves that connect zero and
poles of the Strebel differential, together with the cell
decomposition $\square_\phi$ induced on $M$ by the non-compact horizontal
leaves, form a canonical triangulation (in the sense described above)
$\Delta_\phi$ of $M$.

With these remarks, in \cite{mulase} it had been shown haw to
construct a canonical coordinate system on a Riemann surface $M$:
once they are given a $N$ marked points $M$ and an $N$-tuple of real
numbers, we simply have to introduce the local expression of the
associated Strebel differential, where local means valid in each
0,1,2-cell of $\square_\phi$.
 
\begin{figure}[!t]
  \centering
  \includegraphics[width=.7\textwidth]{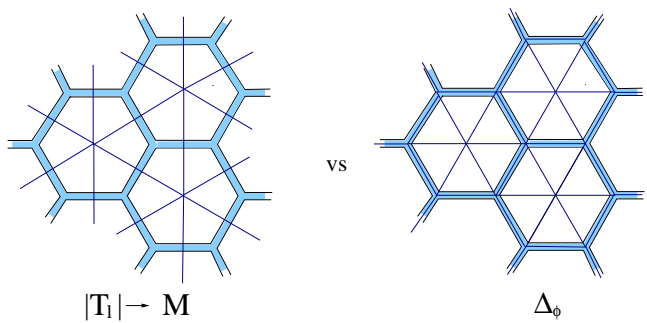}
  \caption{Differences between $|T_l| \rightarrow M$ and $\Delta_\phi$}
  \label{fig:trvstr}
\end{figure}

In \cite{Carfora1}, this construction had been directly applyed to the
triangulation $|T_l \rightarrow M|$,
once they have been performed the following identifications:\\
\begin{center}
  \begin{tabular}{|ccc|}
    \hline
    (M,$\phi$) &{} & RRT\\
    \hline
    $(p_1,\,\ldots,\,p_N)$ 
    & \quad$\longleftrightarrow$\quad &
    $\{\sigma_0(i)\}_{i=1}^{N_0(T)}$ 
    \\
    $(a_1,\,\ldots,\,a_N)$ 
    & \quad$\longleftrightarrow$\quad &
    $\{L(i) = L(\partial\rho^2(i))\}_{i=1}^{N_0(T)}$
    \\
    $\square_\phi$ 
    & \quad$\longleftrightarrow$\quad &
    $|P_{T_l}| \rightarrow M$\\
    \hline
  \end{tabular}
\end{center}
Let us stress that it does not hold the identification between
$\Delta_\phi$ and $|T_l| \rightarrow M$: the canonical
triangulation $\Delta_\phi$ is obtained via the triangulation of each
polygon of $\square_\phi$, while  $|T_l| \rightarrow M$ is the Regge
dual of $|P_{T_l}| \rightarrow M$ (see fig. \ref{fig:trvstr}).

\begin{figure}[!t]
  \centering
  \includegraphics[width=.7\textwidth]{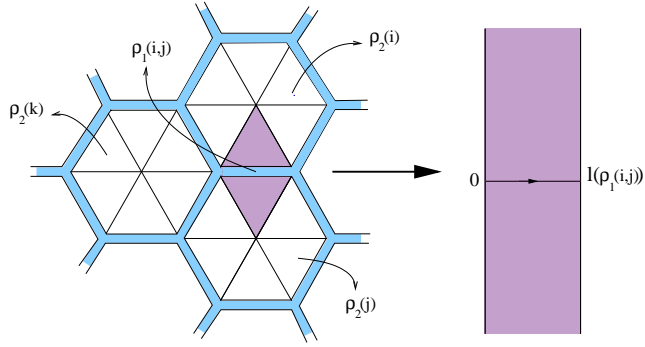}
  \caption{Dual Regge polytope associated to $|T_l| \rightarrow M$}
  \label{fig:dual_pol}
\end{figure}

Let us consider the dual Regge polytope $|P_{T_l}| \rightarrow M$ and
its underlying ribbon graph $\Gamma$. Let $\rho^2(i) \in |P_{T_l}|
\rightarrow M$, $\rho^2(j) \in |P_{T_l}| \rightarrow M$ and $\rho^2(k)
\in |P_{T_l}| \rightarrow M$ be three pairwise adjacent two-cell
barycentrically dual to the vertexes $\sigma_0(i) \in |T_l|
\rightarrow M$,$\sigma_0(j) \in |T_l| \rightarrow M$,$\sigma_0(k) \in
|T_l| \rightarrow M$ (see fig \ref{fig:dual_pol}).  
Each edge of the ribbon graph $\Gamma$ is shared by two triangles of
$\Delta_\phi$ which can be gluing together to get the purple
diamond shape in fig.  \ref{fig:dual_pol}. This is the set of vertical
leaves intersecting $\rho_1(i,j)$. Let $\rho_0(i,j,k)$ and
$\rho_0(j,l,i)$ be the two endpoints of the strip and let us assign
it an arbitrary direction, for example from $\rho_0(i,j,k)$ to
$\rho_0(j,l,i)$. For an arbitrary point P the diamond, the canonical
coordinate $z(i,j)\,=\,\int_{\rho_0(i,j,k)}^P \sqrt{\phi|_{\rho_1(i,j)}}$
maps the diamond to the strip
\begin{equation}
  \label{eq:strip}
  U_{\rho_1(i,j)} \,=\,
  \left\{
    z(i,j) \,\in\, \mathbb{C} 
    \quad\vert\quad
    0\,\leq\,z(i,j)\,\leq,\,l(\rho^1(i,j))
  \right\}
\end{equation}
of infinite height and width $l(\rho^1(i,j))$,where $l(\rho^1(i,j))$
is the length of the considered edge, \ie{} the strip is the local
uniformization on the Riemann surface of the union of the two
triangles. The local expression $\phi|_{\rho_1(i,j)}$ of the Strebel
differential is
\begin{equation}
  \label{eq:stredge}
  \phi|_{\rho_1(i,j)} \,=\, \left(d z(i,j)\right)^2
\end{equation}

On the other hand, the vertexes of the ribbon graphs are, according to
the cellular decomposition induced by the Strebel differential, the
zero of the differential itself. Every quadratic differential has an
expression $ \phi\,=\, \frac{m^2}{4}\, \omega^{m-2}\, (d \omega)^2 $
around a zero of degree $m - 2$. Thus we can use it  as
the expression of the Strebel differential on an open neighborhood
\begin{equation}
  U_{\rho_0(i,j,k)} \,=\,  
  \left\{
    \omega(i,j,k)   \,\in\, \mathbb{C} 
    \quad\vert\quad
    |\omega(i,j,k)| \,\leq\,\delta,\,\omega(i,j,k)(\rho_0(i,j,k)) = 0
  \right\}
\end{equation}
surrounding  $\rho_0(i,j,k)$. Since the 1-skeleton of
$|P_{T_l}\rightarrow M$ is a trivalent graph, $m = 3$ for each vertex,
thus:
\begin{equation}
  \label{eq:strver}
  \phi|_{\rho_0(i,j,k)}\,=\, 
  \frac{9}{4}\, 
  \omega(i,j,k)\, 
  (d \omega(i,j,k))^2
\end{equation}

The transformation law \eqref{eq:strlaw} on the intersection
$U_{\rho_0(i,j,k)}\,\cap\,U_{\rho_1(i,j)}$ gives the transition
function between the two coordinates patches:
\begin{equation}
  \label{eq:edge_vertex_trans}
  \omega(h,j,k) \,=\,
  \begin{cases}
    z(h,\,j)^{\frac{2}{3}} \\
    e^{\frac{2\pi}{3}i}z(j,\,k)^{\frac{2}{3}}\\
    e^{\frac{4\pi}{3}i}z(k,\,h)^{\frac{2}{3}}.
  \end{cases}
\end{equation}

\begin{figure}[!t]
  \centering
  \includegraphics[width=.7\textwidth]{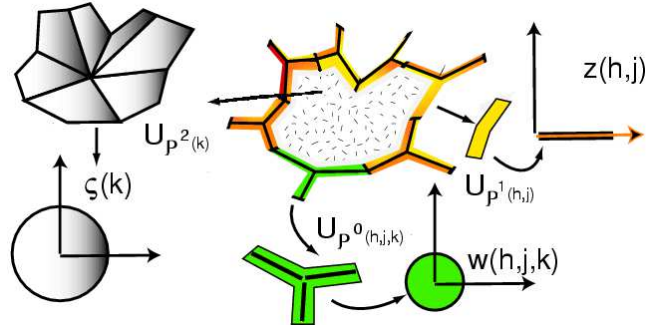}
  \caption{The complex coordinate neighborhoods associated to the dual
    polytope\cite{Carfora1}}
  \label{fig:uniform}
\end{figure}

Due to the conical geometry retained by the deficit angles, the
extension the uniformization on the 2-cells of the cellular
decomposition (\ie{} on the Regge polytopes $\rho_2(i)$) is a little
more subtle.  We will show that is possible to keep track of the
conical geometry of the polytopes in two different ways, thus leading
to two dual uniformizations.

Since the condition \eqref{eq:per}, we can choose a local coordinate
$\zeta(i)$ on an open disk $U_{\rho_2(i)}$ defined by the union of all
the compact horizontal leaves homotopic to $\sigma_0(i)$ such
that\footnote{Let us remember the identification
  $a_i\,\leftrightarrow\,L(i),\,i = 1,\,\ldots,\,N_0(T)$}
\begin{equation}
  \label{eq:strcell}
  \phi|_{\rho_2(i)}\,=\, 
  -\frac{L(i)^2}{4\pi^2}
  \frac{d \zeta(i)^2}{\zeta(i)^2}
\end{equation}

If $z(i,j_\nu),\,\nu=1,\,q(i)$ are the coordinates defined on the
edges surrounding the given 2-cell $\rho_2(i)$, it holds on the
intersections $U_{\rho_2(i)}\,\cap\,U_{\rho_1(i,j_nu)}$:
\begin{equation}
  (d z(i,j_\nu))^2 \,=\,   
  -\frac{L(i)^2}{4\pi^2}
  \frac{d \zeta(i)}{\zeta(i)^2}
\end{equation}
If we fix the integration constants such that the uniformizing unit
disk
\begin{equation}
  U_{\rho_2(i)}\,=\,  \left\{
 \zeta(i) \in \mathbb{C}
    \quad\vert\quad
    |\zeta(i)| \,<\,1,\,\zeta(i)(\sigma_0(i)) = 0
  \right\}
\end{equation}
covers the full polytope $\rho_2(i)$, we get the following transition
functions on $U_{\rho_2(i)}\,\cap\,U_{\rho_1(i,j_nu)}$\cite{Carfora1}:
\begin{equation}
  \zeta(i)\,=\,
  \exp{\left(\frac{2 \pi i}{L(i)} \sum_{\beta = 1}^{\nu - 1}\left(
        L(i, j_\beta) \,+\, z(i, j_\nu)
      \right) \right)} \qquad \nu\,=\,1, \ldots, q(i),
\end{equation}
where $L(i,j)\,=\,l(\rho^1(i,j))$ and $\sum_{\beta = 1}^{\nu - 1}
\doteq 0$ if $\nu = 1$.

Since for any closed curve $c:\;\mathbb{S}^1 \rightarrow
U_{\rho_2(i)}$ homotopic to the boundary of $\overline{U}_{\rho_2(i)}$
it holds
\begin{equation}
  \oint_c \sqrt{\phi|_{\rho_2(i)}} \,=\, L(i),
\end{equation}
then the geometry associated with $\phi|_{\rho_2(i)}$ is described by
a cylindrical metric associated with a quadratic differential with a
second order pole,\ie{}
\begin{equation}
\label{eq:cylmet}
\vert\phi(i)\vert\,\doteq\,
  \vert\phi|_{\rho_2(i)}\vert\,=\, 
  \frac{L(i)^2}{4\pi^2}
  \frac{\vert d \zeta(i)\vert^2}{\vert\zeta(i)\vert^2}.
\end{equation}

We are dealing with such a limiting situation like those described at
the end of section \ref{sec:RRT}: the punctured disk $\Delta^*_i
\,\subset\, U_{\rho^2(i)}$
\begin{equation}
  \Delta^*_i    \,=\,  \left\{
    \zeta(i) \in \mathbb{C}
    \quad\vert\quad
    0  \,<\, |\zeta(i)| \,<\,1
  \right\}
\end{equation}
endowed with the cylindrical metric \eqref{eq:cylmet} is isometric to
a flat semi-infinite cylinder.

To keep track of the conical geometry associated to the polygonal cell
$\rho^2(i)$ we can recall that the conformal factor in
\eqref{eq:metric} defines the metric on the polytopes only up to
conformal symmetry. Thus, for a given deficit angle $\varepsilon(i)
\,=\, 2\pi - \theta(i)$ we can conformally relate the cylindrical
geometry \eqref{eq:cylmet} and the conical one \eqref{eq:metric}
fixing the conformal factor $u$ such that:
\begin{equation}
  \label{eq:metr_cone}
  d s^2_{(i)} \,\doteq\,
  \frac{L(i)^2}{4\pi^2} 
  |\zeta(i)|^{- 2 \frac{\varepsilon(i)}{2 \pi}}
  |d \zeta(i)|^2
\end{equation}

We can eventually state:
\begin{prop}{\cite{Carfora1}}
  If $\{p_{k}\}_{k=1}^{N_{0}}\in M$ denotes the set of punctures
  corresponding to the decorated vertexes $\{\sigma
  ^{0}(k),\frac{\varepsilon (k)}{2\pi } \}_{k=1}^{N_{0}}$ of the
  triangulation $|T_{l}|\rightarrow M$, let $RP_{g,N_{0}}^{met}$ be
  the space of conical Regge polytopes $|P_{T_{l}}|\rightarrow {M}$,
  with $ N_{0}(T)$ labelled vertexes, and let
  ${\mathfrak{P}}_{g},_{N_{0}}$ be the moduli space
  ${\mathfrak{M}}_{g},_{N_{0}}$ decorated with the local metric
  uniformizations $(\zeta(k),ds_{(k)}^{2})$ around each puncture
  $p_k$. Then the map
  $$\Upsilon
  :RP_{g,N_{0}}^{met}\longrightarrow{\mathfrak{P}}_{g},_{N_{0}}$$
  provided by
\begin{multline}
  \Upsilon :(|P_{T_{l}}|\rightarrow {M)\longrightarrow
  }((M;N_{0}),\mathcal{C}
  );\{ds_{(k)}^{2}\}) = \\\label{riemsurf} 
  =\bigcup_{\{\rho^0(i,j,k)\}}^{N_2(T)} U_{\rho^0(i,j,k)}
  \bigcup_{\{\rho^1(i,j)\}}^{N_1(T)}U_{\rho^1(i,j)}
  \bigcup_{\{\rho^2(k)\}}^{N_0(T)}(\Delta^*_k,ds_{(k)}^{2}),
\end{multline}
defines the decorated, $N_{0}$-pointed, Riemann surface $((M;N_{0}),
\mathcal{C})$ canonically associated with the conical Regge polytope $
|P_{T_{l}}|\rightarrow {M}$.
\end{prop}

\subsubsection{The dual decoration: trading curvature for moduli}

The dual decoration, which was introduced in \cite{Carfora1}, is quite
more subtle and arises when we open the cone over $\rho^2(i)$ into its
constituent conical sectors.  To show how it works, let $W_{\alpha
}(k)$, $\alpha =1,...,q(k)$ be the barycenters of the edges $\sigma
^{1}(\alpha )\in |T_{l}|\rightarrow {M}$ incident on $\sigma ^{0}(k)$,
and intercepting the boundary $\partial(\rho^2(k))$ of the polygonal
cell $\rho^2(k)$. Let us denote with $\widehat{L}_{\alpha }(k)$ the
length of the polygonal lines between the points $W_{\alpha }(k)$ and
$W_{\alpha +1}(k)$ (with $\alpha $ defined $\text{mod}q(k)$).  Let us
go back for a while. In the uniformization \eqref{eq:ball}, endowed
with the metric \eqref{eq:metric} (where we have set
$\zeta(k)(\sigma_0(k))=0$) the points $\{W_{\alpha }(k)\}$
characterize a corresponding set of points on the circumference
$\{\zeta (k)\in \mathbb{C} \;|\;\left| \zeta (k)\right| =l(\partial
(\rho ^{2}(k)))\}$, and an associated set of $q(k)$ generators
$\{\overline{W_{\alpha }(k)\sigma ^{0}(k)}\}$ on the cone $C|lk(\sigma
^{0}(k))|$.  Such generators mark $q(k)$ conical sectors
\begin{equation}
  S_\alpha(k)\doteq 
\left( 
c_\alpha(k),\,
\frac{L(k)}{\theta (k)},\,\vartheta _\alpha(k)\right) ,
\end{equation}
with base 
\begin{equation}
  \label{consect}
  c_\alpha(k)\doteq \left\{ 
    \left| \zeta (k)\right| 
    \,=\, L(k),\,
    \arg W_\alpha (k)\leq \arg \zeta(k) \leq \arg W_{\alpha+1}(k)
  \right\}, 
\end{equation}
slant radius $\frac{L(k)}{\theta (k)}$, and with
angular opening 
\begin{equation}
  \label{angles}
  \vartheta _{\alpha }(k)\doteq 
  \frac{\widehat{L}_{\alpha }(k)}{L(k)}\theta (k),
\end{equation}
where $\theta (k)=2\pi -\varepsilon (k)$ is the given conical angle.
The cone
$(D_k(r(k)),ds_{(k)}^{2})$ (see eq. \eqref{eq:ball}) can be formally
represented as:
\begin{equation}
  (D_k(r(k)),ds_{(k)}^{2})=\cup _{\alpha =1}^{q(k)}S_{\alpha }(k).
\end{equation}
If we split the vertex of the cone and of the associated conical
sectors $S_{\alpha }(k)$, then the conical geometry of
$(D_k(r(k)),ds_{(k)}^{2}),ds_{(k)}^{2})$ can be equivalently described
by a cylindrical strip of height $\frac{l(\partial (\rho
  ^{2}(k)))}{\theta (k)}$ decorating the boundary of $\rho ^{2}(k)$.
Each sector $S_{\alpha }(k)$ in the cone gives rise, in such a
picture, to a rectangular region in the cylindrical
strip\cite{Carfora1}.  Introducing a complex variable $v(k) = x(k) +
i\,y(k) \,\in\,\mathbb{C}$, we can explicitly parametrize any such
region as:
\begin{equation}
  R_{\vartheta _{\alpha }(k)}(k)
  \,\doteq\, 
  \left\{ 
    v(k)\in \mathbb{C}|\;0\leq x(k)\leq 
     \widehat{L}_{\alpha }(k),\,
    0\leq y(k)\leq \frac{L(k)}{\theta (k)}
  \right\}.
\end{equation}

In this connecion, we can go a step further. If we consider the full
cylinder, we can think to parametrize it with the complex coordinate
$z$ introduced above. In this case the associated rectangular region
is:
\begin{equation}
\label{rec}
  R_{\theta(k)}(k)
  \,\doteq\, 
  \left\{ 
    v(k)\in \mathbb{C}|\;0\leq x(k)\leq 
     L_{\alpha }(k),\,
    0\leq y(k)\leq \frac{L(k)}{\theta (k)}
  \right\}.
\end{equation}

By means of the canonical conformal transformation:
\begin{equation}
  \label{eq:cyl-to-ann}
  \zeta(k) \,=\,
  e^{\frac{2 \pi i}{L(k)}\,v(k)}
\end{equation}
we can conformally map the cylinder into the annulus:
\begin{equation}
  \label{annul}
  \text{\cyl{k}} \,\doteq\,
  \left\{ 
    \zeta(k) \in \mathbb{C} 
    \quad\vert\quad
    e^{- \frac{2\pi}{2\pi - \varepsilon(k)}}
    \,\leq\, \vert \zeta(k) \vert \,\leq\, 1
  \right\} 
  \,\subset 
  \overline{U_{\rho_2(k)}}. 
\end{equation}

\begin{figure}[!t]
  \centering
  \includegraphics[width=\textwidth]{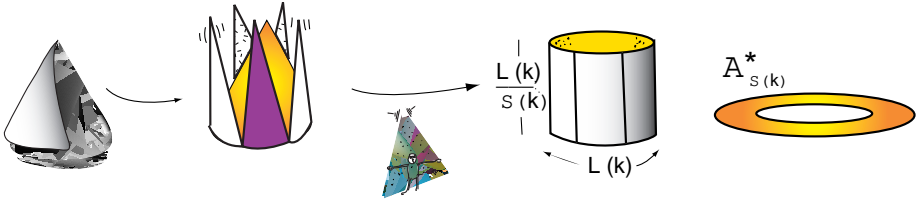}
  \caption{Opening the cones\cite{Carfora1}}
  \label{fig:opening2}
\end{figure}

Then the surface with boundary:
\begin{equation}
  \label{eq:bndry_surf}
  M_\partial \,\doteq\, ((M_\partial,\,N_0),\,\mathcal{C})
\,=\,
\bigcup_{\{\rho^0(i,j,k)\}}^{N_2(T)} U_{\rho^0(i,j,k)}
  \bigcup_{\{\rho^1(i,j)\}}^{N_1(T)}U_{\rho^1(i,j)}
  \bigcup_{\{\rho^2(k)\}}^{N_0(T)}(\text{\cyl{k}},|\phi(k)|)
\end{equation}
defines the blowing up of the conical geometry of
$((M,\,N_0),\,\mathcal{C}, d s_{(k)}^2)$ along the ribbon graph
$\Gamma$\cite{Carfora1}.

The metrical geometry of $(\Delta _{\varepsilon (k)}^{\ast },|\phi
(k)|)$ is that of a flat cylinder with a circumpherence lenght given
by given by $L(k)$ and height given by $\frac{L(k)}{\theta{k}}$
\footnote{Let us recall that this is the slant radius of the
  generalized euclidean cone $(\Delta^*_k, d s^2_{(k)}$ of base circumpherence
  $L(k)$ and vertex conical angle $\theta(k)$}.
We also have 
\begin{equation}
  \partial M_\partial \,=\, \bigsqcup_{k = 1}^{N_0(T)} S_{\theta(k)}^+
  \qquad\quad
  \partial \Gamma \,=\, \bigsqcup_{k = 1}^{N_0(T)} S_{\theta(k)}^-
\end{equation}
where the circles
\begin{subequations}
  \label{eq:boundaries}
  \begin{gather}
    \label{eq:bound+}
    S_{\theta(k)}^+ 
    \,=\,
    \left\{
      \zeta(k) \,\in\, \mathbb{C}
      \quad\vert\quad
      |\zeta(k)| \,=\, e^{- \frac{2\pi}{2\pi - \varepsilon(k)}}
    \right\} \\
    \label{eq:bound-}
    S_{\theta(k)}^- 
    \,=\,
    \left\{
      \zeta(k) \,\in\, \mathbb{C}
      \quad\vert\quad
      |\zeta(k)| \,=\, 1
    \right\}
  \end{gather}
\end{subequations}
denote respectively the inner and outer boundary of the annulus
\cyl{k}. Note that collapsing $S_{\theta(k)}^+$ to a point we get the
original cones $(\Delta^*_k, d s^2_{(k)}$. Thus, the surface with
boundary $M_\partial$ naturally correspond to the ribbon graph
$\Gamma$, naturally associated with the 1-skeleton of the dual Regge
polytope $|P_{T_l}| \rightarrow M$, decorated with the \emph{finite}
cylindrical ends \cyl{k} \cite{Carfora1}.
  
This latter uniformization had proven invaluable in discussing the modular
aspects of simplicial quantum gravity, in particular, it allowed a
detailed analysis of the connection between the Weil-Petersson volume
of the (compactified) moduli space of genus $g$ Riemann surfaces with
$N_0$ punctures $\overline{\mathfrak{M}}_{g},_{N_{0}}$ and the dynamical
triangulation partition function for pure gravity \cite{Carfora1}.

According to the above description, it follows that a metric
triangulation can be seen either as the (singular Euclidean)
uniformization $((M,N_0),\mathbb{C};ds^2_{(k)})$ of a punctured
Riemann surface, or as the uniformization of an open Riemann surface
$((M,\partial{M}_{(k)}),\mathbb{C};|\phi (k)|)$ with finite
cylindrical ends $\Delta _{\varepsilon (k)}^{\ast }$\cite{Carfora1}.
Underlying this geometric duality there is the peculiar role played by
the deficit angles $\varepsilon(k)$ associated with the vertexes of
the triangulation.  They represent the localized curvature degrees of
freedom of the triangulation, and together with the perimeter length
$L(k)$ of the 2-cell dual to the given vertex (puncture), they provide
the relevant geometric information for uniformizing (with a conical
Euclidean metric) the pointed Riemann surface around each puncture.
Conversely, in the open Riemann surface with ($N_0$) boundaries
$\partial{M}_{(k)}$, the deficit angles (again together with the
corresponding 2-cell perimeters) are directly turned into geometric
moduli of the associated annuli\cite{Carfora2}: 
\begin{equation}
\label{curv-moduli}
m(k)
\,\doteq\, 
\frac{1}{2\pi }
\ln{\frac{1}{e^{-\frac{2\pi }{2\pi -\varepsilon (k)}}}}
\,=\,
\frac{1}{2\pi -\varepsilon (k)}.
\end{equation}
In other words, \emph{localized curvature can be traded into modular
  data}. These two points of view merge one into the other when the
deficit angles $\varepsilon(k)$ all degenerate into $2\pi$. This
corresponds to the limiting situation when the triangles incident on
the given vertex of the triangulation are no longer Euclidean but
rather ideal hyperbolic triangles: each finite cylindrical end turns
into a hyperbolic cusp. From a modular point of view we may say that
the two dual singular Euclidean uniformization representing a given
random Regge triangulation merge in a unique hyperbolic uniformization
of the same underlying closed (pointed) Riemann surface.

%% file: bcft_langlands.tex
\chapter{BCFT on the annuli: Langlands' boundary states}
\label{ch:bcft_langlands}

Starting from this chapter, we develop the natural framework to
describe the coupling of a matter field theory with two dimensional
quantum gravity in the framework described in the previous chapter,
\ie{} with the peculiar geometry defined the the dual uniformized open
Riemann surface $M_\partial$ (see eq. \eqref{eq:bndry_surf})
associated with the random Regge triangulation $\vert T
\vert\,\rightarrow\,M$.

The aim of investigating string dualities in such a discretized
approach, leads us to deal with non-critical Polyakov string theory.
Thus, let us consider $D$ scalar fields
$X^\alpha,\,\alpha,=\,1,\,\ldots,\,D$, defined as injection maps from
the Riemann surface $M_\partial$ to an unspecified target space
$\mathcal{T}$. In this connection, at fixed $N_0$ and fixed genus $g$,
the strategy we will follow to quantize the bosonic fields on the open
Riemann surface with boundaries defined in equation
\eqref{eq:bndry_surf} calls into play Boundary Conformal Field Theory
(BCFT). As a matter of fact, we will discuss the quantization of
bosonic fields on each annular region \cyl{k}, defined in
\eqref{annul}, then we shall glue together the resulting BCFTs along
the intersection pattern defined by the ribbon graph $\Gamma$
associated with the triangulation. Finally, as usual in (dynamical or
random) triangulation, the coupling with two-dimensional gravity on
the Riemann surface will be (at least formally) obtained by summing
over all possible triangulation of the original Riemann surface $M$,
\ie{} in the dual picture, over all possible ribbon graphs $\Gamma$.

When we describe the metric geometry of the triangulation
$|T_{l}|\rightarrow M$ as the dual open Riemann surface $M_\partial$,
we are actually unwrapping each conical 2-cell $\rho ^{2}(k)$ of the
dual polytope into a corresponding finite cylindrical end.  The
restriction of each bosonic field to the outer boundary, $X^{\alpha
}(k)|_{+}$, viewed as embedding variables, a priori injects $\boup$ on
a circle of radius $R(k)$ whose length $2 \pi R(k)$ is not necessarily
equal to the circumference of the boundary itself, \emph{i.e.} $L(k)$.
The same is true for the behavior of the fields (injection maps) on
the inner boundary $\boum$, so we can extend this behavior to the
fields defined on the whole cylinder and assume that in general they
are compactified on a circle of circumference $2 \pi
\frac{R^\alpha(k)}{l(k)}$, where $l(k)$ is a lenght parameter built on
the characteristic data of the triangulation.  In order to not impose
any restrictions over this ``unwrapping'' process, we will adopt the
general condition:
\begin{equation}
  \label{winding}
  X^\alpha(k)(e^{2 \pi i}\zeta,\,e^{-2 \pi i}\overline{\zeta}) \,=\, 
  X^\alpha(k)(\zeta,\,\overline{\zeta}) 
  \,+\, 2 \nu^\alpha(k) \pi \frac{R^\alpha(k)}{l(k)}, \qquad \quad 
  \nu^\alpha(k)\,\in\,\mathbb{Z}
\end{equation} 
stating that each field $X^\alpha(k)$ winds $\nu(k)$ times around the
toric cycles of the compact target space $\mathcal{T}$.

Once we are given a Conformal Field Theory (CFT from now on) on a
closed Riemann Surface (\ie{} the Riemann sphere), the problem to
extend this theory to a theory defined on surfaces with boundary is
encoded into find which boundary conditions can be imposed
consistently at the boundaries (in a more stringy language, this is the
question of how many consistent open string theories can be added to a
given closed string one).  In our connection, we have to specify
consistent boundary conditions for the fields
$X^\alpha(\zeta(k),\bar{\zeta}(k)),\, \alpha\,=\,1,\,\ldots,\,D$ on
the boundaries of the cylindrical region \cyl{k}:
\begin{subequations}
\begin{align}
\boup & \doteq\, 
\left\{
   \zeta(k) \,\in\, \mathbb{C} 
    \,\vert\,
   | \zeta(k) | \,=\, e^{\frac{2\pi}{2\pi - \varepsilon(k)}}
\right\}
 \\
\boum & \doteq\, 
\left\{
  \zeta(k) \,\in\, \mathbb{C} 
    \, \vert\,
  | \zeta(k) | \,=\, 1
\right\})
\end{align}
\end{subequations}

This problem is rather similar to the closed conformal field theory
problem arising when we ask if, given a CFT on the Riemann sphere,
this is sufficient to define uniquely the same CFT on higher genus
surfaces. The answer to this question is known exactly: the theory on
the sphere determines uniquely the CFT on higher genus surfaces
because of locality of the Operator Product Expansions (OPE): their
coefficients are uniquely defined on the sphere and do not ``see''
the different genus of the surrounding surface. However, asking for
consistency on higher genus surfaces implies one more condition: the
correlation functions must transform under the action of the modular
group $SL(2,\mathbb{Z})$ at genus one.

An analogous general result holding for BCFT question is not really
known.  Once given a theory on the full complex plane, usually
referred to as \emph{the bulk theory}, what we know is the complete
list of \emph{sewing constraints} that have to be satisfied on if we
introduce a boundary\cite{Lewellen:1991tb,Cardy:1991tv}. However, it
is not known \emph{a priori} for which theory it is possible to find a
complete set of solutions of these constraints. The set of examples in
which a solution has been found shows that, even in this case, modular
invariance plays a fundamental role: this is due to the similarity
between the classification of modular invariant partition functions
and the classification of non-negative integer matrix representation
of the fusion algebra (see \cite{Gaberdiel:2002iw} and references
therein).  To some extents, the theory of the compactified boson
belongs to this set.

The OPE coefficient, which are fully determined by the bulk theory,
define an algebra of fields which the boundary conditions must
respect. We will show that each boundary condition can be described by
a coherent state built on the states space of the bulk theory.
However, not every such a boundary states describes an admissible
extension of the bulk theory. The main point is that boundary states
must preserve the bulk algebra of fields stating a precise gluing
relation between their left and right modes.

Aiming to expose results in a self-contained way, we start the chapter
with a brief introduction to the fundamental concepts in (boundary)
conformal field theory. The expert reader can skip this part, and jump
directly to section \ref{sec:amplitude}.

\section{Bulk and Boundary Conformal Field Theory}
\label{sec:bulkCFT}

To some extents, the CFT associated to the compactified free boson
belongs to the set of conformal field theories whose extension on an
arbitrary surface with boundary can be completely work out once we
know properties of the same theory defined on the full complex plane,
this latter being usually referred to as the \emph{bulk theory}.

Let us recall that, given a vector space $\mathcal{H}$ and the vector
space of linear endomorphism of $\mathcal{H}$ which are finite linear
combination of homogeneous endomorphism, we can define a \emph{field
  of conformal dimension $(h,\bar{h}) \in \mathbb{Z}_+^2$} an
$\text{End} V$-valued formal power series in $\zeta,\bgz$:
\begin{equation}
  \phi(\zeta,\bgz) \,=\, \sum_{m,n \in \mathbb{Z}^2} \phi_{m,n} 
  \zeta^{- m - h}\,\bgz^{- n - \bar{h}},
\end{equation}
\ie{}the coefficient of this formal Laurent expansion are operators in
$\text{End} V$

A bulk conformal theory is defined as an Hilbert space of states
$\mathcal{H}^{(C)}$, endowed with the action of an Hamiltonian
$H^{(C)}$ and of a vertex operation, \ie{} a formal map
\begin{equation}
  \label{eq:vertex_map}
  \Phi^{(C)}(\circ;\,\zeta,\,\bgz): \;  
  \mathcal{H}^{(C)} 
  \,\rightarrow\,
  \text{End} V [\zeta,\bgz]
\end{equation}
associating to each vector $\vert\phi\rangle \in \mathcal{H}^{(C)}$ a
conformal field $\phi(\zeta,\,\bgz)$ of conformal dimension
$h,\bar{h}$.

The bulk theory is completely worked out once they are known the
coefficients of the Operator Product Expansion (OPE) for all fields in
the theory.  Actually, this task is tractable for most of the CFTs,
since the states in $\mathcal{H}^{(C)}$ (and consequently the
associated vertex operators) can be organized into irreducible
representation of the observable algebra generated by the chiral
fields.

Chiral fields depend on only one coordinate, $\zeta$ or $\bgz$, thus
they are either holomorphic or antiholomorphic:
\begin{equation}
  \label{eq:exp}
  W^a(\zeta) \,=\, \sum_m W^a_m \zeta^{- m - h^a}
\qquad
  \overline{W}^a(\bgz) \,=\, \sum_m \overline{W}^a_m \bgz^{- m - \bar{h}^a}
\end{equation}
where the index $a$ labels the different chiral fields in the theory
and $h^a$ and $\bar{h}^a$ are respectively the conformal weights of
$W^a(\zeta)$ and $\overline{W}^a(\bgz)$.  Their Laurent modes $W_m^a$
and $\overline{W}^a_m$ generate two commuting chiral algebras,
$\mathcal{W}$ and $\overline{\mathcal{W}}$, which we will assume to be
isomorphic.

The Virasoro fields $T$ and $\overline{T}$ play a special role among
the chiral fields of a CFT. Their modes $L_n$ and $\overline{L}_n$
close two copies of the Virasoro algebra with central extension $c$:
\begin{gather*}
  \left[L_m\,,\,L_n\right] 
  \,=\, (m - n)\,
  L_{m + n} \,+\, \frac{c}{12} n (n^2 - 1) \delta_{m + n , 0}\\
  \left[\overline{L}_m\,,\,\overline{L}_n\right] 
  \,=\, (m - n)\,
  \overline{L}_{m + n} \,+\, \frac{c}{12} n (n^2 - 1) \delta_{m + n , 0}
\end{gather*}

The Virasoro algebra belongs to the enveloping algebra of the chiral
algebra $\mathcal{W}$. As a matter of fact, since we can represent the
holomorphic and antiholomorphic components of the stress-energy tensor
as the ordered (creator-annihilator) product $T(z) = \gamma
\ordprod{W(\zeta)^a W(\zeta)^a}$, we get a representation of the their
Laurent modes via the Sugawara construction:
\begin{subequations}
  \label{virasoro}
  \begin{gather}
    \label{virasoro0}
    L_0
    \,\doteq\, 
    \sum_{n>0}
    W_{-n} \cdot
    W_n 
    \,+\,
    \frac{1}{2}(W_0)^2 \\
    \label{virasoron}
    L_n
    \,\doteq\, 
    \frac{1}{2}\sum_{m \in \mathbb{Z}}
    \ordprod{W_{n - m}\cdot W_m} 
  \end{gather}
\end{subequations}

The action of the chiral algebra modes $W_m$ and $\overline{W}_m$ determine
a decomposition of the Hilbert space into all the irreducible modules
of the two commuting chiral algebra:
\begin{equation}
  \label{eq:chirdec}
  \mathcal{H}^{(C)} \,\doteq\,
  \bigoplus_{\lambda\,\overline{\lambda}}
  \mathcal{H}^\lambda
  \,\otimes\, 
  \overline{\mathcal{H}}^{\overline{\lambda}},
\end{equation}
where the vectors (highest weights) $\left\{\lambda^a\right\}$ and
$\left\{\overline{\lambda}^a\right\}$ are the charges (\ie{}
respectively the $W_0^a$ and $\overline{W}_0^a$ eigenvalues) associated to
all the possible ground-states (highest weight
states)$|\lambda\rangle$ and $|\overline{\lambda}\rangle$\footnote{Let
  us recall that such ground-states are defined via the relations
  (holomorphic sector):
  \begin{gather*}
  W_n^a  |\lambda\rangle \,=\, 0 \qquad \forall a =
  1,\,\ldots,\,\text{dim} \mathcal{W} \;\text{and}\; \forall \, n > 0
  \\
  \overline{W}_n^a  |\overline{\lambda}\rangle \,=\, 0 \qquad \forall a =
  1,\,\ldots,\,\text{dim} \mathcal{W} \;\text{and}\; \forall \, n > 0
  \end{gather*}}, 
and where each chiral module is defined as: 
\begin{equation}
  \mathcal{H}_{\lambda} \,\doteq\, 
  \mbox{Span}\left\{
    W_{-m_1}^a\ldots W_{-m_N}^a |\lambda\rangle 
    \quad|\quad
    m_i \,>\, 0 \; \forall\,i=1,\,\ldots,N 
  \right\}.
\end{equation}
An equivalent definition holds for the antiholomorphic sector. 

From
these definition we can introduce the vertex operator associated to a
holomorphic ground-state $|\lambda\rangle$:
\begin{equation}
  \label{eq:gs_vertex}
  \varphi_\lambda(\zeta)\,\doteq\,
  \Phi^{(C)}(|\lambda\rangle;\,\zeta) 
  \,=\,
  c_\lambda 
  e^{\int \lambda_-(\zeta) d \zeta}
  e^{i \lambda W_0 }
  z^{i \lambda q}
  e^{\int \lambda_+(\zeta) d \zeta}
\end{equation}
where $\lambda_\pm(\zeta) = \sum_{m>0} \lambda \,W_{\pm m} z^{\mp m -
  1}$, $c_\lambda$ is an arbitrary cocycle and $q$ is a position
operator.

The vertex operator associated with a single oscillator acting on the
conformal vacuum is defined by:
\begin{equation}
  \label{eq:onemode_vertex}
  \Phi^{(C)}(W^a_{-m}\vac;\,\zeta)
  \,=\,
  \frac{1}{(m - 1)!} 
  \frac{d^{m - 1}}{d \zeta}
  W^a(\zeta)  
\end{equation}

For a general homogeneous element $|\phi\rangle \,=\,
W^{a_1}_{-m_1}\ldots W^{a_N}_{-m_N} |\lambda\rangle$, linearity of $\mbox{End}
\mathcal{H}$ ensures: 
\begin{equation}
  \label{eq:gen_vertex}
  \Phi^{(C)}(|\phi\rangle;\,\zeta)\,=\,
  \ordprod{
    \Phi^{(C)}(W^{a_1}_{-m_1}\vac;\,\zeta)
    \,\ldots\,
    \Phi^{(C)}(W^{a_N}_{-m_N}\vac;\,\zeta)
    \Phi^{(C)}(\vert\lambda\rangle;\,\zeta)
  }
\end{equation}

In the definition \eqref{eq:chirdec}, the module $\mathcal{H}^0$
refers to the vacuum representation. From the reconstruction equations
\eqref{eq:gs_vertex} and \eqref{eq:onemode_vertex}, we see that it is
mapped directly into $\mathcal{W}$ by the vertex operation defined
by $\Phi^{(C)}$.

Each irreducible representation $\mathcal{H}^\lambda$ receive a
$\mathbb{Z}$-grading by the action of $L_0$. As a matter of fact the
$L_0$ eigenvalues (\ie{} the conformal weights) 
\begin{equation*}
  L_0 (W_{-m_1}\ldots
W_{-m_N}) |\lambda\rangle \,=\, m_1 \,+\, \ldots \,+\, m_N \,+\, h
|\lambda\rangle
\end{equation*}
define the level of a state as: $l\,=\,\sum_{i=1}^N m_i$ so that each
irreducible representations of $\mathcal{W}$ can be decomposed as 
\begin{equation}
\label{eq:grad}
 \mathcal{H}^\lambda
\,=\, \bigoplus_{l \geq 0} \mathcal{H}^\lambda_l. 
\end{equation}
 
With the above remarks, we see that each copy of the chiral algebra
fulfills all the requirements of a vertex algebra\footnote{Let us
  recall that given a $\mathbb{Z}$-graded vector space $V =
  \bigoplus_m V_m$, an endomorphism E of $V$ is homogeneous of degree
  $j$ if $\text{E}(V_m) \in V_{m + j}$.  Thus we can
  define\cite{Frenkel:2004jn}: 
  \begin{defin}
    A \textsc{VERTEX ALGEBRA} is a collection of data:
    \begin{itemize}
    \item a $\mathbb{Z}$-graded vector space of states $V =
      \bigoplus_{m=0}^\infty V_m$;
    \item a vacuum vector $\vac \in V_0$ 
    \item a translation operator $\omega\,:\; V \rightarrow
      V$ of degree one;
    \item a vertex operation, \ie{} a linear map $\Phi(\circ,\,z)\,:\;
      V \rightarrow \mbox{End} V [[z,z^{-1}]]$ taking each state
      $\vert\phi\rangle \in V_n$ into a field of conformal dimension $n$ 
    \end{itemize}
    subjected to the following constraints:
    \begin{itemize}
    \item \textsc{vacuum axiom}: $\Phi(\vac, z) \doteq \mathbb{I}_V$ 
    \item \textsc{translational axiom}: For any $\vert\phi\rangle \in V$, 
      $\left[\omega , \Phi(\vert\phi\rangle, z)\right] = \partial_z
      \Phi(\vert\phi\rangle, z)$. 
    \item \textsc{locality axiom}: all fields $\Phi(\vert\phi\rangle, z)$
      are mutually local with each other.
    \end{itemize}
  \end{defin}}: for the holomorphic sector, the grading of the Hilbert
space is the one induced by the action of $L_0$ described above. The
conformal vacuum is identified with the ground state $\vac = \vert
\lambda = 0 \rangle$, whose associated vertex operator is indeed
$\mathbb{I}_\mathcal{H}$. Finally, the role of a traslation operator
is played by $L_{-1}$, which satisfies
$\Phi(L_{-1}\vert\phi\vert,\,\zeta) \,=\, \frac{d}{d \zeta}
\Phi(\vert\phi\vert,\,\zeta)$ and $\Phi(\vac,\,\zeta) \,=\, 0$, as we
can see directly applying the Sugawara representation
\eqref{virasoro}.

Let us consider, for each irreducible representations of the
holomorphic\footnote{Obviously, the same holds for the antiholomorphic
  sector} chiral algebra $\mathcal{H}^\lambda$, the eigenspace of
0-level states $V^\lambda_0$, \ie, in the case that
$\mathcal{H}^\lambda$ are the highest weight representations of the
chiral algebra, the subspace containing only the (eventually
degenerate) highest weight state. It will carry an irreducible action
of the horizontal subalgebra of $\mathcal{W}$ (\ie the subalgebra
closed by all the zero modes $W_0^a$) via the action of the associated
linear map:
\begin{equation}
  \label{eq:hor_ac}
  \mathbb{X}^\lambda_{W^a} \,\doteq\, 
  W^a_0\vert_{V^\lambda_0} 
  \,:\; V^\lambda_0 \,\longrightarrow\,V^\lambda_0 
\end{equation}

As a consequence, $\varphi_\lambda(\zeta)$ defined in equation
\eqref{eq:gs_vertex} is the holomorphic part of the (Virasoro primary)
fields which arise as $\mathcal{W}$-primary via the action of the
horizontal subalgebra $\{W_0^a\}$. We can gather the holomorphic and
antiholomorphic part in a single $\mathcal{W}$-primary field $
\varphi_{\lambda,\bar{\lambda}}(\zeta\,\bgz)$ associated to states in
$V^\lambda_0\otimes\overline{V}^\lambda_0$.

\subsection{Glueing conditions and boundary fields}
\label{sec:bo}

Dealing with open string theory naturally impose to study the
consequences of conformal invariance on surfaces delimited by one or
more boundaries, on which we have to define suitable field behaviour.
The prototype example, mostly investigated in literature, is the BCFT
defined on the Upper Half complex Plane (UHP from now on). Its results
are easily extended to other restricted geometries via suitable local
conformal transformations.

Thus, let us consider the complex coordinate $z = x + i y$ and let us
restrict ourselves to $\Im{z} \geq 0$. In the interior part of the
domain, the structure of the boundary CFT is fixed by the requiring
local equivalence with the associated bulk theory: thus, locally,
there is a one-to-one correspondence between fields of the boundary
theory and fields of the parent bulk theory.  Their local structures
coincide in the sense that both theories have identical equal time
commutators with chiral fields and identical operator product
expansions.  In particular, it exists a stress energy tensor
$T_{x_1,x_2}(x,y)$ defined in \cyl{k} and, in this connection, the
presence of a boundary for $\Im{z} = 0$ requires that no energy flows
across the boundary itself.  Introducing the holomorphic and
antiholomorphic components, $T(z) = 2(T_{x x} + i T_{x y})$ and
$\overline{T}(\bar{z}) = 2(T_{x x} - i T_{x y})$, this implies the
condition
\begin{equation}
  \label{eq:no_momflow}
  T(z)\,=\,\overline{T}(\bar{z})|_{y=0}.
\end{equation}

When the bulk theory contains other chiral fields $W$ and $\overline{W}$ of
conformal dimension $h_W$ and $\bar{h}_{\overline{W}}$, we have also to
require the following continuity condition:
\begin{equation}
  \label{eq:cont_cond}
  W(z)\,=\,
  \Omega
  \overline{W}(\bar{z})
  |_{y=0} 
\end{equation}
where $\Omega$ is an automorphism of the Dynkin diagram associated to
the extended chiral algebra \cite{Kato:1996nu}. \textbf{To keep
  contact with the notation mostly used in literature, we use $\Omega$
  to define the glueing authomorphism on the boundary. It has not to
  be confused with the compactification radius $\Omega^\alpha(k)$}.

Since they are defined only on half complex plane, $W$ and
$\overline{W}$ are not sufficient to construct the action of two
commuting copies of the chiral algebra. However, the assumption of the
existence of continuity conditions in \eqref{eq:no_momflow} and
\eqref{eq:cont_cond} has the powerful consequence to give rise to the
action of only one copy of the chiral algebra on the state space
$\mathcal{H}^{(H)}$ of the boundary theory. As a matter of fact, we
can combine $W$ and $\overline{W}$ into a single object
$\mathbf{W}(z)$ which, considering the conformal field theory defined
on the UHP domain can be chosen as:
\begin{equation}
  \label{eq:1al}
  \mathbf{W}(z) \,\doteq\,
  \begin{cases}
    W(z) & \text{when}\, \Im{z}\,\geq\,0\\
    \Omega\overline{W}(\bar{z}) & \text{when}\, \Im{z}\,<\,0
  \end{cases}
\end{equation}
Obviously, different domains allow for different choices of
$\mathbf{W}(z)$. The key point is that, thanks to the gluing condition
on the boundary, the field we are now dealing with is analytic on the
whole complex plane; thus, it can be expanded in a Laurent series: $
\mathbf{W}(z) \,=\, \sum_n W_n^{(H)} z^{- n - h}$, introducing the
modes:
\begin{equation}
\label{eq:open_modes}
  W_n^{(H)} \,=\, \frac{1}{2 \pi i }
  \int_C d z \,z^{n + h_W - 1} \,W(z) 
  \,-\,
  \frac{1}{2 \pi i }
  \int_C d \bar{z} \,\bar{z}^{n + \bar{h}_{\bar{W}} - 1} 
  \,\Omega\overline{W}(\bar{z}) 
\end{equation}

Ward identities for correlators of the boundary theory can be
retrieving from the singular part of the OPE of chiral fields with
bulk fields $\varphi (z, \bar{z})$.  OPE are fixed by requirement of
local equivalence between the bulk theory and the bulk of the boundary
theory. In we define the singular part of $\mathbf{W}$ with
$\mathbf{W}_>(z) = \sum_{n > -h} W_n z^{-n-h}$, The singular part of
the OPE is then given by
\begin{multline}
\label{eq:formalope}
\mathbf{W}(w) \,\varphi (\zeta, \bar{z})
\, \sim \,
 \left[
\mathbf{W}_>(w)
\,,\, 
\varphi (\zeta, \bar{z})
\right] 
\,=\, \\
\sum_{n > -h} \left( \frac{1}{(w-z)^{n+h}} \,
  \Phi\bigl(W^{(C)}_n \varphi; z, \bar z\bigr) + \frac{1}{(w-\bar
    z)^{n+h}} \, \Phi\bigl(\Omega\overline{W}^{(C)}_n \varphi; z,\bar
  z\bigr)\right)
\end{multline}
We have placed a superscript $(C)$ on the modes $W_n^{(C)},
\overline{W}_n^{(C)}$ to display clearly that they act on the elements
$\varphi \in \mathcal{H}^{(C)}$ labeling the bulk fields in the
theory. Let us drop superscript $(H)$ from now on: objects without
superscripts are intended to belong to the boundary theory, while we
will put the superscript $(C)$ when we will have to consider objects
belonging to the bulk theory.

The sum on the right hand side of equation \eqref{eq:formalope} is
always finite because $\varphi$ is annihilated by all modes
$W^{(C)}_m, \overline{W}^{(C)}_m$ with sufficiently large $m$.  For
$\Im{w} >0$ only the first terms involving $W^{(C)}_n$ can become
singular and the singularities agree with the singular part of the OPE
between $W(w)$ and $\varphi(z,\bar z)$ in the bulk theory.  In the
same way, the singular part of the OPE between $\Omega
\overline{W}(w)$ and $\varphi(z,\bar z)$ in the bulk theory is
reproduced by the terms which contain $\overline{W}^{(C)}_n$, if $\Im
w < 0$.

For future utility, let us specialize formula \eqref{eq:formalope} to
the case of a chiral current $\mathbf{J}(w)$ (thus $h_J=1$) acting on
a primary field $\varphi_{\lambda, \bar{\lambda}}\,\in\,
\mathcal{H}_\lambda \otimes \overline{\mathcal{H}}_{\bar{\lambda}}$.
In this case, equation \eqref{eq:formalope} reduces to
\begin{equation}
  \label{eq:currope}
  \mathbf{J}(w)\, \varphi_{\lambda, \bar{\lambda}}(z,\bar z)
\,\sim\,
\frac{ X_J^\lambda}{w-z} \, \varphi_{\lambda, \bar{\lambda}}(z,\bar z) \, - \,
\varphi_{\lambda, \bar{\lambda}}(z,\bar z) \, 
\frac{X_{\Omega \bar{J}}^{\overline{\lambda}}}{w-\bar z}
\end{equation}
The linear maps $X_W^\lambda$ and $X_{\Omega
  \bar{J}}^{\overline{\lambda}}$ has been introduced in equation
\eqref{eq:hor_ac}; they act on $\varphi_{\lambda, \bar{\lambda}}$ by
contraction in the first tensor component.

Ward identities for arbitrary $n$-point functions of fields
$\varphi_{ij}$ follow directly from equation \eqref{eq:formalope}.
They have the same form as those for chiral conformal blocks in a bulk
CFT with $2n$ insertions of chiral vertex operators with charges
$i_1,\dots,i_n,\omega(j_1),..,\omega(j_n)$, where $\omega$ is the
automorphism of the fusion rules algebra induced by $\Omega$. In many
concrete examples, one has rather explicit expressions for such chiral
blocks.  So we see that objects familiar from the construction of bulk
CFT can be used as building blocks of correlators in the boundary
theory (``doubling trick''). Note, however, that the Ward identities
depend on the gluing map $\Omega$.

Thanks to Ward identities and bulk fields OPE we can reduce the
computation of correlators of $n$ bulk fields to a product of 1-point
functions $\langle \phi_{ij} \rangle_\alpha$. These no longer vanish
close to a boundary because, traslation invariance towards the
boundary itself is broken, and they can depend both on their distance
from the boundary itself and, more in general, from the boundary
condition.  Transformation properties of bulk fields with respect to
$L_0$, $L_{\pm 1}$ and  $W_0$ fix the form of the one-point function
to
\begin{equation}
\label{eq:bulk1p}
  \langle \phi_{ij} (z,\bz) \rangle_\alpha
\,=\, \frac{A^\alpha_{i j}}{(z\,-\,\bz)^{h_i + h_j}}
\end{equation}
where $A^\alpha_{i j}\,:\; \mathcal{H}_0^j \rightarrow
\mathcal{H}_0^j$ obeys $\mathbf{X}^i_W A^\alpha_{i j} \,=\,
A^\alpha_{i j} \mathbf{X}^j_{\Omega W}$. This last relation implies $j
= \omega(i^+)$, thus $h_i = h_j$. 

These parameters are not completely free. Sewing constraint has been
worked out by several authors (see \cite{Gaberdiel:2002iw},
\cite{Recknagel:1998ih} and references therein). In particular, if we
consider the two point function of bulk primary fields $\langle
\phi_{ij} \rangle_\alpha$, cluster properties of CFT and conformal Ward
identities (see \cite{Recknagel:1998ih} and \cite{Cardy:1991tv} for
further details) allows to write the following relation for 1-point
function coefficients:
\begin{equation}
\label{eq:ca}
A^\alpha_i\,A^\alpha_j  \,=\,
\sum_k \Xi_{i j k} A^\alpha_0\, A^\alpha_k, 
\end{equation}
where we have defined $A^\alpha_i \doteq A^\alpha_{i
  \omega(i^+)}$. The coefficients $\Xi_{i j k}$ can be expressed as
combination of the coefficients of bulk OPE and of the fusing matrices
relating the conformal blocks of the bulk theory. 

The existence on the boundary of the gluing map $\Omega$, defined in
\eqref{eq:cont_cond}, give rise to the action of one chiral algebra
$\mathcal{W}$ on the state space of the boundary theory and inducing a
decomposition of the open CFT Hilbert Space $\mathcal{H}$ into
irreducible representations of $\mathcal{W}$ \cite{Gaberdiel:2002iw}:
\begin{equation}
\label{eq:decomp}
\mathcal{H} \, = \,
\bigoplus_\lambda \mathcal{V}_\lambda.  
\end{equation}

This allows to introduce a new one-to-one state-field correspondence
$\Phi$ between states $\vert \psi \rangle \in
\mathcal{H}$ and so-called boundary fields $\psi(x)$, which are
defined (at least) for $x$ on the real line \cite{Cardy1}.  Let us
suppose that boundary condition does not jump on the real axis (we
will understand the meaning of this assertion later).  Thus, it exists
a unique $sl_2$ invariant vacuum state $\vert 0 \rangle$. Then, for
any state $\vert \psi \rangle \in \mathcal{H}$, there exists a
boundary operator $\psi(x) \doteq \Phi(\psi ; x)$ such that:
\begin{equation}
 \psi(x)\, \vert 0 \rangle
\,=\, e^{x L_{-1}} \vert \psi \rangle 
\end{equation}
for all $x$ in the real line. In particular, the operator $\psi(0)$
creates the state from the conformal vacuum state $\vert 0 \rangle$.
Note also that $L_{-1}$ generate translations along the boundary.

The conformal dimension of a boundary field
$\psi(x)$ can be read off from the $L_0$-eigenvalue of the
corresponding state $\vert \psi \rangle \in \mathcal{H}$. 
If $\vert \psi \rangle$ is a primary state of conformal weight $h$,
definition of conformal modes analogue to \eqref{eq:open_modes} leads
to:
\begin{equation}
  \label{eq:losuphi}
  \left[
    L_n\,,\,
    \psi(x)
  \right]\,=\,
  x^n\,\left(
    x\,\frac{d}{dx} 
    \,+\,
    h\,
    (n + 1)
    \psi(x)
  \right)
\end{equation}
The boundary fields assigned to elements in the vacuum sector
$\mathcal{H}^0$ coincide with the chiral fields in the theory, i.e.\
$\mathbf{W}(x) = \Phi(w;x)$ for some $w \in \mathcal{H}^0$ and $\Im x
= 0$.  These fields can always be extended beyond the real line and
coincide with either $W$ or $\Omega \overline{W}$ in the bulk.  If
other boundary fields admit such an extension, this suggests an
enlargement of the chiral algebra in the bulk theory.

Decomposition \eqref{eq:decomp} also implies that
the partition function may be expressed as a sum of characters
$\chi_i(q)$ of the chiral algebra, 
\begin{equation}
Z_{(\Omega,\a)}(q) \ := \
\mbox{Tr}_\mathcal{H} (q^{L_0-c/24}) \ = \ \sum_i n^{\Omega\a}_i \,
\chi_i(q),
\end{equation}
where $n^{\Omega\a}_i \ \in \ \mathbb{N}$ are multiplicity
coefficients.

The existence of boundary extra fields motivates to consider more
general correlation functions in which the bulk-field on the interior
of the UHP appear together with boundary fields $\psi(x)$.  Following
the standard reasoning in CFT, it is easy to conclude that the bulk
fields $\varphi_{ij}(z,\bar z)$ give singular contributions to the
correlation functions whenever $z$ approaches the real
line\cite{Cardy:1991tv}.  This can be seen from the fact that the Ward
identities describe a mirror pair of chiral charges $i$ and
$\omega(j)$ placed on both sides of the boundary.  Therefore, the
singularities in an expansion of $\varphi(z,\bar z) \equiv
\varphi_{ij}(z,\bar z)$ around $x = {\Re} z$ are given by primary
fields which are localized at t $x$ on the the real line, i.e.\ the
boundary fields $\psi(x)$. In other words, the observed singular
behavior of bulk fields $\varphi(z,\bar z)$ near the boundary with
associated boundary condition $\alpha$ may be expressed in terms of a
bulk-boundary OPE \cite{Cardy:1991tv}
\begin{equation}
     \varphi(z,\bar z) \ = \ \sum_k \ (2 y)^{h_k - h - \bar h} 
        \ C^{\,\alpha}_{\varphi\; k}\  \psi_k(x), 
\label{bbOPE}
\end{equation}
where $\psi_k(x)$ are primary fields of conformal weight $h_k$. Which
$\psi_k$ can possibly appear on the rhs.\ of (\ref{bbOPE}) is
determined by the chiral fusion of $i$ and $\omega(j)$, but some of
the coefficients $C^{\,\alpha}_{\varphi\, k}$ may vanish for some
$\alpha$.

\section{The quantum amplitude on the annulus}
\label{sec:amplitude}

To work out an explicit expression for an (open) string amplitude over
$M_\partial$, let us approach to the problem dealing first with
building blocks of such an amplitude, \ie{} the partition functions
over the annular domains  \cyl{k}, $k = 1, \ldots, N_0$.

As first remark, let us notice that the computation can be performed
independently for each component $X^\alpha,\,\alpha=1,\,\ldots,D$. As
a matter of fact, in a stringy language the world-sheet action for the
bosonic fields (string coordinates) is% 
%\footnote{Since we are dealing
%with a given cylindrical end, we drop the polytope index (k); it
%will be restored when necessary}
:
\begin{equation}
  \label{eq:action}
  S
\,=\,
\frac{1}{4\pi}
\int d \zeta(k) d \bgz(k)\, 
G_{\alpha\beta}(k) \partial X^\alpha(k)
\bar{\partial} \bar{X}^\beta(k) \,+\,
B_{\alpha\beta}(k) \partial X^\alpha(k)
\bar{\partial} \bar{X}^\beta(k)  \,-\,
\frac{1}{2} \Phi(k)\,R^{(2)}
\end{equation}
where $G(k)$, $B(k)$ and $\Phi(k)$ are respectively the background metric, the
background Kalb-Ramond field and the dilaton.  Sometimes in string
theory literature the $D^2$ components of $G$ and $B$, which encode
the geometrical data of the background, are gathered into a matrix
$E$, whose symmetric part is $G$ and whose antisymmetric part is $B$:
\begin{equation}
  \label{eq:bm}
  E \,=\, G \,+\, B.
\end{equation}
$E$ is usually called \textsc{background matrix}.

We will deal with flat toroidal backgrounds,\ie{} we will consider a
string moving in a background in which $D$ dimensions are compactified
and the metric $G$, the Kalb-Ramond field $B$ and the dilaton are $X$
independent. In particular, in the Polyakov background, we will put to
zero the $B$-field and we will encode all data about the background
metric in the compactification radius value, so that the background
metric can be chosen diagonal and with $G_{\alpha\alpha}(k)=1\,\forall
\alpha =1,\,\ldots,\,D$. Then, the world-sheet action on \cyl{k} is:
\begin{equation}
  \label{eq:action1}
  S\,=\,
  \frac{1}{8\pi} \int d\zeta(k) d\bgz(k)\,
  \partial X^\alpha(k)\, \overline{\partial} \bar{X}^\alpha(k)  
\end{equation}

For the compactified boson, the dynamical degrees of freedom are
defined via the modes of the bosonic fields
$X^\alpha(\zeta(k),\bar{\zeta}(k))$ obeying the equation of
motion $\partial\bar{\partial} X^\alpha(\zeta(k),\bar{\zeta}(k))
\doteq \Delta(k) = 0$, where $\Delta(k)$ is the Laplacian defined over
\cyl{k}.  

The bulk theory sees a free boson as a closed string injection map
propagating from the remote past ($ y \rightarrow -\infty $) to the
remote future ($ y \rightarrow -\infty $) and, after the conformal
mapping in equation \eqref{eq:cyl-to-ann}, we can write the following
mode expansion: 
\begin{multline}
\label{fieldexp}
  X^\alpha(\zeta (k),\overline{\zeta }(k)) 
  \,=\,  
  X_0^\alpha(k) \,-\, 
  i p_L^\alpha(k) \ln\zeta(k)
  \,-\,
  i p_R^\alpha(k) \ln\overline{\zeta }(k)
  \,+\,\\
  i\, \sum_{n\neq 0}
  \frac{1}{n}
  \left(
    \fa^\alpha_n(k)\zeta^{-n} 
    \,+\,
    \ofa^\alpha_n(k)\overline{\zeta}^{-n} 
  \right),
\end{multline}
where  we can identify the center of mass momentum
components with the chiral zero modes:
\begin{equation}
  p_L^\alpha(k) \,=\, \fa_0^\alpha(k) \qquad
  p_R^\alpha(k) \,=\, \ofa_0^\alpha(k) 
\end{equation}

The only chiral fields are the holomorphic and antiholomorphic
currents $ J^\alpha(\z) \,=\, i\,\partial X^\alpha(\z) \,=\, \sum_n
\fa^\alpha_n \, \z^{-n-1}$ and $\overline{J}^\alpha(\bgz) \,=\,
i\,\overline{\partial} X^\alpha(\bgz) \,=\, \sum_n \ofa^\alpha_n \,
\bgz^{-n-1} $ which generate, for each field
$\alpha\,=\,1,\ldots,\,D$, $N_0$ copies of the affine
$\hat{u}(1)_L\,\otimes\,\hat{u}(1)_R$ algebra:
\begin{gather}
  \label{heis}
  \left[ \fa_n^\alpha(k) \,,\, \fa_m^\beta(k) \right] \,=\, 
  n \delta_{n+m,0} \delta^{\alpha,\beta} 
  \qquad
  \left[ \ofa_n^\alpha(k) \,,\, \ofa_m^\beta(k) \right] \,=\, 
  n \delta_{n+m,0} \delta^{\alpha,\beta} 
  \notag \\
  \left[ \fa_n^\alpha(k) \,,\, \ofa_m^\beta(k) \right] \,=\, 0.
\end{gather}

At this level we can think that the theory on each cylindrical end
decouples from the others. Thus, from now on, we will suppress the
polytope index $(k)$. We will assume to have $N_0$ copies of the
same theory, and we will explain later how data propagate over the
ribbon graph, \emph{i.e.} how these different copies glue over it.

The bulk Hamiltonian is $H = \frac{2\pi}{L}
\left(\frac{1}{2} \fa_0^2 + \frac{1}{2}\ofa_0^2 + \frac{1}{2}\sum_{m
    \neq 0} \fa_{-n} \cdot \fa_{n} + \ofa_{-n} \cdot
  \ofa_{n}\right)$, from which, replacing 
the expression of Virasoro generators obtained via the Sugawara
construction in this particular case,eq. \eqref{virasoro} with central
extension $c=D$, we get:
\begin{equation}
  H^{(C)} \,=\, \frac{2\pi}{L} 
  \left( 
    L_0 \,+\, \bar{L}_0 \,-\, \frac{D}{12}
  \right). 
\end{equation}

In the diagonal decomposition \eqref{eq:chirdec} of the Hilbert space
into irreducible representations of the two chiral algebras, $\lambda$
and $\bar{\lambda}$ are the (real numbers) $\hat{u}(1)_L$ and
$\hat{u}(1)_R$ charges respectively.

Due to the winding condition \eqref{winding}, highest weight states
carries different charges w.r.t the action of the zero modes $\fa_0$
and $\ofa_0$.  The total Norther's momentum is \mbox{$p^\alpha
  \,=\,\frac{1}{2} \left( \lambda^\alpha \,+\,
    \overline{\lambda}^\alpha \right)$}, while the variation of
$\alpha$-th field along a closed cycle is \mbox{$X^\alpha(k)
  \left(e^{2 \pi i}\zeta,\,e^{-2 \pi i}\overline{\zeta} \right) \,-\,
  X^\alpha(k) \left(\zeta,\,\overline{\zeta} \right) \,=\, 2 \pi
  \left( \fa_0^\alpha \,-\, \ofa_0^\alpha \right)$}. Thus, the
holomorphic and antiholomorphic $u(1)$ charges
\begin{subequations}
\label{eq:momentum}
\begin{gather}
  \label{eq:momentum_left}
  \lambda^\alpha_{(\mu,\nu)} \,=\, 
  \mu^\alpha \frac{l(k)}{R^\alpha(k)} \,+\,
  \frac{1}{2} \nu^\alpha(k) \frac{R^\alpha(k)}{l(k)} \\
  \label{eq:momentum_right}
  \overline{\lambda}^\alpha_{(\mu,\nu)} \,=\, 
  \mu^\alpha \frac{l(k)}{R^\alpha(k)} \,-\,
  \frac{1}{2} \nu^\alpha(k) \frac{R^\alpha(k)}{l(k)}
\end{gather}
\end{subequations}
We can collect the holomorphic and antiholomorphic
ground-states into a single object $\vert
\lambda^\alpha,\overline{\lambda}^\alpha\rangle = \vert
\lambda^\alpha_{(\mu,\nu)} \rangle \,\otimes\, \vert
\overline{\lambda}^\alpha_{(\mu,\nu)} \rangle$.  The action of
Heisenberg algebra on this vacuum is obvious:
\begin{subequations}
\label{eq:ha}
  \begin{gather}
    \fa_n^\alpha \,
    \vacuum{\alpha}  
    \,\doteq\, 
    \fa_n^\alpha \,\otimes\,\mathbb{I}  \, \vacuum{\alpha} \,=\,
    \fa_n^\alpha 
    \vert 
    \lambda^\alpha_{(\mu,\nu)}
    \rangle \,\otimes\,
    \vert 
    \overline{\lambda}^\alpha_{(\mu,\nu)}
    \rangle \\
    \ofa_n^\alpha \, 
    \vacuum{\alpha}  
    \,=\,
    \mathbb{I} \,\otimes\, \ofa_n^\alpha \, \vacuum{\alpha} \,=\,
    \vert 
    \lambda^\alpha_{(\mu,\nu)}
    \rangle \,\otimes\,
    \ofa_n^\alpha 
    \vert 
    \overline{\lambda}^\alpha_{(\mu,\nu)}
    \rangle 
  \end{gather}
\end{subequations}
and the associated chiral primary field is $\ordprod{e^{i\lambda_\alpha
    X_L^\alpha(\zeta)}\,e^{i\overline{\lambda}_\alpha X_R^\alpha(\bgz)}}$. 

Chiral Fock spaces are constructed by applying the rising operators on
the chiral vacua. From the Sugawara construction of Virasoro
generators, we see that $L_n \vacuum{\alpha} = 0 \,\forall\,n>0 $,
thus the vacua form a continuous set of highest weight states with
holomorphic (resp. antiholomorphic) highest weights $h =
\frac{1}{2}\lambda^2$ (resp. $\overline{h} =
\frac{1}{2}\overline{\lambda}^2$).

The decomposition of the Fock spaces in term of Verma modules is
straightforward: since the descendant states are obtained by the
repeated application of $\fa_{-n},\, n>0$, which is equivalent to the
repeated application of $L_{-n},\, n>0$, because $\fa_{-n}$ also
raises the conformal dimension of $n$, the Fock spaces and the Verma
modules coincide, $\mathcal{H}^\lambda \,=\, \mathcal{V}^h$. However,
this happens only as long as $\lambda \neq \sqrt{2} m$, with $m \in
\frac{1}{2}\mathbb{Z}$. On the other hand, when $\lambda = \sqrt{2} m$
with $m \in \frac{1}{2}\mathbb{Z}$, the presence of a null
state\footnote{Let us recall the definition and the main properties of
  a null state\cite{DiFrancesco}:
  \begin{defin}\textbf{Null state}\\
    Any descendant state state $\vert\chi\rangle \,=\,
    L_{-k_1}\ldots L_{k_N}$, $k_1,\,\ldots,\,k_n > 0$,which is
    annihilated by all the Virasoro generators $L_n$ with $n>0$ is
    called a \textsc{singular vector} (or \textsc{null vector} or
    \textsc{null state}). 
  \end{defin}
  Such a state generates its own Verma module
  $\mathcal{V}^\chi \in \mathcal{V}^\lambda$: as a matter of fact,
  singular vectors are orthogonal to the whole Verma module, and
  this property extends to all its own descendants.} 
at level $2|m| + 1$, causes each Fock space to be a reducible
representation of the Virasoro algebra and to decompose into an
infinite sum of Verma modules:
\begin{equation}
\label{eq:dec}
 \mathcal{H}^{\sqrt{2} m} \,=\, 
\bigoplus_{l=1}^\infty \mathcal{V}^{(|m| + l)^2}
\end{equation}
Let us stress that this decomposition has nothing to do with the
grading induced by the zero modes of the Virasoro algebra and shown in
equation \eqref{eq:grad}: this last decomposition holds for each value
of the left and right chiral algebra charges. 

This decomposition shows an hidden $SU(2)$ symmetry. As a matter of
fact, the Virasoro highest weight state with highest weight $h = j^2$
spans in the set $\{\mathcal{H}^{\sqrt{2} m}$ a $(2|m| +
1)$-dimensional multiplet built with all the $\mathfrak{u}_1$ highest
weight defined by a value of $m$ fulfilling the condition $(|m| + l) =
j^2, l\,\in\,\mathbb{Z}_+$. 

The same holds for the antiholomorphic sector, thus we can
write\cite{Recknagel:1998ih}:
\begin{multline}
  \label{eq:resh}
  \mathcal{H}^{(C)} \,=\, 
  \int_{\substack{\lambda \neq \sqrt{2} m \\ 
      \overline{\lambda} \neq \sqrt{2} n }}
  \mathcal{H}^{\lambda} 
  \otimes 
  \overline{\mathcal{H}}^{\overline{\lambda}} 
  \,\oplus\,  
  \bigoplus_{m,n \in \frac{\mathbb{Z}}{2}}  
  \mathcal{H}^{\sqrt{2} m} 
  \otimes 
  \overline{\mathcal{H}}^{\sqrt{2} n} 
  \,=\,\\
  \int_{\substack{\lambda \neq \sqrt{2} m \\ 
      \overline{\lambda} \neq \sqrt{2} n }}
  \mathcal{V}_{\frac{\lambda^2}{2}} 
  \otimes 
  \overline{\mathcal{V}}_{\frac{\overline{\lambda}^2}{2}} \,   
  \,\oplus\,  
  \bigoplus_{j,\bar{j} \in \frac{\mathbb{Z}_+}{2}}  
  \left(
    {\mathcal{V}_{j^2}}^{\oplus 2j + 1} \otimes 
    {\overline{\mathcal{V}}_{\bar{j}^2}}^{\oplus 2\bar{j} + 1} 
  \right)
\end{multline}

As outlined in the introduction, from a microscopic point of view,
adding boundaries implies to define suitable boundary values
assignations. In a stringy language, it means to work out which open
string theory can be naturally related to closed string. As well
known, closed string theory is naturally coupled to interacting open
string. In this connection, our cylindrical end can be viewed both as
an open string one-loop diagram or as a tree level closed string
diagram, with a closed string propagating for a finite length path.
In the open string (direct) channel, time flows around the cylinder.
The associated quantization scheme defines functions of the modular
parameter $\tau = i \theta(k) = i (2 \pi - \varepsilon(k))$ (see eq.
\eqref{curv-moduli}).  The close string (transverse) channel, sees
time flowing along the cylinder, and the associated quantization scheme
is related to the open string one via the modular transformation $\tau
\rightarrow -\frac{1}{\tau}$. In the following, we will switch back
and forth between these two points of view and the associated
quantization schemes.

The amplitude associated to this diagram is a deep-investigated object
when the boundary assignation are a priori defined as the usual
specification of Neumann or Dirichlet boundary conditions (for a
review see \cite{Saleur:1998hq} and references therein). However, when
we describe the metric geometry of the triangulation $|T_l|
\rightarrow M$ with the Riemann surface $M_\partial$, we are actually
unwrapping each conical two-cell of the dual polytope $|P_{T_l}|
\rightarrow M$ into a finite cylindrical end, in such a way that the
barycenter behavior of the eventually coupled matter gets
``distributed'' over the full outer boundary $\boum$. Thus, in our
description, we well follow the procedure outlined by Charpentier ans
Gawedzky in \cite{Charpentier:1990gn}, which allows to specify the
amplitudes on an arbitrary Riemann surface with boundaries and with an
arbitrary specification of fields on them.  In a general framework,
let us consider a two-dimensional compact (not necessary connected)
Riemann surface $\Sigma$ with boundary $\partial \Sigma \,=\,
\bigsqcup_{I=1}^N S_I$. If we parametrize the boundary loops $S_I \in
\Sigma$ by analytical real maps $p_I\,:\;S^1 \rightarrow S_I$, then
the field theoretical propagation amplitude over the Euclidean surface
$\Sigma$ is formally given by the following functional integral over
the injection maps $X\,;\;\Sigma\rightarrow \mathcal{T}$:
\begin{equation}
  \label{eq:form_ampl}
  \int_{\{ X \circ p_I \,=\, X_I\}} \mathcal{D} X\,
  e^{- S[X]} 
\end{equation}
where $S[X]$ is the euclidean action of the bulk CFT, $\mathcal{D} X$
is the formal measure on the target space and the field $X$ is
restricted to take prescribed values $X_I$ over the boundary loops
$S_I$.

The rater abstract and formal expression \eqref{eq:form_ampl} acquires
a precise sense when we deal with a bosonic field defining a flat
toroidal background. As a matter of fact, if we call $\Xi$ the full
set of possible boundary assignations, is is always possible to fix
some real map $X_{cl}$
\begin{equation}
  \label{eq:xcl}
  X_{cl} \,:\; \Sigma \,\rightarrow\mathbb{R}
\end{equation}
which maps $\partial\Sigma$ diffeomorphically and with the prescribed
orientation in $\Xi$ (\ie{} it exists a real map $X_{cl}$ such that
$ X_{cl} \circ p_I \,=\, X_I$) and such that $X_{cl}$ is an harmonic
function w.r.t $\Delta_\Sigma$, the Laplacian operator defined over
$\Sigma$. Then, for a generic field, we can write:
\begin{equation}
  \label{eq:cl_modes}
  X \,=\, X_{cl} + \tilde{X} 
\end{equation}
where $X_{cl}$ is the harmonic function described above, while
$\tilde{X}\,:\;\Sigma\rightarrow\mathbb{R}$ is the collection of the
off-shell modes of $X$ satisfying
\begin{equation}
  \tilde{X} \circ p_I \,=\, 0
\end{equation}

This last condition implies the diagonal decomposition of the bulk
action, $S[X] \,=\, S[X_{cl}] + S[\tilde{X}]$. Moreover, expanding
$\tilde{X}\,=\,\sum c_M \tilde{X}_M$, where $\tilde{X}_M$ are the non
zero modes of the Laplacian ($\Delta_\Sigma \tilde{X}_M = \Lambda_M
\tilde{X}_M$, $\Lambda_M \neq 0\,\forall M$), we finally get:
\begin{equation}
  \label{eq:form_ampl2}
  \mathcal{A}_\Sigma \,=\, \sum_{X_{cl}} e^{- S[X_{cl}]} 
  \,\times\, 
  \text{det}'[\frac{8\pi^2}{N \Delta_\Sigma}]^\frac{1}{2}
\end{equation}
where the prime means the exclusion of the zero mode, while $N$ is the
normalization of the Laplacian eigenfunctions.  This last expression
clearly diverges and needs to be regularized. If $\Sigma \equiv
\text{\cyl{k}}$ and $N$ is the cylindrical end area, the standard
Riemann Zeta-function regularization technique gives
\begin{equation}
  \label{eq:reg_res}
  \text{det}'\left[
    \frac{8\pi^2 \theta(k)}{L^2(k)\Delta_\Sigma}
  \right]^\frac{1}{2}
  \,=\,
  \frac{1}{4\pi} \frac{1}{\sqrt{\text{Im}\tau} \,\eta(q)}
\end{equation}
where is the Dedekind-$\eta$ function with argument
$q\,=\,e^{2\pi i \tau}$.

To define a suitable parametrization of the on-shell mode, let us
switch to the transverse channel. The modular transformation $\tau
\rightarrow - \frac{1}{\tau}$ redefines
\begin{equation}
  \label{eq:off_tilde}
  \frac{1}{\sqrt{\text{Im}\tau} \,\eta(q)} \,=\, \frac{1}{\eta(\tilde{q})}.
\end{equation}

The Laurent expansion
of the classical mode $X^{\alpha }(\zeta(k),\overline{\zeta(k)})_{cl}$, $\alpha
=1,\ldots,D$  on the
annular region \cyl{k} is:
\begin{align}
  \label{class}
  X^\alpha(\zeta (k),\overline{\zeta }(k))_{cl} &
  \,=\, X^\alpha(\zeta (k))_{cl} 
  \,+\, 
  X^\alpha(\overline{\zeta }(k))_{cl} \notag \\
  X^\alpha(\zeta (k))_{cl} & \,=\, 
  \frac{X_0^\alpha}{2} \,+\, 
  a^\alpha(k) \ln\zeta(k)
  \,+\,
  \sum_{n\neq 0}X_{n}^{\alpha }(k)\zeta ^{n}(k) \notag \\
  X^\alpha(\overline{\zeta }(k))_{cl} & \,=\, 
  \frac{X_0^\alpha}{2} \,+\, 
  b^\alpha(k) \ln\overline{\zeta }(k)
  \,+\,\sum_{n\neq 0}
  \overline{X}_{n}^\alpha(k)\overline{\zeta}^n(k)
\end{align}

Following a prescription introduced in \cite{Charpentier:1990gn} and
specialized to the compactified boson in \cite{Langlands}, it is
possible to parametrize the classical field \eqref{class} in term of
its restrictions to the boundaries $\boup$ and $\boum$:
\begin{multline}
  X^{\alpha}(k)_{cl}|_+
  \,\doteq\, 
  X^\alpha(\zeta (k),\overline{\zeta} (k))|_{S_{\varepsilon (k)}^{(+)}}
  \,=\,
  X_0^\alpha \,-\,
  \frac{2\pi}{2\pi - \varepsilon(k)}
  \left( a^\alpha(k) \,+\, b^\alpha(k) \right) \,+\,  \\
  i \theta(k) 
  \left( a^\alpha(k) \,-\, b^\alpha(k) \right) \,+\,
  \sum_{n\neq 0} b^\alpha_n(k) e^{ i n \vartheta(k)} 
  \qquad  b^\alpha_{-n}(k) \,=\, \overline{b}_n(k)
\end{multline}
\begin{multline}
  X^{\alpha }(k)_{cl}|_-
  \,\doteq\, 
  X^\alpha(\zeta (k),\overline{\zeta} (k))|_{S_{\varepsilon (k)}^{(-)}}
  \,=\,
  X_0^\alpha \,+\, \\
  i \theta(k) 
  \left( a^\alpha(k) \,-\, b^\alpha(k) \right) \,+\,
  \sum_{n\neq 0} a^\alpha_n(k) e^{ i n \vartheta(k)}
  \qquad  a^\alpha_{-n}(k) \,=\, \overline{a}_n(k)
\end{multline}
where we have introduced the two sets of coefficients $\left\{
  a^\alpha_n \right\}$ and $\left\{ b^\alpha_n \right\}$ having
the following expression in terms of the Laurent's modes $X_n^\alpha(k)$ and
$\overline{X}_n^\alpha(k)$:
\begin{subequations}
\begin{align}
a^\alpha_n \,\doteq\, &
e^{-\frac{2\pi}{2\pi - \varepsilon(k)} n} X_n^\alpha(k) \,+\, 
e^{\frac{2\pi}{2\pi - \varepsilon(k)} n} \overline{X}_{-n}^\alpha(k) \\
b^\alpha_n \,\doteq\, &
X_n^\alpha(k) \,+\, 
\overline{X}_{-n}^\alpha(k)
\end{align}
\end{subequations}

Redefining the compactification radius as $\Omega^\alpha(k)
\,=\,\frac{R^\alpha(k)}{l(k)}$, the winding boundary conditions impose
the consequent parametrization of the difference between the
coefficients $a^\alpha(k)$ and $b^\alpha(k)$
\begin{equation}
  a^\alpha(k) \,-\, b^\alpha(k) 
  \,=\, 
  - i \nu(k) \Omega^\alpha(k), \qquad \nu(k)\,\in\,\mathbb{Z},
\end{equation}

The last term to parametrize is the difference between $X^{\alpha
}(k)_{cl}|_+$ and $X^{\alpha }(k)_{cl}|_-$. This can be achieved
\cite{Langlands} via an integer number $\mu(k)\,\in\,\mathbb{Z}$ and a
real parameter $t^\alpha(k)\,\in\,(0,\,2\pi\Omega^\alpha(k)]$\cite{Langlands} 
\begin{equation}
  a^\alpha(k) \,+\, b^\alpha(k) \,=\,
  t^\alpha(k) \,+\,
  2 \pi \mu(k) \Omega^\alpha(k).
\end{equation}

Then, the classical solution is fully parametrized by the two set of
complex numbers $\left\{a^\alpha_n \right\}$ and $\left\{ b^\alpha_n
\right\}$, obeying the reality conditions $a^\alpha_{-n}(k) \,=\,
\overline{a}_n(k)$ and $b^\alpha_{-n}(k) \,=\, \overline{b}_n(k)$, the
couple of integers $(\mu(k),\nu(k)) \,\in\,\mathbb{Z}^2$ and the real
number $t^\alpha(k)\,\in\,(0,\,2\pi\Omega^\alpha(k)]$\cite{Langlands}:
\begin{multline}
  X^\alpha(\zeta (k),\overline{\zeta }(k))_{cl} \,=\,
  X_0^\alpha \,+\, \\
\left[ \frac{t^\alpha(k)}{2} \,+\, \pi\mu(k)\Omega^\alpha(k) \,-\,
  i\frac{\nu(k)}{2}  \Omega^\alpha(k)
\right] 
\ln\zeta(k) \,+\,\\
\left[ \frac{t^\alpha(k)}{2} \,+\, \pi\mu(k)\Omega^\alpha(k) \,+\,
  i\frac{\nu(k)}{2}  \Omega^\alpha(k)
\right] 
  \ln\overline{\zeta}(k) \,+\,\\
  \sum_{n \neq 0} \frac{a^\alpha_n(k)}{\tilde{q}^n \,-\, \tilde{q}^{-n}}
  \left(
    \zeta^n \,-\, \overline{\zeta}^{-n}
  \right)\,+\,
  \sum_{n \neq 0} \frac{b^\alpha_n(k)}{\tilde{q}^{-n} \,-\, \tilde{q}^n}
  \left(
    \left(\frac{\zeta}{\tilde{q}}\right)^n \,-\,
    \left(\frac{\overline{\zeta}}{\tilde{q}}\right)^{-n} 
  \right)
\end{multline}

With these parametrization the ``classical term'' of the full
amplitude will be completely determined in terms of the boundary
values of the Laurent expansion of the classical configuration, modulo
the pair of integers $\left(\mu(k),\nu(k)\right)$ which parametrizes
the coefficients $a^\alpha(k)$ and $b^\alpha(k)$. Thus, the sum over
all the independent ``classical configurations'' in equation
\eqref{eq:form_ampl2} is worked out summing over all values of the
couple of integers $\left(\mu(k),\nu(k)\right)$.  This leads to the
further factorization\cite{Langlands}:
\begin{multline}
  \sum_{\left(\mu,\nu\right)}
  e^{-S[X^\alpha(\zeta (k),\overline{\zeta }(k))_{cl}]}\,=\, \\
  \mathcal{A}_{cl}(\left\{ a^\alpha_n(k) \right\}\left\{
    b^\alpha_n(k) \right\}) 
  \sum_{\left(\mu,\nu\right)}
  \mathcal{A}_{cl}(\mu^\alpha(k),\nu^\alpha(k),t^\alpha),
\end{multline}
where:
\begin{multline}
  \mathcal{A}_{cl}(\left\{ a^\alpha_n(k) \right\}\left\{
    b^\alpha_n(k) \right\}) \,=\,\\ 
  \prod_{n \,>\, 0} 
  \mbox{exp}
  \left[
    -\frac{k}{8}
    \left( 
      a^\alpha_n(k) a^\alpha_{-n}(k) \frac{1 \,+\, \tilde{q}^{2n}}{1
        \,-\, \tilde{q}^{2n}} \,-\, 
      a^\alpha_{-n}(k) b^\alpha_n(k) \frac{2 \tilde{q}^n}{1 \,-\,
        \tilde{q}^{2n}} \,+\, 
    \right.\right. \\ \left.\left. -\,
      a^\alpha_n(k) b^\alpha_{-n}(k) \frac{2 \tilde{q}^n}{1 \,-\,
        \tilde{q}^{2n}} \,+\, 
      b^\alpha_n(k) b^\alpha_{-n}(k) \frac{1 \,+\, \tilde{q}^{2n}}{1
        \,-\, \tilde{q}^{2n}}  
    \right)
  \right]
\end{multline}
and
\begin{equation}
  \sum_{\left(\mu,\nu\right)}\mathcal{A}_{cl}(\mu(k),\nu(k),t^\alpha) \,=\,
  \sum_{\left(\mu,\nu\right)} 
e^{i\frac{\mu^\alpha}{\Omega^\alpha}t^\alpha}
\tilde{q}^{\frac{1}{4} 
    \left(\frac{\mu(k)^2}{\Omega(k)^2} \,+\, \frac{1}{2}
      \nu(k)^2\Omega(k)^2\right)}  
\end{equation}
The full amplitude is then
\begin{multline}
  \label{amptr}
  \mathcal{A}(\left\{ a^\alpha_n(k) \right\}\left\{
    b^\alpha_n(k), t^\alpha(k) \right\}) \,=\,
  \frac{1}{\eta(\tilde{q})}\,
  \mathcal{A}_{cl}(\left\{ a^\alpha_n(k) \right\}\left\{
    b^\alpha_n(k) \right\})\,\\
  \sum_{(\mu,\nu)(k)} 
  e^{i\frac{\mu^\alpha}{\Omega^\alpha}t^\alpha}
  \tilde{q}^{\frac{1}{4} 
    \left(\frac{\mu(k)^2}{\Omega(k)^2} \,+\, \frac{1}{2}
      \nu(k)^2\Omega(k)^2\right)}  
\end{multline}

\section[Generalized Langlands boundary states]{Generalized Langlands
  boundary states and their conformal properties}
\label{sec:conf_bs}

On the annular domain \cyl{k} boundaries are concentric circles
centered on the origins. In radial quantization, boundary conditions
are imposed by propagating states from an initial (coherent) state
$\left| \mathfrak{r}_{(\mu ,\nu )}^{\alpha }(S_{\varepsilon
    (k)}^{(-)})\right\rangle$ residing on the inner boundary $\boum$
to a final (coherent) boundary state $\left| \mathfrak{r}_{(\mu ,\nu
    )}^{\alpha }(S_{\varepsilon (k)}^{(+)})\right\rangle$ naturally
associated to the outer boundary $\boup$. The Hamiltonian generating
this translation is $H^{(C)} = \frac{2\pi}{L(k)}\left(L_0 +
  \overline{L}_0 + \frac{c}{12}\right)$, while the propagator is
$\tilde{q}^{\left(L_0 + \overline{L}_0 + \frac{c}{12}\right)}$.
Therefore, the amplitude on the cylindrical end can be formally
written as:
\begin{equation}
  \label{eq:ampl_bs}
  \mathcal{A}(\left\{ a^\alpha_n(k) \right\}\left\{
    b^\alpha_n(k)\right\}, t^\alpha(k)) \,=
  \sum_{(\mu,\nu)(k)}
  \left\langle\mathfrak{r}_{(\mu ,\nu )}^{\alpha }(S_{\varepsilon
      (k)}^{(+)})\right\vert 
  \tilde{q}^{L_0 \,+\, \bar{L}_0 \,-\, \frac{c}{12}} 
  \left| \mathfrak{r}_{(\mu ,\nu )}^{\alpha }(S_{\varepsilon
      (k)}^{(-)})\right\rangle
\end{equation}

The main goal of \cite{Langlands} was to find a formal map which
associates to each boundary assignation
$\left\{(\mu,\nu),\,\{a_n\}\right\}$ (resp.
$\left\{(\mu,\nu),\,\{b_n\}\right\}$) a well defined coherent state
$\left| \mathfrak{r}_{(\mu ,\nu )}^{\alpha }(S_{\varepsilon
    (k)}^{(-)})\right\rangle$ (resp. $\left| \mathfrak{r}_{(\mu ,\nu
    )}^{\alpha }(S_{\varepsilon (k)}^{(+)})\right\rangle$) $\in
\mathcal{H}_{\mu,\nu} \otimes \overline{\mathcal{H}}_{\mu,\nu}$. 

The procedure, on one hand, exploits the exact correspondence between
the couple of integer parameters of the classical part of the
partition function and the integers which parametrize the left and
right zero modes of the compactified bosonic field, while on the other
hand it introduces a couple of functions
$A_{m_{1},m_{2}}^{n}(a_{n},a_{-n})$ and
$B_{m_{1},m_{2}}^{n}(b_{n},b_{-n})$ of only two parameters, whose
explicit expression can be extracted rewriting
$\mathcal{A}_{cl}(\left\{ a_n(k) \right\}\left\{ b_n(k) \right\})$ in
the form\cite{Langlands}:
\begin{equation}
  \label{eq:bil}   
  \mathcal{A}_{cl}(\left\{ a^\alpha_n(k) \right\}\left\{
    b^\alpha_n(k) \right\}) \,=\,
  \prod_{n=1}^\infty \sum_{m_{1},m_{2}}
  B_{m_{1},m_{2}}^{n} \tilde{q}^{n(m_1 + m_2)} 
  A_{m_{1},m_{2}}^{n} 
\end{equation}

Let us introduce the following generalization of Langlands boundary states:
\begin{multline}
\label{eq:lbs-}
  \left| \mathfrak{r}_{(\mu ,\nu )}^{\alpha }(S_{\varepsilon
      (k)}^{(-)})\right\rangle =e^{i t_{-}^{\alpha }(\lambda _{(\mu ,\nu
      )}^{\alpha }+\overline{\lambda }_{(\mu ,\nu )}^{\alpha })}
  \times \\ \times
  \prod_{n=1}^{\infty
  }\sum_{m_{1},m_{2}}
  A_{m_{1},m_{2}}^{n}(a_{n}^{\alpha },a_{-n}^{\alpha })%
  \frac{\left( \mathfrak{a}_{-n}^{\alpha }\right)^{m_{1}}
    \otimes
    \left( 
      \overline{\mathfrak{a}}_{-n}^{\alpha }
    \right)%
    ^{m_{2}}}{\sqrt{n^{m_{1}+m_{2}}m_{1}!m_{2}!}}
  \left|
    (\mu ^{\alpha },\nu ^{\alpha })
  \right\rangle 
\end{multline}
and
\begin{multline}
\label{eq:lbs+}
  \left| \mathfrak{r}_{(\mu ,\nu )}^{\alpha }(S_{\varepsilon
      (k)}^{(+)})\right\rangle =e^{i t_{+}^{\alpha }(\lambda _{(\mu ,\nu
      )}^{\alpha }+\overline{\lambda }_{(\mu ,\nu )}^{\alpha })}
  \times \\ \times
  \prod_{n=1}^{\infty
  }\sum_{m_{1},m_{2}}
  B_{m_{1},m_{2}}^{n}(b_{n}^{\alpha },b_{-n}^{\alpha })%
  \frac{\left( \mathfrak{a}_{-n}^{\alpha }\right)^{m_{1}}
    \otimes
    \left( 
      \overline{\mathfrak{a}}_{-n}^{\alpha}
    \right)^{m_{2}}}{\sqrt{n^{m_{1}+m_{2}}m_{1}!m_{2}!}}
  \left|
    (\mu^{\alpha },\nu ^{\alpha })
  \right\rangle
\end{multline}
with $t_{-}^{\alpha }-t_{+}^{\alpha }=t^{\alpha }$ and
\begin{multline}
  A_{m_{1},m_{2}}^{n}(a_{n}^{\alpha },a_{-n}^{\alpha })
  \,=\, e^{i \pi s(m_1 + m_2)} \times\\
  \begin{cases}
    (2i\sqrt{n}a_{n}^{\alpha })^{m_{1}-m_{2}}\sqrt{\frac{m_{2}!}{m_{1}!}}%
    e^{-2n|a_{n}^{\alpha }|^{2}}L_{m_{2}}^{(m_{1}-m_{2})}(4n|a_{n}^{\alpha
    }|^{2}) & m_{1}\geq m_{2} \\
    (2i\sqrt{n}a_{n}^{\alpha })^{m_{2}-m_{1}}\sqrt{\frac{m_{1}!}{m_{2}!}}%
    e^{-2n|a_{n}^{\alpha }|^{2}}L_{m_{1}}^{(m_{2}-m_{1})}(4n|a_{n}^{\alpha
    }|^{2}) & m_{2}\geq m_{1}
  \end{cases}
\end{multline}
with $L_{m_{2}}^{(m_{1}-m_{2})}(\circ )$ the $m_{2}$-th Laguerre
polynomial of the $(m_{1}-m_{2})$-th kind.  By replacing the
$a_{n}^{\alpha }$ with the $b_{n}^{\alpha }$, and by conjugation
(induced by the orientation of the boundary), we get also
$\mathcal{B}_{m_{1},m_{2}}^{n}(b_{n}^{\alpha },b_{-n}^{\alpha
})=\overline{\mathcal{A}_{m_{1},m_{2}}^{n}(b_{-n}^{\alpha
  },b_{n}^{\alpha })}$.

Let us stress that the above formula is a generalization of the
original Langland's boundary states. As a matter of fact, the explicit
expression in \cite{Langlands} did not include the phase factor $e^{i
  \pi s(m_1 + m_2)}$ for the $\mathcal{A}_{m_{1},m_{2}}^{n}$ and
$\mathcal{B}_{m_{1},m_{2}}^{n}$ coefficients. The introduction of such
a phase factor is natural since these coefficients enter in the
partition function via the combination in equation
\eqref{eq:bil}. Moreover, its presence is crucial if we want the
boundary states to describe the most general kind of boundary condition.

From a macroscopic point of view, we can assume the presence of a
brane emitting and adsorbing the boundary states via non-perturbative
processes. In the same way, in the open string scheme, branes are
those objects open strings endpoints lay on.  Boundary conformal field
theory allows to define branes in a microscopic CFT language, without
any referring to target space geometry: the boundary CFT describes the
perturbative properties of the open string-brane system. In this
connection, the role of boundary states is obvious: they contain the
complete informations about the static D-branes.  However, as stressed
in the introduction to this chapter, not every linear map such that one
defining generalized Langlands boundary state actually describes an
admissible boundary state, \ie a coherent static D-brane
configuration. To restrict the set of boundary states to an admissible
one, we have to implement the glueing relations \eqref{eq:no_momflow}
and \eqref{eq:cont_cond} directly on \eqref{eq:lbs-} and
\eqref{eq:lbs+}.  Through radial quantization, \ie{} mapping the
cylinder into the annulus via \eqref{eq:cyl-to-ann}, the
\eqref{eq:no_momflow} and \eqref{eq:cont_cond} translates into
\begin{equation}
\label{eq:cont_rq}
  \zeta^2\, T(\zeta)\,=\,
  \bgz^2\bar{T}(\bgz)|_{|\zeta|=1,\,|\zeta|=\tilde{q}}
\qquad
  \zeta^{h_W}\, W(\zeta)\,=\,
  \bgz^{\bar{h}_{\bar{W}}}\bar{W}(\bgz)|_{|\zeta|=1,\,|\zeta|=\tilde{q}}
\end{equation}

Introducing the Laurent expansion of involved fields, we get the
following conditions applied to boundary states:
\begin{equation}
  \label{eq:cont_bs}
  \left( W_n \,-\, (-1)^{\bar{h}_{\bar{W}}}\Omega(\overline{W}_{-n}) \right) \,
  \Vert B \Rangle \,=\, 0 
  \qquad
  \left( L_n \,-\, \overline{L}_{-n} \right) \,
  \Vert B \Rangle \,=\, 0
\end{equation}
where $\Vert s \Rangle$ is a generic conformal boundary state.

In our connection, the only chiral fields involved besides Virasoro
fields are the holomorphic and antiholomorphic currents $J(\zeta)$ and 
$\bar{J}(\bgz)$ generating the Heisenberg algebra \eqref{eq:ha}. Being
the current algebra abelian, the possible gluing maps reduces to:
\begin{subequations}
\begin{gather}
 J(\zeta) \,=\,\bar{J}(\bgz)\\
 J(\zeta) \,=\, -\, \bar{J}(\bgz)
\end{gather}
\end{subequations}
which are respectively mapped into
\begin{subequations}
  \label{cond}
  \begin{gather}
    \label{cpiu}
    \left( \mathfrak{a}_n \,+\, \overline{\mathfrak{a}}_{-n} \right) \,
    \Vert B \Rangle \,=\, 0, \\
    \label{cmeno}
    \left( \mathfrak{a}_n \,-\, \overline{\mathfrak{a}}_{-n} \right) \,
    \Vert B \Rangle \,=\, 0.
  \end{gather}
\end{subequations}

The Sugawara construction \eqref{virasoro} ensures that this
conditions are sufficients to enforce conformal invariance encoded in
\eqref{eq:no_momflow}.

With a careful computation, which exploits recursion relation of
Laguerre polynomials, and which is reported in appendix
\ref{sec:lang_proj}, we reduced generalized Langlands boundary states 
to the Dirichlet and Neumann Ishibashi states associated with the
free bosonic field $X^\alpha(k)$:
\begin{equation}
  \left| \mathfrak{r}_\mu^{\alpha }(S_{\varepsilon
      (k)}^{(-)})\right\rangle ^{(D)}=e^{\sqrt{-1}t_{+}^{\alpha }\mu ^{\alpha }
    \frac{L}{R^{\alpha }}}\exp \left( \sum_{n=1}^{\infty }\frac{1}{n}\left( 
      \mathfrak{a}_{-n}^{\alpha }\right) 
    \left( \overline{\mathfrak{a}}_{-n}^{\alpha}
    \right) 
  \right) 
  \left| (\mu ^{\alpha },0)\right\rangle ,  \label{IshiDir}
\end{equation}
and 
\begin{equation}
  \left| \mathfrak{r}^{\alpha }(S_{\varepsilon
      (k)}^{(-)})\right\rangle ^{(N)}=\exp \left( \sum_{n=1}^{\infty }
    -\frac{1}{n}
    \left( \mathfrak{a}_{-n}^{\alpha }\right) \left( \overline{\mathfrak{a}}%
      _{-n}^{\alpha }\right) \right) \left| (0,\nu ^{\alpha })\right\rangle .
  \label{IshiNeu}
\end{equation}

The most general boundary state will be a superposition of these
states, weighted by suitable coefficients:
\begin{subequations}
\label{eq:bscoeff}
  \begin{gather}
  \Vert \mathfrak{r}^{\alpha }(S_{\varepsilon
      (k)}^{(-)})\Rangle ^{(D)} 
  \,=\, 
  \sum_{\mu \in \mathbb{Z}}
\mathbf{C}_\mu
  \left| \mathfrak{r}_\mu^{\alpha }(S_{\varepsilon
      (k)}^{(-)})\right\rangle ^{(D)}\\
  \Vert\mathfrak{r}^\alpha
    (S_{\varepsilon(k)}^{(-)})\Rangle^{(N)}
      \,=\, 
  \sum_{\nu \in \mathbb{Z}}
\mathbf{C}_\nu
  \left| \mathfrak{r}_\nu^{\alpha }(S_{\varepsilon
      (k)}^{(-)})\right\rangle ^{(N)}
  \end{gather}
\end{subequations}

The glueing automorphism $\Omega$ in\eqref{eq:cont_cond} is sufficient
to determine such coefficients uniquely. Strong
constraints on their expression derive directly by the relation
\eqref{eq:ca} relating coefficients of bulk-fields one-point function.
To show how this works, let us introduce a more compact notation. Let
us omit the vector index $\alpha$. Moreover let us introduce:
\begin{itemize}
\item at fixed glueing map $\Omega$, let
  $|I_{\lambda(p)}\Rangle_{\Omega}$ be the collection of all the
  Ishibashi states which are built on the primary states of the
  Hilbert space of the BCFT defined over \cyl{p};
\item let $Y$ be the collection of all  the indexes $\lambda(p)$
  labeling such Ishibashi states: obviously, $Y
\subseteq \mathcal{Y}$, where $\mathcal{Y} = \left\{
  (\mu,\nu)\right\}$ is the collection of labeling associated to the
irreducible representations of $u_L(1) \times u_R(1)$;
\item let us denote with $\mathcal{A}= \{A(p)\}$ the set of all
  possible boundary assignments, including both the glueing condition
  and the dependence from other parameters, which we include in a
  collective index $\alpha$. Thus, each $A(p) \in \mathcal{A}$ will be
  a pair $A(p) = (\Omega, \alpha)$\footnote{We do not assign a
    polytope index to the sets of labels $Y$, $\mathcal{Y}$ and
    $\mathcal{A}$ since the BCFT constructed is the same over each
    cylindrical ends, thus these sets coincide over each \cyl{p}};
\item let
us indicate with $\Vert\mathfrak{r}_{(p)}(\alpha)\Rangle^{\Omega}$ the
collection of all the admissible boundary states on the boundary
$S_{\varepsilon(p)}^{(-)}$ of \cyl{p}.
\end{itemize}

With the statements above, let us write the relation between the
boundary states and the Ishibashi states as:
\begin{equation}
  \label{eq:bs_newnot}
  \Vert\mathfrak{r}_{(p)}(\alpha)\Rangle^{\Omega} 
  \,=\, 
  \sum_{\lambda(p) \in Y}
  B^{\lambda(p)}_{A(p)} \,
  |I_{\lambda(p)}\Rangle^\Omega,
\end{equation}
$B^{l(p)}_{A(p)}$ being the  $\vert I_{l(p)}\Rangle$ projection  on
$\Vert\mathfrak{r}_{(p)}(\alpha)\Rangle^{\Omega}$.

Coefficients $B^{\lambda(p)}_{A(p)}$ are strictly related both to bulk
fields one-point function coefficients $A^{A(p)}_{\lambda(p)}$ in
\eqref{eq:bulk1p} and to coefficients $C^{A(p)}_{\phi(p)\,\lambda(p)}$
entering int the bulk-to-boundary expansion \eqref{bbOPE}
\cite{Cardy:1991tv,Recknagel:1998ih,Gaberdiel:2002iw}.  In particular,
exploiting transformation properties of correlators under the
conformal mapping relating the finite cylinder to the annulus, one can
show \cite{Recknagel:1998ih}:
\begin{equation}
  \label{eq:ba}
  B_{A(p)}^{\lambda(p)} \,=\, A^{A(p)}_{\lambda(p)}
\end{equation}
Replacing this identity into \eqref{eq:ca}, we get the constraint over
the boundary states' coefficients \footnote{This is the specialized
  version of the sewing constraint holding for the projections of
  boundary states $\Vert\alpha\Rangle$ of a Rational Conformal Field
  Theory on the associated Ishibashi states $\vert j \Rangle$:
  \begin{equation*}
    B^i_\alpha \, B^j_\alpha \,=\, \sum_k C_{i j}^k F_{k 1}
    \left[
      \begin{smallmatrix}
        j & j \\
        i & i 
      \end{smallmatrix}
    \right]\, B^k_\alpha, 
  \end{equation*}
where $C_{i j}^k$ and $F_{k 1}
    \left[
      \begin{smallmatrix}
        j & j \\
        i & i 
      \end{smallmatrix}
    \right]$ are respectively the fusion coefficients and fusion
    matrices of the bulk theory}:
\begin{equation}
  \label{eq:bsc}
B_{A(p)}^{\lambda(p)} \cdot B_{A(p)}^{\lambda'(p)} \,=\,
B_{A(p)}^{\lambda(p) + \lambda'(p)}
\end{equation}
from which we get the boundary states:
\begin{equation}
  \Vert \mathfrak{r}^{\alpha }(S_{\varepsilon
      (k)}^{(-)})\Rangle ^{(D)} 
  \,=\, 
  \frac{1}{\sqrt{2 \frac{L(k)}{R^\alpha(k)}}} 
  \sum_{\mu \in \mathbb{Z}}
  e^{i t_{+}^{\alpha }\mu ^{\alpha }%
    \frac{L}{R^{\alpha }}}\exp \left( \sum_{n=1}^{\infty }\frac{1}{n}\left( 
      \mathfrak{a}_{-n}^{\alpha }\right) \left(
      \overline{\mathfrak{a}}_{-n}^{\alpha 
      }\right) \right) \left| (\mu ^{\alpha },0)\right\rangle , 
\end{equation}
\ie{} the Dirichlet boundary state describing a boundary loop lying on a
D-brane positioned at $X^\alpha\vert_+\,=\,t_+$,
and the states
\begin{equation}
  \Vert\mathfrak{r}^\alpha
    (S_{\varepsilon(k)}^{(-)})\Rangle^{(N)}
  =
  \sqrt{\frac{L(k)}{R^\alpha(k)}}
  \sum_{\nu \in \mathbb{Z}}
  e^{\frac{i}{2} \tilde{t}_{+}^{\alpha }\nu^{\alpha }%
    \frac{R^{\alpha }}{L}}
  \exp \left(-\sum_{n=1}^{\infty }\frac{1}{n}%
    \left( \mathfrak{a}_{-n}^{\alpha }\right) \left( \overline{\mathfrak{a}}%
      _{-n}^{\alpha }\right) \right) \left| (0,\nu^\alpha)\right\rangle,
\end{equation}  
\ie{} a Neumann boundary states describing a boundary loop moving in
the associated direction. In the above formula, $\tilde{t}_+$ is the
$U(1)$ holonomy naturally associated to the boundary loop.

%% file: bio.tex
\chapter{Boundary Insertion Operators}
\label{ch:bio}

To complete the description of the BCFT on the full Riemann Surface,
we need to discuss how the $N_0$ distinct copies of the theory, each
one living on a different cylindrical end \cyl{k}, interact along the
underlying polytope $|P_{T_l}|\,\rightarrow\,M$.  In \cite{Carfora2},
a remarkable intuition proposed that this interaction could be
mediated by boundary conditions changing operators, whose presence is
predicted by boundary conformal field theory
\cite{Cardy1,Recknagel:1998ih}(see the general theory presented in the
previous chapter).
  
In this chapter starting from this statement, we slightly modify it.
We describe interaction on the ribbon graph introducing, beside
ordinary boundary condition changing operators, a new class of
operators which mediate the change in boundary conditions which
actually take place non locally on the boundary shared by two adjacent
cylindrical ends.  As a matter of fact, while in ordinary boundary
conformal field theory a jump in boundary conditions on the boundary
is explained by the local action of boundary operators, this feature
does not fit completely our model, both because, after the glueing
process we do not deal with a true boundary, but with a ``separation
edge'' between the two adjacent cylindrical ends, and because we do
not have a jump in boundary conditions taking place in a precise point
of this edge (which we will go on calling boundary, for the sake of
simplicity) but with two different boundary conditions which, in the
adjacency limit, coexist on this shared boundary. From these
considerations, it follows the description of the dynamical glueing of
the $N_0(T)$ copies of the bosonic BCFT which we present in this
chapter.

\section{Boundary Insertion Operators}

In the background defined in section \ref{sec:bo}, it may happen that
a boundary condition changes along the real line. In radial
quantization, this situation is explained with the presence of a
vacuum which is no longer invariant under the action of
$L_{-1}^{(H)}$. In \cite{Cardy1} it was proposed that such states were
obtained by the action of a boundary operator on the true vacuum,
\ie{} it may happen that boundary operators induce transitions on
boundary conditions along the boundary. These boundary condition
changing operators are associated with vectors in the Hilbert space
depending on both the boundary conditions, and they cannot be obtained
from bulk fields through a bulk to boundary OPE. 

In this section we will propose a slight modification of this picture,
introducing a particular class of boundary condition changing
operators which, living on the boundary shared between two adjacent
polytopes and carrying an irreducible action of the chiral algebra,
will mediate the change in boundary conditions between pairwise
adjacent cylindrical ends.
%%%%%%%%%%%%%%%%%%%%%%%%%%%%%%%%%%%%%%%%%%%%%%%%%%%%%%%%%%%
\begin{figure}[!t]
  \begin{center}
    \includegraphics[width=.4\textwidth]{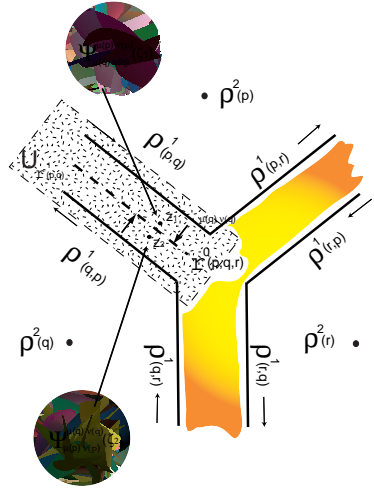}
    \caption{The insertion of boundary operators on the shared
      boundary between pairwise adjacent polytope allows to define
      the superposition of different boundary conditions\cite{Carfora2}} 
    \label{BIO}
  \end{center}
\end{figure}
%%%%%%%%%%%%%%%%%%%%%%%%%%%%%%%%%%%%%%%%%%%%%%%%%%%%%%%%%%%

To fully understand this statement, let us first analyze the geometry
on the neighborhood of a shared ribbon graph edge. Ribbon graph's
strips represent the uniformization of the singular structure of the
RRT in the neighborhood of the 1-skeleton of the original
triangulation. The ribbon graph arises as a consequence of the
presence of a locally uniformizing coordinate $\zeta(k)$, which
provides a counterclockwise orientation in the 2-cells of the original
triangulation: such an orientation gives rise to a cyclic ordering on
the set of the half edges $\left\{
  \rho^1(k)^\pm\right\}_{k=0}^{N_1(T)}$ incident on the vertexes
$\left\{ \rho^0(k)^\pm\right\}_{k=0}^{N_2(T)}$, stating a 1-to-1
correspondence between the 1-skeleton of the original triangulation
and the set o trivalent metric ribbon graphs.  In this connection, let
us consider two adjacent cylindrical ends \cyl{p} and \cyl{q}: these
are dual to the adjacent cells $\rho^2(p)$ and $\rho^2(q)$. The
cylindrical ends are glued to the oriented boundaries
$\partial\Gamma_p$ and $\partial\Gamma_q$ of the ribbon graph, which
are denoted respectively by $\partial\Gamma_p$ and $\partial\Gamma_q$.
Let us consider the oriented strip associated with the edge
$\rho^1(p,q)$ of the triangulation and its uniformized neighborhood
$\left(U_{\rho^1(p,q)},z(p,q)\right)$ and let us take in
$\left(U_{\rho^1(p,q)},z(p,q)\right)$ two infinitesimally neighboring
points $z_1 \,=\, x_1 \,+\ i y_1$ and $z_2 \,=\, x_2 \,+\ i y_2$ such
that $x_1 \,=\, x_2$.  Thus, if $y_1 \,\rightarrow\, 0$ we are
approaching the intersection between the boundary $\partial\Gamma_p$
of the ribbon graph and the oriented edge $\rho^1(p,q)$, while when
$y_2 \,\rightarrow\, 0$ we are approaching the intersection between
the boundary $\partial\Gamma_q$ of the ribbon graph and the oriented
edge $\rho^1(q,p)$.  From a BCFT point of view, this leads to the
conclusion that, in the adjacency limit described above, the
``effective boundary'' between two adjacent cylindrical end is unique
and shares the two different boundary conditions associated to
$S^{(-)}_{\varepsilon(p)}$ and $S^{(-)}_{\varepsilon(q)}$.

In this connection, it is no longer correct to argue the presence of
vacuum state which is not invariant under translations along the
boundary. The shared boundary is obtained glueing two loops, each of
them being part of a domain, on which we have defined a self-contained
BCFT, with its associated Hilbert space and vacuum state invariant
under translation along the boundary loop.

Thus, in order the glueing process to take place coherentely, we have
to require that each state of the Hilbert space melt without breaking
the conformal and chiral symmetry of the model. This leads to the
definition of Boundary Insertion Operators.

A CFT is consistently defined on each cylindrical end once we have
imposed constraints in \eqref{eq:cont_rq}. In particular, specializing
to the Virasoro fields, on \cyl{p} inner boundary we have
\begin{equation}
  \label{eq:nmp}
    \zeta(p)^2\, T(\zeta(p))\,=\,
  \bgz(p)^2\bar{T}(\bgz(p))|_{|\zeta(p)|=1}
\end{equation}
while in the inner boundary of \cyl{q} it holds:
\begin{equation}
  \label{eq:nmq}
    \zeta(q)^2\, T(\zeta(q))\,=\,
  \bgz(q)^2\bar{T}(\bgz(q))|_{|\zeta(q)|=1}
\end{equation}

This conditions allows to combine, on each cylindrical end, $T$ and
$\overline{T}$ in a unique object (well defined conformally mapping
the cylinder into a semi annular domain in the UHP), thus allowing to
associate to each cylindrical end a single copy of the Virasoro
algebra (the same obviously holds for the other chiral fields, see eq.
\eqref{eq:open_modes}). 

In this connection, we can pursue further this last construction, and
implement a non symmetry-breaking glueing of two adjacent cylindrical
end by associating, to each pairwise adjacent couple of them, a unique
copy of both Virasoro and chiral algebras. To achieve this, let us
consider the neighborhood of the $(p,q)$ edge defined in
\eqref{eq:strip} and uniformized with the complex coordinate $z(p,q)$.
Conformal mappings between $z(p,q)$ and the coordinates uniformizing
each cylindrical end are define as:
\begin{equation}
  \zeta(p) \,=\, e^{\frac{2 \pi i}{L(p)}(L(p) - \hat{L}(p,q) + z(p,q))}
\qquad
  \zeta(q) \,=\, e^{\frac{2 \pi i}{L(q)}(L(q) - \hat{L}(q,p) + z(q,p))}
\end{equation}
where it holds $z(q,p) = - z(p,q)$. We can easily express the
holomorphic and antiholomorphic components of the Virasoro fields
defined on each cylindrical end in term of the strip coordinate. For
the holomorphic components we have:
\begin{subequations}
\begin{gather}
  T_{(p)} (z(p,q)) \,=\,  T_{(p)} (\zeta(p)) 
  \left( 
    \frac{d\,z(p,q)}{d\,\zeta(p)}
  \right)^{-2} 
  \\
  T_{(q)} (z(q,p)) \,=\,  T_{(q)} (\zeta(q)) 
  \left( 
    \frac{d\,z(q,p)}{d\,\zeta(q)}
  \right)^{-2}. 
\end{gather}
\end{subequations}
Analogous relations hold for the antiholomorphic sector.

Rescaling $z(q,p) = - z(p,q)$, then we perform the glueing asking for
the following relations to hold:
\begin{subequations}
  \begin{gather}
    T_{(p)} (z(p,q)) \,=\, T_{(q)} (z(p,q))\,\vert_{y(p,q) = 0} \\
    \overline{T}_{(p)} (\bz(p,q)) \,=\, 
    \overline{T}_{(q)} (\bz(p,q))\,\vert_{y(p,q) = 0}     
  \end{gather}
\end{subequations}

\begin{figure}[!t]
  \centering
  \includegraphics[width=.75\textwidth]{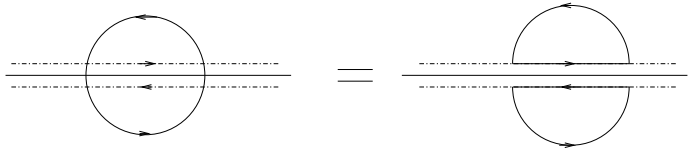}
  \caption{A small integration contour intersecting the $(p,q)$ edge
    of the ribbon graph}
  \label{fig:int_bound}
\end{figure}

These relations allows to associate to each pairwise adjacent couples
of theories a single copy of the Virasoro algebra. As a matter of
fact, these glueing conditions allows to define a unique holomorphic
component of the stress energy-tensor as:
\begin{equation}
  \label{eq:Tpq}
  T_{(p,q)}\,=\, 
\begin{cases}
T_{(p)} (z(p,q)) & \text{if} z(p,q) \,\cup\,\zeta(p) \,\neq\,0\\
T_{(q)} (z(q,p)) & \text{if} z(q,p) \,\cup\,\zeta(q) \,\neq\,0
\end{cases}
\end{equation}
The same holds for the antiholomorphic sector, allowing to define a
unique $\overline{T}(\bz(p,q))$. Moreover,   $T_{(p,q)}$ and
$\overline{T}(\bz(p,q))$ are not independent, because of relations
\eqref{eq:nmp} \eqref{eq:nmq}. Thus, to each pairwise adjacent couples
of cylindrical ends, we can associate a unique copy of the Virasoro
algebra. This can be defined considering a small integration contour
crossing the $(p,q)$ boundary, like that at the rhs of fig. \ref{fig:int_bound}

Thanks to the continuity condition along the boundary, we can actually
write:
\begin{multline}
  \label{eq:Lnpq}
  L_n^{(p,q)}  \,=\, \frac{1}{2 \pi i}\oint_{C(p,q)} z(p,q)^{n + 1}
  T_{(p,q)} (z(p,q)) \,=\,\\
   \frac{1}{2 \pi i}\oint_{C(p)} d z(p,q)\,z(p,q)^{n + 1} T_p
  (z(p,q)) \,+\,\\ \frac{1}{2 \pi i}\oint_{C(q)} d z(q,p) \,z(q,p)^{n +
    1} T_q (z(q,p))
\end{multline}

A similar reasoning can be performed for the other chiral fields of
the model, but performing a particular attention at the gluing maps
$\Omega$ in \eqref{eq:1al}. As a matter of fact, we must keep in
account the fact that this glueing process must relate relate the two
glueing authomorphisms $\Omega_{(p)}$ and $\Omega_{(q)}$ associated to
the BCFTs defined respectively on \cyl{p} and \cyl{q}.

With the above remarks, we can associate to each pairwise coupled
BCFTs a unique Hilbert space carrying a representation of the
$u(1)_{(p,q)}$ algebra.  Consequently the associated collection of
boundary operators will have a primary $u(1)$ (and Virasoro) index
$\lambda(p,q)$. We ask the pair of polytope indexes $(p,q)$ to be
ordered, assuming that this means that the operator in question lives
on the ``$p$-side'' of the boundary of the strip connecting \cyl{p}
with \cyl{q}. Moreover, these operators carry the information about
the conformal boundary condition over $S^{(-)}_{\varepsilon(p)}$, thus
they are labeled by another index $A(p)$:
\begin{equation}
  \psi^{A(p)}_{\lambda(p,q)}(x(p,q)),
\end{equation}
where $x(p,q)\,=\,\Re z(p,q)$.  

In the same way, we can define  with 
\begin{equation}
  \label{eq:bf_1p}
  \psi^{B(q)}_{\lambda(q,p)}(x(q,p))
\end{equation}
(with $x(q,p)\,=\,\Re z(q,p)$) the associated operator, transformed
under the conformal mapping $z(p,q)\,=\,-\,z(q,p)$. According to the
above assumptions, operators in \eqref{eq:bf_1p} live on the $q$- side
of the $(p,q)$ edge, thus carrying the information about the boundary
condition on $S^{(-)}_{\varepsilon(q)}$.

However, the above distinction between $p$-side fields and $q$-side
ones is merely artificial, since the spectrum of boundary operators
associated to a given edge is unique. The natural consequence of this
is that each of them must carry the information about the boundary
conditions adopted on both $S^{(-)}_{\varepsilon(p)}$ and
$S^{(-)}_{\varepsilon(q)}$.

Thus the complete labeling of Boundary Insertion Operators (BIO) which
arises from the BCFT defined on the $(p,q)$-edge is
\begin{equation}
  \label{eq:bio_spectrum}
  \psi_{\lambda(p,q)}^{B(q) A(p)}(x(p,q));
\end{equation}

In this connection, these objects naturally mediate the change in
boundary conditions between pairwise adjacent boundaries.  The labels
ordering, indicates the ``direction'' in which BIO acts, with the
understanding that:
\begin{equation}
  \label{eq:conformal_bio}
  \psi^{\lambda(q,p)}_{A(p) B(q)}(x(q,p))
  \,=\,
  \psi_{\lambda(p,q)}^{B(q) A(p)}(x(p,q));
\end{equation}

At this level of the discussion, the action of BIO is still purely
formal.  Since the switch in boundary conditions (from $A(p)$ to
$B(q)$ and viceversa) will act both on the glueing authomorphism and
on the Wilson line/brane position (see remarks before equation
\eqref{eq:bs_newnot}), we can relate their action to an authomorphism
of the algebra relating the two glueing maps $\Omega_{(p)}$ and
$\Omega_{(q)}$ (see equation \eqref{eq:1al}). Moreover, BIOs are
chiral primary operators of the $U(1)$ algebra: their expression
suggests to adopt a description analogue to that one introduced in
\cite{Behrend:1999bn} for usual boundary operators in Rational
Conformal Field Theory (RCFT). They can be represented as Chiral
Vertex Operators (CVO) which, in every point of the boundary, map the
coupling between states defined by the boundary condition $A(p)$ and
states belonging to the $\lambda(p,q)$ representation of the chiral
algebra to the spectrum of states which defines the boundary condition
$B(q)$.  We can introduce a pictorial description of the interaction
as represented by the (\underline{non local}, because the switch
between boundary condition take place along the full boundary)
interaction vertex:

\vspace{1cm}
%\psgrid[subgriddiv=1,griddots=10,gridlabels=7pt](0,-3)(15,3)%
\psline[linecolor=black]{c-}(5,-1)(6.5,-1)
\psline[linecolor=black]{c-}(6.5,-1)(8,-1)
\psline[linecolor=black]{c-}(6.5,-1)(6.5,.5)

\rput[br](5,-1){$A(p)$}
\rput[bl](8,-1){$B(q)$}
\rput[bl](6,1.1){$\lambda(p,q)$}
\rput[tl](5.9,-.8){$\psi_{\lambda(p,q)}^{B(q) A(p)}$}

\vspace{1.7cm}

This kind of interaction is equivalent to the usual chiral vertex
which represent the map from the fusion of two representation of the
(bulk) Virasoro algebra $\lambda(p)$ and $\lambda'(p)$ to the
representation $\lambda''(p)$. 

The analogy stated with the usual CVO formalism is purely formal,
since the ``fusion'' process described above involve different objects
like boundary states (which are superposition of Ishibashi states) and
irreducible representations of the chiral algebra.  However, this
analogy allows to specify the conformal properties of these operators.
It is possible, a priori, to assign a multiplicity index to BIOs,
$t\,=\,1,\ldots,\,N_{\lambda(p,q)}^{A(p)\,B(q)}$ for each
$\psi^{\lambda(p,q)}_{A(p) B(q)}(x)$ which takes into account the
degeneration which can occur in the map process described before.
Moreover, BIO are primary operators of the boundary chiral algebra,
thus their conformal dimension coincide with the highest weight of the
$V_\lambda(p,q)$ module of the $U(1)$ (Virasoro) algebra:
\begin{equation}
  H(p,q) \,=\,  \frac{1}{2} \lambda(p,q)^2 
\end{equation}

BIOs live on the ribbon graph, thus their interactions are guided by
the trivalent structure of $\Gamma$.  A careful and exhaustive
analysis, which has been entirely reported in appendix
\ref{sec:biobcft}, shows that this structure allows to define all
fundamental coefficients weighting the self-interaction of boundary
insertion operators. Two points functions are well defined on the
edges of the graph, while OPE coefficients naturally defines the
fusion of different boundary insertion operators interacting in
$N_2(T)$ tri-valent vertexes of $\Gamma$. Moreover, we have shown that
thanks to the trivalent structure of the ribbon graph, we can define a
set of sewing constraint these coefficients must satisfy. Remarkably,
these constraints are perfectly analogues to the sewing constraint
introduced in \cite{Lewellen:1991tb} for boundary conditions changing
operator, thus enforcing the analogy between these latters and BIO,
which merge dynamically the $N_0(T)$ copies of the BCFT which are
defined on $M_\partial$.

\section{Investigations at the T-duality self dual radius}
The above description can be pursued further if we explore the
rational limit of the compactified boson BCFT, \ie{} the limit
\begin{equation}
  \label{eq:sdr}
  \Omega(k)\,\doteq\,\Omega(k)_{s.d.}\,=\,\frac{R(k)_{s.d.}}{l(k)}\,=\,\sqrt{2}
\end{equation}
where $\Omega(k)_{s.d.}$ is the self-dual radius, \ie{} the fixed
point of $T$-duality transformation
$\Omega\rightarrow\frac{1}{\Omega}$.  If we substitute
$\Omega\,=\,\sqrt{2}$ into \eqref{eq:momentum}, we obtain
$\lambda_{s.d} \,=\, \frac{ \left( \mu\,+\,\nu \right)}{\sqrt{2}}$ and
$ \overline{\lambda}_{s.d}\,=\,\frac{\left( \mu\,-\,\nu
  \right)}{\sqrt{2}}$, thus we are falling exactly in the situation
described in section \ref{sec:amplitude}. The holomorphic and
antiholomorphic conformal dimensions of primary fields are
respectively $ h_{s.d} \,=\, \left( \frac{\mu\,+\,\nu}{2} \right)^2 $
and $ \overline{h}_{s.d} \,=\, \left( \frac{\mu\,-\,\nu}{2}
\right)^2$, \ie{} they are the square of an integer or semi-integer
number. In full generality, we can write
$h(p)\,=\,j(p)^2,\,j(p)\in\,\frac{1}{2}\mathbb{Z}$. This lead to the
presence of a null state at the level $2j+1$ in the correspondent
$U(1)$ module, which consequently decompose into into an infinite set
of Virasoro ones. Consequences of this can be understood considering
the torus partition function associated the compactified boson with
$\Omega\,=\,\Omega_{s.d.}$ \footnote{In this discussion we will omit
  the polytope index: it will be restored when we will deal again whit
  the interaction of the $N_0$ copies of the BCFT on the ribbon
  graph}:
\begin{equation}
  \label{eq:sd_partfunc}
  Z(\sqrt{2}) \,=\,
  \frac{1}{\left|\eta(\tau)\right|^2}
  \sum_{\mu,\,\nu\,\in\,\mathbb{Z}}
  q^{\frac{1}{4}(\mu\,+\,\nu)^2} 
  \overline{q}^{\frac{1}{4}(\mu\,-\,\nu)^2} 
\end{equation}

If we redefine $m\,=\,\mu\,+\,\nu$ and $n\,=\,\mu\,-\,\nu$ such that
$m\,-\,n\,=\,0\,\text{mod}\,2$, \eqref{eq:sd_partfunc} reads
\begin{equation}
  Z(\sqrt{2}) \,=\,
  \frac{1}{\left|\eta(\tau)\right|^2}
  \sum_{\substack{m,\,n\,\in\,\mathbb{Z} \\ m\,-\,n\,=\,0\,\text{mod}2}}
  q^{\frac{1}{4}m^2} 
  \overline{q}^{\frac{1}{4}n^2}. 
\end{equation}

We can introduce the extended characters\cite{DiFrancesco}:
\begin{subequations}
\label{eq:extchar}
\begin{align}
  \label{eq:extchar1}
  C_0(\tau) 
  & \,=\, 
  \frac{1}{\eta(\tau)}
  \sum_{m \text{even}}q^{\frac{1}{4}m^2} \,=\,
  \frac{1}{\eta(\tau)}
  \sum_{m \in \mathbb{Z}}q^{m^2} \,=\,
  \frac{\theta_3(2\tau)}{\eta(\tau)} \\
  \label{eq:extchar2}
  C_1(\tau) 
  & \,=\, 
  \frac{1}{\eta(\tau)}
  \sum_{m \text{odd}}q^{\frac{1}{4}m^2} \,=\,
  \frac{1}{\eta(\tau)}
  \sum_{m \in \mathbb{Z}}q^{(m\,+\,\frac{1}{2})^2} \,=\,
  \frac{\theta_2(2\tau)}{\eta(\tau)} 
\end{align}
\end{subequations}

where $\theta_i(\tau)$ is the $i$-th Jacobi Theta function. The
partition function now reads:
\begin{equation}
  \label{eq:extpf}
  Z(\sqrt{2}) \,=\,
  |C_0(\tau)|^2 \,+\,
  |C_1(\tau)|^2.
\end{equation}  

We have actually reorganized the infinite Virasoro modules into a
finite number of extended blocks which transforms linearly into each
others by modular transformation. The extended $\mathcal{S}^{ext}$ matrix,
which encodes the modular data via the relation
$C_i(-\frac{1}{\tau})\,=\,\mathcal{S}^{ext}_{ij}C_j(\tau)$, reads\cite{DiFrancesco}:
\begin{equation}
  \label{eq:sext}
\mathcal{S}^{ext} \,=\,
  \frac{1}{\sqrt{2}}
  \begin{pmatrix}
    1 & 1 \\
    1 & -1 
  \end{pmatrix}
\end{equation}

These extended blocks are identified with the two irreducible
representations of the affine algebra $\hat{su}(2)_{k=1}$, namely the
vacuum representation $\mathcal{H}_0$, generated from a state
transforming in the singlet of the horizontal $su(2)$ algebra, and the
representation $\mathcal{H}_\frac{1}{2}$, for which the highest weight
state transform into the fundamental representation of the horizontal
$su(2)$ algebra. Thus, when
$\Omega(k)\,=\,\Omega(k)_{s.d.}$, the BCFT defined over each cylinder
is equivalent to an $\hat{\mathfrak{su}}(2)_{k=1}$-WZW model: it is an
extended non-minimal model generating a Rational Conformal Field
Theory (RCFT), a theory whose possibly infinite Verma modules can be
reorganized into a finite number of extended blocks. These are
irreducible representations of an extended symmetry algebra playing
the role of chiral algebra for the underlining CFT.

In this connection, the set of labels associated to the irreducible
representations of the chiral algebra is finite, 
$\mathcal{Y} = \left\{
  0,\,\frac{1}{2}\right\}$, and the Hilbert space is written as:
\begin{equation}
  \label{eq:su2wzw_hilbspace}
  \mathcal{H}_{bulk} \,=\, 
  \left( 
    \mathcal{H}_{0}^{\hat{su}(2)} 
    \,\otimes\,
    \overline{\mathcal{H}}_{0}^{\hat{su}(2)}
  \right)
  \,\oplus\,
  \left( 
    \mathcal{H}_{\frac{1}{2}}^{\hat{su}(2)} 
    \,\otimes\,
    \overline{\mathcal{H}}_{\frac{1}{2}}^{\hat{su}(2)}
  \right).
\end{equation}

Since they will be useful in the following discussion, let us summarize
the main features of the conformal field theory associated to the
$\hat{\mathfrak{su}}(2)_{k=1}$ WZW model.  $\hat{\mathfrak{su}}(2)_{1}$ (bulk)
primary fields are\cite{DiFrancesco}:
\begin{subequations}
  \label{su2p}
  \begin{align}
    j=0   \,\rightarrow\, & \phi_{(0,0),(0,0)}(z,\bz) \,=\,
    \mathbb{I}\\
    j=\frac{1}{2}  \,\rightarrow\, & 
    \begin{cases}
      \phi_{(\frac{1}{2},\frac{1}{2}),(\frac{1}{2},\frac{1}{2})}(z,\bz)
      \,=\, e^{\frac{i}{\sqrt{2}}X(z)}e^{\frac{i}{\sqrt{2}}X(\bz)}\\
      \phi_{(\frac{1}{2},-\frac{1}{2}),(\frac{1}{2},-\frac{1}{2})}(z,\bz)
      \,=\, e^{-\frac{i}{\sqrt{2}}X(z)}e^{-\frac{i}{\sqrt{2}}X(\bz)}
    \end{cases}
  \end{align}
\end{subequations}

The associated boundary theory is defined by a set of glueing
conditions on the boundary (see formula \eqref{eq:1al}). Let us
remember that, once we are given a glueing authomorphism, we can
associate to each irreducible representation of the chiral algebra an
Ishibashi state\cite{Ishibashi:1988kg},\ie{} a coherent state solution
of the constraint \eqref{eq:cont_bs}. In this connection, let us
denote the two Ishibashi states which are associated with the
$\mathcal{H}_0^{\hat{su}(2)}$ and the
$\mathcal{H}_\frac{1}{2}^{\hat{su}(2)}$ modules respectively with with
$\vert 0 \Rangle$ and $\vert \frac{1}{2} \Rangle$. A remarkable
property defined by Cardy in \cite{Cardy1} states that the set of
boundary conditions of a rational boundary conformal field theory are
labelled exactly by the modules of the chiral algebra, while the
correspondent boundary states are given by\cite{Cardy1}:
\begin{equation}
  \label{eq:cardy_bs}
  \vert\vert A(p)\Rangle \,=
  \sum_{l\,\in\,\mathcal{Y}(p)}
  \frac{\mathcal{S}^{ext}_{A l}}{\sqrt{\mathcal{S}^{ext}_{0 l}}} 
  |l(p)\Rangle
\end{equation}
In the case of $\hat{su}(2)_{k=1}$-WZW model, we can write
$\mathcal{Y}\,\equiv\,\mathcal{A}\,=\,\left\{
  0,\,\frac{1}{2}\right\}$, while formula \eqref{eq:cardy_bs}
specializes to:
\begin{equation}
  \label{eq:cbs}
  \Vert J \Rangle \,=\, 
  2^{- \frac{1}{4}} \,\vert 0 \Rangle \,+\,
  (-1)^{2J}\,2^{- \frac{1}{4}}\,\vert \frac{1}{2} \Rangle, 
\quad\text{with} J = 0, \frac{1}{2}.
\end{equation}

These properties allow for a complete definition of BIO of a  rational
conformal field theory. As a matter of fact, since both boundary states
and boundary operators are identified by primary labels of the
extended chiral algebra, in analogy to what happens for boundary
conditions changing operators of rational minimal models, we can
define the glueing process as the fusion between the representations
associated to the two adjacent boundary states and the one BIO
carries. Thus, let us consider the $(p,q)$ edge $\rho^1(p,q)$. If
$\Vert J(p) \Rangle$ and $\Vert J(q) \Rangle$ are the boundary
conditions shared by $\rho^1(p,q)$, BIOs on $\rho^1(p,q)$ are defined
as:
\begin{equation}
  \label{eq:bio}
  \psi_{j(p,q)}^{J(p)\,J(q)} (x(p,q)) \,=\,
\mathcal{N}_{J(p)\,j(p,q)}^{J(q)}\,\psi_{j(p,q)}(x(p,q)) 
\end{equation}
\ie{} they are the $\hat{\mathfrak{su}}(2)_{k=1}$ primary fields
weighted by hte fusion rule ${N}_{J(p)\,j(p,q)}^{J(q)}$. The latters
are given by a combination of the $\mathcal{S}^{ext}$ matrix entries
via the Verlinde formula
\begin{equation}
  \label{eq:verlfor}
  \mathcal{N}_{J(p)\,j(p,q)}^{J(q)} \,=\,
  \sum_{l\in\mathcal{Y}} 
  \frac{S^{ext}_{J(p)\,l}\,S^{ext}_{j(p,q)\,l}\,
    \overline{S}^{ext}_{l\,J(q)}}{S^{ext}_{0\,l}} 
\end{equation}

As it stands, this construction, valid for the
$\hat{\mathfrak{su}}(2)_{k=1}$ model, does not apply to the
compactified boson at the self-dual radius $\Omega_{s.d}$. As a matter
of fact, even if the Hilbert space of the two models is the same, the
boundary theory associated to the latter one is quite more
complicated. It will lead to the definition of BIOs of the
compactified boson at the self dual radius as truly marginal
deformations of the operators in  \eqref{eq:bio} by the action of
$SO(3)$ elements. 

We will dedicate the remaining part of this section to the
demonstration of this statement.

Due to the Sugawara construction, each highest weight representation
of $\hat{\mathfrak{su}}(2)_{k=1}$ is also an irreducible representation of the
$c=1$ Virasoro algebra, whose generators commute with the horizontal
$su(2)$ ones. Thus, we can decompose each irreducible representation
of $\hat{\mathfrak{su}}(2)_{k=1}$ entering in \eqref{eq:su2wzw_hilbspace} with
respect to $su(2)\,\otimes\,{Vir}$:
\begin{equation}
  \label{eq:su2dec}
  \mathcal{H}_{j}^{\hat{\mathfrak{su}}(2)}
  \,=\, 
  \sum_{n=0}^\infty 
  \mathcal{V}_{(n\,+\,j)^2} 
  \,\otimes\,
  \mathcal{H}_{n\,+\,j}^{su(2)}, 
\end{equation}
where $\mathcal{V}_h$ is the conformal weight $h$ irreducible
representation of the Virasoro algebra, while
$\mathcal{H}_{j}^{\hat{\mathfrak{su}}(2)}$ (resp. $\mathcal{H}_{j}^{su(2)}$) is
the $(2j\,+\,1)$-dimensional representation of $\hat{su}(2)$
(resp. $su(2)$). The bulk Hilbert space then decompose as:
\begin{equation}
  \label{eq:su2bulk_dec}
  \mathcal{H}_{bulk} \,=\, 
  \sum_{\substack{n,\overline{n}\,=\,0\\n\,+\,\overline{n}\,\text{even}}}%
  ^\infty
  \left(
    \mathcal{V}_{\frac{n^2}{4}}
    \,\otimes\,
    \overline{\mathcal{V}}_{\frac{\overline{n}^2}{4}}
  \right)
  \,\otimes\,
  \left(
    \mathcal{H}_{\frac{n}{2}}^{su(2)} 
    \,\otimes\,
    \overline{\mathcal{H}}_{\frac{\overline{n}}{2}}^{su(2)}
  \right).
\end{equation}

As we outlined before, the representations of the Virasoro algebra
which occur in the decomposition are all degenerates. Due to this, the
set of Virasoro primary fields is larger than in the case of generic
compactification radius.  The decomposition \eqref{eq:su2bulk_dec}
shows that, for fixed $(m,\,n)$, the $U(1)_L\,\times\,U(1)_R$
representation of momenta $\lambda_{s.d.}$ and
$\overline{\lambda}_{s.d.}$ contains the Virasoro representation
provided that both $|m|,|n|\,\leq\,j$ and $j\,-\,m$ and $j\,-\,n$ are
integers. Conversely, if $m$ and $n$ satisfies these constraints, the
representation $ \mathcal{V}_{\frac{n^2}{4}} \,\otimes\,
\overline{\mathcal{V}}_{\frac{\overline{n}^2}{4}}$ of the Virasoro
algebra enter exactly once into the representation of
$U(1)\,\times\,U(1)$. Thus the Virasoro primary fields can be labelled
by the triple of indexes $(j;\,m,\,n)$, with the constraint
$|m|,\,|n|\,\leq\,j$. This is the celebrated \textsc{discrete serie of
  states}\cite{Kac:1978ge,Klebanov:1991hx,Callan:1994ub,Kristjansson:2004ny}
$\psi_{j,m} \overline{\psi}_{j,n}$. The associated Ishibashi states
are defined in literature as $\vert
j,\,m,\,n\Rangle$\cite{Callan:1994ub}.

The arising of the discrete serie of states is strictly related to the
deformation of an open bosonic string theory obtained by adding
suitable boundary operators to the action. As a matter of fact,
authors of \cite{Callan:1994ub} showed that the discrete spectrum
generated by the free boson once it is compactified at the $T$-duality
self dual radius can be equivalently explained with an open string
model in which one end of string is subjected to Dirichlet boundary
conditions while, on the other end, the boundary condition is
dynamically defined by adding to the action the following boundary term:
\begin{equation}
  \label{eq:bt_kp}
  S_b \,=\, \int d t \frac{1}{2} 
  \left(
    g e^{i \frac{X(t,0)}{\sqrt{2}}} + 
    \overline{g} e^{- i \frac{\overline{X}(t,0)}{\sqrt{2}}}
  \right)
\end{equation}
They showed that the discrete serie of states cited above saturates
the dynamic of the boundary problem, since Neumann boundary states has
null momentum, and the period of the boundary momenta is such that it
injects momenta which are integral multiples of $\frac{1}{\sqrt{2}}$,
namely the values carried by the discrete states\cite{Callan:1993mw}.

This effects is part of a wider connection in which, once we are given
a boundary conformal field theory, we can associate it different
models considering fluctuations in boundary condensate, where the
condensate is defined by a boundary term added to the bulk action:
\begin{equation}
  S = S_{bulk} + g\int dx \psi(x)
\end{equation}
If the operator $\psi(x)$ is truly marginal (for what this means, see
appendix \ref{ch:bcft_deform}), the deformation does not take the
model away from the renormalization group fixed
point\cite{Recknagel:1998ih}. Thus the bulk theory remains unvaried,
and the perturbation effects involve only a redefinition of the
boundary conditions, thus affecting boundary states and the boundary
operators.

For the sake of completeness, we have included in appendix
\ref{ch:bcft_deform} a comprehensive introduction to the fundamental
concepts and techniques in marginal deformations of a boundary
conformal field theory. These techniques will be used extensively in
the following.

In the connection of the compactified boson, the presence of the
enhanced affine symmetry at special values of the compactification
radius coincide with the presence of new massless \textsc{open} string
states which can be used to deform the theory\cite{Green:1995ga}.
When $\Omega(k) = \sqrt{2}$, the closed affine algebra generators can
be represented in term of the boson field via the Frenkel-Kac-Siegel
construction of the affine algebra. In the closed string channel, the
left moving currents are:
\begin{equation}
  H(\zeta(p)) \,=\, \partial X(\zeta(p)), \qquad
  E^\pm \,=\, \ordprod{e^{\pm i \sqrt{2} X(\zeta(p))}} 
\end{equation}
and the same construction holds for the antiholomorphic sector.
The $\hat{su}(2)_1$ currents modes, $J^a_m$ and $\bar{J}^a_m$
close 2 copies of the affine algebra:
\begin{equation}
  \label{eq:algeb}
  [J^a_m,J^a_n] \,=\,
  \sum_c i f_{abc} J^c_{m+n} \,+\, k m \delta_{a,b} \delta_{m+n,0} 
\end{equation}

The vertex operators for the massless closed string states are given
in term of the currents above as:
\begin{gather}
  \label{eq:massl_vertex}
  V^a_P(p) \,=\, 
  J^a(\zeta(p)) \bar{\partial} \bar{X}(\bar{\zeta}(p)) 
  e^{i P X(\zeta(p)) + \bar{X}(\bar{\zeta}(p))}\\
  \bar{V}^a_P(p) \,=\, 
  \bar{J}^a(\bar{\zeta}(p)) \partial X(\zeta(p)) 
  e^{i P X(\zeta(p)) + \bar{X}(\bar{\zeta}(p))}
\end{gather}
where $P\,=\,p_L \,+\, p_R$ is the total center of mass momentum of
the closed string.  The vertex operators for the new open string
scalar states can be written in the closed string channel as:
\begin{multline}
  S^a_P \,=\, J^a(\zeta(p)) e^{i P X(\zeta(p))} \\\,\equiv\, 
  \frac{1}{2}\left[
    J^a(\zeta(p)) \,-\, 
    \Omega(p)(\bar{J}^a(\bar{\zeta}(p))) e^{i P X(\zeta(p))}
  \right]\vert_{|\zeta|=1,\,|\zeta|=\q}, 
  \label{eq:open_vert}
\end{multline}
where $\Omega(p)$ is the glueing authomorphism on the boundary (see
equation \eqref{eq:1al}).

The occurrence of extra massless open string states in equation
\eqref{eq:open_vert}, indicates that, at this special point in
compactification moduli space, also the chiral algebra of the boundary
theory is enhanced. The associated currents $\mathbf{J}^a(\zeta(p))$
(defined as in equation \eqref{eq:1al}) are truly marginal operators
(see appendix \ref{ch:bcft_deform}) and can actually be combined to
deform the original theory with a boundary action of the form:
\begin{equation}
  \label{def_action}
  S_B \,=\, \int d x(p,q) \sum_a g_a \mathbf{J}^a(\zeta(p))|_{|\zeta|=\q} 
\end{equation}

In this connection, deformations in equation \eqref{eq:bt_kp} are a
particular case of \eqref{def_action}, in which the deformation
involves only generators associated to the simple roots of $su(2)$.
 
Since chiral marginal deformation are truly marginal
\cite{Recknagel:1998ih} the deformed model will change only for a
redefinition of boundary conditions (thus boundary states and boundary
operators).

In particular, in \cite{Green:1995ga,Recknagel:1998ih} it had been
showed that the boundary condition induced by the presence of an
action boundary term like that in \eqref{def_action} is represented by
a boundary state defined as\cite{Green:1995ga,Recknagel:1998ih}
\begin{equation}
  \label{eq:bs_schom}
  \Vert g \Rangle 
  \,=\, 
  g \, \Vert N(0) \Rangle_{s.d.}
  \qquad \text{with} \qquad 
  g \,=\, e^{\sum_a i g_a J^a_0}
\end{equation}
where $J^a_0$ are the horizontal $SU(2)$ algebra generators. 
Thus, According to formula \eqref{eq:bs_schom} boundary states are actually
a rotation of a ``generator'' boundary state, which is that associated
to Neumann boundary conditions with null Wilson line.  This
construction cover the full moduli space of boundary states. Naively,
one can expect to obtain a second branch in the moduli space of
boundary states via the same construction on a ``Dirchlet-like''
unperturbed boundary state $\Vert D(0) \Rangle_{s.d.}$. However,
Dirichlet boundary states are included in the set \eqref{eq:bs_schom}.
They are obtained via a perturbation of the form \eqref{def_action},
with the particular choice $\Gamma\,=\, e^{- i\pi J_0^1}$\cite{Callan:1994ub}:
\begin{equation}
  \label{eq:dir_bs}
  \Vert D(0) \Rangle_{s.d.}\,=\,
e^{- i\pi J_0^1}\Vert N(0) \Rangle_{s.d.} 
\end{equation}

Boundary states in eq. \eqref{eq:bs_schom} satisfy the rotated gluing
condition:
\begin{equation}
  \label{eq:rot_gc}
  \left(
    J^a_m + \Omega\circ\gamma_{\bar{J}}(\bar{J}^a_{-m}) 
  \right)\Vert B(g) \Rangle \,=\,0
\end{equation}
with $\gamma_{\bar{J}}(\bar{J}^a)\,:=\,
e^{-i \sum_b g_b J^b_0}\bar{J}^a e^{i \sum_b g_b J^b_0}$

Cardy's boundary states \eqref{eq:cbs} can be easily retrieved in
the set \eqref{eq:bs_schom}. 
As a matter of fact,
when the unperturbed boundary state is a Neumann one with the Wilson
line parametrized by $\tilde{t}_+$, the general gluing condition 
\begin{equation}
\label{eq:part_gluing}
  E^\pm(\zeta) \,=\, e^{\pm i \sqrt{2} \tilde{t}_+} \bar{E}^\pm(\bar{\zeta}) 
\end{equation}
is invariant under the shift $\tilde{t}_+ \rightarrow \tilde{t}_+ +
\pi\sqrt{2}$. The marginal perturbation implementing this
shift, $\Gamma\,=\,\mbox{exp}(i\frac{\pi}{\sqrt{2}}J^3_0)$ acts non
trivially on a Neumann boundary state $\Vert N(\tilde{t}_+) \Rangle$,
producing exactly a switch of the sign in front of the Ishibashi
state built on the $j=\frac{1}{2}$ $\hat{su}(2)_{1}$ module.  Thus we
can conclude that, in the description of eq. \eqref{eq:bs_schom},
Cardy's boundary corresponds to $\Vert N(0) \Rangle_{s.d.}$ and $\Vert
N(\pi/\sqrt{2}) \Rangle_{s.d.}$:
\begin{subequations}
  \begin{align}
    \Vert 0 \Rangle &
    \,=\, 
    2^\frac{1}{4} 
    \left( 
      \vert 0 \Rangle + \vert \frac{1}{2} \Rangle
    \right)  \,\equiv\, \Vert N(0) \Rangle_{s.d.} \\
    \Vert \frac{1}{2} \Rangle &
    \,=\, 
    2^\frac{1}{4} 
    \left( 
      \vert 0 \Rangle - \vert \frac{1}{2} \Rangle
    \right)  \,\equiv\, \Vert N(\pi/\sqrt{2}) \Rangle_{s.d.}
  \end{align}
\end{subequations}

This result holds for any gluing condition we decide to fix on the
boundary of the cylinder. The independent gluing conditions are then
parametrized by $SO(3)$, because the elements which yield trivial
gluing automorphism are those associated to the center of
$SU(2)$. Cardy's boundary states are then those associated to the
central elements of $SU(2)$.

Authors of \cite{Gaberdiel:2001xm} introduced an alternative
description in which they showed that the boundary states generated by
the perturbation \eqref{def_action} can be directly parametrized by
means of the associated $SU(2)$ elements $g$ and described via the
following formula:
\begin{equation}
  \label{eq:su2_bs_2}
  \Vert g \Rangle 
  \,=\,
  \sum_{j\,\in\,\mathcal{Y}} 
  \sum_{\substack{-j\,\leq\,m\,\leq\,j\\-j\,\leq\,n\,\leq\,j}}
  D^j_{m,\,-n}(g) \vert j;\,m,\,n\Rangle
\end{equation}
where $\vert j;\,m,\,n\Rangle$ is the Ishibashi state associated to
the discrete $h\,=\,j^2$ module of the $su(2)$ algebra and
$D^j_{m,\,-n}(g)$ is the matrix elements of the $j$-representation of
$g\,\in\,SU(2)$, parametrized as $g\,=\,\left(
\begin{smallmatrix}
 a & b \\
 -b* & a*   
\end{smallmatrix}
\right)$.  In this description, the generating boundary state $\Vert
N(0) \Rangle_{s.d.}$ is associated to the $SU(2)$ identity, $g =
\mathbb{I}$ $N(0)\,=\, \left(
  \begin{smallmatrix}
    1 & 0 \\
    0 & 1
  \end{smallmatrix}
\right)$.  

We will not use this construction to define BIO, however we report it
because it allows to write the annulus amplitude with different
boundary conditions associate respectively to $\boum$ and $\boup$ in a
quite simple way. Let us consider the transition amplitude between two
boundary states $|\alpha\rangle$ and $|\beta\rangle$ which can be
defined via the action of $g \in SU(2)$ on $|\alpha\rangle$,
$|\beta\rangle\,=\, g |\alpha\rangle$. From the above definitions of
boundary states, it can be shown that $A_{\alpha, g \alpha}$ depends
only on the conjugacy classes of $g$ (for a detailed demonstration see
\cite{Recknagel:1998ih}, or \cite{Gaberdiel:2001xm}, section
4). Therefore, we can choose to deform the boundary state with an
element in a given torus of $SU(2)$: $t \,=\, h^{-1} g h \,=\,
\left(
\begin{smallmatrix}
 e^{4 \pi  i \lambda} & 0 \\
 0 & e^{- 4 \pi  i \lambda}   
\end{smallmatrix}
\right)$  following the detailed analysis in \cite{Gaberdiel:2001xm}, we
can finally write the amplitude as 
\begin{equation}
  \label{eq:amppp}
  \mathcal{A}(p) \,=\, \frac{1}{\sqrt{2}}
\sum_{j \in \frac{1}{2} \mathbb{Z_+}}
cos(8 \pi j \lambda) \vartheta_{2 \tilde{q}} (\tilde{q})
\end{equation}
where $\tilde{q} = e^{ - \frac{2 \pi i}{\tau}}$ 

A boundary perturbation will affect the boundary operators spectrum
too. Generically, when the perturbing field is truly marginal, the
study of the deformation of a correlator containing both bulk and
boundary fields allows to define the image $\tilde{\psi}_j$ of a
boundary field $\psi_j$ under a rotation generated by the perturbing
field $\psi$ as\cite{Recknagel:1998ih} (see appendix \ref{ch:bcft_deform}):
\begin{equation}
  \label{eq:bf_pert}
  \tilde{\psi}_j\,=\,
  \left[ 
    e^{\frac{1}{2} g \psi} \psi_j
  \right](u)\,:=\,
  \sum_{n=0}^\infty \frac{g^n}{2^n n!}
  \oint_{C_1} \frac{dx_1}{2\pi} \cdots \oint_{C_n} \frac{dx_n}{2\pi}
  \psi_j(u)\psi(x_n)\cdots\psi(x_1)
\end{equation}
where $C_i$ are small circles surrounding the insertion points of the
operators $\psi(x_i)$ on the boundary.

As they stands, representations \eqref{eq:bs_schom} and
\eqref{eq:su2_bs_2} of boundary conditions do not allow to
successfully explain how the transition between pairwise adjacent
boundary conditions take place.  Thus, we have introduced a new
representation for the infinite set of boundary conditions which can be
applied to the compactified boson at the self-dual radius. This
representation merges the infinite choice of boundary conditions with
the the necessity to have a BIO acting ``\emph{a l\'a Cardy}'', ie mediating
between the diffenten boundary conditions exploiting the fusion rules
of the associated chiral algebra.

We can exploit construction \eqref{eq:bs_schom} to
parametrize the generic boundary condition defined over the (inner or
outer) boundary of the $k$-th cylindrical end, represented by the
boundary state $\Vert g(k) \Rangle$, with a couple of elements:
\begin{equation}
  \label{eq:boundary_par}
  \left(
\Vert J(k) \Rangle,\, \Gamma(k)
\right) \qquad \text{with}\quad 
\begin{cases}
J(k) & \,\in\,\mathcal{A} \\
\Gamma(k) & \,\in\,\frac{SU(2)}{\mathbb{Z}_2}
\end{cases}
\end{equation}
being 
$\Vert J(k) \Rangle$ a Cardy's boundary state (thus
corresponding to an element in the center of $SU(2)$) and $\Gamma$ an
$\frac{SU(2)}{\mathbb{Z}_2}$ group element, such that:
\begin{equation}
\label{eq:poly_bs}
  \Vert g(k) \Rangle
  \,=\,
  \Gamma \, \Vert J(k) \Rangle.
\end{equation}

In this connection, the model can be defined not as a truly marginal
deformation of a open string theory by means of the action of elements of the
affine chiral $SU(2)$ , but as a truly marginal deformation of the
$\hat{su}(2)_{k=1}$ by means of $SO(3)$ elements.

Thus, BIO for the compactified boson with $\Omega(p) = \sqrt{2}$ are
actually given by the consequent deformation induced by the two
adjacent boundary conditions on $\hat{\mathfrak{su(2)}}_{k=1}$ WZW
model boundary insertion operators which we introduced in formula
\eqref{eq:bio}

To understand how this deformation affects and define boundary
insertion operators, let us consider the $(p,q)$-edge of the ribbon
graph. Let the two adjacent boundary conditions be defined as:
\begin{subequations}
\label{bcpq}
\begin{align}
  \Vert g_1 (p) \Rangle & \,=\, \Gamma_1(p) \Vert J_1 (p) \Rangle \\
  \Vert g_2 (q) \Rangle & \,=\, \Gamma_2(q) \Vert J_2 (q) \Rangle 
\end{align}  
\end{subequations}

According to the parametrization of boundary condition introduced
above, BIO must mediate both between Cardy's boundary states and
between the $\frac{SU(2)}{\mathbb{Z}_2}$ elements $\Gamma_1(p)$ and
$\Gamma_2(q)$. While the former action is achieved trough the fusion
prefactor $\mathcal{N}_{J(p)\,j(p,q)}^{J(q)}$, the latter can be
understood deforming BIO with the action of both the boundary
potentials which we are adding on the $(p,q)$-edge. As a matter of
fact, according to equation \eqref{bcpq}, the theory on the $(p)$-th
polytope is deformed by the action of the boundary term $S_{B(p)} =
\int dx(p,q) \mathbf{J}_1(\zeta(p))\vert_{|\zeta(p)|=\frac{2 \pi}{2
    \pi - \varepsilon(p)}}$, while the theory on the $(q)$-th polytope
is deformed by the boundary term $S_{B(q)} = \int dx(q,p)
\mathbf{J}_2(\zeta(q))\vert_{|\zeta(q)|=\frac{2 \pi}{2 \pi -
    \varepsilon(q)}}$. Recalling that $x(q,p) = - x(p,q)$, (the
functional part of) boundary insertion operators, we propose
$\psi_{j(p,q)}$ to be deformed by a suitable combination of the
$SO(3)$ operators which are associated to the above boundary terms. We
ask this combination to cancel, on the boundary, the global effect of
the boundary deformation, to let the two ends glue dynamically in such
a way that this dynamic is actually governed by the fusion rules of
the WZW model.  This correspond to a perform over $\psi_{j(p,q)}$ a
rotation induced by the $\frac{SU(2)}{\mathbb{Z}_2}$ element
$\overline{\Gamma} \,=\, \Gamma_2\Gamma_1^{-1}$, with $\Gamma_i =
e^{i\mathbf{J}_i}$. In the same way, $\psi_{j(q,p)}$ will be deformed
by the action of by the action of $\overline{\Gamma}^{-1} \,=\,
\Gamma_1\Gamma_2^{-1}$.

To show how the above rotation alter the functional part of boundary
insertion operators, let us consider the explicit expression of
components of $\hat{\mathfrak{su}}(2)_1$ BIO which we introduced in
\eqref{eq:bio}. Let us drop for a while the dependence from the fusion
rule factor $\mathcal{N}_{J_1(p)\,j(p,q)}^{J_2(q)}$. Such components
are labelled by two (semi-)integers $j = 0, \frac{1}{2}$ and $-j < m <
j$.  For $j = 0$ the unique component is the identity operator
$\psi_{0,0}x(p,q) = \mathbb{I}$, while for $j = \frac{1}{2}$ the two
components are $\psi_{\frac{1}{2}\,\pm\frac{1}{2}} = \left.e^{\pm
    \frac{i}{\sqrt{2}}X[\zeta(p)(z(p,q))]}\right\vert_{y(p,q)=0}$.

Let us consider the action of the deformation on $\psi_{j(p,q)}$
generated by $\overline{\Gamma} =
e^{i\overline{J}}$
(since we are not moving to a definite representation, we can omit the
quantum number $m$). 
According to \eqref{eq:bf_pert}, the rotated boundary operator will
be:
\begin{equation}
\label{eq:part_def}
  \tilde{\psi}_{j(p,q)}(u(p,q))\,=\,
  \left[
    e^{\frac{1}{2} \overline{J}} \psi_{j(p,q)}
  \right](u(p,q))
\end{equation}

We can compute explicitly the expression of $\tilde{\psi}_{j(p,q)}$
thanks to the self-locality of the boundary operators, and to the OPE
between the truly marginal fields in the chiral algebra and a boundary
operator:
\begin{equation}
  \mathbf{J}(x)\psi_j(u) \sim 
\frac{\mathbf{X}_{\overline{J}}^j}{x - u} \psi_j(u) 
\end{equation}
where $\mathbf{X}_{\overline{J}}^j$ is the natural action of the chiral
algebra on a state of the $h=j^2$ $\hat{su}(2)_1$ module (see formula
\eqref{eq:hor_ac}.

An order by order computation in \eqref{eq:part_def} gives:
\begin{equation}
\label{boh}
  \tilde{\psi}_{j(p,q)}(u(p,q))\,=\,
e^{\frac{i}{2} \mathbf{X}_{\overline{J}}^j} \psi_{j(p,q)}  
\end{equation}
\ie{} the natural action of the chiral algebra on the vertex algebra
fields translates into the natural action of an element of
$SU(2)/\mathbb{Z}_2$ on the components of the primary field associated
to a given $\hat{su}(2)_1$'s module. Moving to a specific
representation, equation \eqref{boh} becomes: 
\begin{equation}
\label{eq:rotbio_exp}
  \tilde{\psi}_{(j,\,m)(p,q)}(u(p,q))\,=\,
D^j_{m n}(\Gamma) \, \psi_{(j,n)(p,q)}  
\end{equation}
where $D^j_{mn}$ are the Wigner functions associated to
$\Gamma\,=\,\left(
\begin{smallmatrix}
 a & b \\
 -b* & a*   
\end{smallmatrix}
\right)$:
\begin{multline}
  \label{eq:su2bs_coeff}
  D^j_{m,\,n}(\Gamma) 
  \,=\,
  \sum_{l\,=\,\text{max}(0,n-m)}^{\text{min}(j-m,j-n)} 
  \frac{
    \left[
      (j\,+\,m)!\,
      (j\,-\,m)!\,
      (j\,+\,n)!\,
      (j\,-\,n)!\,
    \right]^{\frac{1}{2}}}
  {(j\,-\,m\,-\,l)!\,
    (j\,+\,n\,-\,l)!\,
    l!\,
    (m\,-\,n\,+\,l)!\,} \\
  \,\times\,
  a^{j\,+\,n\,-\,l}\,
  (a*)^{j\,-\,m\,-\,l}\,
  b^l\,
  (-b*)^{m\,-\,n\,+\,l}.
\end{multline}

Obviously, the spectrum of
$\hat{su}(2)_1$ boundary primary fields must be invariant under the
action of $\mathbb{Z}_2$. A first intuition about this comes once we
remember the gluing condition \eqref{eq:part_gluing}  being invariant
under the shift generating the $\Vert1/2\Rangle$ Cardy's state: the
spectrum of boundary operators is generated the action of a copy of
the chiral algebra:
\begin{equation}
  \mathbb{W}(\zeta)\,=\,
  \begin{cases}
    W(\zeta) & \Im{\zeta}\,\geq\,0\\
    \Omega\circ\gamma_{\bar{\Gamma}}(\bar{W}(\bar{\zeta})) & \Im{\zeta}\,<\,0
  \end{cases}.
\end{equation}
The gluing automorphism itself generates the boundary operators'
spectrum, thus a boundary condensate which leaves invariant the gluing
automorphism automatically leaves invariant the boundary operators'
spectrum too. An explicit computation via formula
\eqref{eq:rotbio_exp} of Cardy's boundary
operators rotated by the action of the boundary condensate
correspondent to the $\Vert N(\pi/\sqrt{2})\Rangle_{s.d.}$ boundary
state confirms this statement: rotated boundary operators are obtained
by multiplication by an inessential phase factor.  

Finally, restoring the fusion multiplicative coefficient, we have the
following expression for boundary insertion operators for the
compactified boson at the self dual radius:
\begin{equation}
  \label{eq:final_bio}
  \psi^{[J_2,\,\Gamma_2](q)\,[J_1,\,\Gamma_1](p)}_{[j,\,m](p,q)}\,=\,
\sum_{n=-j}^j  D^{j(p,q)}_{m\,n(p,q)}(\Gamma_2{\Gamma_1}^{-1}) \, 
  \psi_{[j\,n](p,q)}^{J_2(q)\,J_1(p)}.
\end{equation}
where $\psi_{j(p,q)}^{J(p)\,J(q)} (x(p,q)) \,=\,
\mathcal{N}_{J(p)\,j(p,q)}^{J(q)}\,\psi_{j(p,q)}(x(p,q))$

\section{The algebra of rotated Boundary Insertion Operators}

According to last section remarks, boundary insertion operators for
the compactified boson at enhanced symmetry values of the self-dual
radius have the following expression:
\begin{equation}
  \label{eq:final_bio.}
  \psi^{[J_2,\,\Gamma_2](q)\,[J_1,\,\Gamma_1](p)}_{[j,\,m](p,q)}\,=\,
\sum_{n=-j}^j  D^{j(p,q)}_{m\,n(p,q)}(\Gamma_2{\Gamma_1}^{-1}) \, 
  \psi_{[j\,n](p,q)}^{J_2(q)\,J_1(p)}.
\end{equation}

The aim of this, quite technical, section, is to show that with this
choice, effects of boundary perturbations do not affect the
algebra of boundary operators, thus they do not change the dynamic of
the model. This is a check of consistency of the above choice for
boundary insertion operators, since actually all deformations we have
considered are induced by truly marginal operators, thus they must not
break the $su(2)$ chiral algebra.

Let us deal with the simplest case: the amplitude computed on the open
surface associated with the sphere with three punctures. The amplitude
over each cylindrical end will involve a sum over characters twisted
by the action of conjugacy classes of $\frac{SU(2)}{\mathbb{Z}_2}$.
The full amplitude will involve a sum over the intermediate
channels associated with the three edges of the ribbon graph, each
being associated to an automorphism by the operator $U\,=\, e^{i
  \left[g_1(q)-g_2(p)\right]}, \forall (p,q)\,=\,(1,2),(1,3),(2,3)$. Such an
  automorphism acts at BIOs' level: the BIO associated to the
  $(p,q)$-th edge will be the Cardy's one rotated by the action of the   
 $\frac{SU(2)}{\mathbb{Z}_2}$ element $\Gamma_2{\Gamma_1}^{-1}$

The algebra of rotated BIOs follows from their definition. Let us
notice that rotated BIOs are just a superposition of the different
components of Cardy's $\hat{su}(2)_1$ primary operators (with respect
to the affine chiral algebra). 

Let us focus our attention on the two points function. When we
consider two BIOs both mediating a boundary condition changing in the
$p$-to-$q$ direction, we deal with the following expression: 
\begin{multline}
  \label{eq:rot2points}
  \left\langle 
    \psi^{[J_2,\,\Gamma_2](q)\,[J_1,\,\Gamma_1](p)}_{[j,\,m](p,q)}
    (x_1(p,q)) \,
    \psi^{[J_4,\,\Gamma_4](q)\,[J_3,\,\Gamma_3](p)}_{[j',\,m'](p,q)}
    (x_2(p,q))
  \right\rangle\,=\,\\
\sum_{n\,n'}
  D^{j(p,q)}_{m\,n(p,q)}(\Gamma_2{\Gamma_1}^{-1}) 
  D^{j'(p,q)}_{m'\,n'(p,q)}(\Gamma_4{\Gamma_3}^{-1})
  \left\langle
    \psi^{J_2(q)\,J_1(p)}_{[j,\,n](p,q)}(x_1(p,q)) \,
    \psi^{J_4(q)\,J_3(p)}_{[j',\,n'](p,q)}(x_2(p,q))
\right\rangle
\end{multline}
First of all, we must notice that a coherent gluing impose the two
operators to mediate between the same boundary conditions (see eq.
\eqref{eq:2points}), thus the above expression reduces to:
\begin{multline}
  \label{eq:rot2points_sim}
  \left\langle 
    \psi^{[J_2,\,\Gamma_2](q)\,[J_1,\,\Gamma_1](p)}_{[j,\,m](p,q)}
    (x_1(p,q)) \,
    \psi^{[J_2,\,\Gamma_2](q)\,[J_1,\,\Gamma_1](p)}_{[j',\,m'](p,q)}
    (x_2(p,q))
  \right\rangle\,=\,\\
\sum_{n\,n'}
  D^{j(p,q)}_{m\,n(p,q)}(\Gamma_2{\Gamma_1}^{-1}) 
  D^{j'(p,q)}_{m'\,n'(p,q)}(\Gamma_2{\Gamma_1}^{-1})
  \left\langle
    \psi^{J_2(q)\,J_1(p)}_{[j,\,n](p,q)}
    \psi^{J_2(q)\,J_1(p)}_{[j',\,n'](p,q)}
\right\rangle
\end{multline}
Equation \eqref{eq:rot2points_sim} shows that the net effect of the
rotation on the two points function vanish, because we are actually
implementing the same SU(2) rotation on all boundary fields entering
in the unperturbed correlator. Thus the local $SU(2)$
invariance ensures:
\begin{multline}
  \left\langle 
    \psi^{[J_2,\,\Gamma_2](q)\,[J_1,\,\Gamma_1](p)}_{[j,\,m](p,q)}
    (x_1(p,q)) \,
    \psi^{[J_2,\,\Gamma_2](q)\,[J_1,\,\Gamma_1](p)}_{[j',\,m'](p,q)}
    (x_2(p,q))
  \right\rangle\,=\,\\
  \left\langle
    \psi^{J_2(q)\,J_1(p)}_{[j,\,m](p,q)}(x_1(p,q)) \,
    \psi^{J_2(q)\,J_1(p)}_{[j',\,m'](p,q)}(x_2(p,q)) 
\right\rangle
\end{multline}

Dealing with the two points function between a
$p$-to-$q$ and $q$-to-$p$ mediating operators, the situation is
slightly different. We have to compute:
\begin{multline}
  \label{eq:rot2points2}
  \left\langle 
    \psi^{[J_2,\,\Gamma_2](q)\,[J_1,\,\Gamma_1](p)}_{[j,\,m](p,q)}
    (x_1(p,q)) \,
    \psi^{[J_1,\,\Gamma_1](p)\,[J_2,\,\Gamma_2](q)}_{[j',\,m'](q,p)}
    (x_2(q,p))
  \right\rangle\,=\,\\
\sum_{n\,n'}
  D^{j(p,q)}_{m\,n(p,q)}(\Gamma_2{\Gamma_1}^{-1}) 
  D^{j'(q,p)}_{m'\,n'(q,p)}(\Gamma_1{\Gamma_2}^{-1}) \times \\
  \left\langle 
    \psi^{J_2(q)\,J_1(p)}_{[j,\,n](p,q)}(x_1(p,q)) \,
    \psi^{J_1(p)\,J_2(q)}_{[j',\,n'](q,p)}(x_2(q,p))
\right\rangle
\end{multline}

As a matter of fact, in the previous expression we are dealing with a
representation of diagonal subgroup of the direct product
$\frac{SU(2)}{\mathbb{Z}_2}(p,q)\times\frac{SU(2)}{\mathbb{Z}_2}(q,p)$,
thus it holds (see eq. \eqref{eq:dir_prod}:
 \begin{equation}
\label{eq:dirprod_rep}
D^{j(p,q)}_{m\,n(p,q)}(\Gamma_2{\Gamma_1}^{-1}) 
  D^{j'(q,p)}_{m'\,n'(q,p)}(\Gamma_1{\Gamma_2}^{-1})
\,=\,D^{j \times j'}_{m\,n;\,m'\,n'}(\mathbb{I})
    \end{equation}

The trivial Clebsh-Gordan expansion (eq. \eqref{eq:CG_series}) gives
(we omit the polytope indices writing the Clebsh-Gordan coefficients):
\begin{multline}
  \left\langle 
    \psi^{[J_2,\,\Gamma_2](q)\,[J_1,\,\Gamma_1](p)}_{[j,\,m](p,q)}
    (x_1(p,q)) \,
    \psi^{[J_1,\,\Gamma_1](p)\,[J_2,\,\Gamma_2](q)}_{[j',\,m'](q,p)}
    (x_2(q,p))
  \right\rangle\,=\,\\
\sum_{n\,n'}\,\sum_{J\,N} C_{j_1\,m_1\,j_2\,m_2}^{J\,N}
C_{j_1\,n_1\,j_2\,n_2}^{J\,N}
  \left\langle 
    \psi^{J_2(q)\,J_1(p)}_{[j,\,n](p,q)}(x_1(p,q)) \,
    \psi^{J_1(p)\,J_2(q)}_{[j',\,n'](q,p)}(x_2(q,p))
\right\rangle\,=\,\\
  \left\langle 
    \psi^{J_2(q)\,J_1(p)}_{[j,\,m](p,q)}(x_1(p,q)) \,
    \psi^{J_1(p)\,J_2(q)}_{[j',\,m'](q,p)}(x_2(q,p))
\right\rangle.
\end{multline}
In the last equation we have used the unitarity of Clebsh-Gordan
coefficients (see equation \eqref{eq:CG_unit2}).

\begin{figure}[!t]
  \centering
  \includegraphics[width=.5\textwidth]{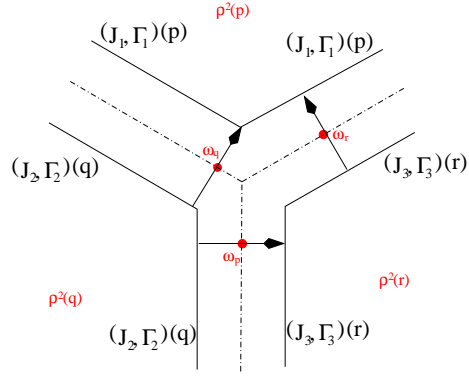}
  \caption{bla bla bla}
  \label{fig:BIO_ope-deformed}
\end{figure}

To calculate the OPE of rotated BIOs, let us notice that the rotation
generated by the boundary condensate does not change them coordinate
dependence. Let us consider the situation depicted in figure
\ref{fig:BIO_ope-deformed}. 

OPE between
$\psi^{[J_1,\,\Gamma_1](p)\,[J_3,\,\Gamma_3](r)}_{[j_1,\,m_1](r,p)}$
and $\psi^{[J_3,\,\Gamma_3](r)\,[J_2,\,\Gamma_2](q)}_{[j',\,m'](q,r)}$
will mediate a change in boundary conditions from
$[J_2,\,\Gamma_2](q)$ to $[J_1,\,\Gamma_1](p)$. In particular,
\begin{multline}
  \notag
  \psi^{[J_1,\,\Gamma_1](p)\,[J_3,\,\Gamma_3](r)}_{[j_1,\,m_1](r,p)}
  (\omega_r) \,
  \psi^{[J_3,\,\Gamma_3](r)\,[J_2,\,\Gamma_2](q)}_{[j',\,m'](q,r)}
  (\omega_q)\,=\, \\
  \sum_{n_1(r,p)\,n_2(q,r)}\,
  D^{j_1(r,p)}_{m_1\,n_1(p,q)}(\Gamma_1{\Gamma_3}^{-1})\, 
  D^{j_2(q,r)}_{m_2\,n_2(q,r)}(\Gamma_3{\Gamma_2}^{-1})\, 
  \psi^{J_1(p)\,J_3(r)}_{[j_1,\,n_1](r,p)}(\omega_r) \,
  \psi^{J_3(r)\,J_2(q)}_{[j_2,\,m_2](q,r)}(\omega_q)
\end{multline}

We are dealing again with a representation of the diagonal subgroup
of the direct product
$\frac{SU(2)}{\mathbb{Z}_2}(r,p)\times\frac{SU(2)}{\mathbb{Z}_2}(q,r)$,
thus applying \eqref{eq:dir_prod} and the Clebsh-Gordan series
expansion \eqref{eq:CG_series} we are left with:
\begin{multline}
\label{interm1}
  \sum_{\substack{n_1(r,p)\\n_2(q,r)}}\,
  \sum_{\substack{j=|j_1-j_2|\\|m|\leq{}j\\|n|\leq{}j}}^{j_1+j_2}
  C^{j\,m}_{j_1(r,p)\,m_1(r,p)\,j_2(q,r)\,m_2(q,r)}\,
  D^j_{m\,n}(\Gamma_1\Gamma_2^{-1})\, \\ \times
  C^{j\,n}_{j_1(r,p)\,n_1(r,p)\,j_2(q,r)\,n_2(q,r)}\, 
  \psi^{J_1(p)\,J_3(r)}_{[j_1,\,n_1](r,p)}(\omega_r) \,
  \psi^{J_3(r)\,J_2(q)}_{[j_2,\,n_2](q,r)}(\omega_q)
\end{multline}

The OPE between Cardy's boundary operators reads:
\begin{multline}
\psi^{J_1(p)\,J_3(r)}_{[j_1,\,n_1](r,p)}(\omega_r) \,
\psi^{J_3(r)\,J_2(q)}_{[j_2,\,n_2](q,r)}(\omega_q)\,=\,
\sum_{j_3\,n_3}\left\vert
  \omega_r\,-\,\omega_q
\right\vert^{H(q,p)\,-\,H(r,p)\,-\,H(q,r)}\\
C^{j_3\,n_3}_{j_1\,n_1\,j_2\,n_2}\,
\mathcal{C}_{j_1\,j_2\,j_3}^{J_1(p)\,J_3(r)\,J_2(q)}\,
\psi^{J_1(p)\,J_2(q)}_{[j_3,\,n_3](q,p)}(\omega_q).
\end{multline}
The Clebsh-Gordan coefficients
$C^{j_3\,n_3}_{j_1\,n_1\,j_2\,n_2}$ compensate the fact that the LHS
 and RHS terms have different transformation behavior under the
action of the horizontal $su(2)$ algebra, while the coefficients 
$\mathcal{C}_{j_1\,j_2\,j_3}^{J_1(p)\,J_3(r)\,J_2(q)}$ reflect the non
trivial dynamic on each trivalent vertex of the ribbon graph.
 
The inclusion of this last OPE into \eqref{interm1} and the Clebsh-Gordan
coefficients' unitarity (equation \eqref{eq:CG_unit2}) leave us with:
\begin{multline}
  \label{eq:rotated_ope}
  \psi^{[J_1,\,\Gamma_1](p)\,[J_3,\,\Gamma_3](r)}_{[j_1,\,m_1](r,p)}
  (\omega_r) \,
  \psi^{[J_3,\,\Gamma_3](r)\,[J_2,\,\Gamma_2](q)}_{[j',\,m'](q,r)}
  (\omega_q)\,=\, \\
\sum_{j_3\,m}
C^{j_3\,m}_{j_1\,m_1\,j_2\,m_2}\,
\mathcal{C}_{j_1\,j_2\,j_3}^{J_1(p)\,J_3(r)\,J_2(q)}\,
  \psi^{[J_1,\,\Gamma_1](p)\,[J_2,\,\Gamma_2](q)}_{[j_1,\,m_1](q,p)}
  (\omega_p) \,
\end{multline}
We demonstrate that  OPE between rotated BIOs is formally equal to OPE between
unrotated BIOs. Thus, on the ribbon graph the non trivial dynamic is
given by the fusion among the three representations entering in each
trivalent vertex.

\section{The action of BIOs at the self-dual radius}

With the above remarks, we can investigate the properties of the four
points functions of BIOs exploiting their crossing properties.

\begin{figure}[!t]
  \centering
  \includegraphics[width=\textwidth]{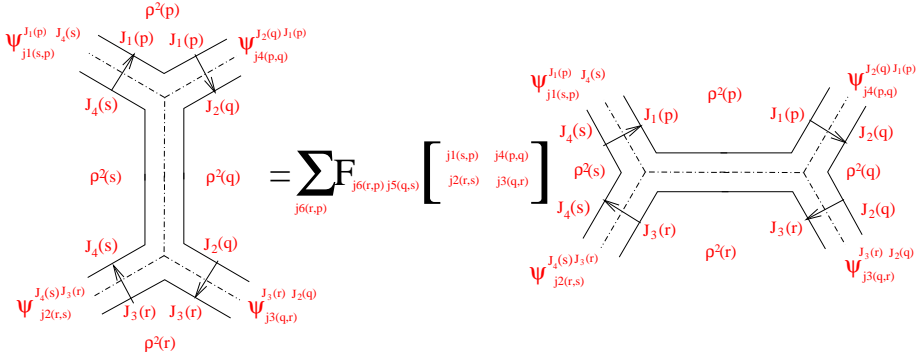}
  \caption{Four points function crossing symmetry}
  \label{fig:BIO_4points}
\end{figure}

First of all let us consider the natural picture in which the
computation of a four points function arises. Let us consider two near
trivalent vertexes. Due to the variable connectivity of the
triangulation, the two configuration shown in figure
\ref{fig:BIO_4points} are both admissible. The transition from the
situation depicted in the lhs and the one in the rhs of the pictorial
identity of figure \ref{fig:BIO_4points}, corresponds exactly to the
transition between the $s$-channel and the $t$-channel of the four
point blocks of a single copy of the bulk theory, thus the two
factorization of the four points function 
$\langle 
\psi_{j_1(s,p)}^{J_1(p)\,J_4(s)}\,
\psi_{j_2(r,s)}^{J_4(s)\,J_3(r)}\,
\psi_{j_3(q,r)}^{J_3(r)\,J_2(q)}\,
\psi_{j_4(p,q)}^{J_2(q)\,J_1(p)}
\rangle$,
pictorially represented in
\ref{fig:BIO_4points}, are related by the bulk crossing matrices:
\begin{equation}
\label{fusion_matrices}
F_{j_6(s,q)\,j_5(r,p)}  
\begin{bmatrix}
  j_4(p,s) & j_1(q,p) \\
  j_3(s,r) & j_2(r,q) 
\end{bmatrix}  
\end{equation} 
The explicit computation of the two factorization leads to the
relation:
\begin{multline}
  \label{eq:fact}
\mathcal{C}^{J_4(s)\,J_3(r)\,J_2(q)}_{j_2(r,s)\,j_3(q,r)\,j_5(q,s)}\,
\mathcal{C}^{J_1(p)\,J_4(s)\,J_2(q)}_{j_1(s,p)\,j_5(q,s)\,j_1(s,p)}\,
\mathcal{C}^{J_1(p)\,J_2(q)\,J_1(p)}_{j_1(s,p)\,j_1(s,p)\,0}
\,=\,\\
\sum_{j_5(r,p)} 
F_{j_6(s,q)\,j_5(r,p)}  
\begin{bmatrix}
  j_4(p,s) & j_1(q,p) \\
  j_3(s,r) & j_2(r,q) 
\end{bmatrix}\,\times\\
\mathcal{C}^{J_1(p)\,J_4(s)\,J_3(r)}_{j_1(s,p)\,j_2(r,s)\,j_6(r,p)}\,
\mathcal{C}^{J_3(r)\,J_2(q)\,J_1(p)}_{j_3(q,r)\,j_4(p,q)\,j_6(p,r)}\,
\mathcal{C}^{J_1(p)\,J_3(r)\,J_1(p)}_{j_6(r,p)\,j_6(p,r)\,0},
\end{multline}
\ie{} the usual BCFT sewing relation among boundary operators' OPEs.

This complete our analysis of the conformal properties of the full
theory arising by glueing together the BCFTs defined over each
cylindrical ends: with the above construction, BIOs play exactly the
role usual boundary operators play in BCFT.

This analogy allows us to apply to BIOs all boundary operators
properties. In particular, we can identify their OPE coefficients with
the fusion matrices \eqref{fusion_matrices} with
the following entries assignation:
\begin{equation}
  \label{eq:OPE_fusion}
  \mathcal{C}^{J_1(p)\,J_2(q)\,J_3(s)}_{j_1(s,p)\,j_2(r,s)\,j_3(q,r)}
  \,=\,
  F_{J_2(q)\,j_3(q,r)}  
  \begin{bmatrix}
    J_1(p)   &  J_3(s)   \\
    j_1(s,p) &  j_2(r,s) 
  \end{bmatrix}  
\end{equation}

Relation \eqref{eq:OPE_fusion}, obtained first in \cite{Runkel:1998pm}
for the $A$-series minimal models exploiting the fact that both the
primary and boundary conditions labels fall in the same set, has been
extended to all minimal models and extended rational conformal field
theories in \cite{Behrend:1999bn} and \cite{Felder:1999ka} noticing the
full analogy between the equation \eqref{eq:fact} and the pentagon
identity for the fusing matrices.

According to \cite{Alvarez-Gaume:1988vr}, WZW-models fusion matrices
coincide with the $6j$-symbols of the corresponding quantum group with
deformation parameter given by the $(k\,+\,h^\vee)$-th root of the
identity, where $k$ and $h^\vee$ are respectively the level and the
dual Coxeter number of the extended algebra. Thus, with $k=1$ and
$h^\vee=2$, the
OPEs coefficients are the $SU(2)_{Q\,=\,e^{\frac{2}{3}\pi i}}$
$6j$-symbols:
\begin{equation}
  \label{eq:OPE_6j}
  \mathcal{C}^{J_1(p)\,J_2(q)\,J_3(s)}_{j_1(s,p)\,j_2(r,s)\,j_3(q,r)}
\,=\,
\begin{Bmatrix}
j_1(s,p) & J_1(p) & J_2(q)  \\
J_3(s) & j_2(r,s) & j_3(q,r)
\end{Bmatrix}_{Q\,=\,e^{\frac{2}{3}\pi i}}
\end{equation}

\section{Open string amplitude on a RRT}

With the computation of OPE's coefficients we have all the building
blocks to construct an open string amplitude on the domain defined by
the open Riemann surface $M_\partial$.

As first step, let us extend results of the previous section to higher
dimensional target spaces.  To this end, let us consider D scalar
fields $X^\alpha, \alpha=1,\,\ldots,\,D$ which, as stated in
\eqref{winding}, wind $\nu^\alpha$ times around the homology cycles of
the compact target space manifold. Let us indicate with
$E_{\alpha\beta} = G_{\alpha\beta} + B_{\alpha\beta}$ the background
matrix on the compact target space manifold, where $G$ is the metric
and $B$ the Kalb-Ramond field. The central charge of the model is
$c=D$.

Let us find out which is the moduli space of inequivalent
compactifications. This can be achieved by considering the torus
partition function of the model: it factorizes into the product of
contributions of each compact direction:
\begin{equation}
  Z_{T_d} \,=\, 
  \frac{1}{|\eta(\tau)|^2} 
  \sum_{\Gamma_{D,D}} 
  q^\frac{p_L^2}{4}
  \overline{q}^\frac{p_R^2}{4}
\end{equation}
Asking for modular invariance let the total momentum 
$\hat{p}=
\left(
  \begin{smallmatrix}
    p_L \\ p_R
  \end{smallmatrix}
\right)$ 
take values
into a self-dual, even-integer, Lorentzian lattice 
$\Gamma_{D,D}$\cite{Johnson:Dbranes}. 
The space of such inequivalent lattices is locally isomorphic to:
\begin{equation}
  \label{lattice_modspace}
  \mathcal{M}
  \,=\, 
  O(d,d,\mathbb{Z}) \backslash O(d,d) /[O(d) \times O(d)],
\end{equation}
which thus is the moduli space of inequivalent toroidal
compactifications in a $D$ dimensional targer
space\cite{Giveon:1994fu,Johnson:Dbranes}.  The different orbits in
this moduli space give rise to different theories in which the
fundamental $U(1)_L \times U(1)_R$ current symmetry can be enhanced to
different symmetry groups of rank at least $D$. This group plays the
role of gauge group in the target space. Our choice is to compactify
each direction at the self dual radius, because this allow to exploit
the previous construction and define a coherent gluing of the
conformal theory along the ribbon graph. This means to choice a
specific orbit into \eqref{lattice_modspace}, \ie{} to fix
definitively the target space modular structure. Moreover, the
background gauge group coming from the closed string sector defined on
the cylindrical ends is $[SU(2)_L \times SU(2)_R]^D$.  On the
contrary, we could choose not to fix the compactification radius as
the self-dual one. This would give us more freedom in choosing the
target space modular and metrical structure, and consequently the
enhanced symmetry group. However, in doing so we can loose information
about the dynamic of the theory on the ribbon graph, because, with the
exception of some particular cases which we will introduce in the
following chapter, combinatorial factor of BIOs would be defined only
as a formal map.

Thus, let us consider each direction compactified at the self dual
radius. 
The amplitude on each \cyl{p} will
receive a contribution from every direction: 
\begin{equation}
  \label{multi_ampl}
  \mathcal{A}_\text{\cyl{p}}
  \,=\,\frac{1}{
    2^\frac{D}{2}
    \left[
      \eta\left(
        e^{-5\frac{4\pi}{\theta(p)}}
      \right)
    \right]^D}
\prod_{\alpha=1}^D
  \sum_{j(p)=0,\frac{1}{2}}
  \cos{(8\pi j(p) \lambda^\alpha(p))}
e^{-\frac{4\pi}{\theta(p)}{j(p)}^{2}}
\end{equation}

Moreover, Boundary Insertion Operators are primaries of the conformal theory,
thus they also factorize into the contribution of each direction. This
means that the full boundary theory factorizes into the contribution
of each compact direction. In this connection, we can exploit a
construction introduced in \cite{Carfora2}, which, exploiting a edge
vertex factorization of the most general correlator we can write on
the ribbon graph, allows to write the contribution to the amplitude
from each compact direction as:
\begin{multline}
  Z(|P_{T_{l}}|) \,=\,\\
  \left( \frac{1}{\sqrt{2}}\right) ^{N_{0}(T)}\sum_{\{j_{p}\in \frac{1}{2}
    \mathbb{Z}_{+}\}}\sum_{\{j_{(r,p)}\}}\prod_{\{\rho
    ^{0}(p,q,r)\}}^{N_{2}(T)}\left\{
    \begin{array}{ccc}
      j_{(r,p)} & j_{p} & j_{r} \\
      j_{q} & j_{(q,r)} & j_{(p,q)}
    \end{array}
  \right\} _{Q=e^{\frac{\pi }{3}i}}\times \\
  \times \prod_{\{\rho ^{1}(p,r)\}}^{N_{1}(T)}\left(
    b_{j_{(r,p)}}^{j_{p}j_{r}}\right) ^{2}L(p,r)^{-2H_{j_{(r,p)}}}
\end{multline}

Collecting the contribution of each direction and applying this
results on the $N_(0)$ channels defined by \eqref{multi_ampl} we
finally have:
\begin{multline}
  Z(|P_{T_{l}}|, D) \,=\, \\
\frac{1}{2^\frac{D N_0(T)}{2}}
\prod_{\alpha=1}^D
\left[\sum_{\{j_{p}\in \frac{1}{2}
    \mathbb{Z}_{+}\}}\sum_{\{j_{(r,p)}\}}\prod_{\{\rho
    ^{0}(p,q,r)\}}^{N_{2}(T)}\left\{
    \begin{array}{ccc}
      j_{(r,p)} & j_{p} & j_{r} \\
      j_{q} & j_{(q,r)} & j_{(p,q)}
    \end{array}
  \right\} _{Q=e^{\frac{\pi }{3}i}}\times \right.\\\left.
  \times \prod_{\{\rho ^{1}(p,r)\}}^{N_{1}(T)}\left(
    b_{j_{(r,p)}}^{j_{p}j_{r}}\right) ^{2}L(p,r)^{-2H_{j_{(r,p)}}}\cos
  (8\pi j_{p}\lambda (i))\frac{e^{-\frac{4\pi }{\theta
        (i)}j_{p}^{2}}}{\eta (e^{- 5\frac{4\pi }{\theta (i)}})}
\right]_{(\alpha)}
\end{multline}
where the subscript $(\alpha)$ indicates the contribution of the
$\alpha$-th direction.

%% file: osgauge.tex
\chapter[Inclusion of Open String gauge degrees of freedom]{Inclusion
of Open String gauge degrees of freedom: a proposal}
\label{ch:os_gauge}

A necessary step to rephrase our model in a gauge/gravity
correspondence connection would be to include open string gauge
degrees of freedom (propagating) along the boundaries of $M_\partial$.
As a matter of fact the model we built in previous chapters presents
an $SU(2)$ gauge symmetry at spacetime level. However, this symmetry
is due to the particular geometry of the system, and it does not allow
to define an appropriate gauge coloring neither of the ribbon graph
underlying the triangulation, nor of the boundary components. Thus, it
seems more appropriate to follow usual techniques in open string
theory, where a non Abelian gauge theory can be naturally included
into an open string model by a suitable assignation of non-Abelian
Chan Paton factors at the open string endpoints \cite{Paton:1969je}.
This will define some modifications in our model.  First of all,
vertex algebra operators (and consequently BIOs) will be valued in the
associated Lie algebra, thus ribbon graph amplitudes will be weighted
by an appropriate group factor.  The natural consequence will be the
coupling of the conformal field theory on each cylindrical end with a
background gauge field.  In this connection, we will show that we can
rephrase the arising of a target space non-Abelian gauge symmetry
which arise with this process with a change in the background matrix,
\ie{} the process is equivalent to move to another point of moduli
space of toroidal compactifications given in \eqref{lattice_modspace}.
According to \cite{Giveon:1994fu,Green:1995ga}, when the chosen
background is a fixed point under the extended $D$-dimensional
T-duality group, the annulus BCFT is equivalent to a
$\mathfrak{g}_{k=1}$ WZW model, where $\mathfrak{g}_{k=1}$ is a
level-one untwisted affine Lie algebra associated to a semisimple
product of simply laced Lie algebras of total rank $D$.

In the last section, we will try to understand how to extend the
gluing technique described in the previous chapters to the current
situation.

\subsubsection{Conventions about $U(N)$ algebra}

Let us represent $U(N) \sim SU(N) \times U(1)$. Thus we can choose a
base of U(N) generators as:
\begin{itemize}
\item The collection of $SU(N)$ generators:
  \begin{equation}
    \label{eq:sung}
    T^a \qquad a \,=\, 1 ,\,\ldots,\,N^2\,-\,1
  \end{equation}
satisfying the traceless condition  $\text{Tr}(T^a) = 0$.
\item $T^0 = C \mathbb{I}$, where $C$ is arbitrary constant. 
\end{itemize}
Let us impose the following normalization condition:
\begin{equation}
  \label{eq:ng}
  \text{Tr}(T^a\,T^b) \,=\, \frac{1}{2}\,\delta^{ab}.
\end{equation}
Equation \eqref{eq:ng} fixes $C\,=\,\frac{1}{\sqrt{2 N}}$.

The completeness relation for generators reads:
\begin{equation}
  \label{eq:gencomp}
  (T^a)^i_j\,(T^b)^k_l
\,=\,
\frac{1}{2}\,
\delta^i_l\,\delta^k_j
\qquad 
\begin{array}{l}
  a\,=\,1,\,\ldots,\,N^2 - 1 \\
i,\,j,\,k,\,l \,=\, 1,\,\ldots,\,N
\end{array}
\end{equation}

The $U(N)$ algebra is defined by:
\begin{subequations}
  \label{sub:alg}
  \begin{gather}
    \label{algas}
    \left[
      T^a\,,\,T^b
    \right]
    \,=\,
    i\,f^{a b c}\,T^c
    \quad \Longrightarrow \quad
    f^{a b c}\,=\,
    \frac{2}{i}\,\text{Tr}
    \left(
      \left[
        T^a\,,\,T^b
      \right]\,T^c
    \right)
    \\
    \left\{
      T^a\,,\,T^b
    \right\}
    \,=\,
    d^{a b c}\,T^c
    \quad \Longrightarrow \quad
    d^{a b c}\,=\,
    2\,\text{Tr}
    \left(
      \left\{
        T^a\,,\,T^b
      \right\}\,T^c
    \right)
    \label{algs}
  \end{gather}
\end{subequations}
where $f^{a b c}$ (resp. $d^{a b c}$) are the antisymmetric
(resp. symmetric) structure constants.

With the above conventions, a rapid computation shows:
\begin{subequations}
  \begin{gather}
    \sum_{b\,c}
    f^{a b c}\,f^{d b c}
    \,=\, 
    N\,(\delta^{a d} \,-\, \delta^{a 0}\,\delta^{d 0})\\
    \sum_{b\,c}
    d^{a b c}\,d^{d b c}
    \,=\, 
    N\,(\delta^{a d} \,+\, \delta^{a 0}\,\delta^{d 0})
  \end{gather}
\end{subequations}

\section{Chan-Paton factors on a RRT}

Chan-Paton factors are non-dynamical degrees of freedom which can be
added to open string endpoints. To show in a few words how it works,
let us consider an open string worldsheet (let us remember that the
cylindrical ends \cyl{k} is an one loop open string worldsheet) with
Neumann boundary conditions on each end, and let us suppose there is a
quark $q$ at one end of the string, ans an antiquark $\bar{q}$ at the
other end, where $q^i, \bar{q}^i = 1, \ldots, N$ are $N$-valued labels
associated to the some representations of a gauge group G.  Equations
of motions for $q_i$ simply require they to be independent from the
coordinate parametrizing the boundary, thus they can be interpreted as
charged, conserved under translations along the boundary. The target
space action automatically has $SO(N)$ symmetry when the charges are
real and $SO(2N)$ when they are complex. As a result, quantization
produces an associated Hilbert space of states on which a spinor
representation of respectively $SO(N)$ or $SO(2N)$ acts.  To restrict
to a subgroup of these two groups, (\ie to a subrepresentation on
which they act) it is necessary to reduce the Hilbert space. This can
be achieved introducing a Lagrangian multiplier in the action.
Unitarity of amplitudes actually restricts the admissible Lie algebras
associated to such terms to belong to the classical series, thus
admissible associated groups are $U(N)$ , $SO(N)$ or $Sp(N)$.

Let us restrict to $G = U(N)$, with $q^i$ in the defining
representation $N$. Quantization of $q^i$ generates $N$ conserved
charges attached to the propagating endpoints. The string for $U(N)$
should be oriented, since the charges at both ends of the string
transform under inequivalent representations of the gauge group. The full
string states now transform under the $N \otimes \overline{N}$
representation, namely the adjoint of $U(N)$. Thus, the generators
$T^a$, $a = 1, \ldots, \text{dim}U(N) = N^2$ labels the string states
belonging to the $N \times \overline{N}$ representation. The matrix
elements $(T^a)^i_j$ specify which charges $\overline{q}^i$ and $q_j$
are created at strings endpoints. The Fock space built over each
ground state $\vert 0, k ; i j \rangle$ is now given by $\mathcal{H}^{k}
\otimes N \otimes \overline{N}$.

For $G = U(N)$, $N \otimes \overline{N}$ transforms in the adjoint
representation, thus massless open string states are natural
candidates for Yang-Mills vector bosons with gauge group $U(N)$. The
same construction might have been done for $q^i$ in another
representations of $U(N)$, but, in these cases, string states would
have not rotated in the adjoint representation of the group, letting
the above interpretation fail.  Thus, the defining representation is
singled out by its relevance in view of Yang-Mills particle
interpretation of string states.

The string for $SO(N)$ or $Sp(N)$ should be unoriented, since charges
at both ends transform under the same representation of the group. If
we consider $f$ to be the defining representation of $SO(N)$, we
obtain string states in the adjoint of $SO(N)$ considering the only
the antisymmetric part of the tensor product of the two
representation, namely $(f \otimes f)_A$.  On the other hand we obtain
the string states in the adjoint of $Sp(N)$ by the symmetrized product
of its defining representation with itself. In each case we recover
exactly the the string states suitable for the Yang-Mills particles
interpretation.

The effect of a general background gauge field is accounted for by
including, for each boundary of the Polyakov path integral, a Wilson
line term $\mbox{Tr\,P\,exp}(-S_A)$, where $S_A$ represents a boundary
condensate of photon vertexes:
\begin{equation}
\label{eq:phot_bc}
S_A \,=\,  \int d \tau\, A_\alpha\, \partial_\tau X^\alpha 
\end{equation}

Preservation of conformal invariance impose the associate
$\beta$-functions to vanish. Equation $\beta_A = 0$ reduce, to the
leading order in the $\sigma$-model expansion, to the Yang-Mills
equation.

In our connection, let us suppose that each cylindrical end's open
string is decored with a suitable assignation of Chan Paton factors.
Thus, we can repeat the previous construction for the theory defined
over each cylindrical end. Let us assign to each open string running
one loop a suitable decoration. In this way, each vertex operator of
the open string spectrum will get as a prefactor a U(N) generator.

Now let us think to glue together the various cylindrical ends along
the pattern defined by the ribbon graph. Some considerations naturally
arise. Let us look again at figure \ref{BIO} on page \pageref{BIO}:
since the two open string worldsheet boundaries which get glued along
one edge of the ribbon graph pattern have opposite orientation, it
seems natural to let the associated quarks to fall into opposite
representations of the gauge group. The same holds for each edge of
the ribbon graph, which thus acquire a well defined gauge coloring.
Thus, we can think to the ribbon graph underling the triangulation as
a sort of image of the gauge coloring of the boundaries of the
surface $\partial M$.

\begin{figure}[!t]
  \centering
  \includegraphics[width=.75\textwidth]{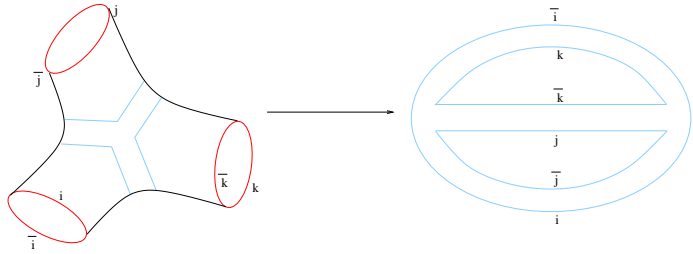}
  \caption{The ribbon graph associated to the open surface dual to
    three-punctured sphere}
  \label{fig:ribbon3p}
\end{figure}

About the spectrum, we can think to define the gluing process once
again introducing a unique spectra of the common boundary theory: the
$(p,q)$ open string will be describe by a unique object being got by a
sort of ``fusion process'' of the Chan-Paton factors of the original
strings and which retains the open string data associated to the the
free ends of the open unglued strings. However, since the fact that
string states rotate into the adjoint representation of the gauge
group is fundamental to their limiting description as particles, we
want an object obtained by the product of the two original Chan-Paton
factors and belonging to the original algebra. The only products
obeying this constraint are the symmetric and antisymmetric products of
generators, thus each BIO belonging to the $(p,q)$ BCFT spectrum will
be decored by a $U(N)$ generator $T^a_{i l}$ defines as:
\begin{equation}
  T^a_{i l}(p,q) \,=\,
  f^{abc}\,\left[T^b_{ij}(p)\,,\,T^c_{jl}(q)\right]
  \,+\,
  d^{abc}\,\left\{T^b_{ij}(p)\,,\,T^c_{jl}(q)\right\}
\end{equation}

Denoting the conformal structure of BIOs in formula
\eqref{eq:final_bio} with a collective index $\Xi(p,q)$, non-Abelian
BIOs will be $\psi_{\Xi(p,q)}^a \,=\, T^a \psi_{\Xi(p,q)}$. As a
consequence, two and three-point functions of BIOs get a prefactor
given by a trace of generators. 

Let us notice that the above construction allows to reissue the
description of the open string gauge data transmitting all the
informations on the associated ribbon graph. Naively speaking, to each
gauge coloring of $\partial M$ boundaries we can uniquely associate a
gauge coloring of the underlying ribbon graph ``displaying'' the
$U(N)$ charges left after the gluing procedure from the outer to the
inner boundary of each cylindrical end (see fig. \ref{fig:ribbon3p}).

In this connection, it is easy to understand that only the two points
function between opposite directed BIOs (in the sense defined in
appendix \ref{sec:biobcft}) makes sense. In this case, the two point
function will be simply weighted by the $U(N)$ generators
normalization constant.  The modified BIOs' algebra can be easily
retrieved by (anti)symmetrizing the product of generators:
\begin{multline}
  \psi^a_{\Xi_1(r,p)}(\omega_r) \,
  \psi^b_{\Xi_2(q,r)}(\omega_q)
  \,\sim\,
  \frac{1}{2}\sum_{\Xi_3,\,c}
  \left\vert
    \omega_r\,-\,\omega_q
  \right\vert^{H(q,p)\,-\,H(r,p)\,-\,H(q,r)} \times \\
  \left(
    i f^{a b c} + d^{a b c}
  \right)
  \mathfrak{C}(\Xi_1(r,p),\Xi_2(p,q),\Xi_3(q,r))
  \psi^c_{\Xi_3(q,p)}(\omega_q),
\end{multline}

Thus, three-point functions will be decored by the sum of the
antisymmetric plus symmetric structure constants.

\section[From triangulations to strings]{Rephrasing the geometrical
  data of the triangulation in term of string quantities}
\label{lalpha}
\begin{figure}[!t]
  \centering
  \includegraphics[width=.50\textwidth]{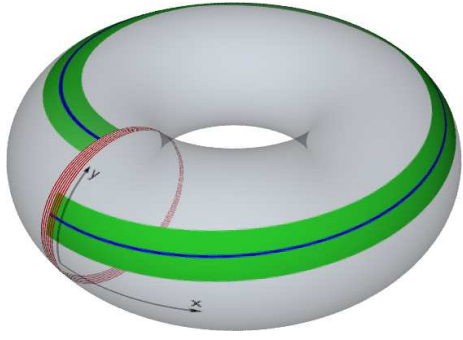}
  \caption{A cylindrical end winding in a two dimensional background}
  \label{fig:toro}
\end{figure}

The overall picture we are dealing with sees a stack of $N$ coincident
D-$p$-branes winding around some (or all) of cycles of the toroidal
background. In this connection, we are considering cylindrical ends as
one-loop open string worldsheets. As prototype picture, we can
consider the first drawing in fig. \ref{fig:toro}. Red circles are the
worldvolumes of N coincident D-0-branes, while the green strip is the
pictorial representation of a cylindrical ends injected by the maps
$X$ and $Y$ on the two dimensional torus. In particular, looking at
the cylindrical end as a one loop open string worldsheet, the
associated open string winds one time around the $X$ direction, while
it does not wind around $Y$. Thus, the injection maps are characterized
by the following values of winding numbers and center of mass
momentum:
\begin{subequations}
  \begin{gather}
    \mu^X_{(O)} = 0 \qquad \nu^X_{(O)} = 1  \\
    \mu^Y_{(O)} = 1 \qquad \nu^X_{(O)} = 0  
  \end{gather}
\end{subequations}

Performing a worldsheet-duality transformation, thus looking at the closed
string which is emitted by the brane and it is reabsorbed after having
run around the $X$ direction once, actually we exchange the role of
total momentum and winding, thus the Hilbert space states are
characterized by the following values of quantum numbers:
\begin{subequations}
  \begin{gather}
    \mu^X_{(C)} = 1 \qquad \nu^X_{(C)} = 0  \\
    \mu^Y_{(C)} = 0 \qquad \nu^X_{(C)} = 1  
  \end{gather}
\end{subequations}

In this connection, computation of the target space spectrum shows
that, when we consider an open string whose endpoint lay on a
D-$p$-brane, the background fields depends only upon coordinates on
the brane world-volume (\ie{} on coordinate $\xi^\mu$, $\mu =
0,\,\ldots\,\,p$ parametrizing the brane embedding in the target
space). Moreover, massless states vertex operators spacetime
components actually divide in two set: those of the first set are
components of a Yang-Mills field living on the D-$p$-brane
world-volume, with $p + 1$ components tangent to the hyperplane,
$A^i(\xi^\mu)$, $\mu, i = 0,\,\ldots,\,p$, while the others
$\phi^m(\xi^\mu) \,=\, \phi^m_b T^b$ are $N \times N$ matrices, whose
entries are scalars field from the brane world volume point of view.
They describe the specific shape of the brane  in the target space.

In ordinary (bosonic) string theory, D-branes are dynamic objects, and such
that, they must be able to respond to the values of the various
background fields in the theory.  This is obvious if one consider that
actually brane's location and shape is controlled by the various open
strings ending on them, and which indeed interact with background
fields.  A world-volume action describing this dynamic has been
derived exploiting $T$-duality in target space (for a comprehensive
review, see \cite{Johnson:Dbranes}, chapter 5). If we consider a
D-$p$-brane in ordinary $D$-dimensional open bosonic string theory,
it turns out to be, in the Abelian case,
\begin{equation}
  \label{eq:dbi}
  S_p \,=\, -T_p \,\int d^{p + 1} \xi \,\e^{-\Phi} \,
\mbox{det}^{\frac{1}{2}} 
\left[G_{i j} \,+\, B_{i j} \,+\, 2 \pi \alpha' F_{i j}
\right]
\end{equation}
In equation \ref{eq:dbi}, $G_{i j}$ and $B_{i j}$ are the
pull-back of background fields to the brane, $\alpha'$ is the Regge
slope and $T_p$ is the D-brane tension defined as:
\begin{equation}
  \label{eq:ten}
  T_p \,=\, \frac{\sqrt{\pi}}{16 \kappa_0} 
(4 \pi^2 \alpha')^{\frac{11 - p}{2}}
\end{equation}

The non-Abelian case is quite more subtle, because the various
background fields typically depends on the transverse coordinates,
which, as stated before, are actually matrix valued in the gauge
algebra. Thus, introduction of non-Abelian quantities, implies to
perform a trace on the integrand, in order to get a gauge invariant
object as action. It has been shown that considering the symmetrized
trace over the gauge indexes, we get a result consistent with various
studies of both scattering amplitudes and non-Abelian soliton
solutions (see \cite{Johnson:Dbranes}, section 5.5, and references
therein). Thus, according to these remark, it has been shown that the
worldvolume action associated to $N$ coincident D-$p$-branes is:
\begin{multline}
\label{eq:nadbi}
S_p \,=\, - T_p \,\int d^{p + 1} \xi \, \mbox{STr} \left\{
  \mbox{det}^{\frac{1}{2}} \left[E_{i j} E_{i m}(Q^{-1} - \delta)^{m
      n}E_{n j} \right.\right.\\\left.\left.  \,+\,
    2\,\pi\,\alpha'\,F_{i j}\right] \mbox{det}^{\frac{1}{2}}[Q^n_m]
\right\}
\end{multline}
where $E$ is the background matrix introduced in \eqref{eq:bm}, and
$Q^n_m \,=\, \delta^n_m + i 2 \pi \alpha' \left[\phi^n , \phi^p\right]
E_{p m}$.

If we expand this Lagrangian to the second order in the gauge field,
and noting that 
\begin{equation}
  \mbox{det}[Q^n_m] \,=\, 1 \,-\,
  \frac{(2\,\pi\,\alpha')^2}{4}\,
  \left[\phi^n,\phi^m\right]\,
  \left[\phi^n,\phi^m\right] \,+\, \ldots
\end{equation}
we can write the leading order of action \eqref{eq:nadbi} as
\begin{equation}
  S_p \,=\, -\frac{T_p(2\,\pi\,\alpha')^2}{4} \int d^{p + 1} \xi\,
e^{- \Phi} \, Tr\left[ F_{a b} F^{a b} \,+\, 2 \mathcal{D}_a \phi^m
  \mathcal{D}^a \phi^m  + \left[\phi^m , \phi^n \right]^2\right]. 
\end{equation}
This is exactly the dimensional reduction of a $D$-dimensional
Yang-Mills Lagrangian, displaying the non trivial scalar potential
involving commutator for the adjoint scalars. This expansion allows
to write the $(p + 1)$-dimensional Yang-Mills coupling for the theory
on the brane:
\begin{equation}
  \label{eq:gym}
  g_{YM,p}^2 \,=\, g_s T^{-1}_p (2\,\pi\,\alpha')^{-2}
\end{equation}
where $g_s = e^\Phi$ is the usual worldsheet string coupling. 

Let us turn to the Polyakov string defined on each cylindrical end via
the action in \eqref{eq:action1}. We can easily rewrite it in a
stringy shape by introducing dimensionful coordinates via the mapping
$X^\alpha(\zeta(k), \bgz(k)) \rightarrow
\frac{l(k)}{2}\,X^\alpha(\zeta(k), \bgz(k))$. Polyakov action becomes:
  \begin{equation}
  \label{eq:act2}
  S(k) \,=\, \frac{1}{2 \pi l(k)^2}
\int d \zeta(k)\,d \bgz(k)\,
\partial\, X^\alpha\, \overline{\partial} \bar{X}_\alpha   
\end{equation}
thus suggesting the parameter $l^2(k)$ to play a role analogue to that
of the Regge slope in the discrete $(k)$-sector. However, from a
classical point of view, since the above analogy has been derived from
a fixed triangulation of the worldsheet, we are not able to associate
it a real physical meaning. Moreover, if we consider the collection of
all the discrete sectors, the above identification seems to lead to a
strange non-sense. As a matter of fact, if we consider the full
surface $M_\partial$, we deal with cylindrical ends having their inner
boundaries glued together along the pattern defined by the ribbon
graph, while outer ones lay on (at least) one stack of $N$ tiled
D-branes.

\begin{figure}[!t]
  \centering
  \includegraphics[width=.45\textwidth]{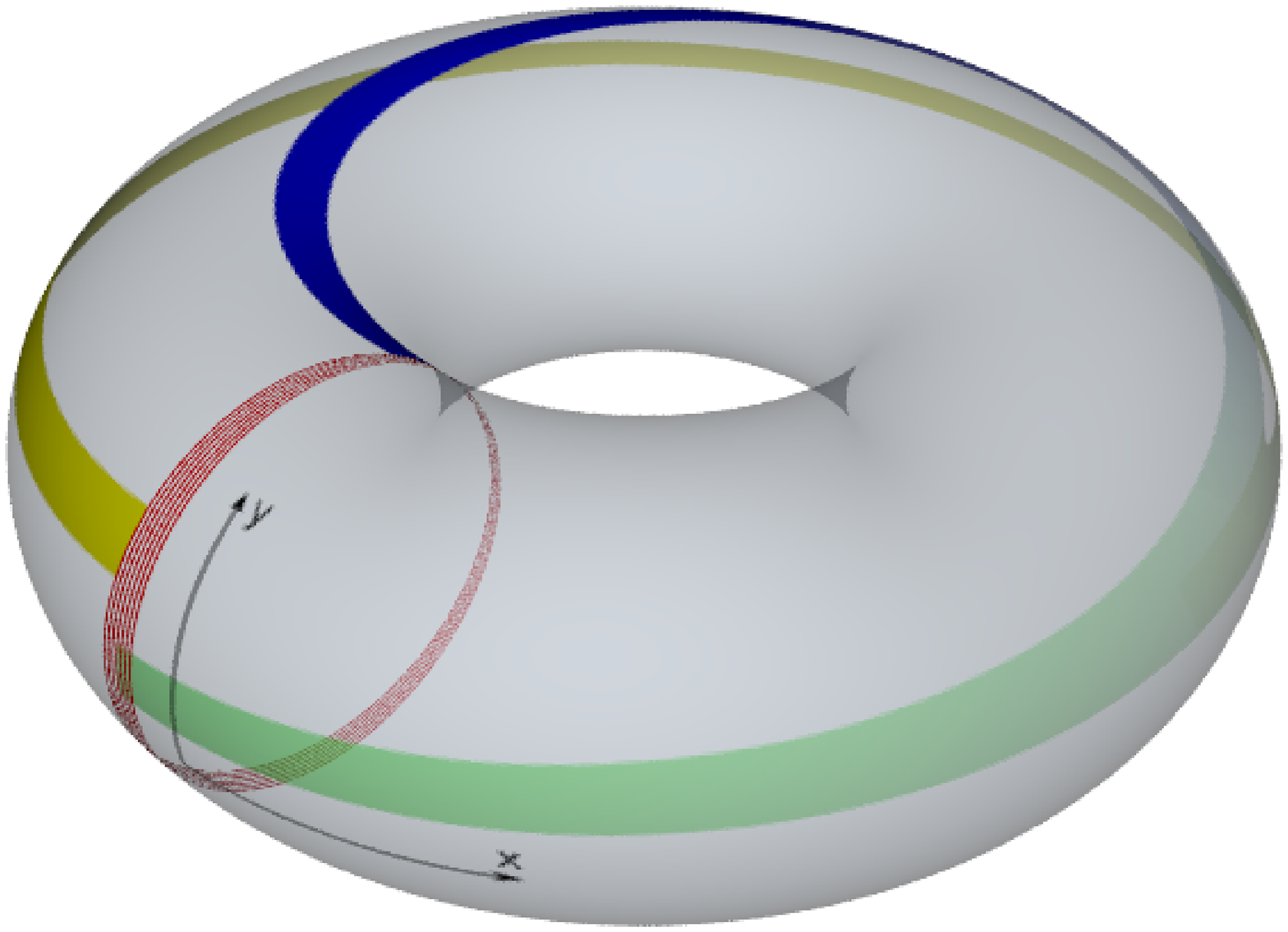}
  \hspace{3mm}
  \includegraphics[width=.45\textwidth]{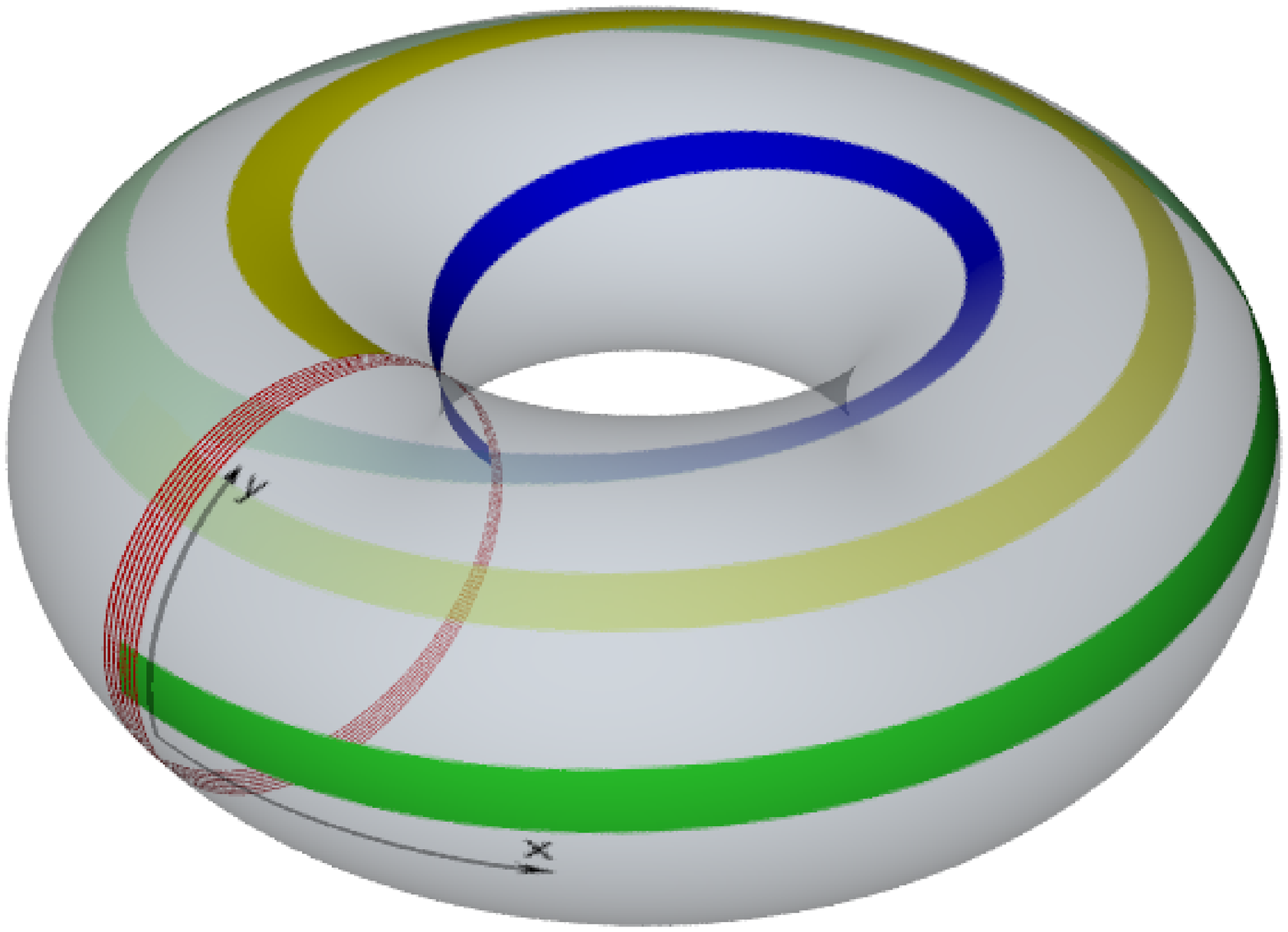}
  \caption{Some simple configurations with three cylindrical ends
    winding in a two dimensional flat toroidal background}
  \label{fig:toro2}
\end{figure}

In particular, let us assume exactly the situation depicted in
fig. \ref{fig:toro2}. We are considering again the open surface dual
to the three-punctured sphere and which has been sketched in
fig. \ref{fig:ribbon3p}. In the left drawing each string has zero
winding number in both direction, while, in the right one, each sting
winds once around $X$ and does not wind around $Y$. Moreover, let us
assume that the outer boundaries of each cylindrical end lays on the
same stack of D-branes. According to the above remarks about the role
played by the parameter $l(k)$, it may seems strange that each copy of
the model, each being defined over one cylindrical end, has associated
an independent value of the ``$\alpha'$-analogue'' parameter. It may
seem meaningless especially if we consider the cylindrical ends
boundaries laying on the same D-brane, whose tension is defined in
term of  $\alpha'$ via the relation \eqref{eq:ten}.

An elegant solution of this puzzle can be achieved if we consider the
problem from a ``quantistic'' point of view: the construction of the
partition function for the bosonic string involve a sum over all
possible configurations \ie{} over all possible ribbon graphs. In this
framework clearly we are considering an average over all possible
parimeters associated to the triangulation, and thus the Regge slope
analogue is not a fixed value associated to each cylindrical end but
it arises as a statistical average over all possible configurations.
A furhter development arise if we performthe continuous limit of such
a discrete model. For discrete planar surfaces this is usually
performed as limit in which the lengths of boundary components go to
infinity, while a suitable cutoff $a$ goes to zero, in order to
maintain the product $L(k) a$ finite.  Thus, if we identify on each
cylindrical end the (statistical average) of the characteristic length
$l(k)$ with:
\begin{equation}
  \label{eq:lk}
  l(k) \,=\, \sqrt{L(k)\,a}
\end{equation}
and we perform the continuum limit requiring that all the (statistical
averages) of finite values of the above products converge to a given
value, we can actually interpret such a result as an analogue of the
Regge slope which, according to relation \eqref{eq:ten}, determine the
tension of the brane naturally included in our model.

Moreover, this defines a precise map between the geometrical data
associated to the triangulation and string quantities.

\section[Extending the
  gluing procedure]{Coupling with background gauge potential: extending the
  gluing procedure}

Let us consider a $D$-dimensional background, in which each direction
$X^\alpha$, $\alpha=0,\,\ldots,\,D-1$ is compactified at the
T-duality self dual radius $\Omega^\alpha = \sqrt{2}$. As explained in
the last chapter, the theory in each direction decouples and the model
present a $(SU(2)_L \times SU(2)_R)^D$ gauge symmetry. In each
direction we can apply previous chapter's constructions. Thus, let us
consider the $k$-th cylindrical end, \cyl{k}, and let us interpret it
as worldsheet associated to one-loop open string diagram. Consistently
with the gluing process, let us assume the injection maps (target
space coordinates) to obey to simple Neumann boundary conditions on
the inner boundary. On the other hand, on the outer boundary, let us
assume to have $p+1$ directions satisfying Neumann boundary conditions
and $D - p - 1$ directions obeying Dirichlet's ones:
\begin{equation}
\left\{ X^\alpha \right\}
\,\doteq\,
\left\{ X^i ,\, X^m\right\},
\text{with} \, 
i\,=\,0,\,\ldots,\,p,\,\, 
m\,=\,p + 1,\,\ldots,\,D - 1 
\end{equation}
According to formulae \eqref{eq:dir_bs} and \eqref{eq:boundary_par},
these can be characterized by:
\begin{align}
  &\text{Neumann directions}:    \qquad   
  \left( 
    \Vert N(0)_{(k)} \Rangle_{sd},\, \mathbb{I} 
  \right)   \\
  &\text{Dirichlet directions}:  \qquad 
  \left( 
    \Vert N(0)_{(k)} \Rangle_{sd},\, e^{-i \pi J_0^1(k)} 
  \right)  
\end{align}

In particular, with the above choice, we are dealing with a
D-$p$-brane laying along the $X^0,\,\ldots,\,X_p$ directions, thus, if
we introduce the D-brane world-volume parameters
$\xi_0,\,\ldots,\,\xi_p$, we
have chosen to parametrize the brane world-volume with 
$\xi_0 = X_0,\,\xi_1= X_1,\,\ldots,\xi_p = X_p$.
 
As described in the previous section, in order to endow the brane with
an interesting dynamic, we have to couple the model to a background
gauge field living on its worldvolume: this can be worked out
introducing non chiral truly marginal deformation defined by the
following boundary action (\cite{Callan:1988wz,Callan:1995xx}):
\begin{equation}
  \label{eq:gaugeba}
  S_A\,=\, \int d \tau 
  \left[
    \sum_{i=0}^{p} A_i(X^i) \partial_\tau X^i  
    \,+\,
    \sum_{m = p + 1}^{D - 1} \phi_m(X^i) \partial_\sigma X^m  
  \right],
\end{equation}
where we have taken the boundary to lay at constant $\sigma$. The $A_i
\,=\, A_i^a T^a $ are Lie algebra valued gauge fields on the D-brane,
while the entries of the $N \times N$ matrices $\phi_m \,=\, \phi_m^a
T^a$ are scalar from the world volume point of view: they describe the
motion of the brane in the transverse space. 
Such a boundary term will act as a non chiral deformation of the
original quantum model by means of a truly marginal  operator.

For the sake of simplicity, let us assume the brane static in the
transverse space imposing $\phi_m = \mathbf{0}_{N \times N}$ $\forall m
= p+1 \ldots D-1$.  Moreover, let us consider constant electric and
magnetic fields along the brane worldvolume, so that the boundary term
reads:
\begin{equation}
  \label{eq:gaugebaF}
  S_A\,=\, 
  F_{i j}
  \int d \tau 
  X^j
  \partial_\tau X^i.  
\end{equation}

Let us move to the Abelian subsector. Such a limiting situation can be
achieved including in each Neumann direction of the T-dual theory a
Wilson line $A_i =
\frac{diag\{\theta_1,\,\ldots,\,\theta_N\}}{\Omega_i}$ which breaks
$U(N)$ to $U(1)^N$. At spacetime level, the global effect will be a
displacement of the position of the N D-branes. Thus, now we are
dealing with N separated D-branes, and we can assume that open strings
generating the cylindrical ends have their outer endpoint attached
over different brane.  

Thus, in each $(k)$-subsector, we can couple the open string with a
different electromagnetic potential $A_i(k)$, allowing the boundary
term to have a simple closed string interpretation. As a matter of
fact, the Lagrangian in \eqref{eq:gaugebaF} is the integral of a total
derivative, thus, after a short computation, $S_A$ can be rewritten as:
\begin{equation}
  \label{eq:gaugebaB}
  S_{A(k)}\,=\, 
  F_{i j}(k)
  \int d \zeta(k) d \bar{\zeta}(k)
  \partial X^i(k) \, 
  \bar{\partial} \bar{X}^j(k).
\end{equation}
Comparing last formula with formula \eqref{eq:action}, we can state
that, in the Abelian subsector, the inclusion of such a boundary term
is equivalent to moving to a different point in flat toroidal
background moduli space (see formula \eqref{lattice_modspace}):
\begin{center}
\begin{tabular}{|ccc|}
\hline
Config. $\mathcal{A}$ & & Config. $\mathcal{B}$\\
\hline
$G_{\alpha \beta} \,=\, \mathbb{I}_{D \times D}$ & & 
$G_{\alpha \beta} \,=\, \mathbb{I}_{D \times D}$ \\
$B_{\alpha \beta} \,=\, 0 \: \forall \alpha,\,\beta$ 
& $\Longleftrightarrow$ &
$B_{\alpha \beta} \,=\, 4 \pi \Lambda_{\alpha \beta}$ \\
$F_{\alpha \beta} \,=\, \Lambda_{\alpha \beta}$ 
& &
$F_{\alpha \beta} \,=\, 0 \: \forall \alpha,\,\beta$\\
\hline
\end{tabular}
\end{center}

On this wise, we can directly choose a peculiar flat toroidal
background, with associated suitable  values of the background matrix
$E$ entries (see formula \eqref{eq:bm}), and rephrase the $B$-field
dependence in term of a gauge field strength associated to a potential
living on the D-brane world-volume.
  
As explained in the previous chapter, the full moduli space of flat
toroidal backgrounds is parametrized by the coset space:
\begin{equation}
  O(D,D,\mathbb{Z}) \,/\, O(D,D,\mathbb{R}) \,\backslash\, O(D) \times O(D).
\end{equation}

In the previous formula  $O(D,D,\mathbb{Z})$ is the generalized
T-duality group.

Thus, higher dimensional toroidal compactifications are described by
non-trivial background fields $B$ and $G$ and, in such a given
background, the maximally enhanced symmetry points are those fixed
under the action of $O(D,D,\mathbb{Z})$, which elements are generated
by a combination of:
\begin{itemize}
\item a conjugations with a matrix $M \in SL(d, \mathbb{Z})$;
\item an integer shift of B by means of antisymmetric, integer valued,
  matrix $\Theta$; 
\end{itemize}
\ie{} the rotated background matrix will be 
\begin{equation}
  \label{eq:rote}
  E'\,=\, M^t \,(E + \Theta)\, M
\end{equation}

In those special points of the moduli space in which $E' = E$, we can
obtain extended target space symmetries with respect to semi-simple
products of simply laced Lie algebras (thus belonging to the A-D-E
decomposition) of total rank $D$.

In particular, the maximally enhanced symmetry background can be
chosen in the following way\cite{Giveon:1994fu}. If
$C_{\alpha\beta},\,\alpha, \beta = 1, \ldots, D$, is the
Cartan matrix of the simply laced Lie algebra of total rank D, then
the background fields are chosen as:
\begin{subequations}
\label{bkBG}
  \begin{gather}
    G_{\alpha \beta} \,=\, \frac{1}{2}  C_{\alpha \beta}  \\
    B_{\alpha \beta} \,=\, G_{\alpha \beta} \,\, \forall\, \alpha \,>\, \beta
    \qquad
    B_{\alpha \beta} \,=\, - G_{\alpha \beta} \,\, \forall\, \alpha \,<\, \beta
    \qquad
    B_{\alpha \alpha} \,=\, 0 \, \forall \alpha
  \end{gather}
\end{subequations}
With these choices, the background matrix $E = G + B$ is an element of
$SL(d,\,\mathbb{Z})$, thus it is fixed under the action of
$O(D,D,\mathbb{Z})$. 

Let us consider, as an example, an $O(D,D,\mathbb{Z})$ transformation
generated by $M = E^{-1}$ and $\Theta = E^\dag + E$. The duality map
states \eqref{eq:rote} tells: $E' \,=\, E^{-1}$ and, if $G =
\mathbb{I}_{D \times D}$ and $B = \mathbf{0}_{D \times D}$, this is
exactly the case of $(SU(2)_L \times SU(2)_R)^D$. 
 
With the above choice of static gauge, this is exactly the situation
for the $D - p - 1$ Dirichlet directions.
Thus, the extended symmetry group associated to the action
\eqref{eq:gaugebaB} will be:
\begin{equation}
\mathbb{G}_D \,=\, ({G}_{p+1} \,\times{G}_{p+1}) \,\times\,  (SU(2)_L \times
  SU(2)_R)^{D - p - 1} 
\end{equation}

\subsection{Chiral currents and boundary states}
As we did for one compactified boson at the self dual radius, the
level-one untwisted Kac-Moody algebra arising in this connection can be
represented in term of the bosonic coordinate by means of the usual
vertex operators construction\cite{DiFrancesco}. In the cylinder
transverse channel (\ie{} in terms of coordinate $\zeta(k)$ and $\bgz(k)$ on
the $k$-th cylindrical end), the left moving and right-moving currents
are, respectively (let us omit the polytopal index $k$)  
\begin{subequations}
  \label{eq:cD}
  \begin{gather}
    \label{eq:lcD}
    H^\alpha(\zeta) \,=\, 
    \partial\,X^\alpha, \qquad 
    E^\lambda (\zeta) \,=\, c(\lambda) \,
    \ordprod{e^{i \lambda_\alpha X^\alpha}}\\
    \label{eq:rcD}
    \overline{H}^\alpha(\bgz) \,=\, 
    M^\alpha_\beta
    \bar{\partial}\,\bar{X}^\beta, \qquad 
    \overline{E}^\lambda (\bgz) \,=\, - \bar{c}(\lambda) \,
    \ordprod{e^{i \lambda_\alpha M^\alpha_\beta \bar{X}^\beta}}.
  \end{gather}
\end{subequations}
 
Here $H^\alpha(\zeta)$ and $ \overline{H}^\alpha(\bgz)$ denotes the
elements in the Cartan subalgebra of ${\mathfrak{g}_D}_{k=1}$, while
$\{\lambda\}$ is the set of roots (positive plus negative) of the
associated semi-simple product of Lie algebras. 
The objects 
$c(\lambda)$ and $\bar{c}(\lambda)$ are $\mathbb{Z}_2$ values
cocycles. They are operators acting on the Fock spaces, and they
depend only upon the momentum part of the free-boson zero modes. Their
inclusion let the combinations of the above currents,
$J^\alpha_L(\zeta)$, to satisfy the correct OPE:
\begin{equation}
  \label{eq:wzwcurr_ope}
  J^a(\zeta) J^b(\zeta') \,\sim\,
\frac{\delta^{a b}}{(\zeta - \zeta')^2}
\,+\,
\sum_c i\,f^{a b c}  
\frac{J^c(\zeta')}{\zeta - \zeta'}
\end{equation}
If we introduce the Laurent expansion, $J^a(\zeta) = \sum_{n \in
  \mathbb{Z}} \zeta^{- n - 1} J^a_n$, we get the commutation relation
for the affine algebra ${\mathfrak{g}_D}_{k=1}$:
\begin{equation}
  \left[ J^a_m\,,\, J^b_n \right]
\,=\,
\sum_c i\,f^{a b c}  \,J^c_{m + n} \,+\,
m\,\delta^{a b}\,\delta_{m + n, \, 0}. 
\end{equation}
Obviously, the same holds for the antiholomorphic sector.

Limiting to the trivial automorphism in equation \eqref{eq:cont_cond},
we get the following constraints on the boundary states:
\begin{equation}
  \left(
    J^a_n + \overline{J}^a_{-n}
  \right)
  \Vert B \Rangle
\end{equation}

Vertex operators for the closed string massless vector states are the
D-dimensional extensions of formula \eqref{eq:massl_vertex}:
\begin{gather}
  \label{eq:massl_vertex_na}
  V^{a \alpha}_P \,=\, 
  J^a(\zeta) \bar{\partial} \bar{X}^\alpha(\bar{\zeta}) 
  e^{i P \cdot X(\zeta) + \bar{X}(\bar{\zeta})}\\
  \bar{V}^{a \alpha}_P \,=\, 
  \bar{J}^a(\bar{\zeta}) \partial X^\alpha(\zeta) 
  e^{i P \cdot X(\zeta) + \bar{X}(\bar{\zeta})}
\end{gather}
where $P\,=\,p_L \,+\, p_R$ is the total center of mass momentum of
the closed string. Once again, vertex operators for new open string scalar
states attached to the boundaries $|\zeta|=1,|\zeta|=\q$ can be
written, in the closed string channel, as:
\begin{equation}
  S^a_P \,=\, J^a(x) e^{i P X(x)},  
  \label{eq:open_vert_na}
\end{equation}
where, as usual, we have parametrized the boundary with $x \doteq
\Re{z(k,\circ)[\zeta(k)]}$.

\subsection{Gluing along the Ribbon Graph: a proposal}
\label{sec:gluingab}

The occurrence of extra massless boundary operators
\eqref{eq:open_vert_na} at the enhanced symmetry point can be
obviously rephrased as the arising of new truly-marginal perturbing
boundary operators, which give rise to the enhanced symmetry also in
the open string spectrum. Such operators came associates with
perturbing terms in the action of the form:
\begin{equation}
  \label{eq:pa}
  {S'}_{g} \,=\,\int d x 
  \,\left(
    \sum_\lambda g_\lambda\,e^{i\lambda_\alpha\,X^\alpha} 
\,+\, 
    \sum_\alpha g_\alpha \partial_x X^\alpha
  \right)\vert_{|\zeta(k)|=1},
\end{equation}
where $(g_\lambda,g_\alpha)$ are real coupling constant, which give
rise to the enhanced symmetry in the open sector when $\lambda_i$
takes special values. In \cite{yegulalp} it was shown that these
particular values arise when we can define a set of vectors:
\begin{equation}
  \hat{\lambda}_\alpha \,=\, 
  (\delta^\alpha_\beta + M^\alpha_\beta) 
  \,\lambda^\beta  
\end{equation}
(where $M=\frac{G + B}{G - B}$), such that the  $\hat{\lambda}$ are
roots of the simply-laced algebra $g_D$. According to deformation
rules of previous chapter, this lead to a modified gluing condition on
the boundary:
\begin{equation}
  J^a(\zeta) \,=\,
 \gamma_{g_a\bar{J}^a_0} (\bar{J}^a)(\bgz), 
\end{equation}
while the effect on the boundary state is a rotation with respect to
the left-moving zero modes of the currents:
\begin{equation}
  \Vert B \Rangle_g
 \,=\,
e^{g_{\hat{\lambda}} E^{\hat{\lambda}}_0 + g_\alpha H^\alpha_0}
  \Vert B \Rangle_g
\qquad\Longrightarrow\qquad
  \Vert B \Rangle_g
 \,=\,
e^{\sum_a g_a J^a_0}
  \Vert B \Rangle
\end{equation}

This construction, perfectly equivalent to the one arising when we
deal with only one injection map, allows us to make a concrete
hypothesis about the possible parametrization of boundary states in
this higher dimensional model.  If we are able to identify a
generating boundary state $\Vert B_0\Rangle$ equivalent to $\Vert N(0)
\Rangle_{sd}$, we can hypothesize to identify Cardy's boundary states
associated to the level one $\mathfrak{g}_D$ WZW-model with the
deformations of $\Vert B_0\Rangle$ with central elements of the group
$\mathbb{G}$. 
Once proved this statement, we could apply the same construction
of previous chapters to write an open string amplitude over $\partial
M$, with the particular choice of background fields in each
cylindrical end given by \eqref{bkBG}.

If we consider the simpler case of $\mathbb{G}$ being a simple Lie
group, strong indications of the above come if we notice that the
number of elements of the (discrete) center of $\mathbb{G}$,
$B(\mathbb{G})$, coincide with the number of level one modules
associated to the (simply-laced) simple Lie algebra $\mathfrak{g}_D$.
Moreover the center of the group is isomorphic to the group of outer
automorphisms associated to the affine simply laced Lie algebra
$\hat{\mathfrak{g}}_D$\footnote{ As a matter of fact, the relation
  between $\mathcal{O}(\hat{\mathfrak{g}}_D)$ and $B(\mathbb{G})$ is
  even stronger because if we consider an element $A \in
  \mathcal{O}(\hat{\mathfrak{g}}_D)$, we can define the associated
  element $b \in B(\mathbb{G})$ as the $\mathcal{S}^{ext}$ dual to
  $A$:
\begin{equation}
  b \,=\, {\mathcal{S}^{ext}}^\dagger\,A\,\mathcal{S}^{ext}
\end{equation}}:
\begin{equation}
  \mathcal{O}(\hat{\mathfrak{g}}_D) \,=\, B(\mathbb{G}).
\end{equation}
Remembering that $\mathcal{O}(\hat{\mathfrak{g}}_D)$ maps the set of
dominant weights which label the irreducible representations of the
affine algebra, this suggests the possibility to generate the full set of Cardy
boundary states associated to the (level one) $\hat{\mathfrak{g}}_D$
WZW model via the action of elements of $B(\mathbb{G})$ on the
boundary state associated to the identity element of $\mathbb{G}$.

% from the same definition of center of a group. As a matter of fact,
% since central elements of $\mathbb{G}$ would commute with all
% elements of the group itself, they are the only ones acting
% trivially on the gluing automorphism, but not trivially on the
% boundary state.

\section{Conclusions and perspectives}

In spirit of understanding the (mathematical and physical) origins of
open/closed string dualities, in previous sections we have
investigated some aspects of coupling of a bosonic string theory with
the peculiar geometry arising when we uniformize the singular
Euclidean structure naturally associated with the Regge polytope
barycentrically dual to a Random Regge Triangulation.

The worldsheet process guiding the holes glueing has been made precise
only at a topological level, while in the more complicated $AdS/CFT$
framework we are still far from understanding the origin of such a
behavior pattern.

In this sense, a remarkable hint has been recently given by Gopakumar,
who intuited that such a process must take place as a change of
variable at level of integrand in the sum over moduli space
parametrizing the free field $\mathcal{N} = 4$ SYM
correlators\cite{Gopakumar:2005fx} and suggested a concrete way of
associating a closed Riemann surface to a gauge theory correlator.
However, the argument introduced by Gopakumar is quite general, since
up to now it has been applied only to the free field theory sector,
thus leaving open the question about a possible application to
interacting theories.

In this connection, it is straightforward to notice that the key
object is represented by the underlying geometrical structure, namely
the skeleton graph naturally associated to gauge correlator and which
can be considered as dual to a triangulation of the Riemann surface
defining the closed string worldsheet. In this connection, in view of
the two dual uniformizations of the Euclidean structure naturally
associated to a RRT and introduced in \cite{Carfora3,Carfora1} which
provide a concrete algorithm to switch from an open to a closed
Riemann surface, it seems that a better understanding of the interplay
between geometrical quantities associated to a discrete realization of
the worldsheet and the analytic quantities of string theory, may
provide a key step towards the understanding of mathematical aspects
of string dualities.

The main goal of previous section has been the description of a very
reliable way of coupling an ordinary bosonic open string theory with
the discrete geometry defined by the uniformization of $M_\partial$. In
particular, exploiting BCFT techniques, we have been able to furnish a
concrete proposal about how to include open string gauge degrees of
freedom in our model. Not less important, we have been able to define
a natural interplay between the geometrical quantities parametrizing
the uniformization defined over $M_\partial$ and the Regge slope
$\alpha'$.

Wide investigation areas are still open. The inclusion of open string
gauge degrees of freedom itself is not completely fixed. In the
abelian case the statement sketched at the end of section
\ref{sec:gluingab} needs some further investigations. Moreover,
nevertheless the nice picture presented in section \ref{lalpha}, the
extension of the above proposal to the non-Abelian case is quite far
from being understood.

Furthermore, a very interesting point to address would be investigating
the possible extending of the above glueing procedure to minimal
string theories\cite{Ginsparg:1993is,DiFrancesco:1993nw}, which,
despite of their simplicity, they are interesting laboratories for the
study of string theories.

%% file: cosmo.tex
\chapter[Geodesics on maximally non compact cosets
$\mathrm{U}/\mathrm{H}$] {Geodesic on maximally non compact cosets
  $\mathrm{U}/\mathrm{H}$ and differential equations}
\label{geodesinomi}

We have recalled how both maximally extended supergravities (of type A and
B) reduce, stepping down from $D=10$ to $D=3$ to the following non
linear sigma model coupled to $D=3$ gravity:
\begin{equation}
\label{lag_sm} \mathcal{L}^{\sigma-model} \,=\,
\sqrt{-\mbox{det} \, g} \, \left[ \, 2 \, R [g] \,+ \,
\ft 12 \, h_{IJ}\left(\phi\right) \partial_{\mu} \phi^I
\partial_{\nu} \phi^J \, g^{\mu\nu} \right]
\end{equation}
where $h_{IJ}$, $I, \, J \,=\, 1,\,\ldots,\,128$, is the metric of
the $128$-dimensional homogeneous coset manifold
\begin{equation}
  \mathcal{M}_{128}=\frac{\mathrm{E_{8(8)}}}{\mathrm{SO(16)}}
\label{m128}
\end{equation}
\par
The above manifold falls in the general category of manifolds
$\mathrm{U/H}$ such that $\mathbb{U}$ (the Lie algebra of
$\mathrm{U}$) is the maximally non-compact real section of a
simple Lie algebra $\mathbb{U}_C$ and the subgroup $\mathrm{H}$ is
generated by the maximal compact subalgebra $\mathbb{H}\subset
\mathbb{U}$. In this case the solvable Lie algebra description of
the target manifold $\mathrm{U/H}$ is universal. The manifold
$\mathrm{U/H}$ is isometrical to the solvable group manifold:
\begin{equation}
\mathcal{M}_{d} =\exp \, \left [  Solv\left( \mathrm{U/H}\right)
\, \right ] \label{geneUH}
\end{equation}
where the solvable algebra $Solv\left( \mathrm{U/H}\right)$ is spanned
by all the Cartan generators $\mathcal{H}_i$ and by the step operators
$E^\alpha$ associated with all positive roots $\alpha >0$ (on the
solvable Lie algebra parametrization of supergravity scalar manifolds
see \cite{Andrianopoli:1996bq,Andrianopoli:1996zg}). On the other hand
the maximal compact subalgebra $\mathbb{H}$ is spanned by all
operators of the form $E_\alpha - E_{-\alpha}$ for all positive roots
$\alpha >0$. So the dimension of the coset $d$, the rank $r$ of
$\mathbb{U}$ and the number of positive roots $p$ are generally
related as follows:
\begin{equation}
 \mbox{dim} \left[ U/H\right] \equiv d \, = \, r +p \quad ; \quad p
 \,
 \equiv \, \# \mbox{positive roots}=  \mbox{dim} \,
 \mathbb{H} \quad ; \quad r \, \equiv \, \mbox{rank} \,
\mathbb{U}
\label{relazionibus}
\end{equation}
\par
In the present chapter we concentrate on studying solutions of a
bosonic field theory of type (\ref{lag_sm}) that are only
time-dependent. In so doing we consider the  case of a generic
manifold $\mathrm{U/H}$ and we show how the previously recalled
algebraic structure allows to retrieve a complete generating
solution of the field equations depending on as many essential
parameters as the rank of the Lie algebra $\mathbb{U}$. These
parameters label the orbits of solutions with respect to the
action of the two symmetries present in (\ref{lag_sm}), namely
$\mathrm{U}$ global symmetry and $\mathrm{H}$ local symmetry.
S
The essential observation is that, as long as we are interested in
solutions depending only on time, the field equations of
(\ref{lag_sm}) can be organized as follows. First we write the
field equations of the matter fields $\phi^I$ which supposedly
depend only on time. At this level the coupling of the sigma-model
to three dimensional gravity can be disregarded. Indeed the effect
of the metric $g_{00}$ is simply that the  field equations for the
scalars $\phi^I$ have the same form as they would have in a rigid
sigma model with just the following proviso. The parameter we use
is proper time rather than coordinate time. Next  in the variation
with respect to the metric we can use the essential feature of
three--dimensional gravity, namely the fact that the Ricci tensor
completely determines also the Riemann tensor. This means that
from the stress energy tensor of the sigma model solution we
reconstruct, via Einstein field equations, also the corresponding
three dimensional metric.
\par
\section{Decoupling the sigma model from gravity}
Since we are just interested  in configurations where the fields
depend only  on time, we take the following ansatz for the three
dimensional metric:
\begin{equation}
  ds^2_{3D} = A^2(t)\, dt^2 - B^2(t)\left(  dr^2 + r^2 d\phi^2 \right)
\label{confpiatto}
\end{equation}
where $A(t)$  and $B(t)$ are undetermined  functions of time. Then
we observe that one of these functions can always be reabsorbed
into a redefinition of the time variable. We fix such a coordinate
gauge by requiring that the matter field equations for the sigma
model should be decoupled from gravity, namely should have the
same form as in a flat metric. This will occur for a special
choice of the time variable. Let us see how.
\par
In general, the sigma model equations, coupled to gravity, have
the following form:
\begin{equation}
  \Box_{cov} \, \phi^I + \Gamma^I_{JK} \partial_\mu \phi^J \,
  \partial_\nu \phi^K \, g^{\mu \nu }= 0
\label{genersigma}
\end{equation}
In the case we restrict dependence only on time the above
equations reduce to:
\begin{equation}
  \frac{1}{\sqrt{-\mbox{det}g}} \,\frac{d}{dt} \left(
  \sqrt{-\mbox{det}g} \, g^{00} \, \frac{d}{dt} \, \phi^I \right)  +
  \Gamma^I_{JK} \frac{d}{dt} \phi^J \,
  \frac{d}{dt}\phi^K \, g^{00 } = 0
\label{sigmamodtime}
\end{equation}
We want to choose a new time $\tau =\tau(t)$ such that  with
respect to this new variable equations (\ref{sigmamodtime}) take
the same form as they would have in a sigma model in flat space,
namely:
\begin{equation}
\ddot{\phi}^I \,+\, \Gamma^{I}_{JK}\, \dot{\phi}^J \, \dot{
\phi}^K \,=\,0 \qquad I, \, J, \, K\,=\, 1,\,\ldots,\,\mbox{dim}
\, \mathcal{M} \label{geodesiaque}
\end{equation}
where $\Gamma^{I}_{JK}$ are the Christoffel symbols for the metric
$h_{IJ}$. The last equations are immediately interpreted as
geodesic equations in the target scalar manifold.
\par
In order for equations (\ref{sigmamodtime}) to reduce to
(\ref{geodesiaque}) the following condition must be imposed:
\begin{equation}
  \sqrt{-\mbox{det}g} \, g^{00} \, \frac{d}{dt} = \frac{d}{d\tau} \,
  \Rightarrow\, dt = \sqrt{-\mbox{det}g} \, g^{00} \, d\tau
\label{timegauge}
\end{equation}
Inserting the metric (\ref{confpiatto}) into the above condition
we obtain an equation for the coefficient $A(t)$ in terms of the
coefficient $B(t)$. Indeed in the new coordinate $\tau$ the metric
(\ref{confpiatto}) becomes:
\begin{equation}
  ds^2_{3D} = B^4(\tau)\, d\tau^2 - B^2(\tau)\left(  dr^2 + r^2 d\phi^2 \right)
\label{confpiatto2}
\end{equation}
 The choice (\ref{confpiatto2}) corresponds
to the following choice of the \textit{dreibein}:
\begin{equation}
  e^0 = B^2(\tau) \, d\tau \quad ; \quad e^1 = B(\tau) dr \quad ; \quad e^2 =
  B(\tau) \, r \, d\phi
\label{dreibeinconfpiatto}
\end{equation}
For such a metric the curvature $2$-form is as follows:
\begin{eqnarray}
R^{01} & = & \frac{2 \dot{B}^2(\tau)- B(\tau) \,
\ddot{B}(\tau)}{B^6(\tau)}\, e^0 \, \wedge \,e^1 \nonumber\\
R^{02} & = & \frac{2 \dot{B}^2(\tau)- B(\tau) \,
\ddot{B}(\tau)}{B^6(\tau)}\, e^0 \, \wedge \,e^2 \nonumber\\
R^{12} & = & -\frac{ \dot{B}^2(\tau)}{B^6(\tau)} e^1 \, \wedge
\,e^2 \label{curvaformd3}
\end{eqnarray}
The Einstein equations, following from our lagrangian
(\ref{lag_sm}) are the following ones, in flat indices:
\begin{equation}
  2 \, G_{ab} \, = \, T_{ab} \quad ; \quad G_{ab} \equiv \mbox{Ric}_{ab}
  - \ft 12 \, R \,\eta_{ab} \quad ; \quad a,b=0,1,2
\label{3einsteino}
\end{equation}
With the above choice of the vielbein, the flat index Einstein
tensor is easily calculated and has the following form:
\begin{equation}
  G_{00}= \frac{\dot{B}^2(\tau)}{2 \,B^6(\tau)} 
  \,;\quad
  G_{0i} = 0 
  \,;\quad 
  G_{ij} = \frac{2\,\dot{B}^2(\tau)
    - B(\tau) \, \ddot{B}(\tau)}{2 \,B^6(\tau)} \,\delta_{ij}
  \,;\qquad i,j = 1,2
  \label{einsteinten}
\end{equation}
On the other hand, calculating the stress energy tensor of the
scalar matter in the background of the metric (\ref{confpiatto2})
we obtain (also in flat indices):
\begin{equation}
  T_{00} = \frac{1}{2 \, B^4(\tau)} \,\left(  \dot{\phi}^I  \dot{\phi}^J
h_{IJ}\right) \quad ;
  \quad T_{0i} =0 \quad ;
  \quad T_{ij} = \frac{1}{2 \, B^4(\tau)} \,\left(  \dot{\phi}^I  \dot{\phi}^J
h_{IJ}\right) \,
  \delta_{ij} \quad ; \quad i,j = 1,2
\label{stresstensore}
\end{equation}
where
\begin{equation}
            \dot{\phi}^I  \dot{\phi}^J h_{IJ} =
  \varpi^2
\label{omegasquare}
\end{equation}
is a constant independent from time as a consequence of the
geodesic equations (\ref{geodesiaque}). To prove this it suffices
to take a derivative in $\tau$ of ${\varpi^2}$ and verify that it
is zero upon use of eq.s (\ref{geodesiaque}).
\par
Hence in order to satisfy the coupled equations of gravity and
matter fields it is necessary that:
\begin{equation}
 2 \,  \frac{\dot{B}^2(\tau)}{2 \,B^6(\tau)} = 2\, \frac{2\,
            \dot{B}^2(\tau)
            - B(\tau) \, \ddot{B}(\tau)}{2 \,B^6(\tau)} = \frac{1}{2 \, B^4(\tau)}
            \,{\varpi^2}
\label{einsteinocon}
\end{equation}
The first of the above equalities implies:
\begin{equation}
  \dot{B} = k \, B \quad \Rightarrow \quad B(\tau) =
  \exp [ k \, \tau]
\label{kdefi}
\end{equation}
where $k$ is some constant. The second equality is satisfied if:
\begin{equation}
  k = \pm \ft 1{\sqrt{2}} \, |\varpi| =
  \pm \ft 1{\sqrt{2}} \,\sqrt{\dot{\phi}^I  \dot{\phi}^J h_{IJ}}
\label{kvalue}
\end{equation}
In this way we have completely fixed the metric of the
three--dimensional space as determined by the solution of the
geodesic equations for the scalar matter:
\begin{equation}
  ds^2_{3D} = \exp \left[ 4 k \,\tau\right]  d\tau^2 - \exp \left[ 2 k
  \,\tau\right]\, \left (dx_1^2 + dx_2^2 \right)
\label{orporpo}
\end{equation}
with the parameter $k$ given by eq.(\ref{kvalue}).
\section{Geodesic equations in target space and the Nomizu
operator} Having clarified how the three dimensional metric is
determined in terms of the solutions of the sigma model, we
concentrate on this latter. We focus on the geodesic equations
(\ref{geodesiaque}) and in order to study them, we rely on the
solvable Lie group description of the target manifold going to an
anholonomic basis for the tangent vectors to the geodesic.
%%%%%%%%%%%%%%%%%%%%%%%%%%%%%
% From Ksenya
%%%%%%%%%%%%%%%%%%%%%%%%%%%%%%%%%%%
Since gravity is decoupled from the scalars, we deal with a rigid
sigma-model where the fields depend only on time
 \begin{equation} \mathcal{L}^{\sigma-model}  \propto
 h_{IJ}(\phi)\dot{\phi}^I\dot{\phi}^J
 \label{lagra3d1}
 \end{equation}
 As was mentioned before, the equations of motion in this case
 reduce to the geodesic equations for the metric $h_{IJ}(\phi)$
 and time plays the role of a parameter along the geodesics
(see eq.(\ref{geodesiaque})).
 Since $h_{IJ}(\phi)$ is the metric of a scalar manifold
 which is a maximally non-compact coset  $\mathcal{M} = \mathrm{U/H}$, we
 can derive this metric from a coset representative
 $\mathbb{L}(\phi)\in U$
 \begin{equation}
 h_{IJ}(\phi) = \mathrm{Tr}(\mathbb{P}_K
 \IL^{-1}\partial_I\IL\IL^{-1}\partial_J\IL)
 \end{equation}
$\mathbb{P}_K$ being a projection operator on the coset directions
of the Lie algebra $\mathbb{U}$ to be discussed in a moment. To
this effect we introduce the following general notation. We make
the orthogonal split of the $\mathbb{U}$ Lie algebra:
\begin{equation}
  \mathbb{U} = \mathbb{H} \oplus \mathbb{{K}}
\label{orthosplitto}
\end{equation}
where $\mathbb{H} \subset \mathbb{U}$ is the maximal compact
subalgebra and $\mathbb{{K}}$ its orthogonal complement.
 We adopt the following normalizations for the generators in each
 subspace:
\begin{eqnarray}
\nonumber && \mathbb{U} = \mathrm{Span}\{H_i, E_{\alpha}, E_{-\alpha}\} \\
&& \mathbb{K}=\mathrm{Span}\{K_A\} = \mathrm{Span}\{H_i,
\ft 1{\sqrt{2}}(E_{\alpha} + E_{-\alpha})\} \\ && \nonumber
\mathbb{H} = \mathrm{Span}\{t_{\alpha}\} =
\mathrm{Span}\{(E_{\alpha} - E_{-\alpha})\}
\end{eqnarray}
The ${\mathbb{U}}$ Lie-algebra valued  left invariant one-form
\begin{equation}
\Omega = \mathbb{L}^{-1} d \mathbb{L} \,=\, V^AK_A +
\omega^{\alpha}t_{\alpha}  \label{solvodecompo}
\end{equation}
is in general expanded along all the generators of $\mathbb{U}$
(not only along $\mathbb{{K}}$) and $V = V^A\, K_A $ corresponds
to the coset manifold vielbein while
 $\omega=\omega^\alpha\, t_\alpha$ corresponds to the
coset manifold $\mathrm{H}$--connection.
\par
 As it is well known neither the
coset representative $\mathbb{L}(\phi)$, nor the one-form $\Omega$
are unique. Indeed $\mathbb{L}$ is defined up to multiplication on
the right by an element of the compact subgroup $h\in H$. This is
a gauge invariance which can be fixed in such a way that the coset
representative lies in the  solvable group ${Solv}(\mathrm{U/H})$ obtained by
exponentiating the solvable subalgebra $Solv(\mathbb{U}/\mathbb{H})$
\begin{equation}
\mathbb{L} \left( \phi \right) \,=\, \exp{\left( Solv(\mathbb{U}/\mathbb{H}) \cdot \phi
\right)} \label{solcoset}
\end{equation}
In the case of $\mathrm{U}/\mathrm{H}$ being maximally non compact
$Solv(\mathbb{U}/\mathbb{H})$ coincides with the Borel subalgebra and
therefore it is spanned by the collection of all Cartan generators
and  step-operators associated with positive roots, as we already
stated, namely:
\begin{equation}
Solv(\mathbb{U}/\mathbb{H}) = \mathrm{Span} \,\left\{ T_A\right\}  =
\mathrm{Span} \, \left \{H_i, E_{\alpha} \right \}
\label{solvospan}
\end{equation}
If the coset representative $ \mathbb{L}$ is chosen to be a
solvable group element, as in eq. (\ref{solcoset}), namely if we
are in the solvable parametrization of the coset, we can also
write:
\begin{equation}
            \Omega = \mathbb{L}^{-1} d\mathbb{L} = \widetilde{V}^i \, \mathcal{H}_i
\, + \,
  \widetilde{V}^\alpha \, E_\alpha = \widetilde{V}^A \, T_A
\label{solvomega}
\end{equation}
since $\Omega$ is contained in the solvable subalgebra
$Solv(\mathbb{U}/\mathbb{H}) \subset \mathbb{U}$. Eq.s (\ref{solvodecompo})
and (\ref{solvomega}) are compatible if and only if:
\begin{equation}
  V^\alpha = \sqrt{2} \, \omega^\alpha
\label{solvocondo}
\end{equation}
In this case we can identify $\widetilde{V}^i={V}^i$ and
$\widetilde{V}^\alpha = \sqrt{2} \, V^\alpha$; eq.(\ref{solvocondo}) is the
solvability condition for a coset
representative.
\par
Hence we can just rewrite the metric of our maximally non--compact
manifold $\mathrm{U} / \mathrm{H}$ as follows:
\begin{equation}
ds^{2}_{\mathrm{U}/\mathrm{H}} \,=\, \sum_{A=1}^{\mbox{dim}
\mathrm{U}}  V^A \,\otimes\, V^A = \tilde{V}^i\oplus \tilde{V}^i +
\ft 12 \tilde{V}^{\alpha}\oplus \tilde{V}^{\alpha}
\label{metricUH}
\end{equation}
It is interesting to discuss what are the residual $\mathrm{H}$--gauge
transformations that remain available after the solvable gauge
condition (\ref{solvocondo}) has been imposed. To this effect we
consider the multiplication
\begin{equation}
  \mathbb{L} \mapsto \mathbb{L} \, h = \overline{\mathbb{L}}
\label{hgauga}
\end{equation}
where
\begin{equation}
  h = \exp \, \left[  \theta^\alpha  \, t_\alpha  \right]
\label{helemo}
\end{equation}
is a finite element of the $\mathrm{H}$ subgroup singled out by generic
parameters theta.
 For any such element we can always write:
\begin{eqnarray}
h^{-1} \, t_\alpha  \, h & = & A(\theta)_\alpha^{\phantom{\alpha }\beta} \,t_\beta
\nonumber\\
h^{-1} \, K_A  \, h &  = & D(\theta)_A^{\phantom{A }B} \,K_B
\label{duematrici}
\end{eqnarray}
where the matrix $A(\theta)$ is the adjoint representation of $h$
and $D(\theta)$ is the $D$--representation of the same group
element. We obtain:
\begin{eqnarray}
            \overline{\Omega} \equiv \overline{\mathbb{L}}^{-1} \,
            d\overline{\mathbb{L}} & = & h^{-1}dh + h^{-1} \, \Omega \, h
            \nonumber\\
            & = & \underbrace{h^{-1}\, dh  + h^{-1} \,
            \omega \, h}_{=\,\,\overline{\omega}} \, + \underbrace{\, h^{-1} \, V \,
            h}_{=\,\,\overline{V}}
\label{omegabarra}
\end{eqnarray}
where
\begin{eqnarray}
\overline{\omega}^\alpha & = & \frac {1}{\mbox{tr}(t_\alpha ^2)}
\,\mbox{tr} \left( h^{-1}\, dh \, t_\alpha \right)  \, + \,
 \omega^\beta  \, A(\theta)_{\beta}^{\phantom{\beta}\alpha}\nonumber\\
\overline{V}^\alpha & = & V^\beta \, D(\theta)_{\beta}
^{\phantom{\beta}\alpha} \, + \, V^i \,
D(\theta)_i^{\phantom{i}\alpha} \, \label{bongo}
\end{eqnarray}
Suppose now that the coset representative $\mathbb{L}$ is
solvable, namely it satisfies eq.(\ref{solvocondo}). The coset
representative $\overline{\mathbb{L}}$ will still satisfy the same
condition if the $h$-compensator satisfies the following
condition:
\begin{equation}
 \frac{\sqrt{2}}{\mbox{tr}(t_\alpha ^2)} \, \,\mbox{tr} \left( h^{-1}(\theta)\,
dh(\theta) \, t_\alpha
  \right)= V^\beta \, \left(  A(\theta)_\beta^{\phantom{\beta}\alpha}
  \, -\, D(\theta)_\beta^{\phantom{\beta}\alpha} \right) \, + \,  V^i \,
  D(\theta)_i^{\phantom{i}\alpha}
\label{daequa}
\end{equation}
The above equations are a set of $n=\#roots = \mbox{dim}H$
differential equations on the parameters $\theta^\alpha$ of the
$h$--subgroup element (\textit{compensator}). In the following we
will use such set of equations as the basis of an algorithm to
produce solutions of the geodesic equations (\ref{geodesiaque}).

Given these preliminaries, we can establish a new notation. We
introduce tangent vectors to the geodesics in the anholonomic
basis:
\begin{equation}
\Phi^A \,=\, \widetilde{V}^A_I\left(\phi\right) \dot{\phi}^I
\label{anholotang}
\end{equation}
which are functions only of time: $\Phi^A \,=\,\Phi^A(t)$. In this
basis the field equations reduce to
\begin{equation}
\label{D=3feqn} \dot{\Phi}^A \,+\, \Gamma^A_{BC} \Phi^B \Phi^C
\,=\,0
\end{equation}
where now $\Gamma^A_{BC}$ are the components of the Levi-Civita
connection in the chosen anholonomic basis. Explicitly they are
related to the components of the Levi Civita connection in an
arbitrary holonomic basis by:
\begin{equation}
\Gamma^A_{BC}\,= \Gamma^I_{JK}V^A_I V^J_B V^K_C - \partial_K
(V^A_J)V^J_B V^K_C \label{capindgamma}
\end{equation}
where the inverse vielbein is defined in the usual way:
\begin{equation}
V^A_I \, V^I_B \,=\, \delta^A_B \label{invervielb}
\end{equation}
The most important point here is that,  the connection
$\Gamma^A_{BC}$ can be identified with the \textit{Nomizu
connection} defined on a solvable Lie algebra, if the coset
representative $\mathbb{L}$ from which we construct the vielbein
via eq.(\ref{solvodecompo}), is solvable, \textit{namely if and
only if} the solvability condition (\ref{solvocondo}) is
satisfied. In fact, as we can read in \cite{alekseevskii}, once we
have defined over $Solv$ a non degenerate, positive definite and
symmetric form:
\begin{eqnarray}
\langle \,,\, \rangle & \; : \; Solv \otimes Solv \longrightarrow \mathbb{R}
\nonumber \\
\langle X \,,\, Y \rangle & \; = \; \langle Y \,,\, X \rangle
\end{eqnarray}
whose lifting to the manifold produces the metric, the covariant
derivative is defined through the \textbf{Nomizu operator}:
\begin{equation}
\forall X \in Solv \,:\, \mathbb{L}_X: Solv \longrightarrow Solv
\end{equation}
so that
\begin{equation}
\forall X,Y,Z \in Solv \,:\, 2 \langle Z \,,\, \mathbb{L}_X Y
\rangle \,=\, \langle Z, \left[ X,Y \right] \rangle \,-\, \langle
X, \left[ Y,Z \right] \rangle \,-\, \langle Y, \left[ X,Z \right]
\rangle \label{Nomizuoper}
\end{equation}
while the Riemann curvature 2-form is given by the commutator of
two Nomizu operators:
\begin{equation}
R^W_{\phantom{W}Z} \left( X,Y \right) \,=\, \langle W \,,\,
\left\{ \left[ \mathbb{L}_X , \mathbb{L}_Y \right] \,-\,
\mathbb{L}_{\left[X,Y\right]} \right\} Z \rangle
\label{Nomizucurv}
\end{equation}
This implies that the covariant derivative explicitly reads:
\begin{equation}
\mathbb{L}_X \,Y \,=\, \Gamma_{XY}^Z \,Z \label{Gammonedefi}
\end{equation}
where
\begin{equation}
\Gamma_{XY}^Z \,=\,\frac{\left(\langle Z, \left[ X,Y \right]
\rangle \,-\, \langle X, \left[ Y,Z \right] \rangle \,-\, \langle
Y, \left[ X,Z \right] \rangle\right)}{2\,<Z,Z>}  \qquad
\forall X,Y,Z \in Solv \label{Nomizuconne}
\end{equation}

In concrete, the non degenerate, positive definite, symmetric form
on the solvable Lie algebra which agrees with equation
(\ref{metricUH}) is defined by setting:
\begin{eqnarray}
\langle \mathcal{H}_i \,,\, \mathcal{H}_j \rangle & \,=\, & 2 \, \delta_{ij}
\nonumber \\
\langle \mathcal{H}_i \,,\, E_\alpha \rangle & \,=\, & 0 \nonumber \\
\langle E_\alpha \,,\, E_\beta \rangle & \,=\, &
\delta_{\alpha,\beta}
\end{eqnarray}
$\forall \mathcal{H}_i ,\, \mathcal{H}_j \,\in\,
\mathrm{CSA}_{\mathrm{E_{8(8)}}} $ and $\forall E_\alpha$, step
operator associated to a positive root $\alpha$ of
$\mathrm{E_{8(8)}}$. Then the Nomizu connection (which is
constant) is very easy to calculate. We have:
\begin{equation}
\begin{array}{lll}
\Gamma^i_{jk}& \,=&\,  0 \nonumber \\
\Gamma^i_{\alpha\beta} &\,=& \,  \ft 12\left(-\langle
E_\alpha,\,\left[E_\beta,\,H^i\right]\rangle - \langle
E_\beta,\,\left[E_\alpha,\,H^i\right]\rangle\right)\,=\,\ft 12 \,
\alpha^i \delta_{\alpha\beta}
\nonumber \\
\Gamma^{\alpha}_{ij}   &\,=&
\, \Gamma^{\alpha}_{i\beta} \,=\, \Gamma^i_{j\alpha} \,=\,0 \nonumber \\
\Gamma^{\alpha}_{\beta i} &\,=& \, \ft 12\left(\langle
E^\alpha,\,\left[E_\beta,\,H_i\right]\rangle -
\langle E_\beta,\,\left[H_i,\,E^\alpha\right]\rangle\right)\,=\,
-\alpha_i\,\delta^{\alpha}_{\beta}\nonumber\\
\Gamma^{\alpha+\beta}_{\alpha\beta}   &\,=&
\, -\Gamma^{\alpha+\beta}_{\beta\alpha}\,=\,\ft 12 N_{\alpha\beta}\nonumber\\
\Gamma^{\alpha}_{\alpha+\beta\,\beta}  &\,=&
 \, \Gamma^{\alpha}_{\beta\,\alpha+\beta} \,=\,\ft 12 N_{\alpha\beta}
\end{array}
 \label{Nomizuconne2}
\end{equation}
where $N_{\alpha\beta}$ is defined by the commutator:
\begin{equation}
\left[ E_\alpha \,,\, E_\beta \right] \,=\,
N_{\alpha\beta}\,E_{\alpha + \beta} \label{nalfabeta}
\end{equation}
which has to be worked out in the algebra.
\footnote{The values of the constants $N_{\alpha\beta}$, that enable to construct
explicitly the representation of $E_{8(8)}$, used in this paper,
 are given in the hidden appendix. To see it, download the source file, delete the
tag $end\{document\}$
 after the bibliography and $LaTeX$.}
Notice that
$\Gamma^Z_{XY}\neq \Gamma^Z_{YX}$ since its expression consists of
the first term which is antisymmetric in $(X,\,Y)$ and the sum of
the last two which is symmetric. The component
$\Gamma^{\alpha}_{\beta i}$ consists of the sum of two equal
contributions from the antisymmetric and symmetric part, the same
contributions cancel in  $\Gamma^{\alpha}_{i\beta}$ which indeed
vanishes.
 By substituting the explicit expression of the Nomizu connection
in (\ref{D=3feqn}) and introducing for the further convenience new
names for the tangent vectors along the Cartan generators
$\chi^i\equiv\Phi^i $ we have the equations:
\begin{eqnarray}
\label{D=3feqn_2} \dot{\chi}^i & \,+\, &  \, \ft 12
\,\sum_{\alpha\in \Delta_+}  \alpha^i \Phi_\alpha^2 \,=\, 0
\nonumber \\
\dot{\Phi}^\alpha &\,+\, &  \,\sum_{\beta\in
\Delta_+}\,N_{\alpha\beta} \Phi^{\beta}  \Phi^{\alpha+\beta} -
\alpha_i\,\chi^i\Phi^\alpha\,=\,0
\end{eqnarray}
Eq.s (\ref{D=3feqn_2} ) encode all the algebraic structure of the
$D=3$ sigma model and due to our oxidation algorithm of the
original supergravity in ten dimensions.
\par

All this means that, thanks to the solvability of the algebra (and
also to the fact that we know the explicit form of the connection
via the Nomizu operator), we have reduced the entire problem of
finding time dependent backgrounds for either type IIA or type IIB
superstrings or M-theory to the integration of a system of
differential equations firmly based on the algebraic structure of
$E_{8(8)}$. This is a system of non-linear differential equations,
and from this point of view it might seem hopeless to be solved.
Yet, due to its underlying algebraic structure, one can use its
isometries to generate the complete integral depending on as many integration
constants as the number of equations in the system. This is the compensator
algorithm we
alluded to above, which we shortly outline. To this effect we
discuss the role of initial conditions for the tangent vectors to
the geodesics. There exist a number of possibilities for such
conditions that can truncate the whole system to  smaller and
simpler ones. The simplest choice is to put all root-vectors to
zero in the origin. This will ensure that root-vectors will remain
zero at all later times and the system will reduce to
\begin{equation}
\dot{\chi}^i = 0 \label{dchi=0}
\end{equation}
The solution of such a reduced system is trivial and consists of a
constant vector $\tilde{V}^A = (\chi^i, 0)$.  If we apply an
$\mathrm{H}$--rotation to this tangent vector
\begin{equation}
  \overline{V}^A = V^B \, D(\theta)_B^{\phantom{B}A}
\label{rotatedV}
\end{equation}
we produce a new one, yet, for generic $\mathrm{H}$--rotations we will
break the solvable gauge, so that the result no longer produces a
solution of eq.s(\ref{D=3feqn_2}). However, if we restrict the
$\theta^\alpha$ parameters of the rotation to satisfy condition
(\ref{daequa}), then the solvable gauge is preserved and the
rotated tangent vector $\overline{V}^A$ is still a solution of
eq.s(\ref{D=3feqn_2}). Hence a general algorithm to solve the
differential system (\ref{D=3feqn_2}) has been outlined. One
starts from the trivial solution in eq.(\ref{dchi=0}) and then
tries to solve the differential equation for the  theta parameter
corresponding to one particular $\mathrm{H}$--generator $t_\alpha =
E_\alpha -E_{-\alpha}$. Applying this rotation to the trivial
solution we obtain a new non trivial one. Then starting from such
a new solution we can repeat the procedure and try to solve again
the differential equation for the theta parameter relative to a
new generator. If we succeed we obtain a further new solution of
the original system and we can repeat the procedure a third time
for a third generator, iteratively. Indeed, considering
eq.(\ref{daequa}) we see that if $h(t)$ is just a general element
of the subgroup $\mathrm{H}$, the system is
 rather difficult to solve, yet if we choose
 a rotation around a single axis $h_{\alpha_0} =
 e^{\theta^{\alpha_0}(t)t_{\alpha_0}}$,  then
$\frac{1}{\mathrm{Tr}(t_{\alpha_0}^2)}\mathrm{Tr}(h^{-1}dh
t_{\alpha_0}) = \dot{\theta^{\alpha_0}}$ and, if all the other
equations for $\alpha \ne \alpha_0$ are identically satisfied, as
it will turn out to be the case in the examples we consider, then
the system reduces to only one first order differential equation
on the angle $\theta^{\alpha_0}(t)$.

We name such an algorithm \textit{the compensator method} and we will
illustrate it in the next chapter with specific examples.

\chapter{The $\mathrm{A_2}$ toy model as a paradigma}
\label{exampsolv} 

In this chapter we consider explicit examples of
solutions of the geodesic problem in the case of an $\mathrm{A_2}$
simple algebra. Later we will consider the possible embeddings of such
an algebra into the $\mathrm{E_8}$ algebra, so that the solutions we
construct here will be promoted to particular solutions of the full
$\mathrm{E_{8(8)}/SO(16)}$ sigma model. The diverse embeddings will
correspond to diverse oxidations of the same three dimensional
configuration to $D=10$ configurations.  In other words there exist
various non abelian solvable subalgebras $S_5 \subset
Solv(\mathrm{E_8}/\mathrm{SO(16)})$ of dimension $5$ which by means of
a linear transformation can be identified as the solvable Lie algebra
of the simple Lie algebra $A_2$, namely the solvable Lie algebra
description of the coset manifold:
\begin{equation}
 \mathcal{M}_5 \equiv \exp \left [Solv(\mathrm{A_2}) \right ] \,
 \cong \, \frac{\mathrm{SL(3,\mathbb{R})}}{\mathrm{SO(3)}}
\label{solv5}
\end{equation}
The  detailed study of this model provides our paradigma for the
general solution of the complete theory based on the coset
manifold $\mathrm{E_{8(8)}/SO(16)}$. We emphasize that the
possibility of choosing a \textit{normal form} for the initial
tangent vector to the geodesic  allows to reduce the system of
first order equations to a much simpler set, as we started to
discuss in the previous chapter in general terms. Such a normal
form can be chosen in different ways. In particular it can always
be chosen so that it contains only Cartan generators. When this is
done the system   is always exactly solvable and in terms of pure
exponentials. The solution obtained in this way provides a
representative for the orbit of geodesics modulo isometries. We
can then generate new solutions of the differential system
(\ref{D=3feqn_2}) by the compensator method we described in the
previous chapter.
  In this chapter we
illustrate such an algorithm in the case of the toy $\mathrm{A_2}$
model. The resulting solutions have not only a tuitional interest,
rather they provide  examples of solutions of the full
$\mathrm{E_{8(8)}}$ system and hence of full supergravity. It
suffices to embed the $\mathrm{A_2}$ Lie algebra in the full
algebra $\mathrm{E_{8(8)}}$. We will discuss such embeddings and
the corresponding oxidations of our sigma model solutions in later
chapters.

\section{Structure of the $\mathrm{A_2}$ system}

Our model consists of 5 scalar fields, which parametrize a coset
manifold $\mathcal{M}_5 = \mathrm{SL(3)/SO(3)}$. Our chosen
conventions are as follows. The two simple roots of
$\mathrm{SL(3)}$ are:
\begin{equation}
\beta_1  = \left\{ \sqrt{2} \, , \, 0 \right\}, \hspace{0.5cm}
\beta_2 = \left\{ - \ft 1{\sqrt{2}}\, ,
\,\sqrt{\ft 32}\right\} \label{simplea2}
\end{equation}
and the third positive root, which is the highest is:
\begin{equation}
  \beta_3 = \beta_1 + \beta_2 = \left\{  \ft 1{\sqrt{2}}\, ,
\,\sqrt{\ft 32}\right\} \label{highroota2}
\end{equation}
Furthermore the step operator $E_{\beta_3}$ is defined through
the commutator:
\begin{equation}
  E_{\beta_3} = \left[ E_{\beta_1} \, , \,
  E_{\beta_2}\right]
\label{steppotre}
\end{equation}
and this completely fixes all conventions for the Lie algebra
structure constants.
\par
The three generators of the maximally compact subgroup are defined
as:
\begin{equation}
  t_1 = E_{\beta _1} -E_{-\beta _1} \quad , \quad t_2 = E_{\beta _2} -E_{-\beta _2}
  \quad , \quad t_3 = E_{\beta _3} -E_{-\beta _3}
\label{generso3}
\end{equation}
and they satisfy the standard commutation relations:
\begin{equation}
  \left[ t_i \, , \, t_j \right] \, = \, \epsilon _{ijk} \, t_k
\label{standeso3}
\end{equation}
In the orthogonal decomposition of the Lie algebra:
\begin{equation}
  \mathrm{A_2} = \mathrm{SO(3)} \, \oplus \, \mathbb{K}_5
\label{orthodeco}
\end{equation}
the $5$-dimensional subspace $\mathbb{K}_5$  is identified with
the tangent space to $\mathcal{M}_5$ and corresponds to the $j=2$
representation of ${\mathrm{SO(3)}}$
 \begin{equation}
[t_{\beta},K_A] = Y_{\beta A}^BK_B
\end{equation}

This subspace is spanned by the following generators:
\begin{equation}
  \mathbb{K}_5 
  \,=\,\mathrm{Span}
  \left\{
    H_1,\,
    H_2,\,
    \frac{E_{\beta_1} + E_{-\beta_1}}{\sqrt{2}},\,
    \frac{E_{\beta_2} + E_{-\beta_2}}{\sqrt{2}},\,
    \frac{E_{\beta_3} + E_{-\beta_3}}{\sqrt{2}}
  \right\}
  \label{K5generi}
\end{equation}

Applying to this case the general formulae (\ref{D=3feqn_2}) based
on the Nomizu connection (\ref{Nomizuconne2}) we obtain the
differential system:
\begin{eqnarray}
 &&\dot{\chi}_1(t) + \ft 1{\sqrt{2}}\Phi^2_1(t) -
  \ft 1{2\sqrt{2}}\Phi_2^2(t) +
  \ft 1{2\sqrt{2}}\Phi_3^2(t)   =  0 \nonumber\\
 &&\dot{\chi}_2(t) + \ft {\sqrt{3}}{2\sqrt{2}}\Phi^2_2(t)
  + \ft {\sqrt{3}}{2\sqrt{2}}\Phi^2_3(t)  =  0
            \nonumber\\
            && \dot{\Phi}_1(t)  + {{\Phi }_2}(t)\,{{\Phi }_3}(t) -
  {\sqrt{2}}\,{{\Phi }_1}(t)\,
            {{\chi }_1}(t)
            =  0 \nonumber\\
&& \dot{\Phi}_2(t) - {{\Phi }_1}(t)\,
                     {{\Phi }_3}(t)   +
  \ft 1{\sqrt{2}}\Phi_2(t)
                     \chi_1(t) -
  \sqrt{\ft 32}
            \Phi_2(t)\chi_2(t)
            = 0\nonumber\\
&& \dot{\Phi}_3(t) -\ft 1{\sqrt{2}}\Phi_3(t)\chi_1(t) -
  \sqrt{\ft 32}\Phi_3(t)\chi_2(t)
            = 0 \label{A2system} 
\end{eqnarray}
          
In order to solve this differential system of equations we recall
their geometrical meaning. They are the geodesics equations for the
manifold (\ref{solv5}) written in flat indices, namely in an
anholonomic frame. Any geodesics is completely determined by two data:
the initial point $p_0 \in M_5$ and the initial tangent vector
$\overrightarrow{t}_0 \in T(M_5)$ at time $t=0$. Since our manifold is
homogeneous, all points are equivalent and we can just choose the
origin of the coset manifold. Since we are interested in determining
the orbits of geodesics modulo the action of the isometry group, the
relevant question is the following: {\it in how many irreducible
  representations} of the tangent group $\mathrm{SO(3)}$ does the
tangent space decompose? The answer is simple: the $5$ dimensional
tangent space is irreducible and corresponds to the $j=2$
representation of $\mathrm{SO(3)}$. The next question is: \textit{
  what is the normal form of such a representation} and how many
parameters does it contain. The answer is again simple. A spin two
representation is just a symmetric traceless tensor $g_{ij}$ in three
dimensions. By means of $\mathrm{SO(3)}$ rotations we can reduce it to
a diagonal form and the essential parameters are its eigenvalues,
namely two parameters, since the third eigenvalue is minus the sum of
the other two, being the matrix traceless. So by means of
$\mathrm{SO(3)}$ rotations a generic $5$-dimensional tangent vector
can be brought to contain only two parameters. This argument is also
evident from the consideration that $5-3=2$, namely by means of the
three $\mathrm{SO(3)}$ parameters we can set three components of the
$5$-dimensional vector to zero.
          
We can also analyze the normal form of the $5$--dimensional
representation from the point of view of eigenstates of the angular
momentum third component $t_3$. This latter has skew eigenvalues $\pm
2, \pm 1 $ and $0$. The transformation of the matrix $g=\left\{ g_{ij}
\right\} $ under any generator $t$ of the $\mathrm{SO(3)}$ Lie algebra
is
\begin{equation}
  \delta \, g \, = \, \left[ t \, , \, g \right]
\label{trasforg}
\end{equation}
so that the pair of skew eigenstates of the generator $t_3$, as given
in eq.  (\ref{tgene3}), pertaining to the skew eigenvalues $\pm 2$ is
provided by the symmetric matrices of the form:
\begin{equation}
g_{(\pm 2)} =  \left(\begin{array}{ccc}
  a & 0 & b \\
  0 & 0 & 0 \\
  b & 0 & -a
\end{array} \right)
\label{eigepm2}
\end{equation}
which can be diagonalized through $\mathrm{SO(3)}$ rotations (actually
$\mathrm{SO(2)}$ in this case) and brought to the normal form:
\begin{equation}
g_{2} =  \left(
\begin{array}{ccc}
  \sqrt{a^2 + b^2} & 0 & 0 \\
 0 & 0 & 0 \\
  0 & 0 & -\sqrt{a^2 + b^2}
\end{array} \right)
\label{eige2}
\end{equation}
which is just one of the two in the pair of skew eigenstates. On
the other hand the symmetric traceless matrix that corresponds to
the null eigenstate of $t_3$ is:
\begin{equation}
  g_{(0)} = \left( \begin{array}{ccc}
             s & 0 & 0 \\
             0 & -2s & 0 \\
             0 & 0 & s \
  \end{array}\right)
\label{eige0}
\end{equation}
A superposition $g_{2} + g_{(0)}$ provides the most general
diagonal traceless symmetric matrix, namely the \textit{normal
form} to which any state in the $j=2$ irreducible representation
can be brought by means of $SO(3)$ rotations.
\par
Alternatively, since the $j=2$ representation is provided by the
tangent space to the $\mathcal{M}_5$ manifold, spanned by the
coset generators of $\mathrm{SL(3,\mathbb{R})}$ not lying in the
compact $\mathrm{SO(3)}$ subalgebra, we can identify the normal
form of a $5$--dimensional vector as one with non vanishing
components only in the directions of the Cartan generators.
Indeed, by means of $\mathrm{SO(3)}$ rotations any vector can be
brought to such a form and the counting of independent parameters
coincides, namely two. This is a completely general statement for
maximally non compact coset manifolds. The rank of the coset is
equal to the number of independent parameters in the normal form
of the $\mathbb{H}$ representation provided by the coset subspace
$\mathbb{K}$.
\par
Relying on these considerations, let us consider the explicit
representation of the group $\mathrm{SO(3)}$ on the tangent space
to our manifold $\mathcal{M}_5$ and how, by means of its
transformation we can bring the initial tangent vector to our
geodesic to our desired normal form. Indeed our  aim is to solve
 the geodesic equations (\ref{A2system}) fixing
initial conditions:
\begin{equation}
  \left\{ \chi_1(0)\, , \, \chi_2(0) \, , \,\Phi_1(0)\, , \, \Phi_2(0) \,
  , \, \Phi_3(0)\right \} \, = \,\tilde{V} \, =
  \, \left\{ \tilde{V}_1, \tilde{V}_2, \dots ,\tilde{V}_5 \right \}
\label{pincus}
\end{equation}
where $\tilde{V}$ is the normal form of the $5$ vector.
To this effect it is convenient to inspect the representative
matrices of $\mathrm{SO(3)}$ on the tangent space. The three
generators of the maximally compact subgroup were  defined in
(\ref{generso3}) and in the basis of $\mathbb{K}_5$ provided by
the generators (\ref{K5generi}) the $ 5 \times 5$ matrices
representing $\mathrm{SO(3)}$ are:
\begin{gather}
  \label{j2generi}
  t_1^{[5]}\,=\,
  \begin{pmatrix}
    0 & 0 & -2 & 0 & 0\\
    0 & 0 & 0 & 0 & 0\\
    2 & 0 & 0 & 0 & 0\\ 
    0 & 0 & 0 & 0 & 1\\
    0 & 0 & 0 & -1 & 0
  \end{pmatrix},\quad
  t_2^{[5]}\,=\,
  \begin{pmatrix}
    0 & 0 & 0 & 1 & 0\\
    0 & 0 & 0 & -\sqrt{3} & 0\\
    0 & 0 & 0 & 0 & -1\\
    -1 & \sqrt{3} & 0 & 0 & 0\\
    0 & 0 & 1 & 0 & 0
  \end{pmatrix}
  \notag\\
  t_3^{[5]} \,=\,
  \begin{pmatrix}
    0 & 0 & 0 & 0 & -1\\
    0 & 0 & 0 & 0 & -\sqrt{3}\\
    0 & 0 & 0 & -1 & 0\\
    0 & 0 & 1 & 0 & 0\\
    1 & \sqrt{3} & 0 & 0 & 0
  \end{pmatrix}
\end{gather}

These matrices have  the expected skew eigenvalues:
\begin{equation}
  \left( \pm 2,\, \pm 1,\, 0 \right)
\end{equation}
For the generator $t^{[5]}_3$ the corresponding eigenvectors are:
\begin{equation}
  \begin{array}{lcl}
             \mbox{eigenvalue $0$} &
             \Rightarrow & \{ -{\sqrt{3}},1,0,0,0\} \\
             \mbox{eigenvalues $\pm 1$} & \Rightarrow &
             \left \{\begin{array}{l}
                        \{ 0,0,1 ,1,0\} \\
                                \{ 0,0,-1 ,1,0\} \\
             \end{array} \right.\\
             \mbox{eigenvalues $\pm 2$ } & \Rightarrow &
             \left \{ \begin{array}{l}
                                \{ \frac{1}{{\sqrt{2}}},
  {\sqrt{\frac{3}{2}}},0,0,
  {\sqrt{2}}\} \\
                                \{ - \frac{1}{{\sqrt{2}}}
                      ,-{\sqrt{\frac{3}{2}}},
  0,0,{\sqrt{2}}\} \\
\end{array} \right.\
  \end{array}
\label{eigevalori}
\end{equation}
So reduced to normal form the $5$-vector of initial condition is
 a linear combination of  the vectors $\overrightarrow{g}_{\pm 2}=\{ \pm
\frac{1}{{\sqrt{2}}},
 \pm {\sqrt{\frac{3}{2}}},0,0,
  {\sqrt{2}}\}$ with the vector $\overrightarrow{g}_{0}=\{ -{\sqrt{3}},1,0,0,0\}$.
In particular writing:
\begin{eqnarray}
  \tilde{V}_{\mbox{normal form}} &=& a \, \overrightarrow{g}_{0}
+ b \left( \overrightarrow{g}_{+2} - \overrightarrow{g}_{-2}\right) \nonumber\\
 &=& \left( -\sqrt{3} \, a \, + \sqrt{2} \, b ,a \, + \, \sqrt{6} \,
 b, 0,0,0 \right)
\label{normalforma}
\end{eqnarray}
we obtain an initial tangent vector that has non vanishing
components only in the directions of the Cartan generators. \footnote{Indeed,
starting from the Cartan subalgebra,
we can generate the whole $\mathbb{K}$ space by applying the adjoint action of the
$\mathbb{H}$
subalgebra $Ad_h H_i = h^{\alpha}[H_i,t_{\alpha}] = \sqrt{2}\alpha_i
h^{\alpha}K_{\alpha}$.} For
reasons of later convenience we parametrize the initial normal
tangent vector as follows:
\begin{equation}
  \tilde{V}_{\mbox{normal form}} =\left (\frac
  {\omega  - \kappa }{4\sqrt{2}}, \frac{3\omega +\kappa }{4\sqrt{6}}, 0, 0, 0\right )
\label{generatvecto}
\end{equation}
and we conclude that we can find a generating solution of the
geodesic equations if we solve the first order system for the
tangent vectors (eq.s (\ref{A2system})) with the initial
conditions given by eq.(\ref{generatvecto}).
With such  conditions the differential system (\ref{A2system}) is
immediately solved  by:
\begin{eqnarray}
  \Phi^{(gen)}_1(t) &=& 0 , \quad \Phi^{(gen)}_2(t)=0 , \quad \Phi^{(gen)}_3(t) =0
  \nonumber\\
            \chi^{(gen)}_1(t) &=& \frac
  {\omega  - \kappa }{4\sqrt{2}} , \quad \chi^{(gen)}_2(t) = \frac{3\,\omega +\kappa
}{4\sqrt{6}}
\label{A2genersolut}
\end{eqnarray}
From this generating solution we can obtain new ones by performing
$\mathrm{SO(3)}$ rotations such that they keep the solvable
parametrization of the coset stable. In particular by rotating
along the three possible rotation axes we can switch on the root
fields $\Phi_\beta(t)$, one by one. This procedure is discussed
in  chapter \ref{diffecompe}.

\section{Scalar fields of the $A_2$ model}
In order to find the solutions for the scalar fields $\phi^I$, we
have to construct explicitly the $\mathrm{SL(3,\mathbb{R})/SO(3)}$
coset representative $\mathbb{L}$. First, we fix the parametrization
of the coset representative as follows
\begin{equation}
  \mathbb{L}=\exp\left[ \varphi^3(t) \, E_3\right]\,
\exp[\varphi^1(t) \, E_1 \, + \,\varphi^2(t) \, E_2] \,
  \exp\left[ h^1(t) \, H_1\,+\, h^2(t) \,
  H_2\right] \,
\label{cosettus3}
\end{equation}
Note that here we have ordered the exponentials by height grading,
first the highest root of level two, then the simple roots of
level one, finally the Cartan generators of level zero. As we will appreciate
in eq.s (\ref{formeident}), this is
crucial in order to interpret the scalar fields $\varphi_i$ as the
components of the corresponding $p$-forms, in oxidation. Choosing
the following normalizations for the generators of the fundamental
defining representation of the group $\mathrm{SL(3,\mathbb{R})}$:
\begin{eqnarray}
H_1  =  \left(\begin{matrix} \frac{1}{{\sqrt{2}}} & 0 & 0 \cr 0 & -
\frac{1}{{\sqrt{2}}}   & 0 \cr 0 & 0 & 0 \cr
             \end{matrix} \right),\quad
H_2  = \left(\begin{matrix} \frac{1}{{\sqrt{6}}} & 0 & 0 \cr 0 &
\frac{1}
            {{\sqrt{6}}} & 0 \cr 0 & 0 & -{\sqrt{\frac{2}
                                {3}}} \cr   \end{matrix} \right)
\label{cartanini}
\end{eqnarray}
and
\begin{eqnarray}
E^1  =  \left( \begin{matrix} 0 & 1 & 0 \cr 0 & 0 & 0 \cr 0 & 0 & 0 \cr
\end{matrix} \right), \quad 
E^2  = \left(\begin{matrix} 0 & 0 & 0 \cr 0 & 0 & 1 \cr 0 & 0
& 0 \cr \end{matrix}  \right), 
\quad E^3  =  \left(\begin{matrix} 0 & 0 & 1 \cr 0 & 0 & 0
\cr 0 & 0 & 0 \cr  \end{matrix}\right) \label{Esteppini}
\end{eqnarray}
we construct a coset representative $\mathbb{L} \in
\mathrm{SL(3,\mathbb{R})/SO(3)}$ explicitly as the following upper
triangular matrix:
\begin{equation}
\mathbb{L} = 
\begin{pmatrix}
  e^{\frac{h_1(t)}{\sqrt{2}} +
    \frac{h_2(t)}{\sqrt{6}}} &
  e^{- \frac{{h_1}(t)}{\sqrt{2}} +
    \frac{h_2(t)}{\sqrt{6}}}\varphi_1(t) &
  e^{-\sqrt{\frac{2}{3}}h_2(t)}
  (\ft 12\varphi_1(t)\varphi_2(t) +
  \varphi_3(t)) \cr 0 & e^
  {-\frac{{h_1}(t)}{{\sqrt{2}}} + \frac{{h_2}(t)}{{\sqrt{6}}}} &
  e^{-\sqrt{\frac{2}{3}}h_2(t)}\varphi_2(t) \cr 0 & 0 & e^
  {- \sqrt{\frac{2}{3}}h_2(t)}  
\end{pmatrix}
\label{explicoset}
\end{equation}

Then we calculate the vielbein components through the formula:
\begin{equation}
  V^I = \mbox{Tr}\left[\mathbb{L}^{-1} \frac{d}{dt}\mathbb{L} \,
  \mathbf{K}_5^I\right]
\label{Vivielbe}
\end{equation}
where $\mathbf{K}_5^I$ are the generators of the coset defined in
eq.(\ref{K5generi}). The vielbein $V^A$ can be found explicitly as a
function of time, recalling that in the solvable gauge it is connected
with the solutions of the eq.s (\ref{A2system}) by the formula
$\tilde{V}^i={V}^i \,,\,\tilde{V}^\beta = \Phi^\beta =
\sqrt{2}V^\beta$.  
We obtain the following equations:
\begin{equation}
  \begin{array}{lclcl}
    V^1 &=&\dot{h}_1(t) & = & \chi_1(t) \\
    V^2 &=&\dot{h}_2(t) & = & \chi_2(t) \\
    V^3 &=& e^{-{\sqrt{2}}\,{h_1}(t)} \, 
    \frac{1}{\sqrt{2}}\,{\dot{\varphi}_1}(t)&=&
    \frac{1}{\sqrt{2}}\Phi_1(t)\\
    V^4 &=& e^{\frac{{h_1}(t) - {\sqrt{3}}\,{h_2}(t)}{{\sqrt{2}}}}\,
    \frac{1}{{\sqrt{2}}}\,{\dot{\varphi }_2}\,(t)
    &=& \frac{1}{\sqrt{2}}\Phi_2(t)\\
    V^5 &=& e^{-\frac{{h_1}(t) +
        {\sqrt{3}}\,{h_2}(t)}{{\sqrt{2}}}}\,
    \frac{\left( {{\varphi}_2}(t)
        \,{\dot{\varphi }_1}\,(t) -
        {{\varphi }_1}(t)\,{\dot{\varphi }_2}\,(t) + 2\,{\dot{\varphi
          }_3}\,(t)\right)}{2\,
      {\sqrt{2}}\,} &=&
    \frac{1}{\sqrt{2}}\Phi_3(t)\
  \end{array}
  \label{seconequaz}
\end{equation}
where in the last column we are supposed to write whatever
functions of the time $t$ we have found as solutions of the
differential equations (\ref{A2system}) for the tangent vectors. For future use in
the oxidation procedure it is convenient to give a name to the
following combination of derivatives:
\begin{equation}
  W(t) =  {{\varphi }_2}(t)\,{{\varphi }_1}\,'(t) -
             {{\varphi }_1}(t)\,{{\varphi }_2}\,'(t) + 2\,{{\varphi }_3}\,'(t)
\label{Wdefi}
\end{equation}
and rewrite the last of equations (\ref{seconequaz}) as follows:
\begin{equation}
\Phi_3(t) = \ft 12 e^{-\frac{{h_1}(t) +
{\sqrt{3}}\,{h_2}(t)}{{\sqrt{2}}}}\,W(t)  \
\label{fi3equaw}
\end{equation}
In particular, the generating solution for the tangent vectors
(inserting $\chi^1 = \frac{\omega - \kappa}{4\sqrt{2}}$, $\chi^2 =
\frac{3\omega + \kappa}{4\sqrt{6}}$, $\Phi^1 = 0$, $\Phi^2 = 0$,
$\Phi^3 = 0$) gives, up to irrelevant integration constants, the following scalar
fields:
\begin{eqnarray}
&& \nonumber h_1(t) = \frac{(\omega - \kappa)t}{4\sqrt{2}},\hspace{0.5cm}
h_2(t) = \frac{(3\omega + \kappa)t}{4\sqrt{6}}, \\
&& \varphi_1 (t) =0, \hspace{0.5cm} \varphi_2 (t) =0,
\hspace{0.5cm} \varphi_3 (t) = 0
\end{eqnarray}
%%%%%%%%%%%%%%%%%%
\section[Generation of new solutions via the compensator
method]{Differential equations for the $\mathrm{H}$-compensators and
  the generation of new solutions}
\label{diffecompe}
Non trivial solutions of the system (\ref{A2system}) can now be  obtained
from the generating solution (\ref{A2genersolut}) by means of a
suitable $\mathrm{H}$-subgroup compensating transformation, applying to the present case
the general procedure of the compensator method outlined at the end of chapter
\ref{geodesinomi}.
In previous paragraphs we have already collected all the ingredients which are
necessary to
construct the explicit form of eq.s (\ref{daequa}). Indeed from
eq.s (\ref{cartanini}), (\ref{Esteppini}), by recalling the definition
(\ref{generso3}), we immediately obtain the three generators $t_i$ of
the compact subgroup $\mathrm{SO(3)}$ in the $3$--dimensional
representation which is also the adjoint:
\begin{equation}
  t_1^{[3]}=\left(\begin{array}{ccc}
             0 & 1 & 0 \\
             -1 & 0 & 0 \\
             0 & 0 & 0 \
  \end{array} \right), \quad t_2^{[3]}=\left(\begin{array}{ccc}
             0 & 0 & 0 \\
             0 & 0 & 1 \\
             0 & -1 & 0 \
  \end{array} \right),\quad t_3^{[3]}=\left(\begin{array}{ccc}
             0 & 0 & 1 \\
             0 & 0 & 0 \\
             -1 & 0 & 0 \
  \end{array} \right)
\label{tgene3}
\end{equation}
On the other hand in eq. (\ref{j2generi}), we constructed the
generators $t_i$ in the $5$--dimensional $j=2$ representation,
spanned by the vielbein. Hence introducing
a compensating group element $h \in \mathrm{SO(3)}$, parametrized by three
time dependent angles in the following way:
\begin{equation}
  h=\exp \left[ \theta _3(t) \, t_3 \right] \, \exp \left[ \theta _2(t) \, t_2 \right]
  \, \exp \left[ \theta _1(t) \, t_1 \right]
\label{compensah}
\end{equation}
we immediately obtain the explicit form of
the adjoint matrix $A(\theta)$ and of the matrix $D(\theta)$, by
setting:
\begin{eqnarray}
  A(\theta)&=&\exp \left[ \theta _3(t) \, t_3^{[3]} \right] \, \exp \left[ \theta
_2(t) \, t_2^{[3]} \right]
  \, \exp \left[ \theta _1(t) \, t_1^{[3]} \right]\nonumber\\
            D(\theta)&=&\exp \left[ \theta _3(t) \, t_3^{[5]} \right] \, \exp \left[
\theta
_2(t) \, t_2^{[5]} \right]
  \, \exp \left[ \theta _1(t) \, t_1^{[5]} \right]
\label{compensahrep}
\end{eqnarray}
Inserting the normal form vector (\ref{generatvecto}) and the above
defined matrices $A(\theta)$ and $D(\theta)$ into the differential system
(\ref{daequa}) we obtain the following explicit differential
equations for the three time dependent $\theta$-parameters:
\begin{eqnarray}
\nonumber &&{\dot{\theta} _3}(t)=\ft 14\omega \,
                                 \sin 2\,\theta _3(t) \\
\nonumber && {\dot{\theta} _2}(t) = \ft 18 \left[ \kappa  + \omega
\,\cos 2\,\theta _3(t) \right] \,
            \sin 2\,\theta _2(t)\\ \nonumber
&& {\dot{\theta} _1}(t) = - \ft 1{16} [ \kappa  + \kappa \,\cos 2\,{{\theta
}_2}(t) + \omega [\cos2\theta_2(t) - 3]\cos2\theta_3(t)]
\,\sin 2\,{{\theta }_1}(t) + \\ && + \ft 12\,\omega \,{\sin^2
{{\theta }_1}(t)}\,\sin {{\theta }_2}(t)\,
            \sin 2\,{{\theta }_3}(t)
            \label{A2thetas}
\end{eqnarray}
At the same time the rotated tangent vector reads as follows in terms
of the chosen angles:
\begin{eqnarray}
V_{\mbox{rot}}  & \equiv & \overrightarrow{v}_{\mbox{n.f.}}\,D(\theta)
  \nonumber\\
V_{\mbox{rot}}^1 &=& \ft 1{16\sqrt{2}}\left\{- \cos 2\theta_1
                                \left[2\kappa + 2\kappa \cos 2\theta_2  +
                                          \omega\cos 2(\theta_2
-\theta_3)\right.\right.\nonumber\\&& 
\left.\left. -
6\omega \cos 2\theta_3  + \omega \cos 2(\theta_2 + \theta_3)  \right]
 + 8\omega \sin 2\theta_1 \sin \theta_2
                     \sin 2\theta_3  \right \} \nonumber\\
V_{\mbox{rot}}^2 &=& \ft 1{16\sqrt{6}}\left \{-2\,\kappa  + 6\,\kappa \,\cos
2\,{{\theta }_2}  +
             3\,\omega \,\cos 2\,\left( {{\theta }_2}  - {{\theta }_3}  \right)
                     \right.
\nonumber\\
&& \left.
             +
             6\,\omega \,\cos 2\,{{\theta }_3}  +
             3\,\omega \,\cos 2\,\left( {{\theta }_2}  + {{\theta }_3}  \right) 
\right\}
                        \nonumber\\
V_{\mbox{rot}}^3 &=& \ft 1{16\sqrt{2}}\left\{-\left[ 2\,\kappa
 + 2\,\kappa \,\cos 2\,{{\theta }_2}  +
                                          \omega \,\cos 2\,\left( {{\theta }_2}  -
{{\theta }_3}  \right)  -
                                          6\,\omega \,\cos 2\,{{\theta }_3} 
\right.\right.\nonumber\\
&&\left.\left.+
                                          \omega \,\cos 2\,\left( {{\theta }_2}  +
{{\theta }_3}  \right)
                                          \right] \,\sin 2\,{{\theta }_1}  -
             8\,\omega \,\cos 2\,{{\theta }_1} \,\sin {{\theta }_2} \,
                     \sin 2\,{{\theta }_3} \right\}  \nonumber\\
V_{\mbox{rot}}^4 &=&\ft 1{4\sqrt{2}} \left \{\cos {{\theta }_1} \,\left( {\kappa } +
                                \omega \,\cos 2\,{{\theta }_3} \right) \,\sin
2\,{{\theta }_2}  +
             2\,\omega \,\cos {{\theta }_2} \,\sin {{\theta }_1} \,
                     \sin 2\,{{\theta }_3} \right \}\nonumber\\
V_{\mbox{rot}}^5 &=& \ft 1{4\sqrt{2}}\left \{- \left( \kappa  + \omega \,\cos
(2\,{{\theta }_3} ) \right) \,
                                \sin {{\theta }_1}\,\sin 2\,{{\theta }_2}
                                \right.\nonumber\\
                                &&\left. +
             2\,\omega \,\cos {{\theta }_1} \,\cos {{\theta }_2} \,
                     \sin 2\,{{\theta }_3} \right \}
\label{ruotatone}
\end{eqnarray}
In this way finding solutions of the original differential system for
tangent vectors is reduced to the problem of finding solutions of the
differential system for the compensating angles (\ref{A2thetas}).
The main property of this latter system is that it can be solved
iteratively. By inspection we see that the first of eq.s
(\ref{A2thetas}) is a single differential equation in separable
variables for the angle $\theta_3$. Inserting the resulting solution
into the second of eq.s (\ref{A2thetas}) produces a new differential
equation in separable variables for $\theta_2$ which can also be
solved by direct integration. Inserting these results into the last
equation produces instead a non--linear differential equation for
$\theta_1$ which is not with separable variables and reads as
follows:
\begin{equation}
  p_1(t)\sin 2\,{{\theta }_1}(t) + p_2(t){\sin^2 {{\theta }_1}(t)} +
  {{\theta }_1}'(t)=0
\label{generth1}
\end{equation}
In eq.(\ref{generth1}) $p_i(t)$ are two functions of time determined by the previous
solutions for $\theta_{2,3}(t)$.
Explicitly they read:
\begin{eqnarray}
p_1(t) & = & \ft 1{32}\left\{ 2\,\kappa  + 2\,\kappa \,\cos 2\,{{\theta }_2}(t) +
            \omega \,\cos 2\,\left[ {{\theta }_2}(t) - {{\theta }_3}(t) \right]
\right.\nonumber\\
            &&\left. -
            6\,\omega \,\cos 2\,{{\theta }_3}(t) +
            \omega \,\cos 2\,\left[ {{\theta }_2}(t) + {{\theta }_3}(t) \right]
\right\}
\nonumber\\
p_2(t) & = & -\ft 12 \omega \sin \theta_2(t)\sin 2\theta_3(t)
\label{expp12}
\end{eqnarray}
and we can evaluate them using the general solutions of the first two
equations in (\ref{A2thetas}), namely:
\begin{eqnarray}
\theta_3(t) & = & -\arcsin \left[ \frac{e^{\frac{t \, \omega}{2}}}
                     {{\sqrt{e^{t\,\omega } + e^{\omega \,{{\lambda }_3}}}}} 
\right]\nonumber\\
\theta_2(t) & = &  -\arcsin\frac{e^
                                {\frac{t(\kappa  + \omega)  + \lambda_2}{4}}}{\sqrt{e^
                                 {t\omega} + e^{\omega\lambda_3}+e^
                                {\frac{t(\kappa  + \omega)  + \lambda_2}{2}}}}
\label{th23}
\end{eqnarray}
where $\lambda_{2,3}$ are two integration constants.
Equation (\ref{generth1}) is actually an integrable differential
equation. Indeed
multiplying (\ref{generth1}) by $1/ \sin ^{2}\theta_1$
and introducing the new depending variable $y(t) = \cot \theta_1$,
(\ref{generth1}) becomes actually the following linear differential equation
for $y(t)$
\begin{equation}
 2y(t) p_1(t) + p_2(t) - y(t)^\prime =0
 \label{trasformata}
 \end{equation}
which can easily be solved.
Hence the general integral of (\ref{generth1}) reads as follows:
\begin{equation}
  {{{{\theta }_1}(t)}\rightarrow
          {-\mbox{arccot}\left [e^{2\,\int {p_1}(t)\,dt}\,
                      \left( -\int \frac{{p_2}(t)}{e^{2\,\int {p_1}(t)\,dt}}\,dt +
                                {{\lambda }_1} \right) \right]}}
\label{generint1}
\end{equation}
where $\lambda_1$ is a third integration constant.
\par
In this way the system of eq.s (\ref{A2thetas}) has obtained a fully
general solution containing three integration constants. By inserting
this general solution into equation (\ref{ruotatone}) one also
obtains a complete general solution of the original differential
system for the tangent vectors containing five integration constants
$\omega,\kappa, \lambda_1 , \lambda_2 , \lambda_3$, as many as the
first order equations in the system.
\par
Let us consider for instance the choice $\lambda_2=\lambda_3=0$. In
this case the solution (\ref{th23}) for the rotation angles $\theta_{3,2}$
reduces to:
\begin{eqnarray}
\theta_3(t) & \hookrightarrow & -\arcsin \frac{e^{\frac{t\omega}{2}}}
            {{\sqrt{
                                 1 + e^{t\,\omega } }}} \nonumber\\
\theta_2(t) & \hookrightarrow & -\arcsin \frac{e^{\frac{t\,\left( \kappa  + \omega
\right) }{4}}}
            {{\sqrt{1 + e^{t\,\omega } +
                                  e^{\frac{t\,\left( \kappa  + \omega  \right) }{2}}}}}
\label{thet32spec}
\end{eqnarray}
and by replacing this result into the integrals we get:
\begin{eqnarray}
  2\,\int {p_1}(t)\,dt & = & \frac{t\,\left( \kappa  - \omega  \right)  + 4\,\log (1
+ e^{t\,\omega }) -
            2\,\log (1 + e^{t\,\omega } +
                      e^{\frac{t\,\left( \kappa  + \omega  \right) }{2}})}{4}
\label{integral1}
\end{eqnarray}
and
\begin{equation}
  -\int \frac{{p_2}(t)}{e^{2\,\int {p_1}(t)\,dt}}\,dt = \frac{1}{1 + e^{t\,\omega }}
\label{integral2}
\end{equation}
Substituting the above explicit integrations into eq.(\ref{generint1}) we obtain:
\begin{eqnarray}
  \theta_1 & \hookrightarrow & {{
            {\mbox{arccot}\left[\frac{e^{\frac{t\,\left( \kappa  - \omega  \right)
}{4}}\,
                                \left( 1 + \left( 1 + e^{t\,\omega } \right)
\,{{\lambda }_1} \right) }
                                {{\sqrt{1 + e^{t\,\omega } +
                                          e^{\frac{t\,\left( \kappa  + \omega 
\right) }{2}}}}}\right]}}}
\label{the3spec}
\end{eqnarray}
that together with eq.s (\ref{thet32spec}) provides an explicit
solution of equations (\ref{A2thetas}). We can replace such a result
in eq.(\ref{ruotatone}) and obtain the tangent vectors after three
rotations. Yet as it is evident form eq.(\ref{ruotatone}) the first two
rotations are already sufficient to obtain a solution where all the
entries of the $5$--dimensional tangent vector are non vanishing and
hence all the root fields are excited. In the sequel we will consider
the two solutions obtained by means of the first rotation and by
means of the first plus the second. They will constitute our
paradigma of how the full system can be eventually solved. These
solutions however, as we discuss in later chapters, are not only
interesting as toy models and examples. Indeed through oxidation they
can be promoted to very interesting backgrounds of ten dimensional
supergravity that make contact with the physics of $S$--branes.

\subsection{Solution of the differential equations for the tangent vectors with
  two Cartan and one nilpotent field} 

Let us consider the system (\ref{A2thetas}) and put
\begin{equation}
  \theta_1=\theta_2= \mbox{const}=0
\label{th1th2=0}
\end{equation}
This identically solves the last two equations and we are left with
the first whose general integral was already given in
eq.(\ref{th23}). By choosing the integration constant $\lambda_3=0$
we can also write:
\begin{equation}
\theta_3(t) = \arccos \frac{1}{{\sqrt{1 + e^{t\omega  }}}}
\label{sol3th}
\end{equation}
By inserting (\ref{sol3th}) and
(\ref{th1th2=0}) into (\ref{ruotatone}) we obtain the desired
solution for the tangent vectors:
\begin{eqnarray}
\chi_1(t) & = & -\frac{ \kappa  + \omega \,\tanh \frac{t\,\omega }{2}  }
  {4\,{\sqrt{2}}}, \hspace{0.5cm}
\chi_2(t)  =  \frac{\kappa  - 3\,\omega \,\tanh \frac{t\,\omega
}{2}}{4\,{\sqrt{6}}},\nonumber\\
                                 \Phi_1(t) & = & 0, \hspace{0.5cm}
\Phi_2(t)  =  0, \hspace{0.5cm}  \Phi_3(t) =
\frac{\omega}{\sqrt{(1 + e^{- t\omega})(1 + e^{t\omega})}}
\label{generetsolu}
\end{eqnarray}
where one root field is excited.
\par
Next we address the problem of solving the equations for the
scalar fields, namely eq.s(\ref{seconequaz}), which are
immediately integrated, obtaining:
\begin{eqnarray}
h_1(t) & = &-\frac{t\kappa + 2
             \log ({\cosh\frac{t\,\omega }{2}})}{4\,{\sqrt{2}}}, \hspace{0.5cm}
             h_2(t)  =  \frac{t\kappa  - 6\,\log (\cosh \frac{t\,\omega
}{2})}{4\,{\sqrt{6}}},\nonumber\\
\varphi_1(t) &=& 0, \hspace{0.5cm} \varphi_2(t) = 0,
\hspace{0.5cm}
  \varphi_3(t) = \frac{{\sqrt{1 + \tanh (\frac{t\,\omega }{2})}}}
            {e^{\frac{t\,\omega }{2}}\,{\sqrt{1 + e^{t\,\omega }}}}
\label{finsol}
\end{eqnarray}
We can now insert eq.s (\ref{finsol}) into the form of the coset
representative (\ref{explicoset}) and we obtain the geodesic as a
map of the time line into the solvable group manifold and hence
into the coset manifold depending on your taste for
interpretation:
\begin{equation}
  \mathbb{R}_t \, \hookrightarrow \, \exp \left [Solv(\mathrm{A_2}) \right
  ]\simeq \frac{\mathrm{SL(3,\mathbb{R})}}{\mathrm{SO(3)}}
  \label{mojna}
\end{equation}

In chapter (\ref{oxide1a2}) the oxidation of this sigma--model
solution to a full fledged supergravity background in $D=10$ is studied.

\subsection{Solution of the differential equations for the tangent vectors with
  two Cartan and
  three nilpotent fields}

Then we continue the hierarchical solution of the (\ref{A2thetas})
differential system by considering the next rotation $\theta_2$.  We
set $\theta_1= \mbox{const} = 0$ and we replace in eq.s
(\ref{A2thetas}) the solution (\ref{sol3th}) for $\theta_3$, with
$\lambda_3=0$. The first and the last differential equations are
identically satisfied. The second equation was already solved in
eq.(\ref{th23}).  By choice of the irrelevant integration variable
$\lambda_2=0$ a convenient solution of the above equation is provided
by the following time dependent angle:
\begin{equation}
  \theta_2(t) = -\arcsin \frac{e^{\frac{t\,\left( \kappa  + \omega
        \right) }{4}}}
  {{\sqrt{1 + e^{t\,\omega } +
        e^{\frac{t\,\left( \kappa  + \omega  \right) }{2}}}}} \label{sol2th}
\end{equation}
Inserting (\ref{sol3th}) and (\ref{sol2th}) into (\ref{ruotatone}) we obtain :
\begin{eqnarray}
  \chi_1(t) & = &  \frac{- {\left( 1 + e^{t\,\omega } \right) }^2\,\kappa    -
    \left( -1 + e^{t\,\omega } \right) \,
    \left( 1 + e^{t\,\omega } +
      2\,e^{\frac{t\,\left( \kappa  + \omega  \right)
        }{2}} \right) \,\omega }
  {4\,{\sqrt{2}}\,\left( 1 + e^{t\,\omega } \right) \,
    \left( 1 + e^{t\,\omega } + e^
      {\frac{t\,\left( \kappa  + \omega  \right) }{2}}
    \right) } \nonumber\\
  \chi_2(t) & = & \frac{\left( 1 + e^{t\,\omega } -
      2\,e^{\frac{t\,\left( \kappa  + \omega  \right)
        }{2}} \right) \,\kappa  -
    3\,\left( -1 + e^{t\,\omega } \right) \,\omega }{4\,{\sqrt{6}}\,
    \left( 1 + e^{t\,\omega } + e^
      {\frac{t\,\left( \kappa  + \omega  \right) }{2}}
    \right) }\nonumber\\
  \Phi_1(t) & = & - \frac{e^{\frac{t\,\left( \kappa 
          + 3\,\omega  \right)
      }{4}}\,\omega }
  {\left( 1 + e^{t\,\omega } \right) \,
    {\sqrt{1 + e^{t\,\omega } +
        e^{\frac{t\,\left( \kappa  + \omega  \right)
          }{2}}}}} \nonumber\\
  \Phi_2(t) & = & \frac{e^{\frac{t\,\left( \kappa  + \omega  \right) }{4}}\,
    \left( \kappa  + e^{t\,\omega }\,\left( \kappa  - \omega  \right)  +
      \omega  \right) }{2\,{\sqrt{1 + e^{t\,\omega }}}\,
    \left( 1 + e^{t\,\omega } + e^
      {\frac{t\,\left( \kappa  + \omega  \right) }{2}} \right) }
  \nonumber\\
  \Phi_3(t) & = & \frac{e^{\frac{t\,\omega }{2}}\,\omega }
  {{\sqrt{\left( 1 + e^{t\,\omega } \right) \,
        \left( 1 + e^{t\,\omega } +
          e^{\frac{t\,\left( \kappa  + \omega  \right) }{2}}
        \right)
      }}}
  \label{2rotasolu}
\end{eqnarray}
Integrating  eq.s(\ref{seconequaz}) with this new choice of the
left hand side we obtain:
\begin{eqnarray}
  h_1(t) & = & \frac{t\,\left( -\kappa  + \omega  \right)  
    - 4\,\log (1 + e^{t\,\omega
    }) +
    2\,\log (1 + e^{t\,\omega } +
    e^{\frac{t\,\left( \kappa  + \omega  \right)
      }{2}})}{4\,{\sqrt{2}}} \nonumber\\
  h_2(t) & = & \frac{t\,\left( \kappa  + 3\,\omega  \right)  -
    6\,\log (1 + e^{t\,\omega } +
    e^{\frac{t\,\left( \kappa  + \omega  \right)
      }{2}})}{4\,{\sqrt{6}}}\nonumber\\
  \varphi_1(t) &=&\frac{1}{1 + e^{t\,\omega }}\nonumber\\
  \varphi_2(t) &=& - \frac{\left( 1 + e^{t\,\omega } \right) }
  {1 + e^{t\,\omega } + e^{\frac{t\,\left( \kappa  + \omega  \right) }{2}}}
  \nonumber\\
  W(t) &=& \frac{2\,e^{t\,\omega }\,\omega }
  {\left( 1 + e^{t\,\omega } \right) \,
    \left( 1 + e^{t\,\omega } + e^
      {\frac{t\,\left( \kappa  + \omega  \right) }{2}}
    \right) }
  \label{2finsol}
\end{eqnarray}
In chapter \ref{oxide2a2} we will see how this $\sigma$-model
solution can be oxided, among other choices, to an interesting
$S3/S1$-brane solution of type IIB supergravity.

\chapter[The $\mathrm{E_8}$ Lie algebra]{The $\mathrm{E_8}$ Lie
  algebra: Reduction, Oxidation and subalgebra embeddings}
\label{genoxide}

We come now to a close examination of the $\mathrm{E_8}$ Lie algebra
and we show how the hierarchical dimensional
\textit{reduction/oxidation}
\cite{Keurentjes:2002xc}--\cite{Julia:1980gr} of supergravity
backgrounds is algebraically encoded in the hierarchical embedding of
subalgebras into the $\mathrm{E_8}$ algebra. Similarly the structure
of the bosonic lagrangians of type II A/B supergravities in $D=10$
\cite{Campbell:zc,Schwarz:1983qr,Howe:1983sr,Castellani:1993ye} is
encoded in the decomposition of the solvable Lie algebra
$Solv(\mathrm{E_{8(8)}}/\mathrm{SO(16)})$ according to irreducible
representations of two inequivalent subgroups
$\mathrm{GL(7,\mathbb{R})_{A/B} }\subset \mathrm{E_{8(8)}}$,
respectively associated with the moduli space of flat metrics on a
torus $T^7$ in compactified type II A or type II B theory
\cite{Andrianopoli:1996bq,Andrianopoli:1996zg,dualiza1}.

In order to carry out our programme we begin by spelling out the
$E_{8}$ Lie algebra in our chosen conventions.

Using the Cartan--Weyl basis the Lie algebra can be written in the
standard form:
\begin{eqnarray}
\left[ \mathcal{H}_i \, ,\, \mathcal{H}_j \right] &=& 0 \nonumber\\
\left[ \mathcal{H}_i \,,\, E_\alpha \right] &=&\alpha_i \,
E_\alpha \quad \quad \quad \quad\quad
\forall \alpha \in \Delta_+ \nonumber\\
\left[ \mathcal{H}_i \,,\, E_{-\alpha} \right] &=&- \alpha_i \,
E_{-\alpha} \nonumber\\
\left[ E_\alpha \,,\, E_\beta \right] &=&
N_{\alpha\beta}\,E_{\alpha + \beta}\quad \quad \quad \mbox{if
$\alpha + \beta \in
\Delta_+$}\nonumber\\
\left[ E_\alpha \,,\, E_\beta \right] &=& 0\quad \quad \quad \quad
\quad \quad \mbox{if $\alpha + \beta
\notin \Delta_+$}\nonumber\\
\left[ E_\alpha \,,\, E_{-\beta} \right]& = & \delta_{\alpha\beta}
\, \, \alpha^i \, {H}_i \label{cartaweyl}
\end{eqnarray}
where $\mathcal{H}_i$ are the $8$ Cartan generators, $E_{\alpha}$ are
the $120$ step operators associated with the positive roots $\alpha
\in \Delta_+$.\footnote{The values of the constants $N_{\alpha\beta}$,
  that enable to construct explicitly the representation of
  $E_{8(8)}$, used in this paper, are given in the hidden appendix. To
  see it, download the source file, delete the tag $end\{document\}$
  after the bibliography and $LaTeX$.}

Our choice of the  simple roots as vectors in an Euclidean
$\mathbb{R}^8$ space is the following one
\begin{eqnarray}
  \alpha_1 & = & \{ 0,
  1, -1, 0, 0, 0, 0, 0\}  \nonumber\\
  \alpha_2 & = & \{ 0,
  0, 1, -1, 0, 0, 0, 0\}  \nonumber\\
  \alpha_3 & = & \{ 0,
  0, 0, 1, -1, 0, 0, 0\}  \nonumber\\
  \alpha_4 & = & \{ 0,
  0, 0, 0, 1, -1, 0, 0\}  \nonumber\\
  \alpha_5 & = & \{ 0,
  0, 0, 0, 0, 1, -1, 0\}  \nonumber\\
  \alpha_6 & = & \{ 0,
  0, 0, 0, 0, 1, 1, 0\}  \nonumber\\
  \alpha_7 & = & \{ - \ft 12, - \ft 12, - \ft 12 , 
  - \ft 12 , - \ft 12  , - \ft 12,
  -\ft 12, -\ft 12\}  \nonumber\\
  \alpha_8 & = & \{ 1, -1, 0, 0, 0, 0, 0 , 0 \}  \nonumber
  \label{simplerutte}
\end{eqnarray}

The Dynkin diagrams corresponding to
$\mathrm{GL(7,\mathbb{R})}_{A/B} $ are defined by the following
simple roots:
\begin{eqnarray}
\mathrm{GL(7,\mathbb{R})}_{A}&\leftrightarrow
&\{\alpha_1,\,\alpha_2,\,\alpha_3,\,\alpha_4,\,\alpha_6,\,\alpha_8\}\nonumber\\
\mathrm{GL(7,\mathbb{R})}_{B}&\leftrightarrow
&\{\alpha_1,\,\alpha_2,\,\alpha_3,\,\alpha_4,\,\alpha_5,\,\alpha_8\}
\label{sl7ab}
\end{eqnarray}
\begin{figure}[!t]
  \centering
\includegraphics[width=12cm]{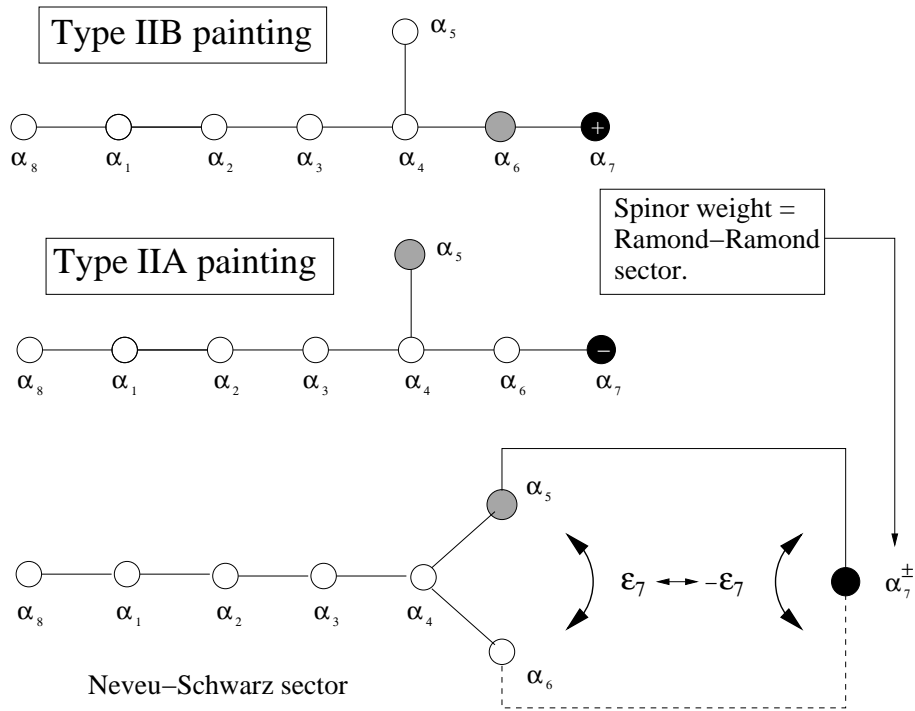}
  \caption{There are two different ways of embedding the $\mathrm{SL(7,R)}$ Lie
algebra in $E_{8(8)}$ which correspond to the type IIA and type IIB
interpretation of the same sigma model. This can also be seen as a
different way of painting the same Dynkin diagram with blobs that are
either associated with the metric (white) or with the $B$--field
(gray) or with the Ramond--Ramond field (black). Furthermore the
T--duality transforming the A painting into the B one is just the
change in sign of the $\epsilon_6$ vector in Euclidean space. Indeed
this corresponds, physically to inverting one of the torus radii $R_9
\rightarrow \alpha^\prime /R_9$.}
  \label{pittodinko}
\end{figure}

These two choices are illustrated in fig.\ref{pittodinko} where the
roots belonging to the $SL(7,\mathbb{R}) \subset GL(7,\mathbb{R})$
subgroup of the metric group are painted white. The roots eventually
corresponding to a B--field are instead painted black, while the root
eventually corresponding to a RR state are painted gray. As one sees
the difference between the A and B interpretation of the same Dynkin
diagram, named by us a \textit{painting} of the same, resides in the
fact that in the first case the RR root is linked to a metric, while
in the second it is linked to a B--field.

In order to motivate the above identifications, let us start recalling
that the $T^7$ metric--moduli parametrize the coset ${\mathcal
  M}^{A/B}_g\,=\, \mathrm{GL}(7,\mathbb{R})_{A/B}/\mathrm{SO(7)}$ in
the type IIA or B frameworks. If we describe ${\mathrm {\mathcal
    M}^{A/B}_g}$ as a solvable Lie group generated by the solvable Lie
algebra $Solv( {\mathcal M}^{A/B}_g)$ \cite{dualiza1,dualiza2} then
its coset representative $L^p{}_{\hat{q}}$ (in our notation the hatted
indices are rigid, i.e. are acted on by the compact isotropy group)
will be a solvable group element which, in virtue of the Iwasawa
decomposition can be expressed as the product of a matrix ${\mathcal
  N}^{-1T}$, which is the exponent of a nilpotent matrix, times a
diagonal one ${\mathcal H}^{-1}$: $L\,=\,{\mathcal N}^{-1
  T}\,{\mathcal H}^{-1}$. Indeed the matrix ${\mathcal N}^{-1T}$ is
the exponential of the subalgebra $\mathcal{A}$ of $Solv( {\mathcal
  M}^{A/B}_g)$ spanned by the shift operators corresponding to the
$\mathrm{GL}(7,\mathbb{R})_{A/B}$ positive roots, while ${\mathcal
  H}^{-1}$ is the exponential of the six--dimensional
$\mathrm{GL}(7,\mathbb{R})_{A/B}$ Cartan subalgebra. The vielbein
$E_p{}^{\hat{q}}$ corresponding to the $T^7$ metric $g_{pq}$ will have
the following expression :
\begin{eqnarray}
E&=&L^{-1 T}\,=\,{\mathcal N}\,{\mathcal H}\,,\nonumber\\
g&=& E\,E^T\,=\,{\mathcal N}\,{\mathcal H}\,{\mathcal
H}^T\,{\mathcal N}^T\,. \label{NHHtNt}
\end{eqnarray}

The matrix ${\mathcal N}$ is non--trivial only if $T^7$ has
off--diagonal metric--moduli. In the case of a straight torus, namely
when $g_{pq}\,=\, e^{2\sigma_p}\,\delta_{pq}\,=\, R_p^2\,\delta_{pq}$
the diagonal entries of ${\mathcal H}$ are just the radii $R_p$:
${\mathcal H}_p{}^{\hat{q}}\,=\,R_p\,\delta_p{}^{\hat{q}}$.

The decomposition of $Solv(E_{8(8)}/SO(16))=Solv_8$ with respect to\\
$Solv(GL(7,\mathbb{R})/SO(7))=Solv_7^{A/B}$ has the following form:
\begin{eqnarray}
  Solv_8 &=& Solv_7^{A/B}+ o(1,1)+ {\mathcal  A}^{[1]}+{\mathbf
    B}^{[1]}+{\mathbf B}^{[2]}+ \sum_{k} {\mathbf C}_{A/B}^{[k]}
\end{eqnarray}
where $o(1,1)$ denotes the Cartan generator $H_{\alpha[7]}$
parametrized by the ten dimensional dilaton $\phi $, ${\mathbf
B}^{[2]}\,=\, B_{2+p \, , \,2+q}\, {\mathbf B}^{p\, \, q}$ is the subalgebra
parametrized by the internal components of the Kalb--Ramond field
and \\${\mathbf C}_{A/B}^{[k]}\,=\, C_{2+p_1\,\dots\, 2+p_k}\,{\mathbf
C}^{p_1\,\dots\, p_k}$ the subalgebra spanned by the internal
components of the R--R k--form (in our conventions $C_{2+p_1,\dots
2+p_k}$ for $k>4$ are the {\em dualized} vectors $C_{2+q_1\,\dots
\, 2+q_{7-k},\mu}$, with $\epsilon^{p_1\dots p_k q_1\dots
q_{7-k}}\,\neq\,0$). Finally the spaces ${\mathcal{A}}^{[1]}$ and
${\mathbf  B}^{[1]}$ are parametrized by the dualized
Kaluza--Klein and Kalb--Ramon vectors: $g_{p\,\mu},\,B_{2+p,\,\mu}$. It
can be verified that the shift generators corresponding to
${\mathrm E}_{8(8)}$ positive roots decompose into order--$k$
antisymmetric tensorial representations ${\mathbf
  T}^{[k]}\,=\,\{{\mathbf T}^{{p_1\dots p_k}}\} $ with respect to
the adjoint action of $\mathrm{GL}(7,\mathbb{R})_{A/B}$:
\begin{eqnarray}
  {\mathbf E}&\in & \mathrm{GL}(7,\mathbb{R})_{A/B}\,:\,\,\,{\mathbf
    E}\cdot {\mathbf T}^{{p_1\dots p_k}}\cdot {\mathbf
    E}^{-1}\,=\,E^{p_1}{}_{q_1}\dots E^{p_k}{}_{q_k}{\mathbf
    T}^{{q_1\dots q_k}}
\end{eqnarray}
From the definitions (\ref{sl7ab}) we see that the shift generators
corresponding to positive roots decompose with respect to
$\mathrm{GL}(7,\mathbb{R})_{A}$ 
into the subspaces 
${\mathbf B}^{[2]}$
and 
${\mathbf C}_{A}^{[k]}$, 
$k\,=\,1,\,3,\,5,\,7$ 
and with respect to
$\mathrm{GL}(7,\mathbb{R})_{B}$ 
into 
${\mathbf B}^{[2]}$ 
and 
${\mathbf C}_{B}^{[k]}$, 
$k\,=\,0,\,2,\,4,\,6$. 

As far as the R--R scalars are
concerned, these representations correspond indeed to the tensorial
structure of the type IIA spectrum 
\begin{equation*}
  C_{2+p},\,C_{2+p \, ,\, 2+q \,
    ,\, 2+r},\,C_{\mu},\,C_{2+p\, , \, 2+q\, , \, \mu}  
\end{equation*}
and type IIB
spectrum 
\begin{equation*}
  C,\,C_{2+p, 2+q},\,C_{2+p, 2+q, 2+r, 2+s},\,C_{2+p,\mu}.
\end{equation*}

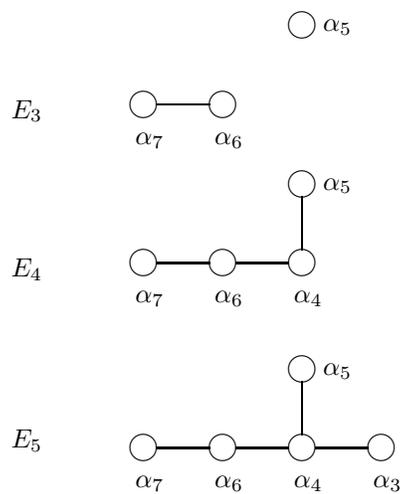
\begin{figure}[!p]
  \centering
  \begin{picture}(100,245)
    \put (-70,160){$E_3$} \put (-20,165){\circle {10}} \put
    (-23,150){$\alpha_7$} \put (-15,165){\line (1,0){20}} \put
    (10,165){\circle {10}} \put (7,150){$\alpha_6$}
    \put (40,195){\circle {10}} \put (48,193){$\alpha_5$}
%%%%%%%%%%%%%%%%%%%%%%%%%%
%%%%%%%%%%%%%%%%%%%%%%%%%%%%%%%%%
    \put (-70,100){$E_4$} \put (-20,105){\circle {10}} \put
    (-23,90){$\alpha_7$} \put (-15,105){\line (1,0){20}} \put
    (10,105){\circle {10}} \put (7,90){$\alpha_6$} \put (15,105){\line
      (1,0){20}} \put (40,105){\circle {10}} \put (37,90){$\alpha_4$}
    \put (40,135){\circle {10}} \put (48,132.8){$\alpha_5$} \put
    (40,110){\line (0,1){20}}
%%%%%%%%%%%%%%%%%%%%%%%%%%
%%%%%%%%%%%%%%%%%%%%%%%%%%%%%%%%%
    \put (-70,35){$E_5$} \put (-20,35){\circle {10}} \put
    (-23,20){$\alpha_7$} \put (-15,35){\line (1,0){20}} \put
    (10,35){\circle {10}} \put (7,20){$\alpha_6$} \put (15,35){\line
      (1,0){20}} \put (40,35){\circle {10}} \put (37,20){$\alpha_4$}
    \put (40,65){\circle {10}} \put (48,62.8){$\alpha_5$} \put
    (40,40){\line (0,1){20}} \put (45,35){\line (1,0){20}} \put
    (70,35){\circle {10}} \put (67,20){$\alpha_{3}$}
%%%%%%%%%%%%%%%%%%%%%%%%%%
%%%%%%%%%%%%%%%%%%%%%%%%%
  \end{picture}
  \caption{The Dynkin diagrams of $E_{3(3))}\subset
    E_{4(4)}\subset E_{5(5)}$ and the labeling of simple roots }
  \label{except1}
\end{figure}

We can now define a one-to-one correspondence between axions and
${\mathrm E}_{8(8)}$ positive roots. The $T^7$ moduli space is ${\rm
  SO}(7,7)_T/[{\rm SO}(7)\times {\rm SO}(7)]\,=\, {\rm Exp}(Solv_T)$
parametrized by the scalars $g_{2+p \, ,\, 2+q}$ and $B_{2+p\, ,\,
  2+q}$, where:
\begin{eqnarray}
  Solv_T &=&Solv_7^{A/B}+ {\mathbf  B}^{[2]}
\end{eqnarray}
In three dimensions the scalar fields deriving from the
dualization of $g_{p\mu}$ and $B_{p \,\mu}$ together with the dilaton
$\phi$ enlarge the manifold ${\rm SO}(7,7)_T/[{\rm SO}(7)\times
{\rm SO}(7)]$ to ${\rm SO}(8,8)/[{\rm SO}(8)\times {\rm
SO}(8)]\,=\, {\rm Exp}(Solv_{NS})$ where now:
\begin{eqnarray}
Solv_{NS} &=&Solv_7^{A/B}+o(1,1)+{\mathcal A  }^{[1]}+{\mathbf
B}^{[1]} +{\mathbf  B}^{[2]}\,,
\end{eqnarray}
This manifold is parametrized by the 64 NS scalar fields. If we
decompose $Solv_8$ with respect to $Solv_{NS}$ we may achieve an
intrinsic group--theoretical characterization of the NS and R--R
scalars. From this point of view the R--R scalar fields span the
64--dimensional subalgebra $Solv_8/Solv_{NS}$ which coincides with
a spinorial representation of ${\rm SO}(7,7)_T$ with a definite
chirality. Therefore the corresponding positive roots have grading
one with respect to the ${\rm SO}(7,7)_T$ spinorial root
$\alpha[7]$. Finally the higher--dimensional origin of the three
dimensional scalar fields can be determined by decomposing
$Solv_8$ with respect to the solvable algebra $Solv_{11-D}$
generating the scalar manifold ${\rm E}_{11-D(11-D)}/{\rm H}$ of
the D--dimensional maximal supergravity. This decomposition is
defined by the embedding of the higher--dimensional duality groups
${\rm E}_{11-D(11-D)}$ inside the three dimensional one. The
Dynkin diagrams of the $E_{11-D(11-D)}$ nested Lie algebras are
arranged according to the the pictures displayed in Fig.
\ref{except1} and Fig. \ref{except2}.

Let us now comment on the geometrical relation between the Type
IIA and IIB representations. The two ${\rm
SL}(7,\mathbb{R})_{A/B}$ Dynkin diagrams are mapped into each
other by the ${\rm SO}(7,7)$ outer authomorphism
$\epsilon_7\rightarrow -\epsilon_7$ which corresponds,  in the
light of our parametrization of the ${\rm E}_{8(8)}$ Cartan
generators, to a T--duality along the direction $x^9$.  To show
that this operation is indeed a T--duality (see \cite{Lu:1996ge}
and also \cite{Bertolini:1999uz} for a geometrical definition
of T--duality in the solvable Lie algebra formalism) let us recall
the parametrization of the Cartan subalgebra in our setup:
\begin{subequations}
  \label{typeIIAB}
  \begin{align}
    & \mbox{\bf Type IIB:} \notag\\
    \vec{h}\cdot\vec{H} & \,=\,
    \sum_{p=1}^7\,\sigma_{p}\,
    (\epsilon_p-\epsilon_8)-\frac{\phi}{2}\,\alpha_7\,=\,
    \sum_{p=1}^7\,\tilde{\sigma}_{p}\,
    (\epsilon_p-\epsilon_8)+2\phi\,\epsilon_8\\
    & \mbox{\bf Type IIA:}\notag\\
    \vec{h}\cdot\vec{H} & \,=\,
    \sum_{p=1}^7\,\sigma_{p}\,
    (\epsilon^\prime_p-\epsilon^\prime_8)-\frac{\phi}{2}\,
    (\alpha_7+\epsilon_7)\,=\,
    \sum_{p=1}^7\,\tilde{\sigma}_{p}\,
    (\epsilon^\prime_p-\epsilon^\prime_8)+2\phi\,\epsilon^\prime_8\\
    &\text{with}\quad\epsilon^\prime_w\,=\,\notag
    \begin{cases}
      \epsilon_w &\qquad \text{if}\,w\neq7\\
      \epsilon_w &\qquad \text{if}\,w=7
    \end{cases}
  \end{align}
\end{subequations}
where, in the case of a compactification on a straight torus,
$\sigma_p\,=\,\log{(R_p)}$ and
$\tilde{\sigma}_p\,=\,\log{(\tilde{R}_p)}$, $R_p$ and
$\tilde{R}_p$ being the $T^7$ radii in the ten--dimensional
Einstein-- or string--frame respectively.  Let us consider a
T--duality along directions $x^{i_1},\dots,\,x^{i_k} $:
$\tilde{R}_{i_r}\,\rightarrow\, 1/\tilde{R}_{i_r}$
($r=1,\dots,k,\,\,\alpha^\prime\,=\,1$). The transformation
$\epsilon_{i_r}\,\rightarrow\, -\epsilon_{i_r}$ in the expression
of $\vec{h}\cdot \vec{H}$ can be absorbed by the transformation
$\tilde{\sigma}_{i_r}\,\rightarrow\, -\tilde{\sigma}_{i_r}$ and
$\phi\,\rightarrow\, \phi-\sum_{r=1}^k\, \tilde{\sigma}_{i_r}$
which is indeed the effect of the T--duality. 

As a result of this analysis the precise one--to--one
correspondence between axions and positive roots can now be given
in the following form:

\begin{eqnarray}
  \mbox{Type IIB:}&&\nonumber\\
  &&C_{2+p_1 \, \dots \,2+p_k}\,\leftrightarrow\, \alpha_7+\epsilon_{p_1}+\dots
  \epsilon_{p_k} \,,\,\,\,(k\,=\,2,\,4)\,,\nonumber\\
  &&C_{2+p\, , \,\mu}\,\leftrightarrow\, \alpha_7+\epsilon_{q_1}+\dots
  \epsilon_{q_{6}} \,,\,\,\,(\epsilon^{p q_1\dots q_{6}}\neq 0)\,,\nonumber\\
  %% C_{(0)}\,\leftrightarrow\, \alpha_7\,,\nonumber\\
  &&B_{2+p\, , \, 2+q}\,
  \leftrightarrow\, \epsilon_{p}+ \epsilon_{q}\,,\nonumber\\
  &&B_{2+p\, , \,\mu}\,
  \leftrightarrow\, -\epsilon_{p}- \epsilon_{8}\,,\nonumber\\
  &&\gamma_{2+p}{}^{2+q}\,
  \leftrightarrow\,\epsilon_{p}- \epsilon_{q}\nonumber\\
  &&\gamma_\mu{}^{2+q}\,
  \leftrightarrow\,\epsilon_{q}- \epsilon_{8}\nonumber\\
  \mbox{Type IIA:}&&\nonumber\\
  &&C_{2+p_1\dots 2+p_k}\,\leftrightarrow\,
  \alpha_7+\epsilon_7+\epsilon^\prime_{p_1}+\dots \epsilon^\prime_{p_k}
  \,,\,\,\,(k\,=\,1,\,3)\,,\nonumber\\
  &&C_{2+p_1 \, , \, 2+p_2\, , \,\mu}\,\leftrightarrow\,
  \alpha_7+\epsilon_7+\epsilon^\prime_{q_1}+\dots \epsilon^\prime_{q_{5}}
  \,,\,\,\,(\epsilon^{p_1 p_2 q_1\dots q_{5}}\neq 0)\,,\nonumber\\
  &&C_{\mu}\,\leftrightarrow\, \alpha_7+\epsilon_7+\epsilon^\prime_{1}+\dots
  \epsilon^\prime_{7}\,,\nonumber\\
  &&B_{2+p\, , \, 2+q}\,\leftrightarrow\, \epsilon^\prime_{p}+
  \epsilon^\prime_{q}\,,\nonumber\\
  &&B_{2+p\, , \,\mu}\,\leftrightarrow\, -\epsilon^\prime_{p}-
  \epsilon^\prime_{8}\,,\nonumber\\
  &&\gamma_{2+p}{}^{2+q}\,\leftrightarrow\,\epsilon^\prime_{p}-
  \epsilon^\prime_{q}\nonumber\\&&
  \gamma_\mu{}^{2+q}\,\leftrightarrow\,\epsilon^\prime_{q}-
  \epsilon^\prime_{8}\nonumber\\
\end{eqnarray}
where $\gamma_{2+p}{}^{2+q}$ are the parameters entering the matrix
${\mathcal N}$ and which determine the off--diagonal entries of
the $T^7$ vielbein $E_p{}^{\hat{q}}$:
\begin{eqnarray}
  \mathcal{N}&\equiv& \exp{(\gamma_{2+p}{}^{2+q}\,\mathcal{A}_q{}^p)}
\end{eqnarray}
for a precise definition of the above  exponential representation
see \cite{dualiza1}. The fields $\gamma_\mu{}^{q}$ denote the
scalars dual to the Kaluza--Klein vectors.

\begin{figure}[!p]
  \centering
\begin{picture}(100,245)
\put (-70,160){$E_6$} \put (-20,165){\circle {10}} \put
(-23,150){$\alpha_7$} \put (-15,165){\line (1,0){20}} \put
(10,165){\circle {10}} \put (7,150){$\alpha_6$} \put
(15,165){\line (1,0){20}} \put (40,165){\circle {10}} \put
(37,150){$\alpha_4$} \put (40,195){\circle {10}} \put
(48,193){$\alpha_5$} \put (40,170){\line (0,1){20}} \put
(45,165){\line (1,0){20}} \put (70,165){\circle {10}} \put
(67,150){$\alpha_{3}$} \put (75,165){\line (1,0){20}} \put
(100,165){\circle {10}} \put (97,150){$\alpha_{2}$}
%\put (105,100){\line (1,0){20}}
%\put (130,100){\circle {10}}
%\put (127,85){$\alpha_\ell$}
%%%%%%%%%%%%%%%%%%%%%%%%%%
%%%%%%%%%%%%%%%%%%%%%%%%%%%%%%%%%
\put (-70,100){$E_7$} \put (-20,105){\circle {10}} \put
(-23,90){$\alpha_7$} \put (-15,105){\line (1,0){20}} \put
(10,105){\circle {10}} \put (7,90){$\alpha_6$} \put (15,105){\line
(1,0){20}} \put (40,105){\circle {10}} \put (37,90){$\alpha_4$}
\put (40,135){\circle {10}} \put (48,132.8){$\alpha_5$} \put
(40,110){\line (0,1){20}} \put (45,105){\line (1,0){20}} \put
(70,105){\circle {10}} \put (67,90){$\alpha_{3}$} \put
(75,105){\line (1,0){20}} \put (100,105){\circle {10}} \put
(97,90){$\alpha_{2}$} \put (105,105){\line (1,0){20}} \put
(130,105){\circle {10}} \put (127,90){$\alpha_1$}
%%%%%%%%%%%%%%%%%%%%%%%%%%
%%%%%%%%%%%%%%%%%%%%%%%%%%%%%%%%%
\put (-70,35){$E_8$} \put (-20,35){\circle {10}} \put
(-23,20){$\alpha_7$} \put (-15,35){\line (1,0){20}} \put
(10,35){\circle {10}} \put (7,20){$\alpha_6$} \put (15,35){\line
(1,0){20}} \put (40,35){\circle {10}} \put (37,20){$\alpha_4$}
\put (40,65){\circle {10}} \put (48,62.8){$\alpha_5$} \put
(40,40){\line (0,1){20}} \put (45,35){\line (1,0){20}} \put
(70,35){\circle {10}} \put (67,20){$\alpha_{3}$} \put
(75,35){\line (1,0){20}} \put (100,35){\circle {10}} \put
(97,20){$\alpha_{2}$} \put (105,35){\line (1,0){20}} \put
(130,35){\circle {10}} \put (127,20){$\alpha_1$} \put
(135,35){\line (1,0){20}} \put (160,35){\circle {10}} \put
(157,20){$\alpha_8$}
%%%%%%%%%%%%%%%%%%%%%%%%%%
%%%%%%%%%%%%%%%%%%%%%%%%%
\end{picture}
\caption{The Dynkin diagrams of $E_{6(6))}\subset
E_{7(7)}\subset E_{8(8)}$ and the labeling of simple roots}
\label{except2}
\end{figure}

\chapter{Oxidation of the $\mathrm{A_2}$ solutions}
\label{occidoa2}

In this chapter, as a working illustration of the oxidation process we
derive two full fledged $D=10$ supergravity backgrounds corresponding
to the two $A_2$ sigma model solutions derived in previous chapters.
As we already emphasized in our introduction the correspondence is not
one-to-one, rather it is one-to-many. This has two reasons. First of
all we can either oxide to a type IIA or to a type IIB configuration.
Secondly, even within the same supergravity choice (A or B), there are
several different oxidations of the same abstract sigma model
solution, just as many as the different ways of embedding the solvable
$Solv(\mathrm{A_2})$ algebra into the solvable
$Solv(\mathrm{E_8}/SO(16))$ algebra. This embeddings lead to quite
different physical interpretations of the same abstract sigma model
solution.
\par
Our first task is the classification of these inequivalent embeddings.

\section{Possible embeddings of the $A_2$ algebra}
In order to study the possible embeddings it is convenient to rely on
a compact notation and on the following graded structure of the
Solvable Lie algebra $Solv(\mathrm{E_8}/SO(16))$ characterized by the
following non vanishing commutators:
\begin{subequations}
  \label{gradastructa}
  \begin{align}
    \left[ \mathcal{A} \, ,\, \mathcal{A}\right]
    &\,=\,
    \mathcal{A}\\
    \left[\mathcal{A} \, ,\, \mathcal{A}^{[1]}\right]
    &\,=\,  
    \mathcal{A}^{[1]}\\
    \left[ \mathbf{B}^{[2]} \, , \,\mathbf{B} ^{[1]}\right]
    &\,=\,
    \mathcal{A}^{[1]}\\
    \left[ \mathcal{A} \, ,\, \mathbf{B}^{[2]}\right]     
    &\,=\,
    \mathbf{B}^{[2] }\\
    \left[ \mathcal{A} \, ,\, \mathbf{C}^{[k]}\right]
    &\,=\,
    \mathbf{C}^{[k]} \\
    \left[ \mathbf{B}^{[2]} \, ,\, \mathbf{C}^{[k]}\right] 
    &\,=\,
    \mathbf{C}^{[k+2]}\\
    \left[ \mathbf{C}^{[k]} \, ,\, \mathbf{C}^{[6-k]}\right]
    &\,=\,
    {\mathbf{B}}^{[1]}\\
    \left[ \mathbf{C}^{[4]} \, ,\, \mathbf{C}^{[4]}\right]  
    &\,=\,
    {\mathbf{A}}^{[1]}
\end{align}
\end{subequations}

In eq. (\ref{gradastructa}) $\mathcal{A}^{[1]}$, $\mathbf{B}^{[2]}$,
$\mathbf{B}^{[1]}$ and $\mathbf{C}^{[k]}$ are the spaces of nilpotent
generators defined in the previous chapter. While $\mathcal{A}$ is the
$Solv_7^{A/B}$ Lie algebra. In view of the above graded structure
there are essentially $8$ physically different ways of embedding the
$A_2$ algebra into $\mathrm{E_{8(8)}}$.
\begin{description}
\item[1] Every root $\beta_{1,2,3}$ is a metric generator ${\mathcal{A}}$.
  In this case the $A_2$ Lie algebra is embedded into the
  $\mathrm{SL(7,\mathbb{R})}$ subalgebra of
  $\mathrm{E_{8(8)}}$ and the corresponding oxidation leads to a
  purely gravitational background of supergravity which is identical
  in the type IIA or type IIB theory.
\item [2] The two simple roots $\beta_{1,2}$ are respectively
  associated with a metric generator ${\mathcal{A}}$ and a $B$-field
  generator $\mathbf{B}$.  The composite root $\beta_3$ is associated
  with a second $B$-field generator. This is so because the
  $\mathbf{B}$-generators span an antisymmetric rank $2$
  representation of $\mathrm{SL(7,\mathbb{R})}$. In this case
  oxidation leads to a purely NS configuration, shared by type II A
  and type II B theory, involving the metric, the dilaton and the
  B-field alone.
\item[3] The two simple roots $\beta_{1,2}$ are respectively associated with  a
  metric generator
  ${\mathcal{A}}$ and with a RR $k$--form generator $\mathbf{C}^{[k]}$.
  The composite root $\beta_3$ is associated with a second RR generator
  $\mathbf{C}^{[k]}$ pertaining to the same $k$. This follows again from
  the fact that the $\mathbf{C}^{[k]}$ generators span an
  $\mathrm{SL(7,\mathbb{R})}$representation. In this case oxidation
  leads to different results in type II A and type II B theories,
  although the metric is the same for the two cases and it has non
  trivial off-diagonal parts.
\item[4] The simple roots $\beta_{1,2}$ are respectively associated
  with a RR $k$-form generator $\mathbf{C}^{[k]}$ and with a B-field generator
  $\mathbf{B}$. The composite root $\beta_3$ is associated with a $k+2$ form
  generator $\mathbf{C}^{{k+2}}$. In this case the metric is purely
  diagonal and we have non trivial B-fields  and RR forms. Type IIA and
  type IIB oxidations are just different in this sector. The NS sector
  is the same for both.
\item[5] The two simple roots $\beta_{1,2}$ are respectively
  associated with a RR generator $\mathbf{C}^{[k]}$ and a RR generator
  $\mathbf{C}^{[6-k]}$.
  The composite root $\beta_3$ is associated with a
  $\mathbf{B}^{[1]}$ generator. The oxidation properties of this
  case are just similar to those of the previous case. Also here the
  metric is diagonal.
\item[6] The root $\beta_1$ corresponds to an off diagonal element of
  the internal metric, namely belongs to $\mathcal{A}$, while
  $\beta_{2,3}$ correspond to scalars dual to the Kaluza--Klein vectors
  $g_{2+p\, ,\,\mu}$ namely belong to $\mathcal{A}^{[1]}$.
\item[7] The root $\beta_1 \, \in \, \mathbf{B}^{[2]}$, namely it
  describes an internal component of the B--field. The root $\beta_2 \, \in \,
  \mathbf{B}^{[1]}$
  namely it corresponds to a B--field with mixed indices. The root
  $\beta_3 \, \in \, \mathcal{A}^{[1]}$ is associated with a mixed
  component of the metric.
\item[8] In type IIB theory the roots $\beta_{1,2}$ belong to
  $C^{[4]}$ namely are associated with two different components of the
  internal 4--form, while $\beta_3 \, \in \,\mathcal{A}^{[1]}$
  describes a mixed component of the metric.
\end{description}
\section{Choice of one embedding example}
As an illustration, out of the above list
we choose one example of embedding that has an immediate and nice
physical interpretation in terms of a brane system. We consider  the
case $4$, with a RR generator $\mathbf{C}^{[2]}$
and a B-field generator respectively associated with $\beta_{1,2}$ and
a $\mathbf{C}^{[4]}$ generator associated with the composite root
$\beta_{3}$.
In particular we set:
\begin{eqnarray}
  \beta_1 & \rightarrow & \mathbf{B}^{34}\nonumber\\
\beta_2 & \rightarrow & \mathbf{C}^{89}\nonumber\\
\beta_3 & \rightarrow & \mathbf{C}^{3489} \, \sim \, \mathbf{C}^{\mu  567}
\label{associati}
\end{eqnarray}

More precisely this corresponds to identifying $\beta _{1,2,3}$ with
the following roots of $E_{8(8)}$ according to their classification
given in the appendix:
\begin{equation}
  \begin{array}{cccclcc}
             \beta_1& \hookrightarrow &  \alpha [69] & = &\epsilon _1 + \epsilon _2 &
\leftrightarrow & B_{34}\\
             \beta_2 & \hookrightarrow & \alpha [15] & = & \alpha[7] +\epsilon _6 +
\epsilon
_7  & \leftrightarrow & C_{89}  \\
             \beta_3& \hookrightarrow & \alpha [80] & = & \alpha [7] +\epsilon _1 +
\epsilon _1 +
             \epsilon _6 + \epsilon _7&  \leftrightarrow & C_{3489} \sim C_{\mu 567} \
  \end{array}
\label{identificazie}
\end{equation}
where $\alpha[7]=\left\{ - \ft 12 ,  - \ft 12 , - \ft 12, - \ft 12, - \ft
12, - \ft 12, - \ft 12, - \ft 12 \right \}$ is the spinorial simple root
of $\mathrm{E_{8(8)}}$.
\par
Next given the explicit form of the two roots $ \beta_1  \hookrightarrow  \alpha
[69]$ and $\beta_2  \hookrightarrow  \alpha [15]$ we construct the $2$--dimensional
subspace of the Cartan
subalgebra which is orthogonal to the orthogonal complement of
$\alpha[69]$ and $\alpha[15]$ in $\mathbb{R}^8$. We immediately see
that this subspace is spanned by all $8$ vectors of the form:
\begin{equation}
  \overrightarrow{h} = \left\{ x,x,y,y,y,-y,-y,y \right\}
\label{hsubspace}
\end{equation}
so that we find:
\begin{equation}
  \overrightarrow{h} \cdot \overrightarrow{\alpha}[69] = 2x \quad ;
  \quad \overrightarrow{h} \cdot \overrightarrow{\alpha}[15] = - (x +
  3y)
\label{halp}
\end{equation}
Then we relate the fields $x$ and $y$ to the diagonal part of the ten
dimensional metric.
\par
To this effect we start from the
general relations between the ten-dimensional metric in the \textit{Einstein
frame} and the fields in three-dimensions evaluated in the $D=3$
Einstein frame, then we specialize such relations to our particular
case.
\paragraph{General relations in dimensional reduction}
The Einstein frame metric in $D=10$ can be written as:
\begin{equation}
  G_{MN}^{(Einstein)} = \left(\begin{array}{c|c}
             \exp [4 \phi_3 -\ft 12 \phi] \, g^{(E,3)}_{\mu\nu} + G_{ij} \,
             \mathcal{A}^i_\mu \, \mathcal{A}^i_\nu & G_{ik} \,
             \mathcal{A}^k_\mu  \\
             \hline
             G_{jk} \,
             \mathcal{A}^k_\nu  &G_{ij} \
  \end{array} \right)
\label{metrica10}
\end{equation}
where $g^{(E,3)}_{\mu\nu}$ is the three dimensional Einstein frame
metric (\ref{confpiatto2}) determined by the solution of the
$D=3$-sigma model via equations (\ref{einsteinocon}) and
(\ref{kdefi},\ref{kvalue}). On the other hand
$G_{ij}$ is  the Einstein frame metric in the internal seven
directions.
It parametrizes the coset:
\begin{equation}
 \frac{\mathrm{GL(7,\mathbb{R})}}{\mathrm{SO(7)}} = O(1,1) \, \times
 \, \frac{\mathrm{SL(7,\mathbb{R})}}{\mathrm{SO(7)}}
\label{7torusmoduli}
\end{equation}
In full generality, recalling eq.s(\ref{NHHtNt}) we can  set
(\cite{dualiza1,dualiza2}):
\begin{equation}
 G = E \,  E^T \quad ; \quad E = \mathcal{N} \, \mathcal{H}
\label{diag+ndiag}
\end{equation}
where, in this case:
\begin{equation}
  \mathcal{N}_{ij} = \delta_{ij}
\label{nilpotentN}
\end{equation}
since there are no roots associated with metric generators, while  the diagonal matrix:
\begin{equation}
  H_{ij} = \exp[\sigma_i] \, \delta_{ij}
\label{sigmadefi}
\end{equation}
parametrizes the degrees of freedom associated with the Cartan subalgebra
of $\mathrm{O(1,1)} \times \mathrm{SL(7,\mathbb{R})}$. The  relation
of the fields $\sigma_i$ with
the dilaton field and the Cartan fields of $E_{8(8)}$ is
obtained through the following general formulae:
\begin{eqnarray}
  \overrightarrow{h} &=&\sum_{p=1}^7 \, \left( \sigma_p + \ft 14 \phi\right)
  \,\overrightarrow{\epsilon}_p + 2 \phi_3  \, \overrightarrow{\epsilon} _8\nonumber\\
  \phi_3 & = & \ft 18 \, \phi - \ft 12 \sum_{p=1}^7 \, \sigma_p
\label{identicarta}
\end{eqnarray}
$\phi$ being the dilaton  in $D=10$ and $\phi_3$ its counterpart in $D=3$. The above formula follows immediately from eq.(\ref{typeIIAB})
\par
We also stress the following general property of the parametrization
(\ref{metrica10}) for the $D=10$ metric:
\begin{equation}
  \sqrt{-\mbox{det}G} \, G^{00} \, = \, \sqrt{-\mbox{det}g} \,\, g^{00}
\label{property}
\end{equation}
having denoted $G$ the full Einstein metric in ten dimension and $g$
the Einstein metric in three dimension.
\par
\paragraph{Specializing to our example}
Hence, in our example  the ansatz for $G_{ij}$ is diagonal
\begin{equation}
  G_{ij} =   \exp \left[ 2
  \sigma_i\right] \, \delta_{ij} \quad ; \quad i=1,\dots,7
\label{sigmafildi}
\end{equation}
and we  obtain the following relation between the fields $x$
and $y$ and the diagonal entries of the metric and the dilaton:
\begin{eqnarray}
\phi = - \overrightarrow{h}\cdot \overrightarrow{\alpha}[7]& =  & x+y \nonumber\\
\sigma_{1,2} & = & \frac {3x-y}{4} \nonumber\\
\sigma_{3,4,5} & = & \frac{3y-x}{4} \nonumber\\
\sigma_{6,7}&=&-\frac{5y+x}{4}
\label{xyident}
\end{eqnarray}
Calling $\widetilde{h}_{1,2}$ the Cartan fields in the abstract $\mathrm{A_2}$
model discussed in chapter \ref{exampsolv}, we have:
\begin{equation}
  {\widetilde{h}}\cdot \beta_1 = \sqrt{2} \,
  \widetilde{h}_1 \quad ; \quad {\widetilde{h}}\cdot
  \beta_2 = - \ft 1{\sqrt{2}} \, \widetilde{h}_1 \, + \,
  \sqrt{\ft 32} \, \widetilde{h}_1
\label{htildeident}
\end{equation}
so that we can conclude:
\begin{equation}
  x =  \ft 1{\sqrt{2}} \, \widetilde{h}_1 \, \quad ; \quad y = -
  \ft 1{\sqrt{6}} \widetilde{h}_1
\label{xyversush}
\end{equation}
We can also immediately conclude that:
\begin{equation}
 Q^2 =  \frac{d}{dt}{h}\cdot \frac{d}{dt}{h}
  =\frac{d}{dt}{\widetilde{h}}\cdot \frac{d}{dt}{\widetilde{h}}
  = |\chi_1|^2 + |\chi_2|^2
\label{hnorme}
\end{equation}
On the other hand the interpretation of $Q^2$ is the following.
Consider the
parameter $\varpi^2$  appearing in the three-dimensional
metric determined from the sigma model by Einstein equations. It is
defined as:
\begin{equation}
  \varpi^2 = h_{IJ} \dot{\phi}^I \, \dot{\phi}^J = \sum_{i=1}^8
  |\chi_i|^2 + \sum_{\alpha=1}^{120} \,|\Phi|^2
\label{varpi}
\end{equation}
If we calculate $ \varpi^2$  using the generating solution or any other
solution obtained from it by compensating $\mathrm{H}$-transformations, its
value, which is a constant, does not change. So we have:
\begin{equation}
  \varpi^2 = \sum_{i=1}^8
  |\chi_i^{(\mbox{gen.sol.})}|^2
\label{varpigensol}
\end{equation}
and in the lifting of our $\mathrm{A_2}$ solutions we can conclude
that $Q^2=\varpi^2$. Let us calculate this crucial parameter for the
case of the non trivial $A_2$ solutions discussed above. By means
of straightforward algebra we get:
\begin{equation}
  \varpi^2 =  |\chi_1^{(\mbox{gen.sol.})}|^2 + 
  |\chi_2^{(\mbox{gen.sol.})}|^2  = \ft
  1{24} \, \left(
    \kappa^2 + 3 \, \omega^2 \right)
  \label{finalvarpi}
\end{equation}
Next we turn to the identification of the $p$-forms. As we will
explicitly verify by checking type IIB supergravity field equations,
the appropriately normalized identifications are the following ones:
\begin{eqnarray}
B_{[2]} & = &  \varphi_1(t) \, dx_3 \, \wedge \, dx_4 \nonumber\\
C_{[2]} & = &  \varphi_2(t) \, dx_8 \, \wedge \,
dx_9 \nonumber\\
C_{[4]} &=& \varphi_3(t) \,dx_3 \, \wedge \,
dx_4\, \wedge \, dx_8 \, \wedge \,
dx_9 \, + \, U
\label{formeident}
\end{eqnarray}
where $U$ is the appropriate $4$--form needed to make the
corresponding field strength self dual.
\par
In this way recalling the normalizations of type IIB field strengths
as given in appendix we get:
\begin{eqnarray}
F^{NS}_{[3]034} & = & \ft 16 \varphi_1 \, '(t)\nonumber\\
F^{RR}_{[3]089} & = & \ft 16 \varphi_2 \, '(t)\nonumber\\
F^{RR}_{[5]03489}&=&\ft 1{240} \, W(t) \, \nonumber\\
F^{RR}_{[5]12567}&=&\ft 1{240} \, W(t) \,
\sqrt{-\mbox{det}g} \, \frac{1}{g_{00} \, g_{33} \, g_{44} \, g_{88} \, g_{99}}
\label{Fformidenti}
\end{eqnarray}
and we recognize that the combination $W(t)$ defined in
eq.(\ref{Wdefi}) is just the self-dual $5$-form field strength
including Chern-Simons factors.

\subsection[Oxidation of the solution with only one
  root switched on]{Full oxidation of the $A_2$ solution with only one
  root switched on} 
\label{oxide1a2} 

Let us now focus on the $A_2$ solution involving only the highest root
(similar solutions were obtained in
\cite{Ivashchuk:1997pk,Cornalba:2002nv,Cornalba:2003ze,Leblond:2003ac,
  Kruczenski:2002ap,Ohta:2003pu,Emparan:2003gg,Buchel:2003xa,pope1,pope2,
  lukas1,lukas2,quevedo}), namely on eq.s (\ref{generetsolu}) and
(\ref{finsol}). Inserting the explicit form of the Cartan fields in
eq.s(\ref{xyversush}) and then using (\ref{xyident}) we obtain the
complete form of the metric
\begin{eqnarray}
  ds^2 &=&-r^2_{[0]}(t) \,dt^2 + r^2_{[1|2]}(t) \, \left( dx_1^2 +
  dx_2^2 \right) + r^2_{[3|4]}(t) \, \left( dx_3^2 +
  dx_4^2 \right) \nonumber\\
  && + r^2_{[5|6|7]}(t) \, \left( dx_5^2 + dx_6^2 +
  dx_7^2\right) +
  r^2_{[8|9]}(t) \, \left( dx_8^2 +
  dx_9^2 \right)
  \label{a2firstmetric}
\end{eqnarray}
which is diagonal and it is parametrized
by five time dependent \textit{scale factors}
\begin{eqnarray}
r^2_{[0]}(t)&=& e^{t\,{\sqrt{\frac{{\kappa }^2}{3} + {\omega }^2}}}\,
  {\sqrt{\cosh \frac{t\,\omega }{2}}} \nonumber\\
r^2_{[1|2]}(t)&=&e^{\frac{t\,{\sqrt{\frac{{\kappa }^2}{3} + {\omega }^2}}}{2}}\,
  {\sqrt{\cosh \frac{t\,\omega }{2}}}\nonumber\\
  r^2_{[3|4]}(t)&=&\frac{1}{e^{\frac{t\,\kappa }{6}}\,{\sqrt{\cosh \frac{t\,\omega
  }{2}}}}\nonumber\\
r^2_{[5|6|7]}(t)&=&{\sqrt{\cosh \frac{t\,\omega }{2}}}\nonumber\\
r^2_{[8|9]}(t)&=&\frac{e^{\frac{t\,\kappa }{6}}}{{\sqrt{\cosh \frac{t\,\omega }{2}}}}
\label{a2firstscafacti}
\end{eqnarray}
We also obtain the explicit form of the dilaton, which turns out to be linear in
time:
\begin{equation}
 \phi= -\ft {1}{6} \, \kappa \, t
\label{a2firstdilaton}
\end{equation}
Calculating the Ricci tensor of the metric (\ref{a2firstmetric}) we
find that it is also diagonal and it has five independent eigenvalues
respectively given by:
\begin{eqnarray}
Ric_{00} & = & \frac{\left( {\kappa }^2 + 9\,{\omega }^2 + {\kappa }^2\,\cosh t\,\omega
                        \right) \,{\mbox{sech}^2(\frac{t\,\omega }{2})}}{288}
\nonumber\\
Ric_{11}=Ric_{22} & = & \frac{{\omega }^2\,{\mbox{sech}^2(\frac{t\,\omega }{2})}}
  {32\,e^{\frac{t\,{\sqrt{\frac{{\kappa }^2}{3} + {\omega }^2}}}{2}}} \nonumber\\
Ric_{33}=Ric_{44} & = & \frac{\omega^2\mbox{sech}^2(\frac{t\omega}{2})}
  {32\,e^{\frac{t\,{\sqrt{\frac{{\kappa }^2}{3} + {\omega }^2}}}{2}}} \nonumber\\
Ric_{55}=Ric_{66}=Ric_{77} & = & \frac{-\omega^2\mbox{sech}^3(\frac{t\omega}{2})}
  {32\,e^{\frac{t\,\left( \kappa  +
                                             6\,{\sqrt{\frac{{\kappa }^2}{3} +
{\omega }^2}} \right) }{6}}}
\nonumber\\
Ric_{88}=Ric_{99} & = & \frac{- \omega^2\mbox{sech}^3(\frac{t\omega}{2}) }
  {32\,e^{\frac{t\,\left( \kappa  +
                                             6\,{\sqrt{\frac{{\kappa }^2}{3} +
{\omega }^2}} \right) }{6}}}
\label{a2firstRicci}
\end{eqnarray}
On the other hand inserting the explicit values of scalar fields (\ref{finsol})
into equations (\ref{formeident}) we obtain:
\begin{align}
  \label{a2firstpform}
  F^{NS}_{[3]} & \,=\, 0 \notag\\
  F^{RR}_{[3]} & \,=\, 0 \\\notag
  F^{RR}_{[5]} & \,=\, 
  \frac{\omega\,\,
    dt\,\wedge\,dx_3\,\wedge\,dx_4\,\wedge\,dx_8\,\wedge\,dx_9
  }{
    1 + \cosh t\,\omega 
  }
  \,+\,
  \frac{\omega\,\,
    dx_1\,\wedge\,dx_2\,\wedge\,dx_5\,\wedge\,dx_6\,\wedge\,dx_7
  }{
    2 
  }
\end{align}

Considering eq.s(\ref{a2firstpform}) and (\ref{a2firstdilaton})
together the physical interpretation of the parameters $\omega$ and
$\kappa$ labeling the generating solution, becomes clear. They are
respectively associated to the charges of the $D3$ and $D5$ branes
which originate this classical supergravity solution. Indeed, as it
is obvious from the last of eq.s (\ref{a2firstpform}), there is a
dyonic $D3$-brane whose magnetic charge is uniformly distributed on
the Euclidean hyperplane 12567 while the electric charge is attached to the
 Minkowskian hyperplane 03489. The magnetic charge per unit volume
is $\omega/2$. With our
choice of the $\mathrm{A_2}$ subalgebra, there should also be a $D5$-brane
magnetically dual to an Euclidean $D$-string extending in the
directions 89. In this particular solution, where $\varphi_{1,2}=0$
the $F^{RR}_{[3]}$ vanishes, yet the presence of the $D5$ brane is
revealed by the dilaton. Indeed in a pure $D3$ brane solution the
dilaton would be constant. The linear behaviour
(\ref{a2firstdilaton}) of $\phi$, with coefficient $-\kappa/6$  is due
to the $D5$ brane which couples non
trivially to the dilaton field. Such an interpretation will become
completely evident when we consider the oxidation of the solution
obtained from this by a further $\mathrm{SO(3)}$ rotation which switches on
all the roots. This we do in the next section. Then we will
discuss how both oxidations do indeed satisfy the field equations of
type IIB supergravity and we will illustrate their physical
properties as cosmic backgrounds.

\subsection[Oxidation of the solution with all three
roots switched on]{Full oxidation of the $\mathrm{A_2}$ solution with
  all three roots switched on} 
\label{oxide2a2} 

Let us then turn to
the $\mathrm{A_2}$ solution involving all the three nilpotent fields,
namely to eq.s (\ref{2rotasolu}) and (\ref{2finsol}). Just as before,
by inserting the explicit form of the Cartan fields in
eq.s(\ref{xyversush}) and then using (\ref{xyident}) we obtain the
complete form of the new metric, which has the same diagonal structure
as in the previous example, namely
\begin{eqnarray}
  ds^2 &=&-\overline{r}^2_{[0]}(t) \,dt^2 + \overline{r}^2_{[1|2]}(t) \, \left(
    dx_1^2 +
    dx_2^2 \right) + \overline{r}^2_{[3|4]}(t) \, \left( dx_3^2 +
    dx_4^2 \right) \nonumber\\
  && + \overline{r}^2_{[5|6|7]}(t) \, \left( dx_5^2 + dx_6^2 +
    dx_7^2\right) +
  \overline{r}^2_{[8|9]}(t) \, \left( dx_8^2 +
    dx_9^2 \right)
  \label{a2secondtmetric}
\end{eqnarray}
now, however, the  \textit{scale factors} are given by:
\begin{eqnarray}
  \overline{r}^2_{[0]}(t)&=&e^{t\,\left( \frac{-\omega }{4} +
      {\sqrt{\frac{{\kappa }^2}{3} + {\omega }^2}} \right)
  }\,
  {\left( 1 + e^{t\,\omega } \right) }^{\frac{1}{4}}\,
  {\left( 1 + e^{t\,\omega } + e^
      {\frac{t\,\left( \kappa  + \omega  \right) }{2}}
    \right) }^{\frac{1}{4}} \nonumber\\
  \overline{r}^2_{[1|2]}(t)&=&e^{\frac{t\,\left( -3\,\omega  +
        2\,{\sqrt{3}}\,{\sqrt{{\kappa }^2 +
            3\,{\omega }^2}} \right) }{12}}\,
  {\left( 1 + e^{t\,\omega } \right) }^{\frac{1}{4}}\,
  {\left( 1 + e^{t\,\omega } + e^
      {\frac{t\,\left( \kappa  + \omega  \right) }{2}}
    \right) }^{\frac{1}{4}}\nonumber\\
  \overline{r}^2_{[3|4]}(t)&=&\frac{e^{\frac{-\left( t\,\kappa  \right) }{6} +
      \frac{t\,\omega }{4}}\,
    {\left( 1 + e^{t\,\omega } +
        e^{\frac{t\,\left( \kappa  + \omega  \right) }{2}}
      \right) }^{\frac{1}{4}}
  }{{\left( 1 + e^{t\,\omega } \right) }^{\frac{3}{4}}}\nonumber\\
  \overline{r}^2_{[5|6|7]}(t)&=&\frac{{\left( 1 + e^{t\,\omega } \right)
    }^{\frac{1}{4}}\,
    {\left( 1 + e^{t\,\omega } +
        e^{\frac{t\,\left( \kappa  + \omega  \right) }{2}}
      \right) }^{\frac{1}{4}}
  }{e^{\frac{t\,\omega }{4}}}\nonumber\\
  \overline{r}^2_{[8|9]}(t)&=&
  \frac{e^{\frac{t\,\left( 2\,\kappa  + 3\,\omega  \right)
      }{12}}\,
    {\left( 1 + e^{t\,\omega } \right) }^{\frac{1}{4}}}{{\left( 1 +
        e^{t\,\omega } + e^{\frac{t\,\left( \kappa  + \omega
            \right) }{2}} \right)
    }^{\frac{3}{4}}}
  \label{a2secondscafacti}
\end{eqnarray}
and the dilaton is no longer linear in time, rather it is given by:
\begin{equation}
  \phi = \frac{-\left( t\,\kappa  \right)  - 3\,\log (1 + e^{t\,\omega }) +
    3\,\log (1 + e^{t\,\omega } +
    e^{\frac{t\,\left( \kappa  + \omega  \right) }{2}})}{6}
  \label{a2secondilaton}
\end{equation}

\begin{figure}[!t]
  \centering
  \includegraphics[width=5cm]{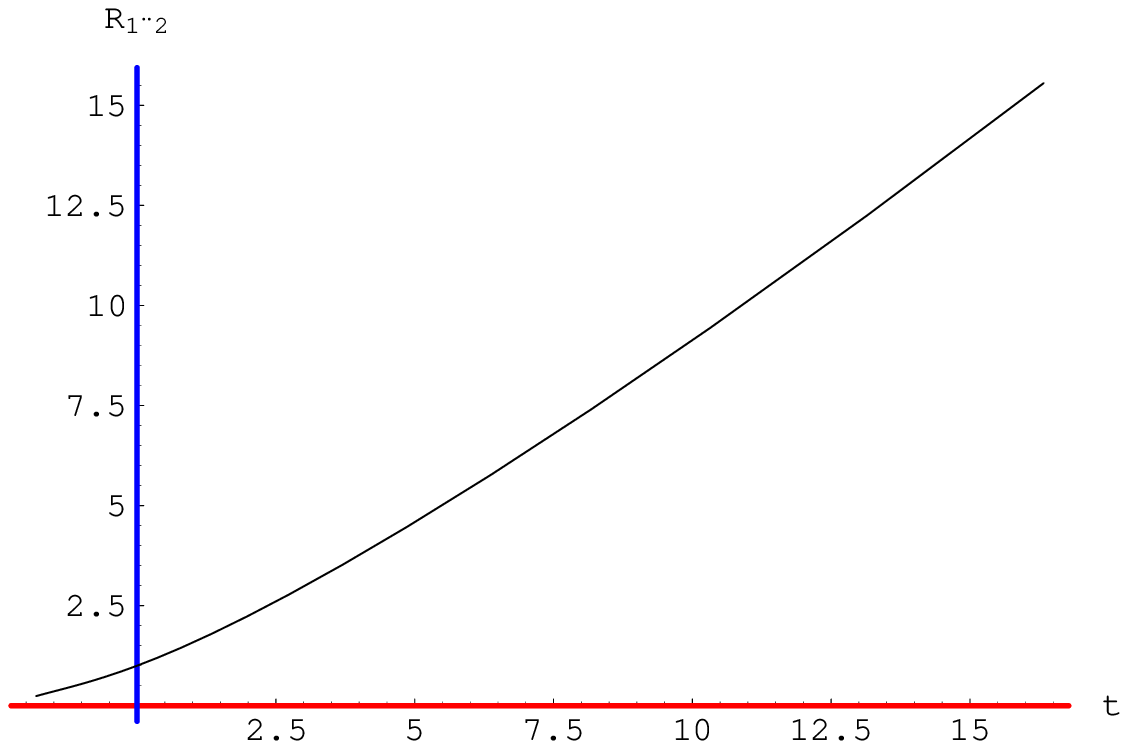}
  \includegraphics[width=5cm]{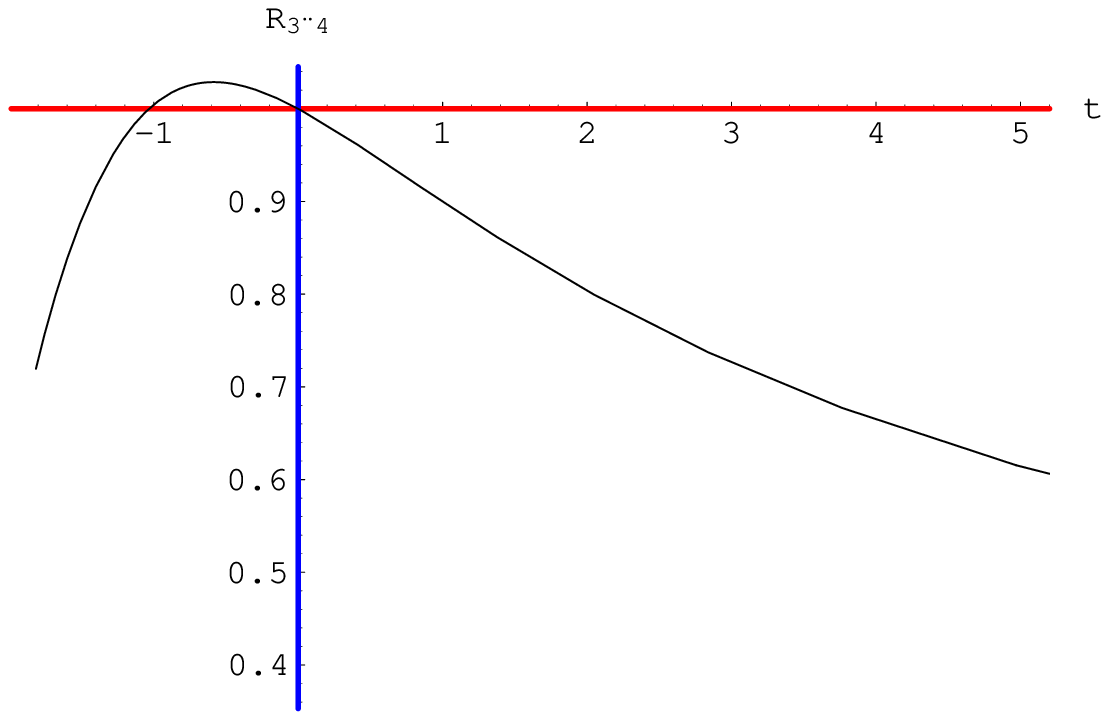}
  \includegraphics[width=5cm]{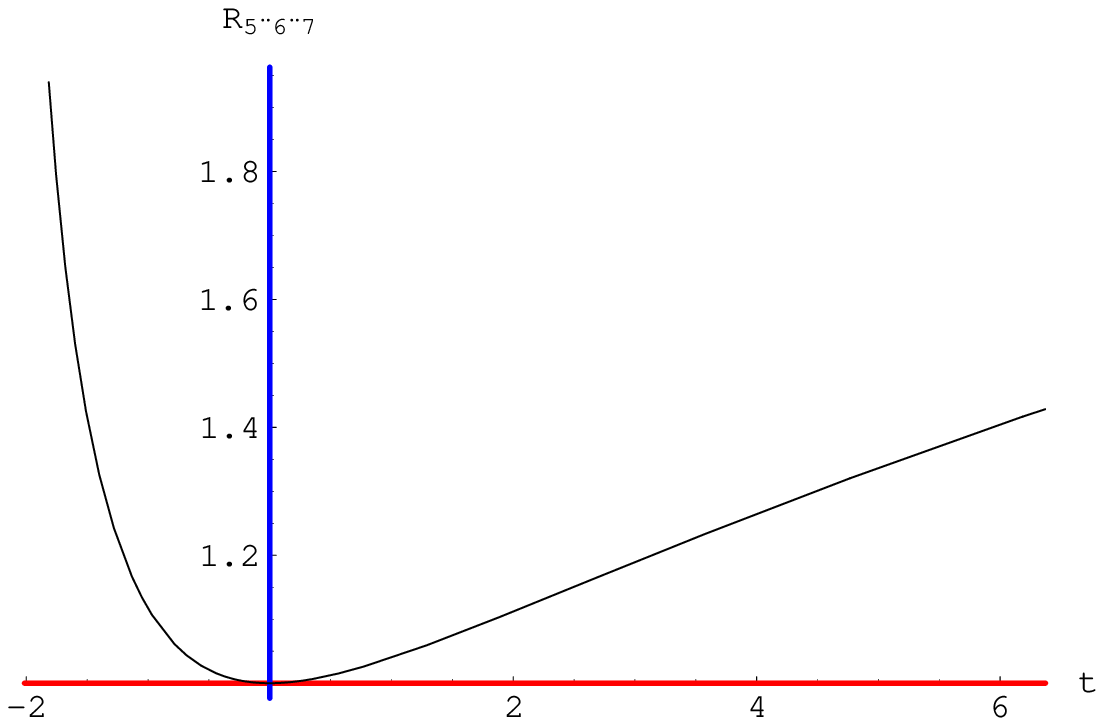}
  \includegraphics[width=5cm]{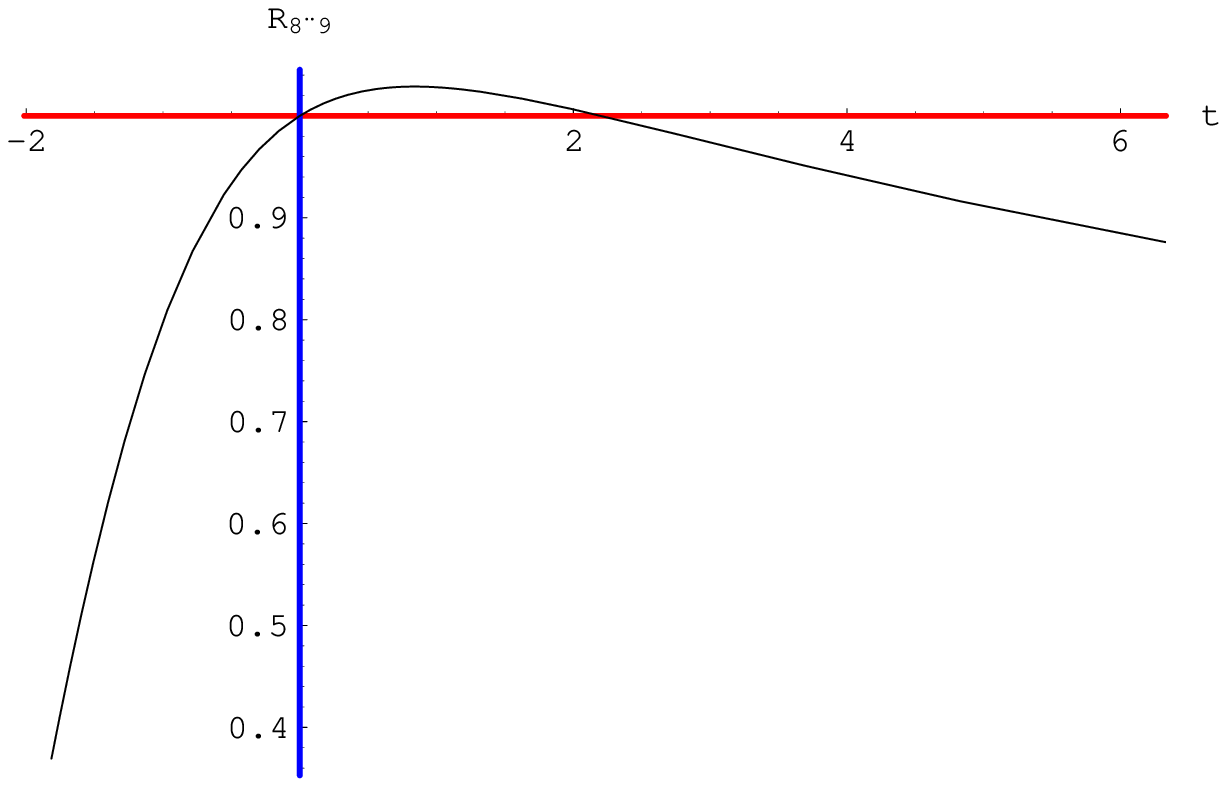}
  \caption{Plots of the scale factors $r^2_{[\alpha]}$, 
  $\alpha= 1|2 \, ,\, 3|4 \, ,\, 5|6|7 \, , \, 8|9$ as functions of
  the cosmic time $t=\tau(T)$ in the case of the choice of parameters
  $\omega =1$, $\kappa=0.5$ and for the $A_2$ solution with only the
  highest root switched on.}
  \label{radiio1k05}
\end{figure}

\begin{figure}[p]
  \centering
  \includegraphics[width=5cm]{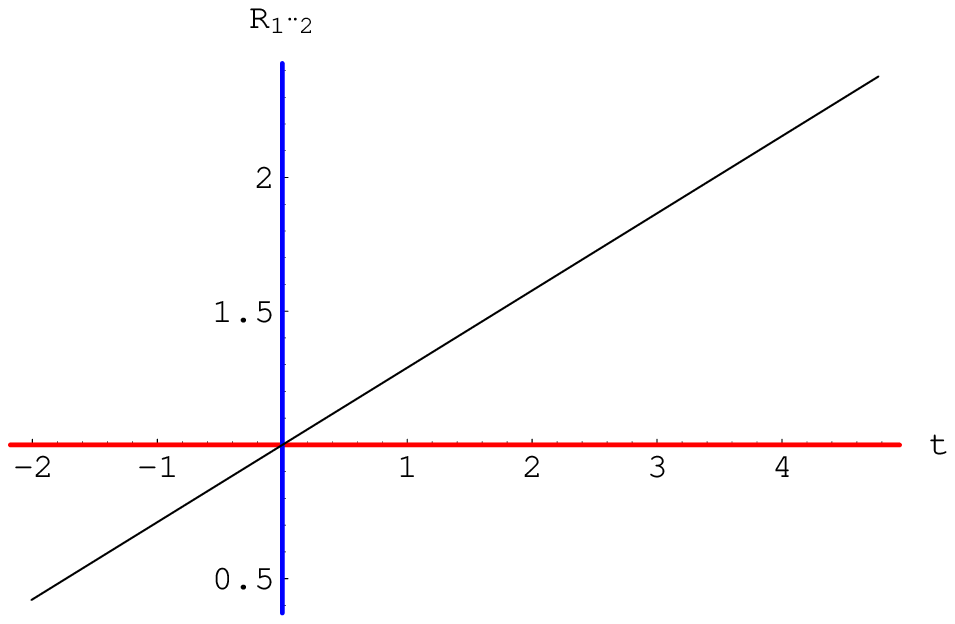}
  \includegraphics[width=5cm]{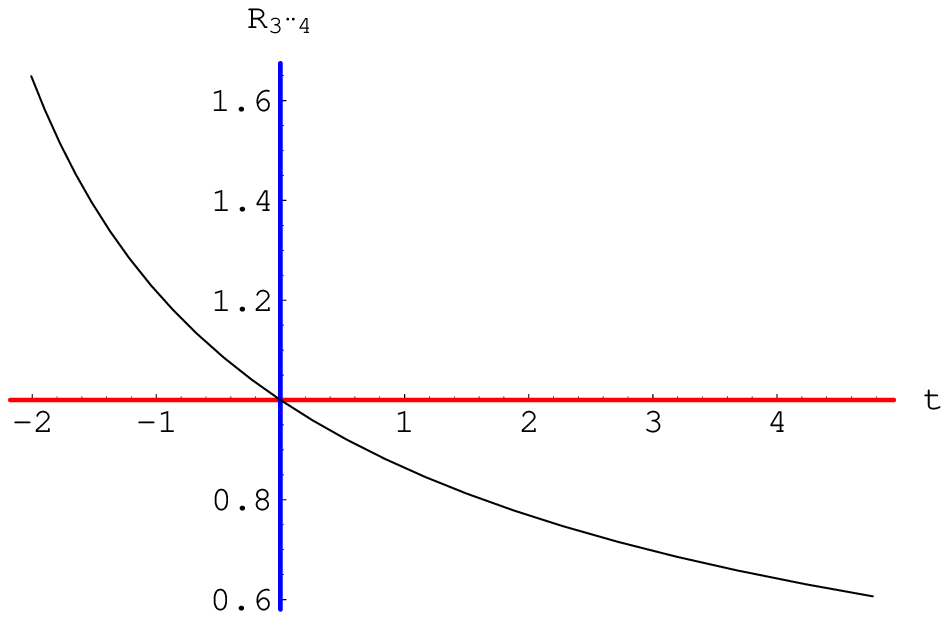}
  \includegraphics[width=5cm]{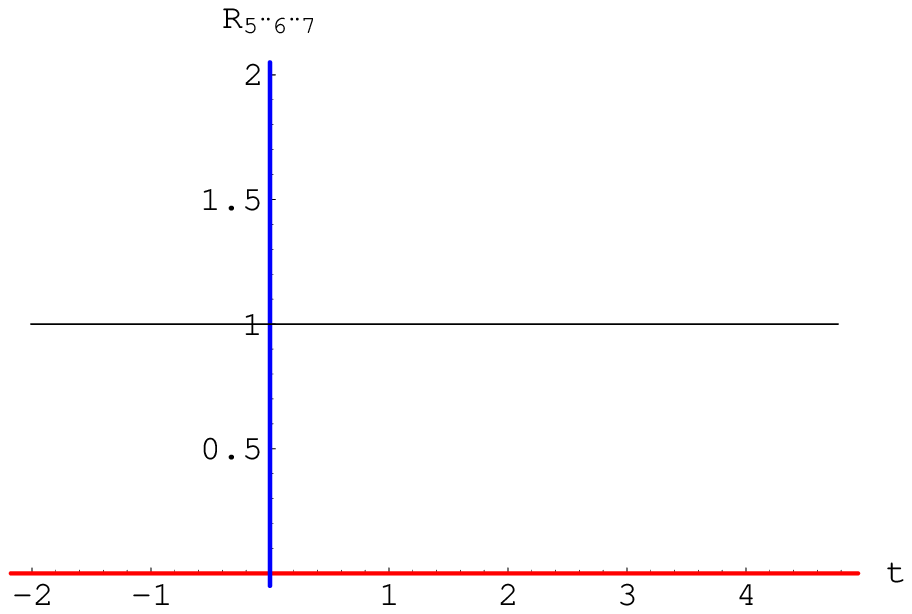}
  \includegraphics[width=5cm]{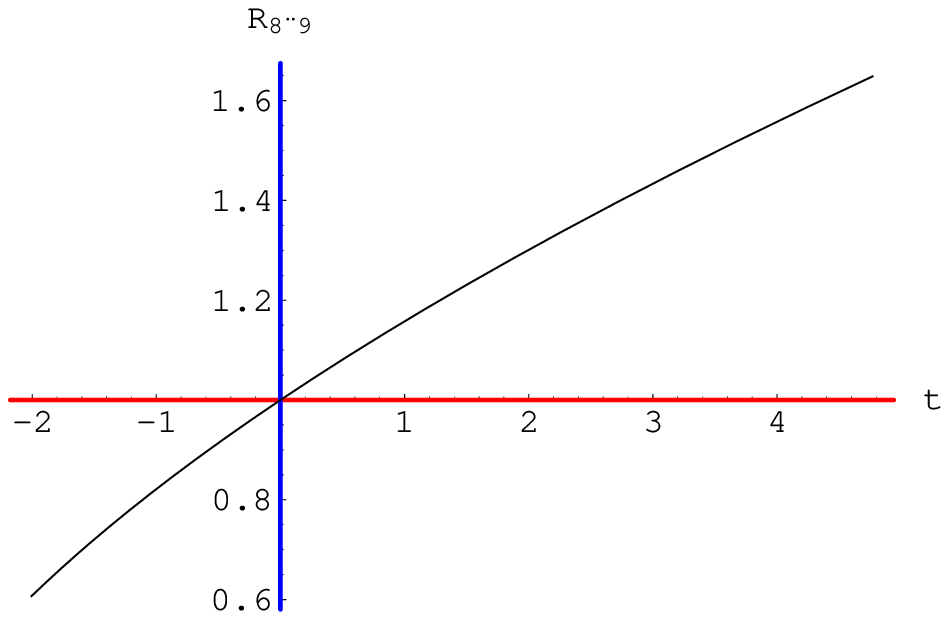}
  \caption{Plots of the scale factors $r^2_{[\alpha]}$, 
    $\alpha= 1|2 \, ,\, 3|4 \, ,\, 5|6|7 \, , \,8|9$ as functions of
    the cosmic time $t=\tau(T)$ in the case of the choice of
    parameters $\omega =0$, $\kappa=1$ and for the $A_2$ solution with
    only the highest root switched on.}
  \label{radiio0k1}
\end{figure}

\begin{figure}[p]
  \centering
  \includegraphics[width=5cm]{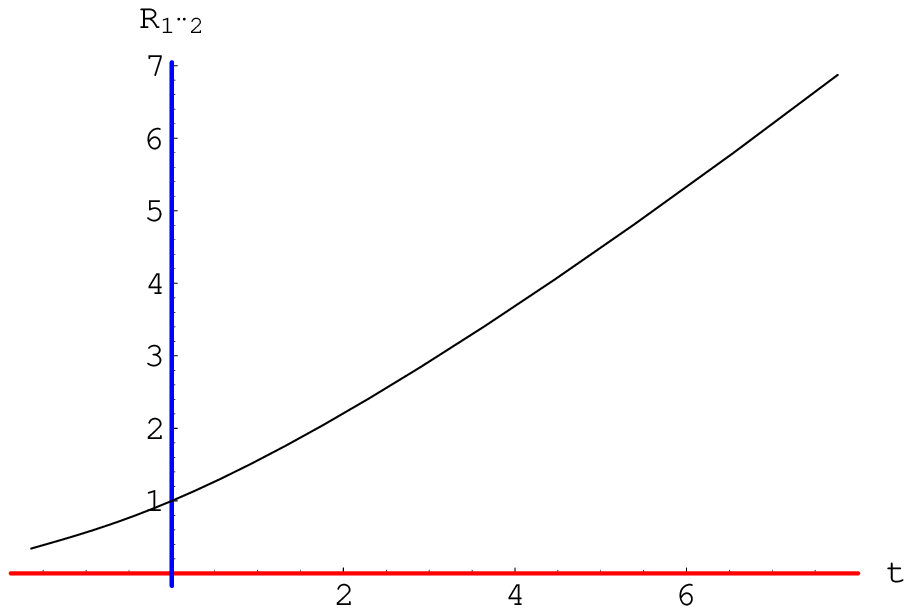}
  \includegraphics[width=5cm]{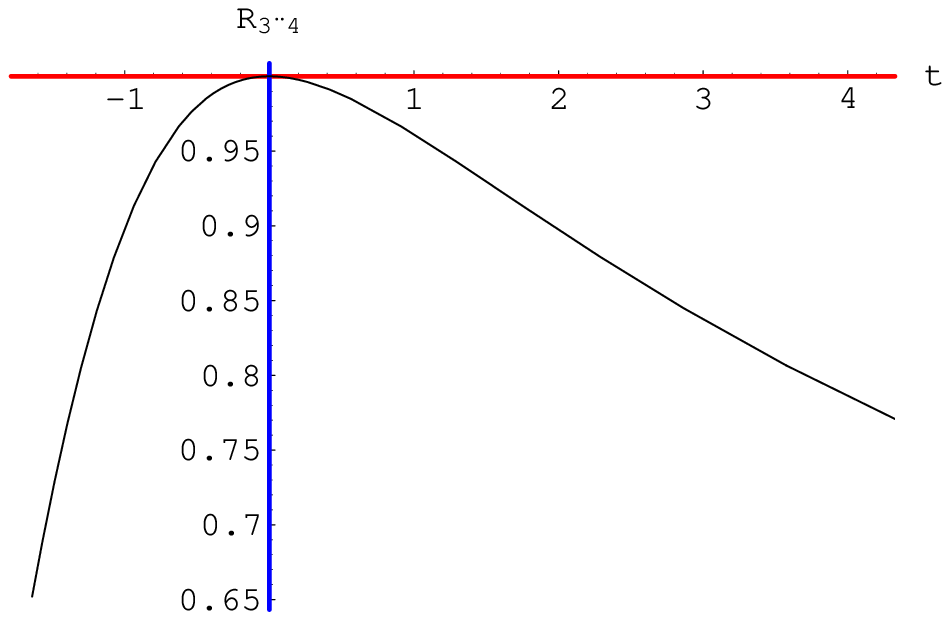}
  \includegraphics[width=5cm]{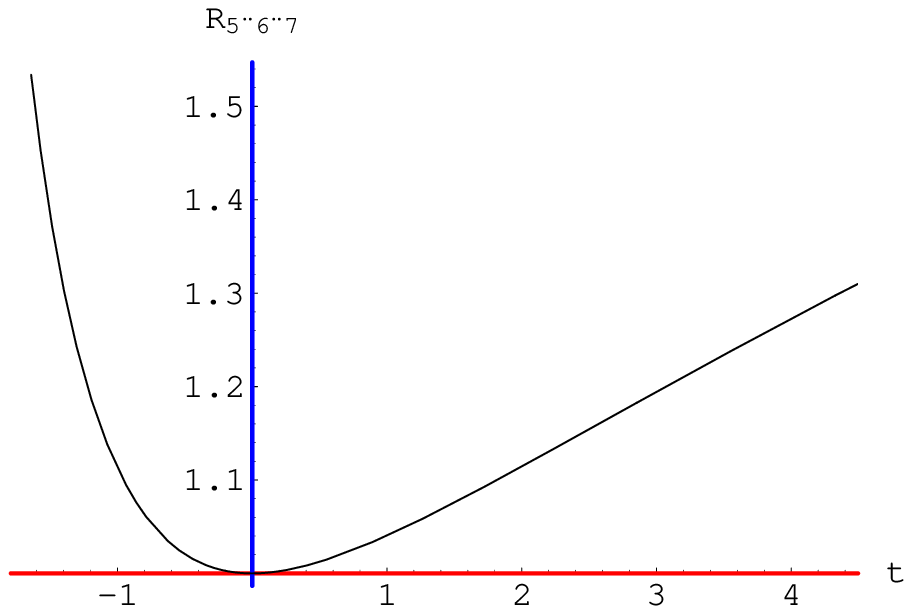}
  \includegraphics[width=5cm]{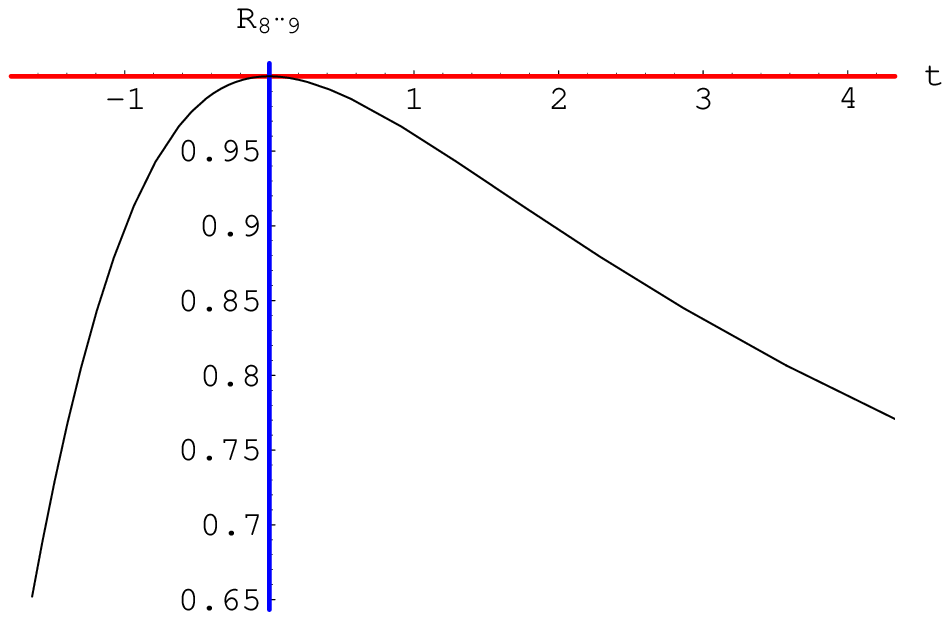}
  \caption{Plots of the scale factors $r^2_{[\alpha]}$, 
  $\alpha= 1|2 \,,\, 3|4 \, ,\, 5|6|7 \, , \,8|9$ as functions of the
  cosmic time $t=\tau(T)$ in the case of the choice of parameters
  $\omega =1$, $\kappa=0$ and for the $A_2$ solution with only the
  highest root switched on.}
  \label{radiio1k0}
\end{figure}

Calculating the Ricci tensor of the metric
(\ref{a2secondtmetric},\ref{a2secondscafacti}) we find it diagonal
with five different eigenvalues, just as in the previous case, but
with a modified time dependence, namely:

\begin{subequations}
  \label{a2secondRicci}
  \begin{multline}
    Ric_{00}\,=\, 
    \frac{1}
    {576\,
      \left( 1 + e^{t\,\omega} \right)^2\,
      \left( 
        1 + e^{t\,\omega } +
        e^\frac{t\,\left( \kappa  + \omega  \right)}{2}
      \right)^2}
\\\times 
    \left[
      \left( 1 + e^{t\,\omega} \right)^2\,
      \left( 
        4 + 
        8\,e^{t\,\omega} + 
        4\,e^{2\,t\,\omega} +
        23\,e^\frac{t\,( \kappa  + \omega )}{2} +
        e^{t\,( \kappa  + \omega )} +
        23\,e^\frac{t\,( \kappa  + 3\,\omega )}{2}
      \right)\,\kappa^2  
    \right. \\ \left.
      - 6\,e^\frac{t\,( \kappa  + \omega)}{2}\,
      \left( 
        -1 + e^{2\,t\,\omega } 
      \right)\,
      \left( 
        7 + 
        7\,e^{t\,\omega}  +
        e^\frac{t\,\left( \kappa  + \omega  \right) }{2} 
      \right)\,\kappa \,\omega
    \right.\\\left. 
      9\,\left( 8\,e^{t\,\omega } + 16\,e^{2\,t\,\omega } + 
        8\,e^{3\,t\,\omega } +
        3\,e^{\frac{t\,\left( \kappa  + \omega  \right) }{2}} +
        e^{t\,\left( \kappa  + \omega  \right) } +
        10\,e^{t\,\left( \kappa  + 2\,\omega  \right)}
      \right.\right.\\\left.\left. +
        17\,e^{\frac{t\,\left( \kappa  + 3\,\omega  \right) }{2}} +
        e^{t\,\left( \kappa  + 3\,\omega  \right) } +
        17\,e^{\frac{t\,\left( \kappa  + 5\,\omega  \right) }{2}} +
        3\,e^{\frac{t\,\left( \kappa  + 7\,\omega  \right) }{2}}
      \right)\,\omega^2
    \right] 
  \end{multline}
  \begin{multline}
    Ric_{11}\,=\,Ric_{22}\,=\, 
    \frac
    {1}
    {64\,
      \left( 
        1 + e^{t\,\omega} 
      \right)^2\,
      \left( 
        1 + 
        e^{t\,\omega } +      
        e^\frac{t\,( \kappa  + \omega ) }{2} 
      \right)^2}\,  \\ \times
    \Biggl[ 
      e^\frac
      {t\,\left( 
          \omega - 
          \sqrt{\frac{\kappa^2}{3} + \omega}^2
        \right)}
      {2}\,
      \left( 
        e^\frac
        {t\,(\kappa + 6\,\omega)}
        {2}\,
        ( \kappa  - \omega )^2 +
        8\,e^\frac{t\,\omega}{2}\,\omega^2 +
        16\,e^\frac{3\,t\,\omega}{2}\,\omega^2 +
        8\,e^\frac{5\,t\,\omega}{2}\,\omega ^2
      \right.\\\left. 
        + 
        4\,e^{t\,\left(\kappa  + \frac{3\,\omega}{2}\right)}\,\omega^2 +
        e^\frac{t\,\kappa}{2}\,(\kappa  + \omega)^2 +
        e^\frac{t\,(\kappa  + 4\,\omega)}{2}\,
        (3\,\kappa^2 - 2\,\kappa\,\omega  + 11\,\omega^2)
      \right.\\\left. +
        e^{\left(\frac{t\,\kappa}{2} + t\,\omega\right)}\,
        (3\,\kappa^2 + 2\,\kappa\,\omega  + 11\,\omega^2)
      \right) 
    \Biggr]
  \end{multline}
  \begin{multline}
    Ric_{33}\,=\,Ric_{44}\,=\,
    \frac
    {1}
    {64\,
      e^\frac{
        t\,
        \left( 
          \kappa - 
          6\,\omega  +
          6\,\sqrt{
            \frac{
              \kappa^2
            }
            {3} 
            + \omega^2
          }
        \right) 
      }
      {6}\,
      \left( 
        1 + e^{t\,\omega } 
      \right)^3\,
      \left( 
        1 + 
        e^{t\,\omega} +
        e^{\frac{
            t\,(\kappa + \omega)
          }
          {2}
        } 
      \right)^2
    }  
    \\ \times
    \left[
      e^\frac{t\,( \kappa  + 6\,\omega)}{2}\,
      (\kappa  - \omega)^2 - 
      8\,e^\frac{t\,\omega}{2}\,
      \omega^2 - 
      16\,e^\frac{3\,t\,\omega}{2}\,
      \omega^2 -
      8\,e^\frac{5\,t\,\omega}{2}\,
      \omega^2 -
      12\,e^{t\,\left(
          \kappa + \frac{3\,\omega}{2} 
        \right)}\,
      \omega^2 
    \right.\\\left. +  
      e^\frac{t\,\kappa}{2}\,
      (\kappa  + \omega)^2 + 
      e^\frac{t\,(\kappa  + 4\,\omega)}{2}\,
      (3\,\kappa^2 - 2\,\kappa\,\omega  - 21\,\omega^2) 
    \right.\\\left. +
      e^\frac{t\,(\kappa  + 2\,\omega)}{2}\,
      (3\,\kappa^2 + 2\,\kappa\,\omega  - 21\,\omega^2) 
    \right]
  \end{multline}
  \begin{multline}
    Ric_{55}\,=\,Ric_{66}\,=\,Ric_{77}\,=\,
    \frac{1}{
      64\,
      \left( 
        1 + e^{t\,\omega} 
      \right)^2\,
      \left( 
        1 + e^{t\,\omega} +
        e^\frac{
          t\,(\kappa + \omega) 
        }
        {2} 
      \right)^2} \\
    \times
\Bigg[ %blue open
e^\frac{
t\,\left( 
\omega - 
2\,\sqrt{\frac{\kappa^2}{3} + 
\omega^2}
\right)
}
{2}
\Big[  %red open
e^\frac{t\,(\kappa + 6\,\omega)}{2}\,
(\kappa  - \omega)^2 + 
8\,e^\frac{t\,\omega}{2}\,
\omega^2 +
16\,e^\frac{3\,t\,\omega}{2}\,
\omega^2 +
8\,e^\frac{5\,t\,\omega}{2}\,
\omega^2 \\ +
4\,e^{t\,\left(
\kappa  + \frac{3\,\omega}{2}
\right)}\,
\omega^2 +
e^\frac{t\,\kappa }{2}\,
(\kappa  + \omega)^2 
e^\frac{t\,(\kappa  + 4\,\omega)}{2}\,
(3\,\kappa^2 - 2\,\kappa\,\omega  + 11\,\omega ^2)  \\ +
e^\frac{t\,(\kappa  + 2\,\omega)}{2}\,
(3\,\kappa^2 + 2\,\kappa\,\omega  + 11\,\omega ^2) +
\Big]  %red closed
\Bigg] %blue closed
  \end{multline}
  \begin{multline}
    Ric_{88}\,=\,Ric_{99}\,=\,
    -\,  
    \frac{1}{
      64\,
      \left(1 + e^{t\,\omega}\right)^2\,
      \left(1 + e^{t\,\omega } +
        e^\frac{t\,\left( \kappa  + \omega 
          \right)}{2} 
      \right)^3} \\ \times
    \Bigg[ %blue open
    e^\frac{
      t\,\left( 
        \kappa + 
        6\,\omega - 
        6\,\sqrt{\frac{\kappa^2}{3} + 
          \omega^2}
      \right)}{6}\,
    \Big[  %red open
3e^\frac{t\,(\kappa + 6\,\omega)}{2}\,
(\kappa  - \omega)^2 + 
8\,e^\frac{t\,\omega}{2}\,
\omega^2 +
16\,e^\frac{3\,t\,\omega}{2}\,
\omega^2 +
8\,e^\frac{5\,t\,\omega}{2}\,
\omega^2 \\ -
4\,e^{t\,\left(
\kappa  + \frac{3\,\omega}{2}
\right)}\,
\omega^2 +
e^\frac{t\,(\kappa  + 4\,\omega)}{2}
(-3\,\kappa  + \omega)^2 +
3\,e^\frac{t\,\kappa}{2}\, 
(\kappa  + \omega)^2 +   
e^\frac{t\,(\kappa  + 2\,\omega)}{2}
(3\,\kappa  + \omega)^2 
\Big]  %red closed
    \Bigg] %blue closed
  \end{multline}
\end{subequations}

On the other hand inserting the explicit values of scalar fields
(\ref{2finsol}) into equations (\ref{formeident}) we obtain:
\begin{subequations}
  \label{a2secondpform}
  \begin{align}
    F^{NS}_{[3]} & \,=\, 
    - \frac{1}{4}\omega\,
    \text{sech}^2\frac{t\,\omega}{2}\,
    dt\,\wedge\,dx_3\,\wedge\,dx_4 \\
    F^{RR}_{[3]} & \,=\, 
    \frac{
      e^\frac{t\,(\kappa  + \omega)}{2}\,
      \left(
        \kappa + 
        e^{t\,\omega}(\kappa  - \omega)  +
        \omega\right)}
    {2\,\left( 
        1 + e^{t\,\omega} +
        e^\frac{t\,(\kappa  + \omega)}{2}
      \right)^2}
    \,dt\,\wedge\,dx_8\,\wedge\,dx_9\\
    F^{RR}_{[5]} & \,=\,
    \frac{
      e^{t\,\omega}\,\omega\,
      dt\,\wedge\,dx_3\,\wedge\,dx_4\wedge\,dx_8\wedge\,dx_9
    }{
      \left(
        1 + e^{t\,\omega}
      \right)
      \left(
        1 + e^{t\,\omega} + \e^\frac{t\,(\kappa + \omega)}{2}
      \right)
    }
    \,-\,
    \omega\,
    dx_1\,\wedge\,dx_2\,\wedge\,dx_5\wedge\,dx_6\wedge\,dx_7
  \end{align}
\end{subequations}

This formula completes the oxidation also of the second sigma model
solution to a full fledged $D=10$ type IIB configuration. As expected
in both cases the ten dimensional fields obtained by oxidation satisfy
the field equations of supergravity as formulated in the appendix. We
discuss this in the next chapter.

\subsection{How the supergravity field equations are satisfied and
  their cosmological interpretation} 

Taking into account that the
Ramond scalar $C_0$ vanishes the effective bosonic field equations of
supergravity reduce to:
\begin{eqnarray}
  d\star d \phi & = & 
  \ft 1 2 \left( e^{-\phi} \,F_3^{NS} \wedge \star F_3^{NS} -
    e^{\phi} \,F_3^{RR} \wedge \star F_3^{RR} \right) \label{dilatequa}\\
  0 & = & F_3^{NS} \wedge \star F_3^{NS} \label{ramscalequa}\\
  d\left( e^{-\phi} F_3^{NS} \right) &=& -F_3^{RR} \wedge \star F_5^{RR}
  \label{Bfielequa}\\
  d\left( e^{\phi} F_3^{RR} \right) &=& F_3^{NS} \wedge \star F_5^{RR}
  \label{Cfielequa}\\
  d\left( \star F_3^{RR} \right) &=& - F_3^{NS} \wedge F_3^{RR}
  \label{Cfielequabis}\\
  -2 Ric_{MN}& =& \hat{T}_{MN} \label{einsteinequazia}
\end{eqnarray}
where the reduced stress energy tensor $\hat{T}_{MN}$ is the
superposition of two contributions that we respectively attribute to
the $D3$ brane and to the $D5$-brane, namely:
\begin{eqnarray}
  \hat{T}_{MN} &=& \hat{T}_{MN}^{[D3]} + \hat{T}_{MN}^{[D5]}\label{totstress}\\
  \hat{T}_{MN}^{[D3]} & \equiv & 150
  {F}_{[5]{M}\cdot\cdot\cdot\cdot}
  {F}_{[5]{N}}^{\phantom{{M}}\cdot\cdot\cdot\cdot}\label{d3stress}\\
  \hat{T}_{MN}^{[D5]} & \equiv & 
  \frac{1}{2}\partial_{{M}}\varphi\partial_{{N}}\varphi
  + 9 \left( e^{-\varphi}F_{[3]{M}\cdot\cdot}^{NS}\,
    F_{[3]{N}}^{{NS}\phantom{{M}}\cdot\cdot} +e^{\varphi}{
      F}_{[3]{M}\cdot\cdot}^{RR}
    { F}_{[3]{N}}^{RR\phantom{{M}}\cdot\cdot}\right)\nonumber\\
  && -\frac{3}{4}\,
  g_{{MN}}\,\left(e^{-\varphi}F_{[3]\cdot\cdot\cdot}^{NS}
    F_{[3]}^{NS\cdot\cdot\cdot}+e^{\varphi}{{F}}_{[3]\cdot\cdot\cdot}^{RR}{
      F}^{RR\cdot\cdot\cdot}_{[3]}\right)\label{d5stress}
\end{eqnarray}
By means of laborious algebraic manipulations that can be easily
performed on a computer with the help of MATHEMATICA, we have
explicitly verified that in both cases, that of chapter \ref{oxide1a2}
and that of chapter \ref{oxide2a2} the field
eq.s(\ref{dilatequa})-(\ref{einsteinequazia}) are indeed satisfied, so
that the oxidation procedure we have described turns out to be well
tuned and fully correct.

\begin{figure}[!t]
  \centering
  \includegraphics[width=5cm]{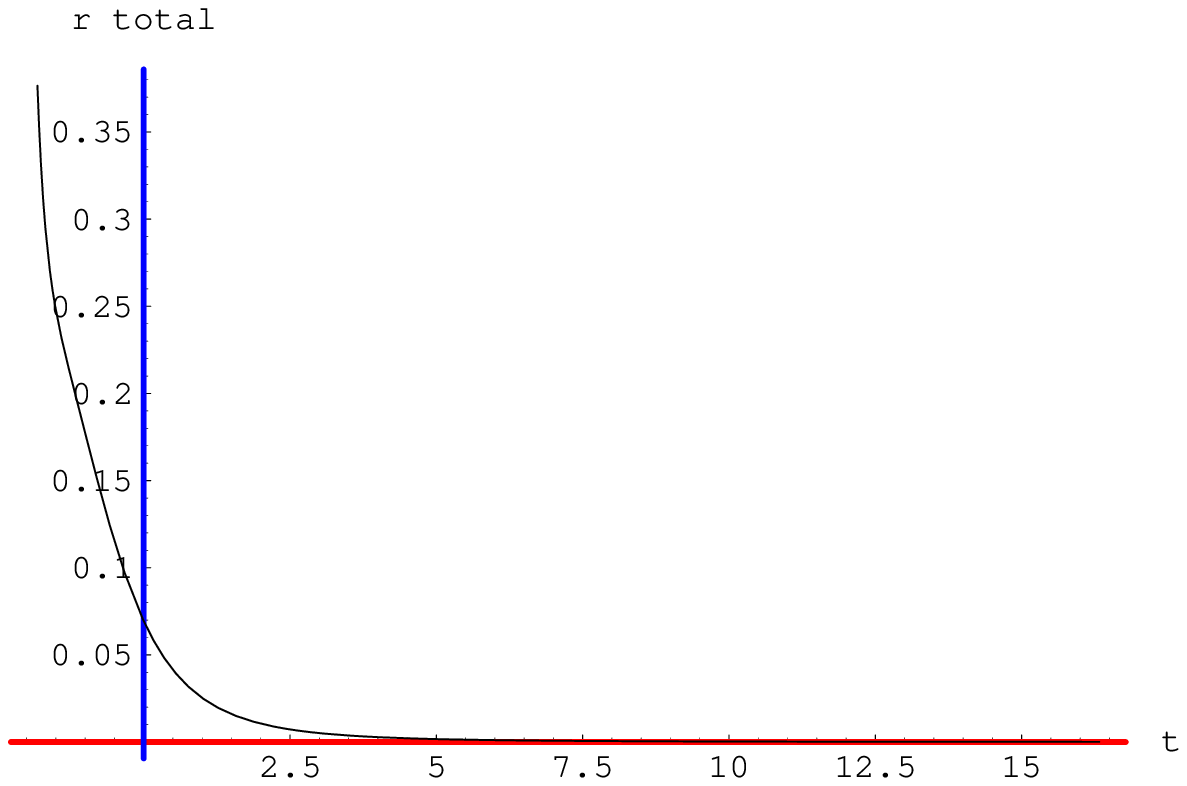}
  \includegraphics[width=5cm]{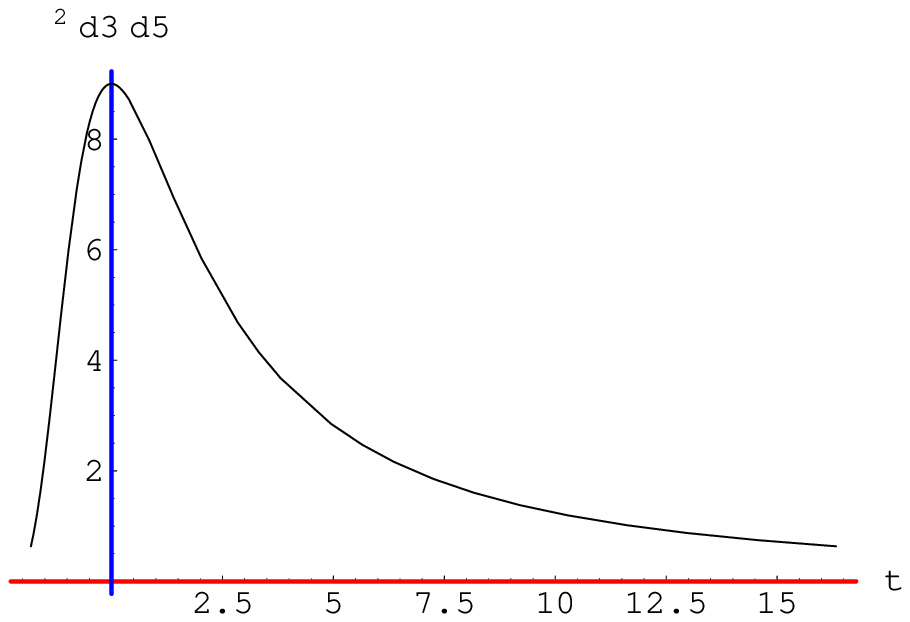}
  \includegraphics[width=5cm]{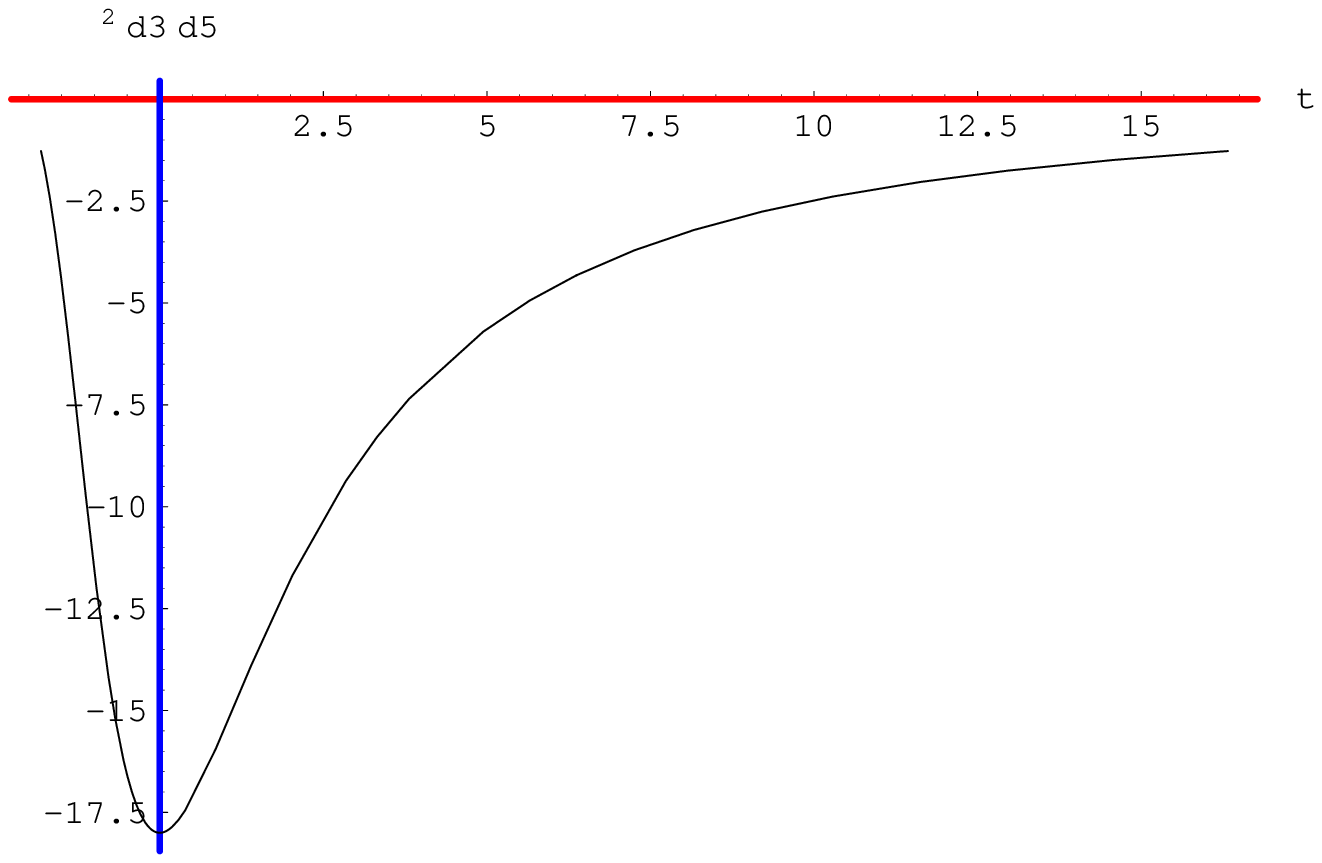}
  \caption{Plots of the energy densities  in the case of the 
    choice of parameters $\omega =1$, $\kappa=0.5$ and for the $A_2$
    solution with only the highest root switched on.  The first
    picture plots the total density $\rho^{tot}(\tau)$. The second
    picture plots the ratio $\rho^{d3}(\tau)/\rho^{tot}(\tau)$ and the
    third plots the ratio $\rho^{d3}(\tau)/\rho^{d5}(\tau)$}
  \label{rhoo1k05}
\end{figure}

\begin{figure}[!t]
  \centering
  \includegraphics[width=5cm]{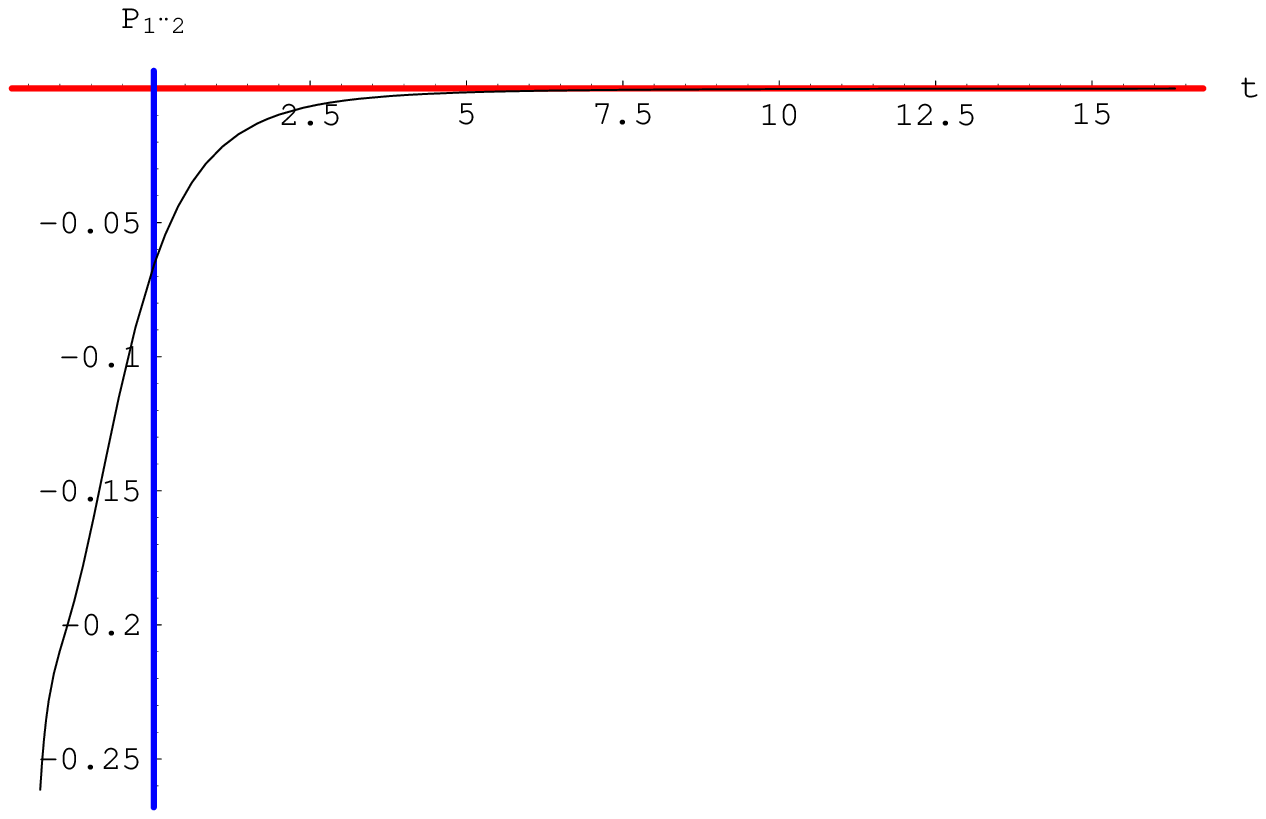}
  \includegraphics[width=5cm]{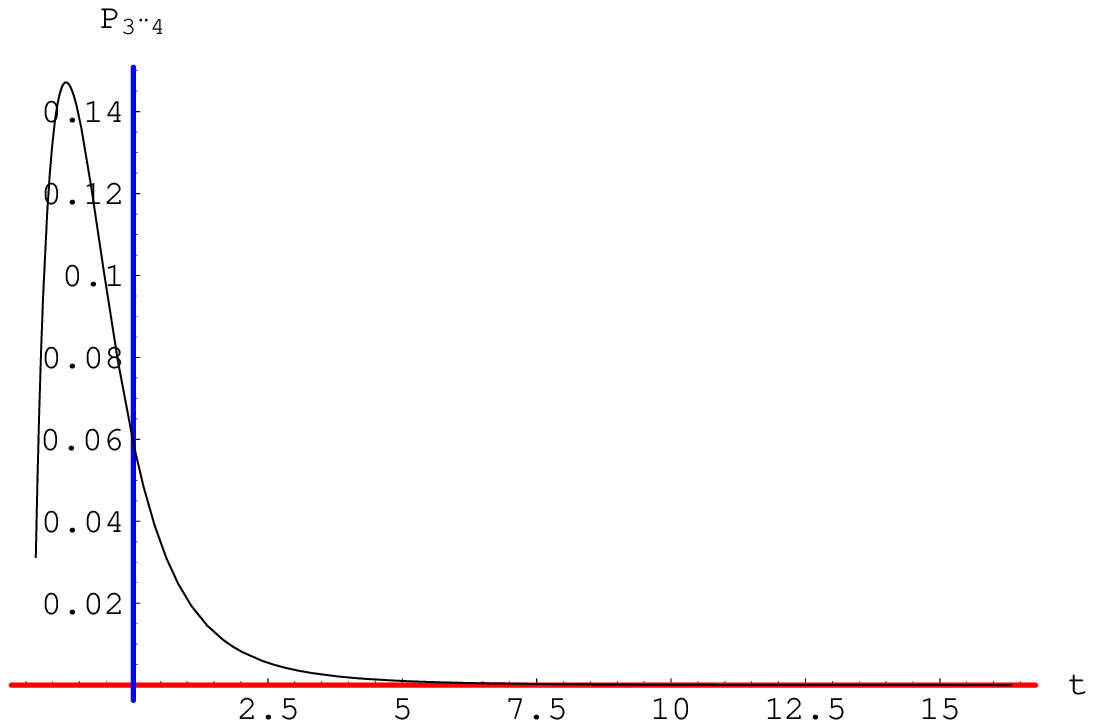}
  \includegraphics[width=5cm]{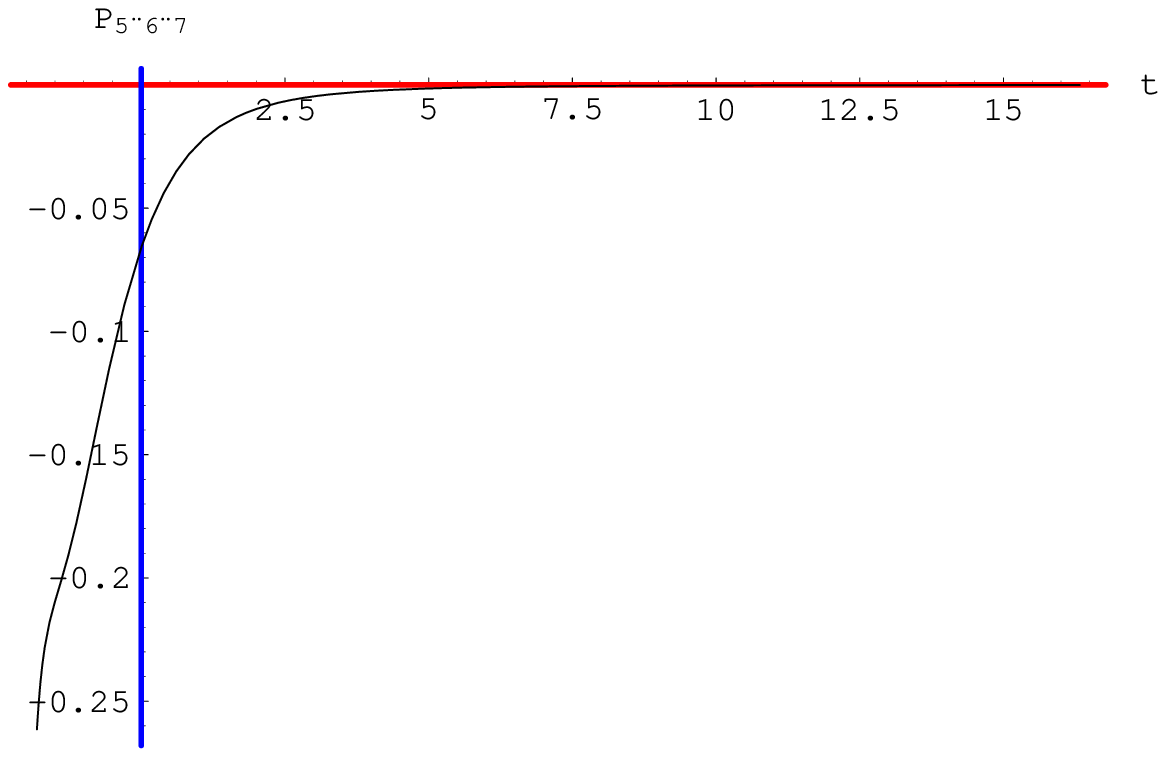}
  \includegraphics[width=5cm]{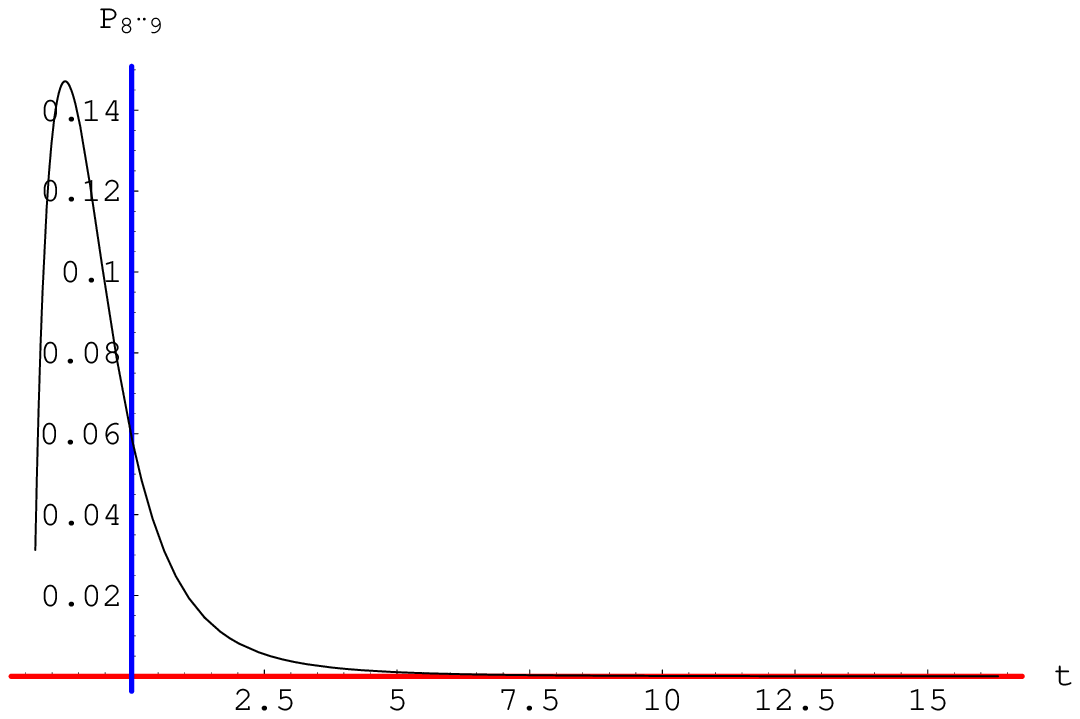}
  \caption{Plots of the pressure eigenvalues 
    $P_{[\alpha]}$, $\alpha= 1|2 \, ,\, 3|4\, , \, 5|6|7 \, , \,8|9$
    as functions of the cosmic time $t=\tau(T)$ in the case of the
    choice of parameters $\omega =1$, $\kappa=0.5$ and for the $A_2$
    solution with only the highest root switched on.}
  \label{pressuro1k05}
\end{figure}

In order to enlighten the physical meaning of the type IIB superstring
backgrounds we have eventually constructed it is worth to analyze the
structure of the stress energy tensor. First, reintroducing the
missing traces we define:
\begin{eqnarray}
T_{MN}^{[D3]} &=& \hat{T}_{MN}^{[D3]}-\ft 12 \, g_{MN} \,
\hat{T}_{RS}^{[D3]}\, g^{RS} \nonumber\\
T_{MN}^{[D5]} &=& \hat{T}_{MN}^{[D5]}-\ft 12 \, g_{MN} \,
\hat{T}_{RS}^{[D5]}\, g^{RS} \nonumber\\
T_{MN}^{tot} &=&T_{MN}^{[D3]} + T_{MN}^{[D5]}
\label{stresscontra}
\end{eqnarray}
It turns out that the stress energy tensors are diagonal, just as the
metric, and have the form of a perfect
fluid, but with different pressure eigenvalues in the various
subspaces. Indeed we can write:
\begin{eqnarray}
T_{00}^{tot,\, D3,\, D5} & = & g_{00} \, \rho^{tot,\, D3,\, D5} \nonumber\\
T_{i_\alpha j_\alpha}^{tot,\, D3,\, D5} & = & -g_{i_\alpha j_\alpha} \,
P_{\alpha}^{tot,\, D3,\, D5}
\label{definizionipress}
\end{eqnarray}
where $\alpha$ denotes the four different submanifolds extending in
directions:
\begin{equation}
  \alpha= 1|2 \, , \, 3|4 \, , \, 5|6|7 \, ,\, 8|9
\label{directions}
\end{equation}
We can now analyze the specific properties of the two example of
solutions.

\subsection{Properties of the solution with just one root switched on}
In the case of the time dependent background described in chapter
(\ref{oxide1a2}) and obtained by oxiding the solution (\ref{finsol})
we obtain for the \textit{energy densities}:
\begin{eqnarray}
  \rho^{tot} & = & \frac{{\left( 1 + e^{t\,\omega } \right) }^2\,{\kappa }^2 +
    9\,e^{t\,\omega }\,{\omega }^2}{36\,
    e^{t\,{\sqrt{\frac{{\kappa }^2}{3} + {\omega }^2}}}\,
    {\left( 1 + e^{t\,\omega } \right) }^2\,
    {\sqrt{\cosh \frac{t\,\omega }{2}}}} \nonumber\\
  \rho^{d3} & = & \frac{{\omega }^2}
  {16\,e^{t\,{\sqrt{\frac{{\kappa }^2}{3} + {\omega }^2}}}\,
    {\cosh^{\frac{5}{2}} \frac{t\,\omega }{2}}}\nonumber\\
  \rho^{d5}    & =&\frac{{\kappa }^2}
  {36\,e^{t\,{\sqrt{\frac{{\kappa }^2}{3} + {\omega }^2}}}\,
    {\sqrt{\cosh \frac{t\,\omega }{2}}}}
  \label{energydensitya2first}
\end{eqnarray}
for the \textit{total pressures}:
\begin{eqnarray}
  P_{1|2}^{tot} & = &-\frac{\kappa^2 + 9\omega^2 + \kappa^2\cosh t\omega}{144
    e^{t\sqrt{\frac{\kappa^2}{3} + \omega^2}}
    \cosh^{\frac{5}{2}}\frac{t\omega}{2}} \nonumber\\
  P_{3|4}^{tot} & = & -\frac{ e^{t\left( \omega  -
        \sqrt{\frac{\kappa^2}{3} + \omega^2}\right) } 
    \left( \kappa^2 - 9\omega^2 +
      \kappa^2\cosh t\omega\right)}{36
    {\left( 1 + e^{t\omega} \right)}^2
    \sqrt{\cosh \frac{t\omega }{2}}}\nonumber\\
  P_{5|6|7}^{tot} & = & -\frac{ \kappa^2 + 9\omega^2 + \kappa^2\cosh t\omega}
  {144e^{t\sqrt{\frac{\kappa^2}{3} + \omega^2}}
    \cosh^{\frac{5}{2}}\frac{t\omega}{2}}\nonumber\\
  P_{8|9}^{tot} & = &-\frac{e^{t\left(\omega -\sqrt{\frac{\kappa^2}{3} + 
          \omega^2} \right) }
    \left(\kappa^2 - 9\omega^2 +\kappa^2\cosh t\omega\right)}{36
    \left( 1 + e^{t\omega} \right)^2\sqrt{\cosh \frac{t\omega}{2}}}
  \label{totpressio}
\end{eqnarray}
for the \textit{pressures associated with the $D3$ brane}:
\begin{eqnarray}
  P_{1|2}^{D3} & = & \frac{-{\omega }^2}
  {16\,e^{t\,{\sqrt{\frac{{\kappa }^2}{3} + {\omega }^2}}}\,
    {\cosh^{\frac{5}{2}}\frac{t\omega }{2}}} \nonumber\\
  P_{3|4}^{D3} & = & \frac{{\omega }^2}
  {16\,e^{t\,{\sqrt{\frac{{\kappa }^2}{3} + {\omega }^2}}}\,
    {\cosh^{\frac{5}{2}}\frac{t\omega }{2}}}\nonumber\\
  P_{5|6|7}^{D3} & = & \frac{-{\omega }^2}
  {16\,e^{t\,{\sqrt{\frac{{\kappa }^2}{3} + {\omega }^2}}}\,
    {\cosh^{\frac{5}{2}}\frac{t\,\omega }{2}}}\nonumber\\
  P_{8|9}^{D3} & = & \frac{{\omega }^2}
  {16\,e^{t\,{\sqrt{\frac{{\kappa }^2}{3} + {\omega }^2}}}\,
    {\cosh^{\frac{5}{2}}\frac{t\,\omega }{2}}}
  \label{d3pressio}
\end{eqnarray}
and for the pressures associated with the $D5$ brane:
\begin{eqnarray}
  P_{1|2}^{D5} & = & \frac{-{\kappa }^2}
  {72\,e^{t\,{\sqrt{\frac{{\kappa }^2}{3} + {\omega }^2}}}\,
    {\sqrt{\cosh \frac{t\,\omega }{2}}}} \nonumber\\
  P_{3|4}^{D5} & = & \frac{-{\kappa }^2}
  {72\,e^{t\,{\sqrt{\frac{{\kappa }^2}{3} + {\omega }^2}}}\,
    {\sqrt{\cosh \frac{t\,\omega }{2}}}}\nonumber\\
  P_{5|6|7}^{D5} & = & \frac{-{\kappa }^2}
  {72\,e^{t\,{\sqrt{\frac{{\kappa }^2}{3} + {\omega }^2}}}\,
    {\sqrt{\cosh \frac{t\,\omega }{2}}}}\nonumber\\
  P_{8|9}^{D5} & = & \frac{-{\kappa }^2}
  {72\,e^{t\,{\sqrt{\frac{{\kappa }^2}{3} + {\omega }^2}}}\,
    {\sqrt{\cosh \frac{t\,\omega }{2}}}}
  \label{d5pressio}
\end{eqnarray}
As we see from its analytic expression the total energy density is an
exponentially decreasing function of time which tends to zero at
asymptotically late times ($t \mapsto \infty$). What happens instead
at asymptotically early times ($t \mapsto -\infty$) depends on the
value of $\kappa$.  For $\kappa=0$ we have $\lim_{t\mapsto-\infty}
\rho^{tot}(t)=0$, while for $\kappa \ne 0$ we always have $
\lim_{t\mapsto-\infty} \rho^{tot}(t)=\infty$. This is illustrated, for
instance, in figs.(8) and (10). This phenomenon is related to the
presence or absence of a $D5$ brane as it is evident from eq.s
(\ref{energydensitya2first}) which shows that the
\textit{dilaton}-($D5$) brane contribution to the energy density is
proportional to $\kappa^2$ and it is always divergent at
asymptotically early times, while the $D3$ brane contribution tends to
zero in the same regime.

We also note, comparing eq.s(\ref{d5pressio}) with
eq.s(\ref{d3pressio}) that the pressure contributed by the
\textit{dilaton}--$D5$--brane system is the same in all directions
1--9, while the pressure contributed by the $D3$--brane system is just
opposite in the direction 3489 and in the transverse directions 12567.
This is the origin of the \textit{cosmic billiard phenomenon} that we
observe in the behaviour of the metric scale factors.
%, as originally envisaged, by the asymptotic analysis of
%Damour et al \cite{biliardi}.
Indeed the presence of the $D3$-brane causes, at a certain instant of
time, a switch in the cosmic expansion. Dimensions that were
previously shrinking begin to expand and dimensions that were
expanding begin to shrink. It is like a ball that hits a wall and
inverts its speed. In the exact solution that we have constructed
through reduction to three dimensions this occurs in a smooth way.
There is a maximum and respectively a minimum in the behaviour of
certain scale factors, which is in relation with a predominance of the
$D3$--brane energy density with respect to the total energy density.
The cosmic $D3$--brane behaves just as an instanton. Its contribution
to the total energy is originally almost zero, then it raises and
dominates for some time, then it exponentially decays again. This is
the smooth exact realization of the potential walls envisaged by
Damour et al.

To appreciate such a behaviour it is convenient to consider some plots
of the scale factors, the energy densities and the pressures.  In
order to present such plots we first reduce the metric
(\ref{a2firstmetric}) to a standard cosmological form, by introducing
a new time variable $\tau$ such that:
\begin{equation}
  r_{[0]}(t) \, dt = d\tau
  \label{taudefini1}
\end{equation}
Explicitly we set:
\begin{equation}
  \tau = \int_{0}^{T} \, r_{[0]}(t) \, dt
  \label{cosmotime1}
\end{equation}
and inserting the explicit form of the scale factor $r_{[0]}(t)$ as
given in eq.(\ref{a2firstscafacti}) we obtain:

\begin{multline}
  \label{tau1}
  \tau(T) \,=\,  
  \frac{4\,2^{\frac{3}{4}}}{\omega  -
    4\,{\sqrt{\frac{{\kappa }^2}{3} + {\omega }^2}}}\,\left [ {_2 F_1}\left
      (-  \ft {1}
      {8}   + \ft {{\sqrt{\frac{{\kappa
              }^2}{3} + {\omega }^2}}}
      {2\,\omega },-  \ft {1}{4}   ,
      \ft {7}{8} + \ft {{\sqrt{\frac{{\kappa }^2}{3} +
            {\omega }^2}}}
      {2\,\omega },-1 \right) 
  \right.\\ \left.
    -\, e^
    {\frac{T\,\left( -\omega  +
          4\,{\sqrt{\frac{{\kappa
                }^2}{3} + {\omega
              }^2}} \right)
      }{8}}\,
    {_2 F_1}\left (-  \ft{1}{8}    +
      \ft {{\sqrt{\frac{{\kappa }^2}{3} +
            {\omega }^2}}}{2\,\omega },
      -  \ft {1}{4}   ,
      \ft {7}{8} + \ft {{\sqrt{\frac{{\kappa }^2}{3} +
            {\omega }^2}}}
      {2\,\omega },-e^{T\,\omega } \right)
  \right]
\end{multline}
which expresses $\tau$ in terms of hypergeometric functions and
exponentials.

\begin{figure}[!t]
  \centering
  \includegraphics[width=5cm]{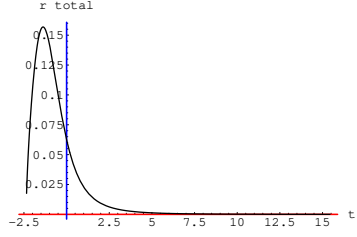}
  \caption{Plot of the  energy density as function of the cosmic time 
    $t=\tau(T)$ in the case of a pure $D3$ brane system, namely for
    the the choice of parameters $\omega =1$, $\kappa=0$ and for the
    $A_2$ solution with only the highest root switched on.}
  \label{rhototo1k0}
\end{figure}

In fig.(\ref{radiio1k05}) we observe the billiard phenomenon in a
generic case where both parameters $\omega$ and $\kappa$ are non
vanishing. Since the value of $\omega$ can always be rescaled by a
rescaling of the original time coordinate $t$, we can just set it to
$1$ and what matters is to distinguish the case $\omega \ne 0$ where
the $D3$ brane is present from the case $\omega = 0$ corresponding to
its absence. Hence fig.(\ref{radiio1k05}) corresponds to the presence
of both a $D3$--brane and a \textit{dilaton}--$D5$--brane system.  A
very different behavior occurs in fig.(\ref{radiio0k1}) where
$\omega=0$. In this case there is no billiard and the dimensions
either shrink or expand uniformly.  On the other hand in
fig.(\ref{radiio1k0}) we observe the pure billiard phenomenon induced
by the $D3$-brane in the case where no \textit{dilaton}-$D5$--brane is
present, namely when we set $\kappa=0$. In this case, as we see, the
parallel directions to the Euclidean $D3$--brane, namely 3489 have
exactly the same behaviour: they first inflate and then they deflate,
namely there is a maximum in the scale factor. The transverse
directions to the $D3$ brane 567 have the opposite behaviour. They
display a minimum at the same point where the parallel directions
display a maximum.  In all cases the directions 12 corresponding to
the spatial directions of the three dimensional sigma model world
suffer a uniform expansion.
\par
Let us now consider the behavior of the energy densities.
%%%%%%%%%%%%%%%%%%%%%%
In fig.(\ref{rhoo1k05}) we focus on the mixed case $\omega=1$,
$\kappa=0.5$ characterized by the presence of both a $D3$ brane and
\textit{dilaton}-$D5$--brane system. As we see the total energy
density exponentially decreases at late times and has a singularity at
asymptotically early times. This is like in a standard Big Bang
cosmological model with an indefinite expansion starting from an
initial singularity. Yet the ratio of the $D3$ energy with respect to
the total energy has a maximum at some instant of time and this is the
cause of the billiard phenomenon in the behaviour of the scale factors
respectively parallel and transverse to the $D3$ brane itself.  The
two contributions to the energy density from the $D3$--brane and from
the dilaton have the same sign and the plot of their ratio displays a
maximum in correspondence with the billiard time.  With the same
choice of parameters $\omega=1$, $\kappa=0.5$ the physical behavior of
the system can be appreciated by looking at the plots of the pressure
eigenvalues. They are displayed in fig.(\ref{pressuro1k05}). We
observe that the pressure is negative in the directions transverse to
the $D3$ brane 12 and 567. Slowly, but uniformly it increases to zero
in these directions.  In the directions parallel to the brane the
pressure is instead always positive and it displays a sharp maximum at
the instant of time where the billiard phenomenon occurs.

\begin{figure}[!t]
  \centering
  \includegraphics[width=5cm]{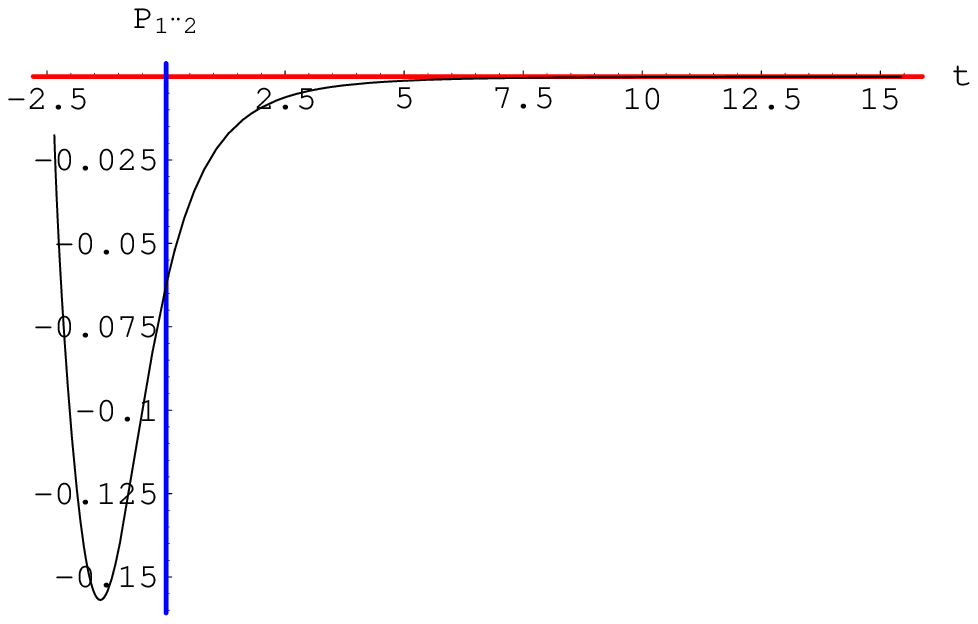}
  \includegraphics[width=5cm]{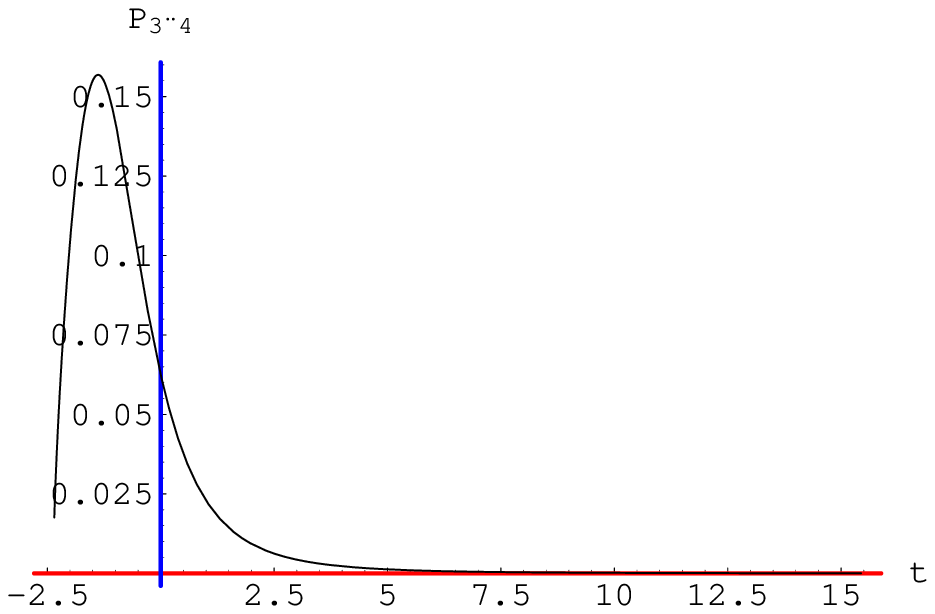}
  \includegraphics[width=5cm]{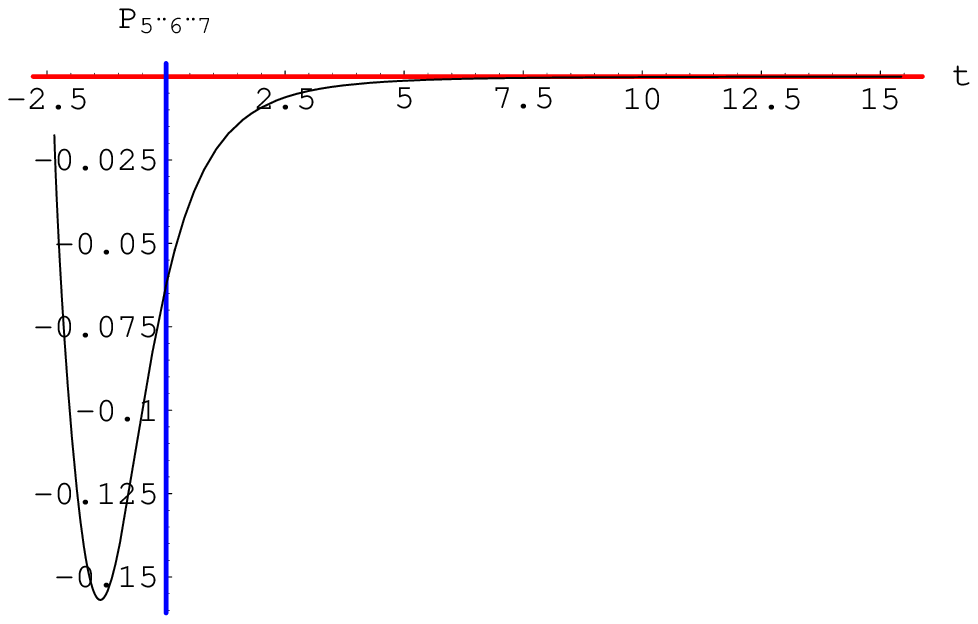}
  \includegraphics[width=5cm]{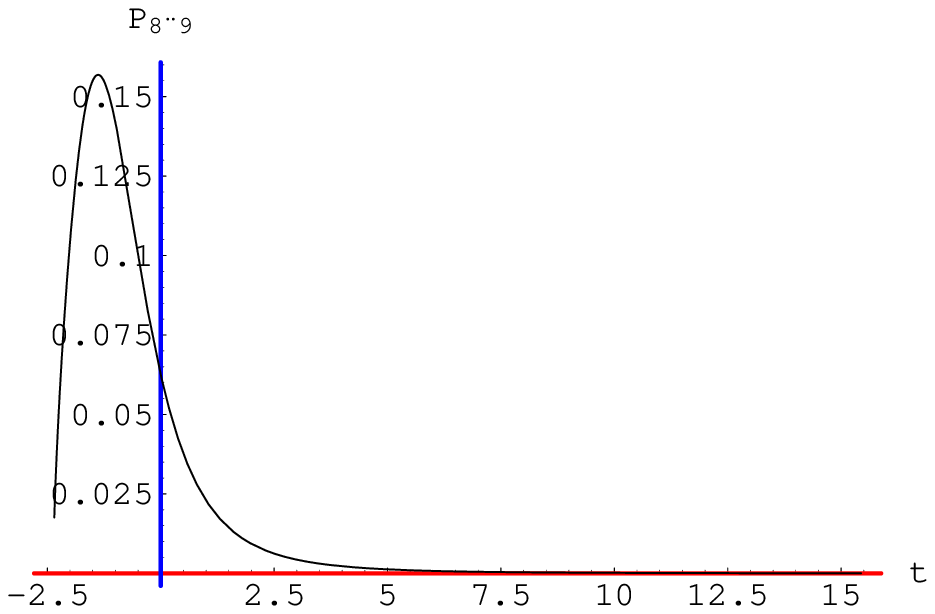}
  \caption{Plots of the pressure eigenvalues $P_{[\alpha]}$, 
    $\alpha= 1|2 \, ,\, 3|4\, , \, 5|6|7 \, , \, 8|9$ as functions of
    the cosmic time $t=\tau(T)$ in the case of the choice of
    parameters $\omega =1$, $\kappa=0$ and for the $A_2$ solution with
    only the highest root switched on.  This case corresponds to a
    pure $D3$ brane system.}
  \label{pressuro1k0}
\end{figure}

For a pure $D3$ brane system, namely for $\kappa=0$ and $\omega=1$ the
energy density starts at zero, develops a maximum and then decays
again to zero. This can be seen in fig.(\ref{rhototo1k0}). The plot of
the pressures is displayed, for this case in fig.(\ref{pressuro1k0}).
In this case the pressure in the directions transverse to the brane,
i.e. 12567 is negative and it is just the opposite of the pressure in
the directions parallel to the brane, namely 3489.  This behavior
causes the corresponding scale factors to suffer a minimum and a
maximum, respectively.

\subsection{Properties of the solution with all roots
  switched on} 

Let us now discuss the properties of the second solution where all the
roots have been excited. In chapter \ref{oxide2a2} we considered the
oxidation of such a sigma model solution and we constructed the
corresponding $D=10$ supergravity background given by the metric
(\ref{a2secondtmetric}, \ref{a2secondscafacti}) and by the field
strengths (\ref{a2secondpform}). Looking at eq.s (\ref{a2secondpform})
we see that the interpretation of the parameter $\omega$ is still the
same as it was before, namely it represents the magnetic charge of the
dyonic $D3$-brane. At $\omega=0$ the $D3$--brane disappears. Yet it
appears from eq.s (\ref{a2secondpform}) that there is no obvious
interpretation of the parameter $\kappa$ as a pure $D5$-brane charge.
Indeed there is no choice of $\kappa$ which suppresses both the NS and
the RR $3$--form field strengths.

Following the same procedure as in the previous case we calculate the
energy density and the pressures and we separate the contributions due
to the $D3$--brane and to the \textit{dilaton}--$D5$--brane system.
After straightforward but lengthy algebraic manipulations, implemented
on a computer with MATHEMATICA we obtain:

\begin{subequations}
  \label{energydensitya2second} 
  \begin{equation}
    \rho^{d3}\,=\,
    \frac{{\omega }^2}
    {4\,e^{t\,\left( \frac{-5\,\omega }{4} +
          {\sqrt{\frac{{\kappa }^2}{3} + {\omega
              }^2}} \right) }\,
      {\left( 1 + e^{t\,\omega } \right) }^{\frac{5}{4}}\,
      {\left( 1 + e^{t\,\omega } +
          e^{\frac{t\,\left( \kappa  + \omega  \right) }{2}}
        \right) }^
      {\frac{5}{4}}}
  \end{equation}
  \begin{multline}
    \rho^{d5} \,=\,
    \frac{e^
      {t\,\left( \frac{\omega }{4} -
          {\sqrt{\frac{{\kappa }^2}{3} + {\omega
              }^2}} \right)}\,}
    {192\,
      \left( 1 + e^{t\,\omega } \right)^\frac{9}{4}\,
      \left( 1 + e^{t\,\omega } +
        e^{\frac{t\,\left( \kappa  + \omega  \right) }{2}}
      \right)^\frac{9}{4}} \, \\\times
 \left [ {\left( 1 + e^{t\,\omega } \right) }^2\,
    \left( 4 + 8\,e^{t\,\omega } + 4\,e^{2\,t\,\omega } +
      20\,e^{\frac{t\,\left( \kappa  + \omega  \right) }{2}}+
      e^{t\,\left( \kappa  + \omega  \right) } +
      20\,e^{\frac{t\,\left( \kappa  + 3\,\omega  \right) }{2}}
    \right) \,\kappa^2 
\right.\\\left.
- 6\,e^
    {\frac{t\,\left( \kappa  + \omega  \right) }{2}}\,
    \left( -1 + e^{2\,t\,\omega } \right) \,
    \left( 6 + 6\,e^{t\,\omega } +
      e^{\frac{t\,\left( \kappa  + \omega  \right) }{2}} \right) \,
    \kappa 
\right.\\\left.
+ 3\,\left( 8\,
      e^{\frac{t\,\left( \kappa  + \omega  \right) }{2}} +
      3\,e^{t\,\left( \kappa  + \omega  \right) } +
      26\,e^{t\,\left( \kappa  + 2\,\omega  \right) } +
      24\,e^{\frac{t\,\left( \kappa  + 3\,\omega  \right) }{2}}
    \right.\right.\\\left.\left. +
      3\,e^{t\,\left( \kappa  + 3\,\omega  \right) } +
      24\,e^{\frac{t\,\left( \kappa  + 5\,\omega  \right) }{2}} +
      8\,e^{\frac{t\,\left( \kappa  + 7\,\omega  \right) }{2}}
    \right) \,
    {\omega }^2\right]
  \end{multline}
  \begin{multline}
    \rho^{tot}\,=\,
    \frac{e^{t\,\left( 
          \frac{\omega }{4} - \sqrt{\frac{\kappa^2}{3} + \omega^2} 
        \right)}}
    {192\,
      \left( 1 + e^{t\,\omega} \right)^\frac{9}{4} \,
      \left( 1 + e^{t\,\omega} +
        e^{\frac{t\,\left(\kappa +
              \omega\right)}{2}}\right)^\frac{9}{4}}\\ \times 
    \left [
      \left(1 + e^{t\,\omega}\right)^2 \,
      \left( 
        4 + 8\,e^{t\,\omega} + 4\,e^{2\,t\,\omega} +
        20\,e^\frac{t\,(\kappa + \omega)}{2} +
        e^{t\,(\kappa  + \omega)} +
        20\,e^\frac{t\,(\kappa + 3\,\omega)}{2}
      \right)\, \kappa^2  
    \right.\\\left.
      - 6\, e^\frac{t\,(\kappa  + \omega)}{2}\,
      \left( -1 + e^{2\,t\,\omega} \right)\,
      \left( 
        6 + 6\,e^{t\,\omega } +
        e^\frac{t\,( \kappa  + \omega )}{2} 
      \right) \,
      \kappa \,\omega
    \right.\\\left.
      + 3\,\left( 16\,e^{t\,\omega } +
        32\,e^{2\,t\,\omega } + 16\,e^{3\,t\,\omega } +
        8\,e^{\frac{t\,\left( \kappa  + \omega  \right) }{2}} +
        3\,e^{t\,\left( \kappa  + \omega  \right) } 
      \right.\right.\\\left.\left.
        +26\,e^{t\,\left( \kappa  + 2\,\omega  \right) } +
        40\,e^{\frac{t\,\left( \kappa  + 3\,\omega  \right) }{2}} +
        3\,e^{t\,\left( \kappa  + 3\,\omega  \right) } +
        40\,e^{\frac{t\,\left( \kappa  + 5\,\omega  \right) }{2}} +
        8\,e^{\frac{t\,\left( \kappa  + 7\,\omega  \right) }{2}}
      \right) \,
      \omega^2
    \right]
  \end{multline}
\end{subequations}

We see from the above formulae that the energy density contributed by
the $D3$--brane system is proportional to $\omega^2$ as before and
vanishes at $\omega=0$. However there is no choice of the parameter
$\kappa$ which suppresses the \textit{dilaton}--$D5$--contribution
leaving the $D3$--contribution non--zero.

The pressure eigenvalues can also be calculated just as in the previous example
but the resulting analytic formulae are quite messy and we do not
feel them worthy to be displayed. It is rather convenient to consider
a few more plots.

Just as in the previous case we define the cosmic time through the
formula (\ref{cosmotime1}). In this case, however, the integral does
not lead to a closed formula in terms of special functions and we just
have an implicit definition:
\begin{equation}
  \tau(T) \, \equiv \, \int _{0}^{T} \, e^
  {\frac{t\,\left( \frac{-\omega }{4} +
        {\sqrt{\frac{{\kappa }^2}{3} +
            {\omega }^2}}
      \right) }{2}}\,
  {\left( 1 + e^{t\,\omega } \right) }^{\frac{1}{8}}\,
  {\left( 1 + e^{t\,\omega } +
      e^{\frac{t\,\left( \kappa  + \omega 
          \right) }
        {2}} \right) }^{\frac{1}{8}}\,dt
  \label{cosmictau2}
\end{equation}

Let us now observe from eq.s (\ref{a2secondscafacti}),
(\ref{a2secondpform}) that there are the following critical values of
the parameters:
\begin{description}
  \item[1] For $\omega=0$ and $\kappa \ne 0$ there is no $D3$--brane
  and there is just a $D$--string dual to a $D5$--brane.
  \item[2] For $\kappa = \pm \ft 3 2 \, \omega$ the scale factor in
  the directions 34 tends to a finite asymptotic value respectively
  at very early or very late times.
\end{description}

\begin{figure}[!t]
  \centering
  \includegraphics[width=5cm]{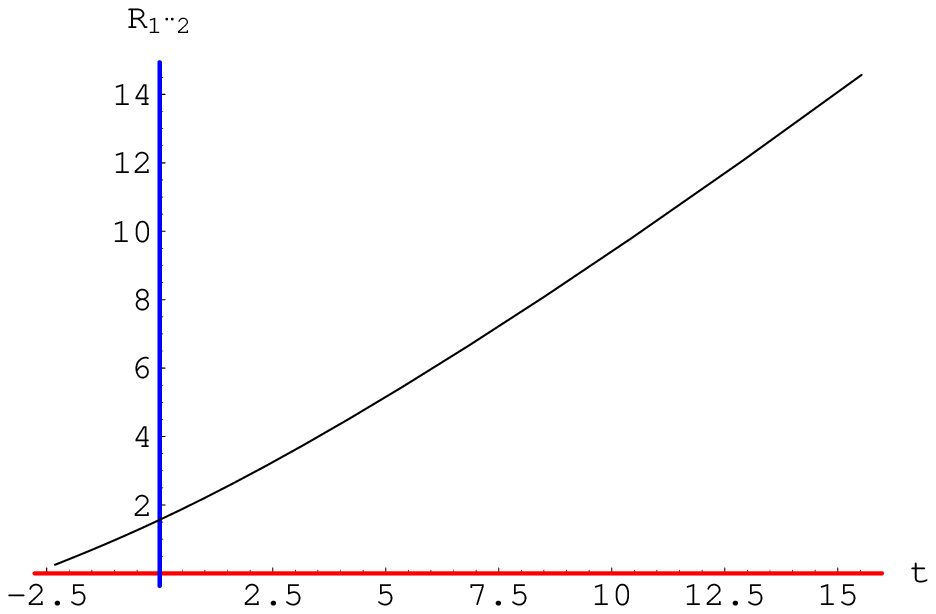}
  \includegraphics[width=5cm]{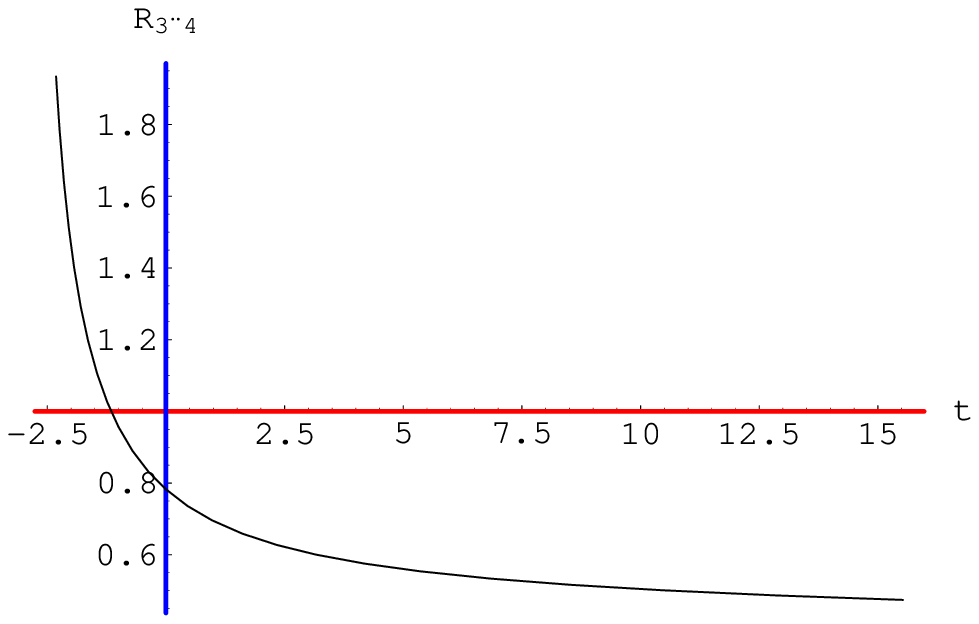}
  \includegraphics[width=5cm]{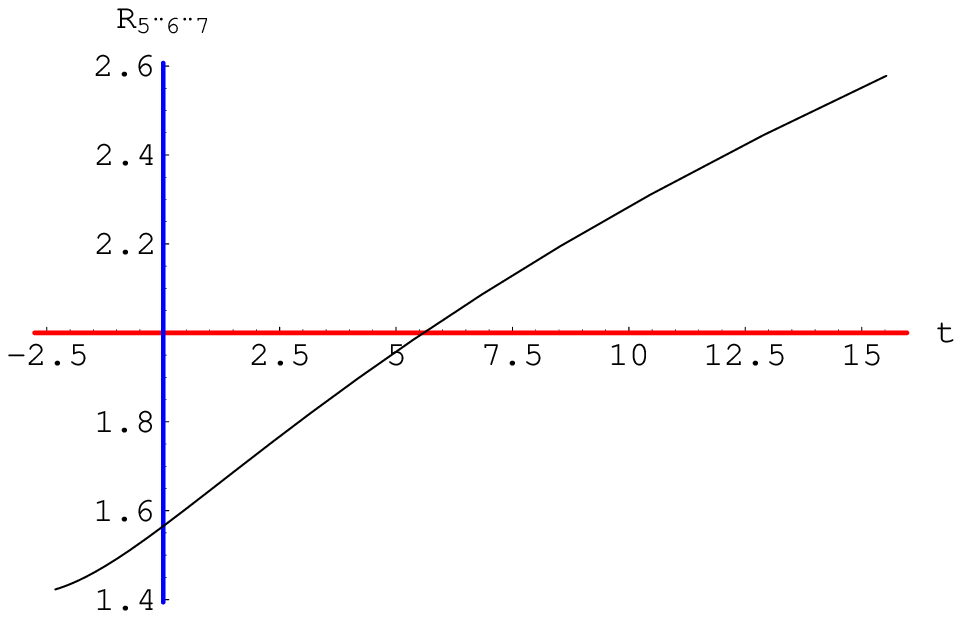}
  \includegraphics[width=5cm]{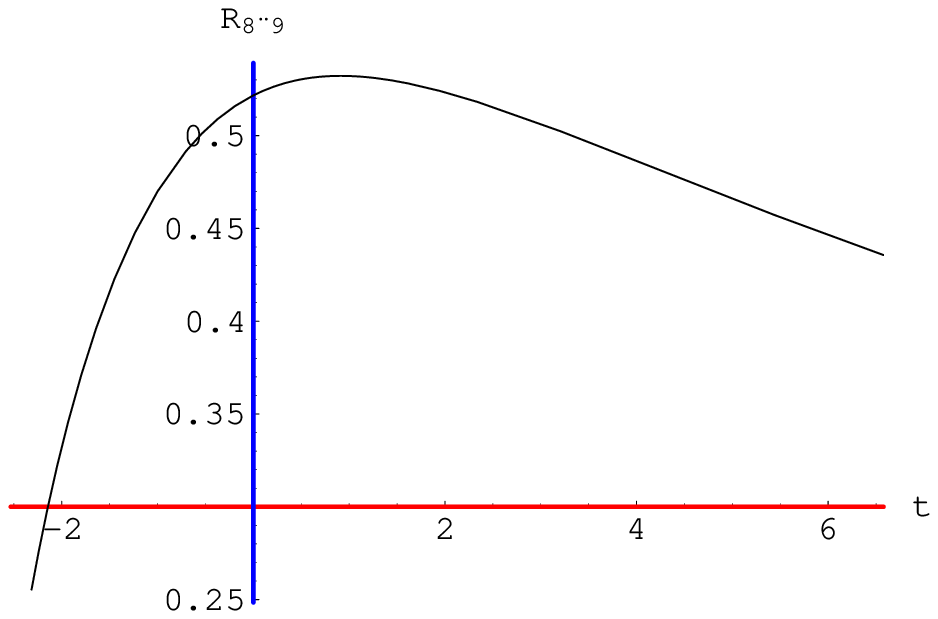}
  \caption{Plots of the scale factors $\overline{r}^2_{[\alpha]}$, 
    $\alpha= 1|2 \, ,\,3|4 \, , \, 5|6|7 \, , \,8|9$ as functions of
    the cosmic time $t=\tau(T)$ with the parameter choice $\omega =0$,
    $\kappa=3/2$ and for the $A_2$ solution with all the roots
    switched on.}
  \label{radii2o0k15}
\end{figure}

The plot of the scale factors for the choice $\omega=0$,
$\kappa=3/2$ is given in fig.(\ref{radii2o0k15})

\begin{figure}[!t]
  \centering
  \includegraphics[width=5cm]{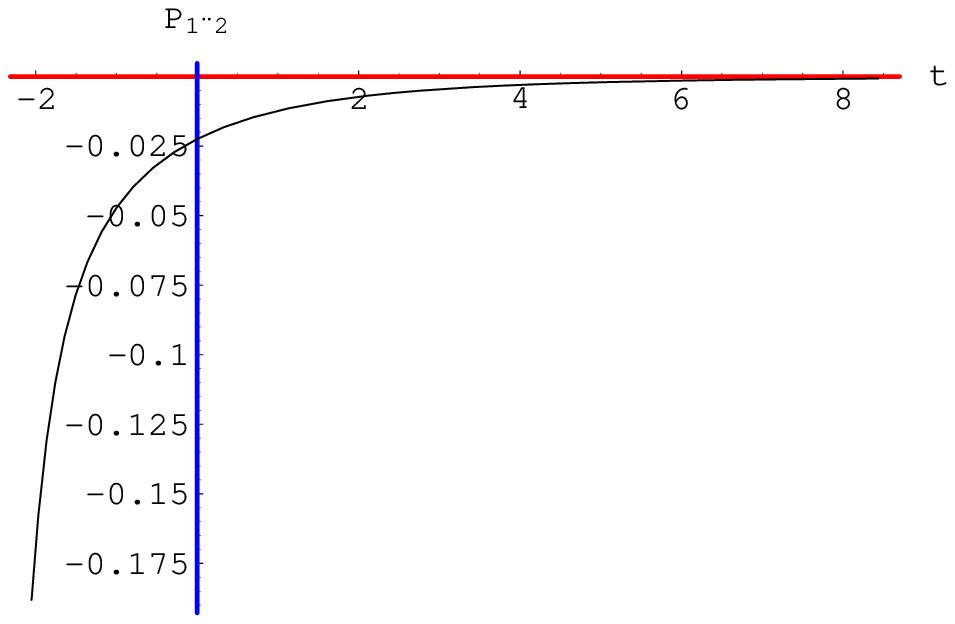}
  \includegraphics[width=5cm]{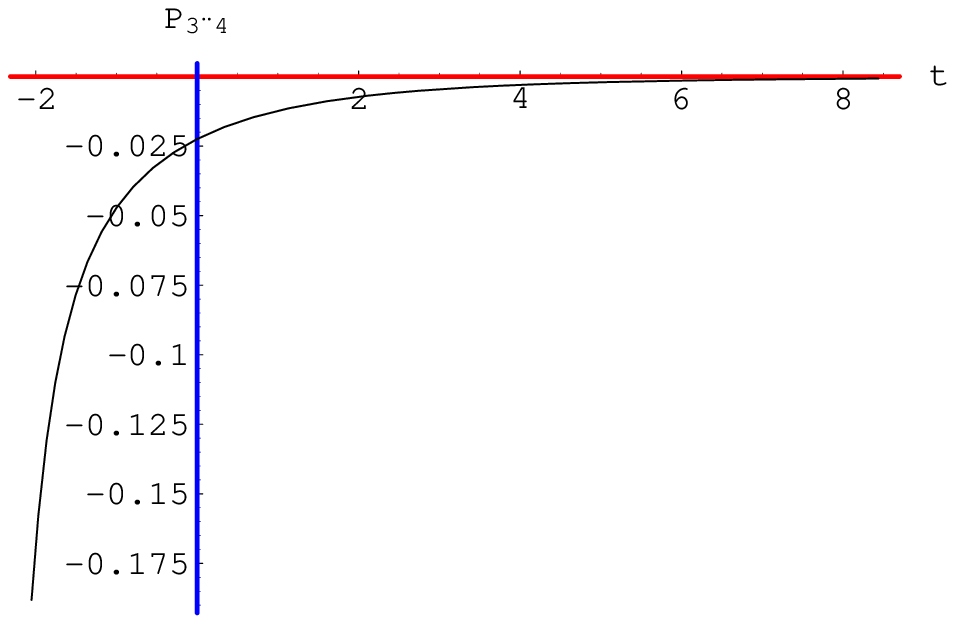}
  \includegraphics[width=5cm]{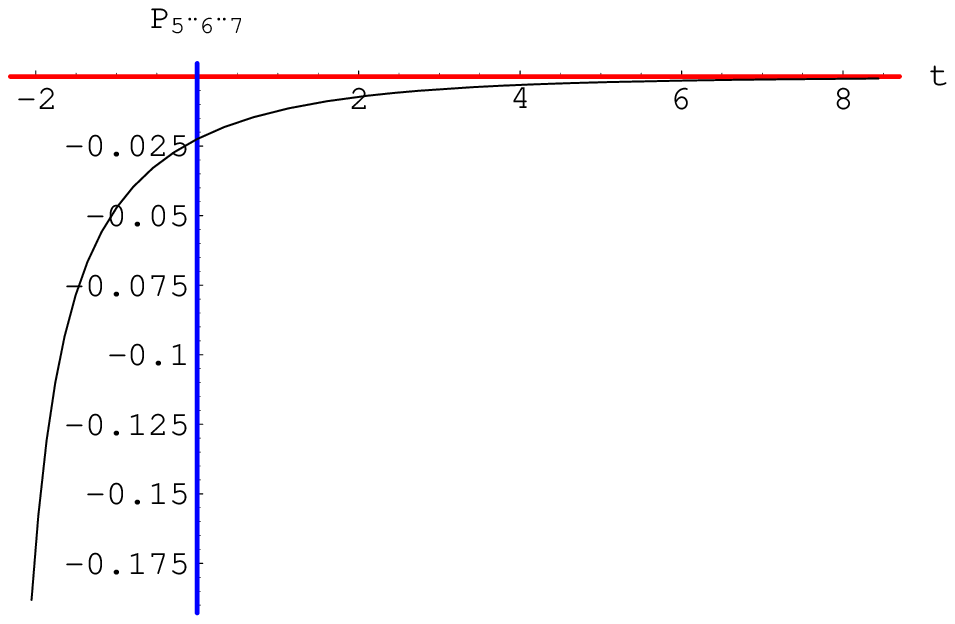}
  \includegraphics[width=5cm]{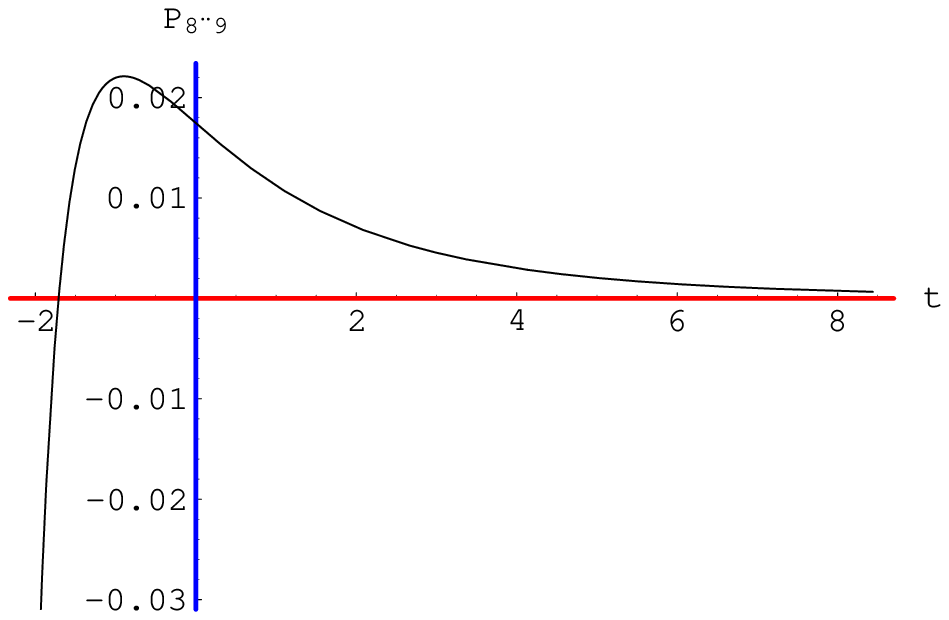}
  \caption{Plots of the pressure eigenvalues $\overline{P}_{[\alpha]}$,
    $\alpha= 1|2 \, ,\, 3|4 \, , \, 5|6|7 \, , \, 8|9$ as functions of
    the cosmic time $t=\tau(T)$ with the parameter choice $\omega =0$,
    $\kappa=3/2$ and for the $A_2$ solution with all the roots
    switched on.}
  \label{press2o0k15}
\end{figure}

As already stressed, this a pure $D$-string system and indeed the
billiard phenomenon occurs only in the directions 89 that correspond
to the euclidean $D$--string world--sheet. In all the other directions
there is a monotonous behavior of the scale factors.  The $D$-string
nature of the solution is best appreciated by looking at the behavior
of the pressure eigenvalues, displayed in fig.(\ref{press2o0k15})

\begin{figure}[!t]
  \centering
  \includegraphics[width=5cm]{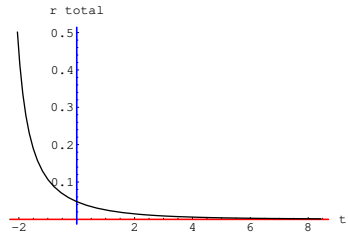}
  \caption{Plot of the  energy density as function of the cosmic time 
    $t=\tau(T)$ with the parameter choice $\omega =0$, $\kappa=3/2$
    and for the $A_2$ solution with all the roots switched on.}
  \label{2o0k15rhotot}
\end{figure}

As we see the positive bump in the pressure now occurs only in the
$D$--string directions 89, while in all the other directions the
pressure is the same and rises monotonously to zero from large
negative values. The pressure bump is in correspondence with the
billiard phenomenon. The energy density is instead a monotonously
decreasing function of time (see fig.(\ref{2o0k15rhotot})).  

\begin{figure}[!t]
  \centering
  \includegraphics[width=5cm]{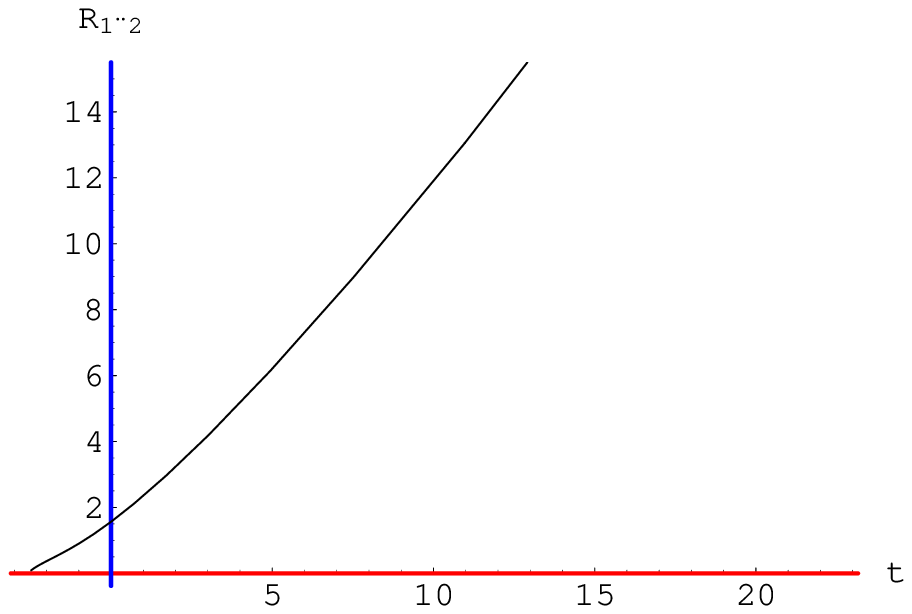}
  \includegraphics[width=5cm]{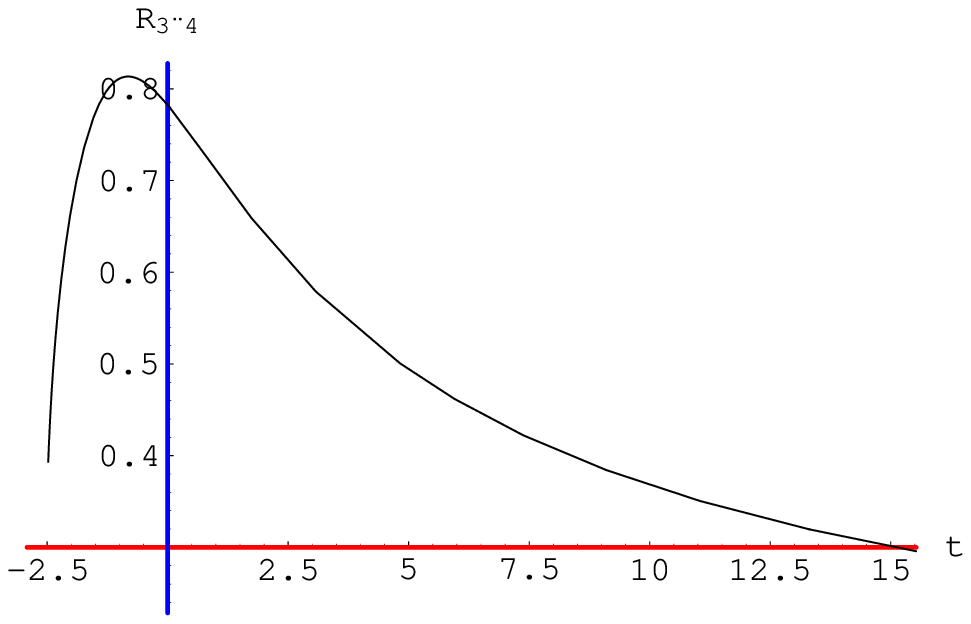}
  \includegraphics[width=5cm]{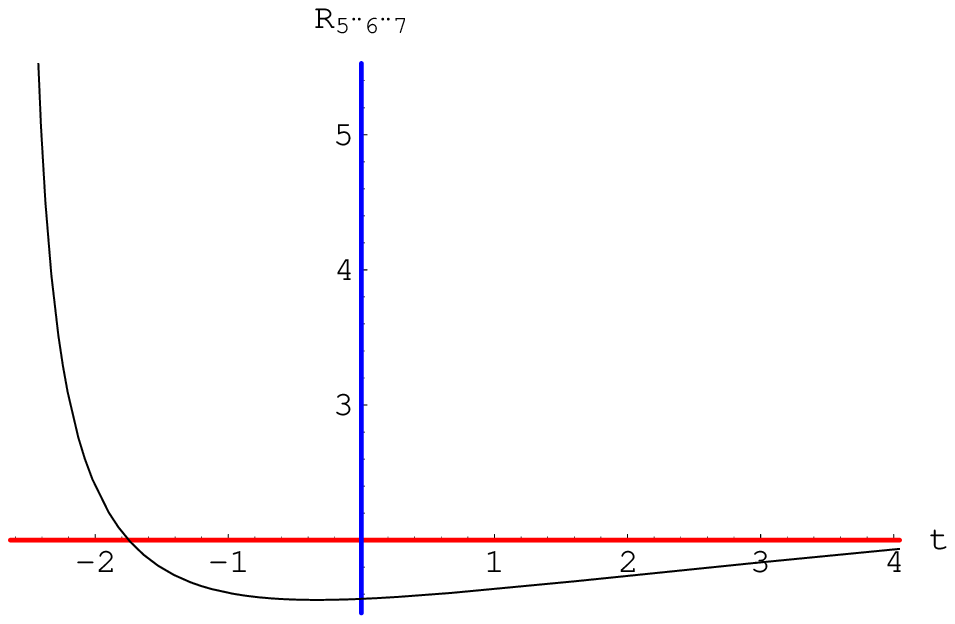}
  \includegraphics[width=5cm]{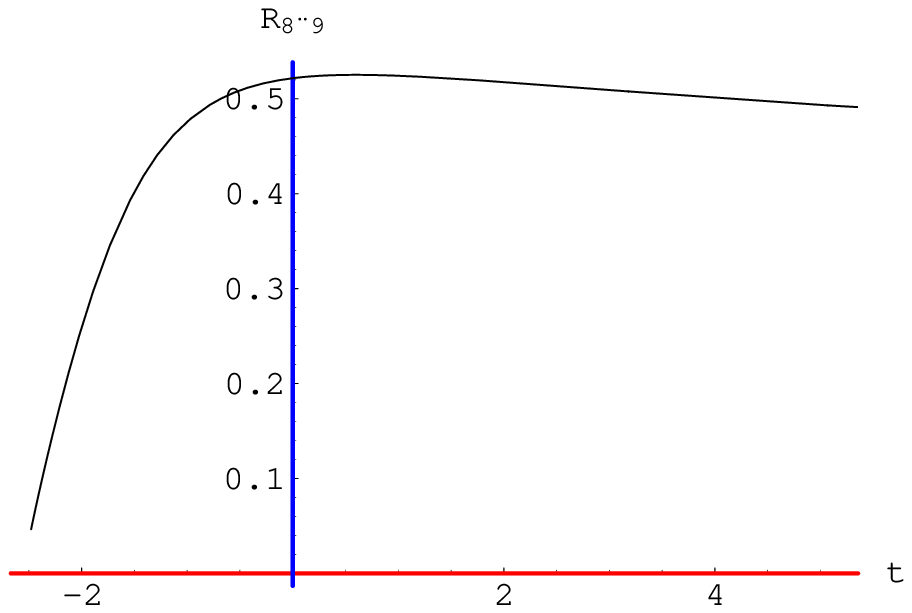}
  \caption{Plots of the scale factors $\overline{r}^2_{[\alpha]}$, 
    $\alpha= 1|2 \, ,\,3|4 \, , \, 5|6|7 \, , \, 8|9$ as functions of
    the cosmic time $t=\tau(T)$ with the parameter choice $\omega =1$,
    $\kappa=0.8$ and for the $A_2$ solution with all the roots switched
    on.}
  \label{radii2o1k08}
\end{figure}

An intermediate case is provided by the parameter choice $\omega=1$,
$\kappa=0.8 < 3/2$. The plots of the scale factors are given in
fig.(\ref{radii2o1k08})

The mixture of $D3$ and $D5$ systems is evident from the pictures.
Indeed we have now a billiard phenomenon in both the directions 34
and 89 as we expect from a $D3$--brane, but the maximum in 34 is much
sharper than in 89. The maximum in 89 is broader because it takes
contribution both from the $D3$ brane and from the $D$--string.

\begin{figure}[!t]
  \centering
  \includegraphics[width=5cm]{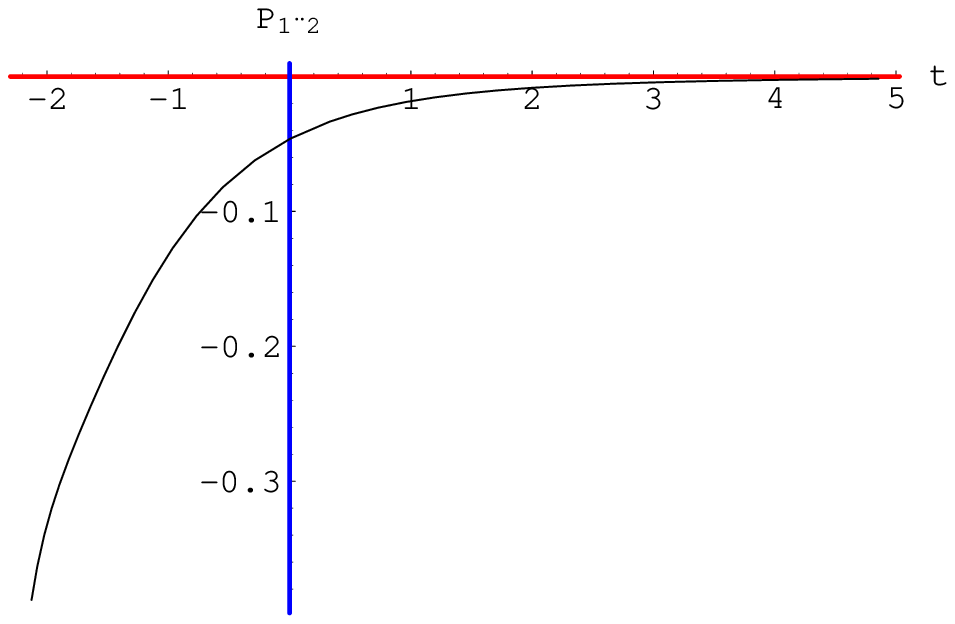}
  \includegraphics[width=5cm]{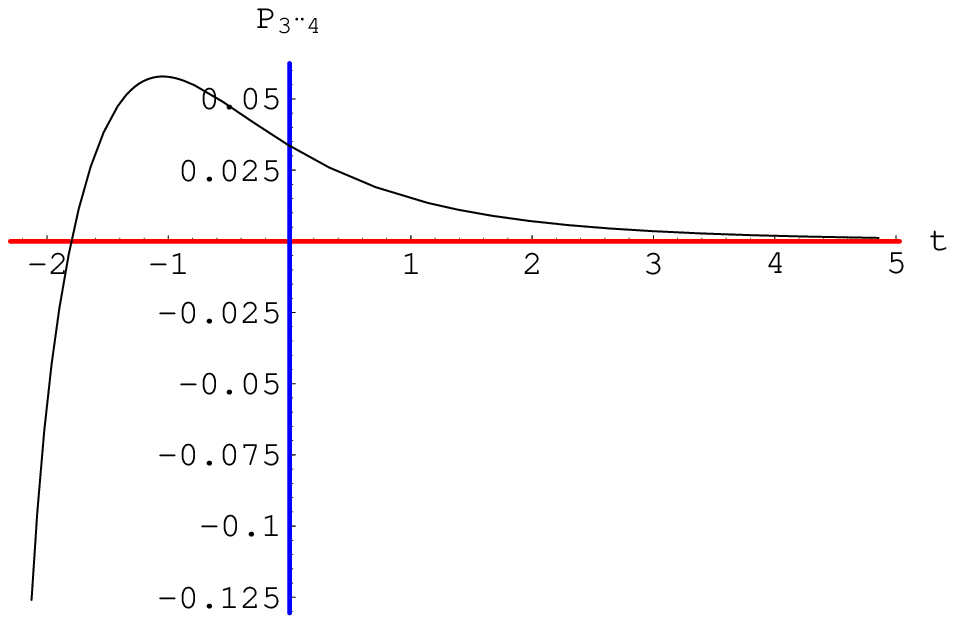}
  \includegraphics[width=5cm]{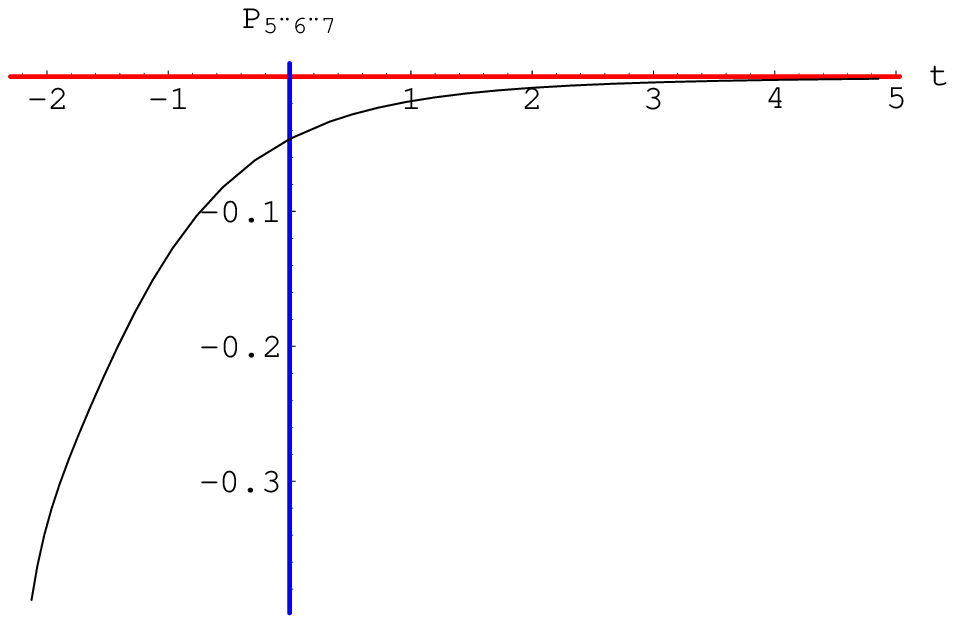}
  \includegraphics[width=5cm]{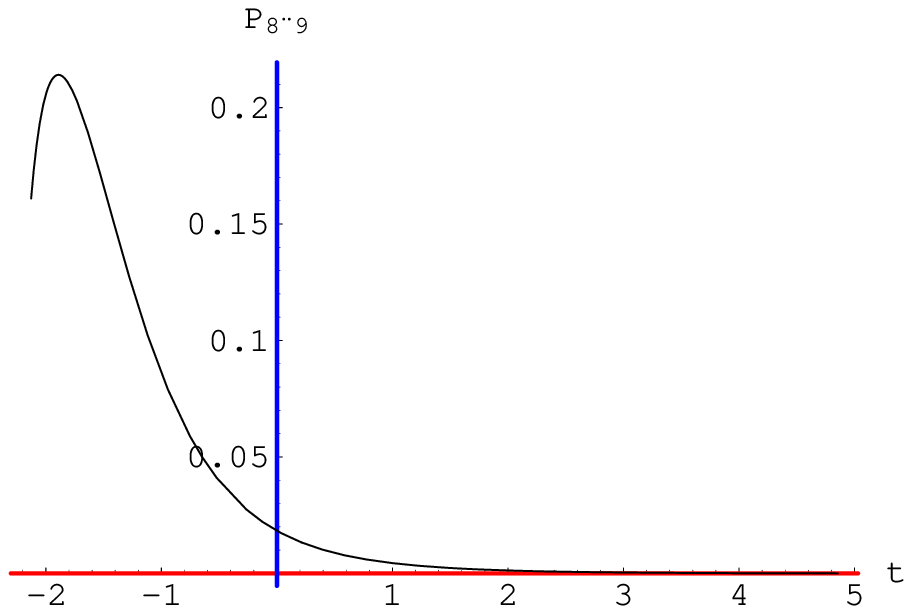}
  \caption{Plots of the pressure eigenvalues
    $\overline{P}_{[\alpha]}$, $\alpha= 1|2\, ,\, 3|4 \, , \, 5|6|7 \,
    , \, 8|9$ as functions of the cosmic time $t=\tau(T)$ with the
    parameter choice $\omega =1$, $\kappa=0.8$ and for the $A_2$
    solution with all the roots switched on.}
  \label{press2o1k08}
\end{figure}

\begin{figure}[!t]
  \centering
  \includegraphics[width=5cm]{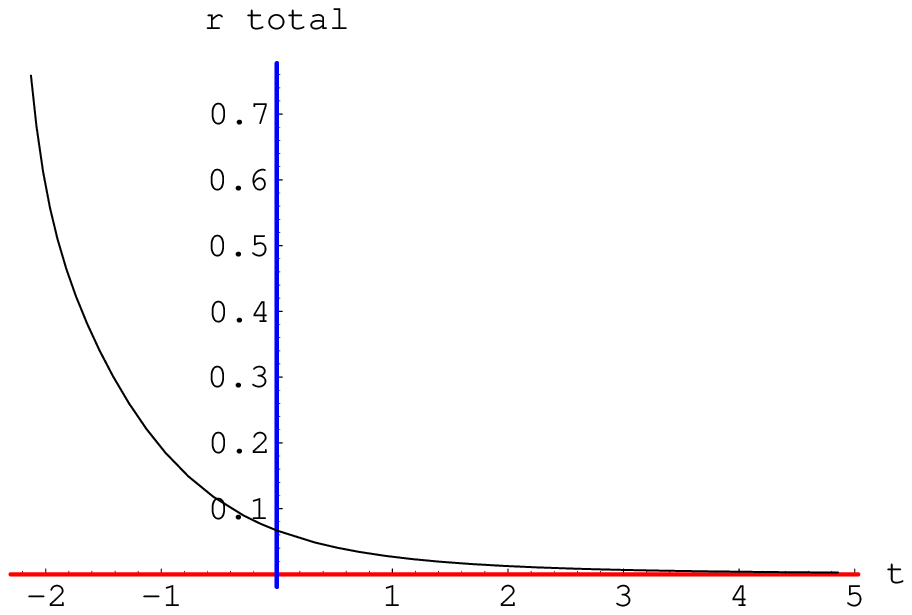}
  \includegraphics[width=5cm]{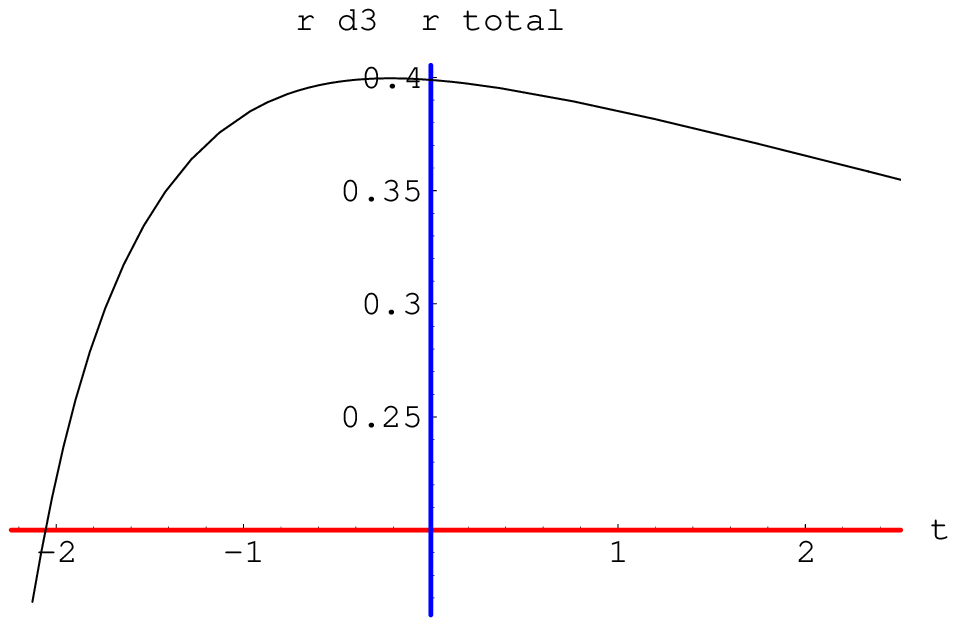}
  \includegraphics[width=5cm]{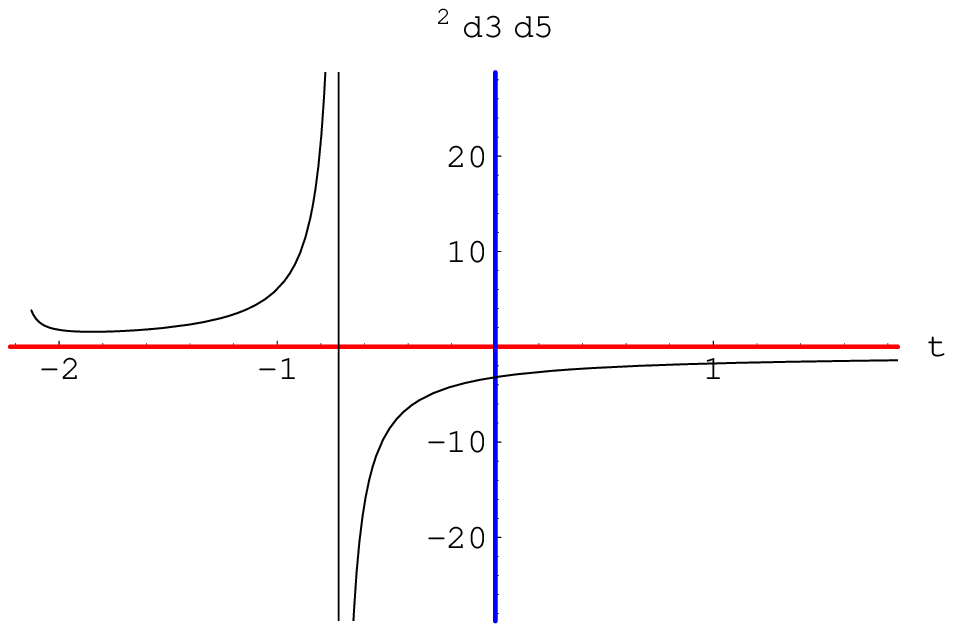}
  \includegraphics[width=5cm]{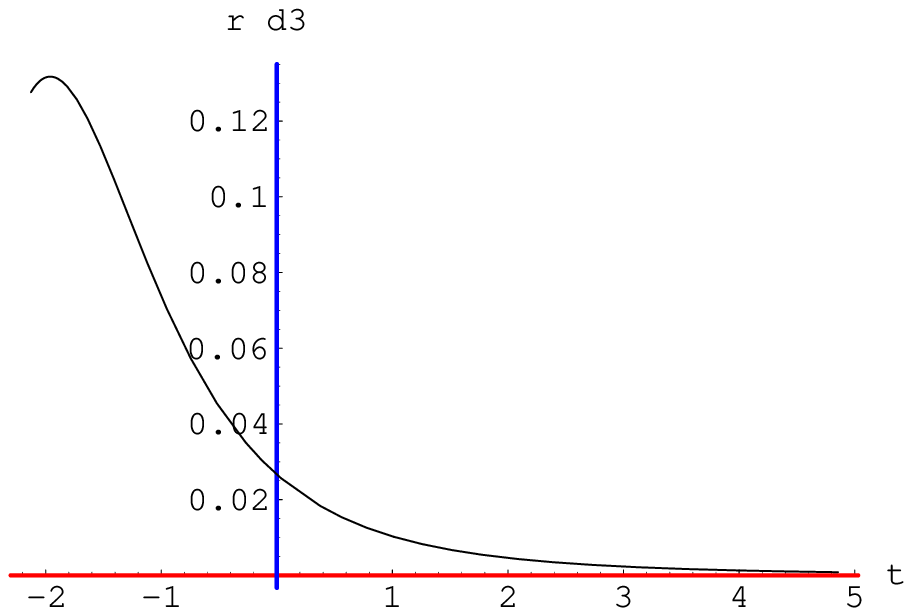}
  \includegraphics[width=5cm]{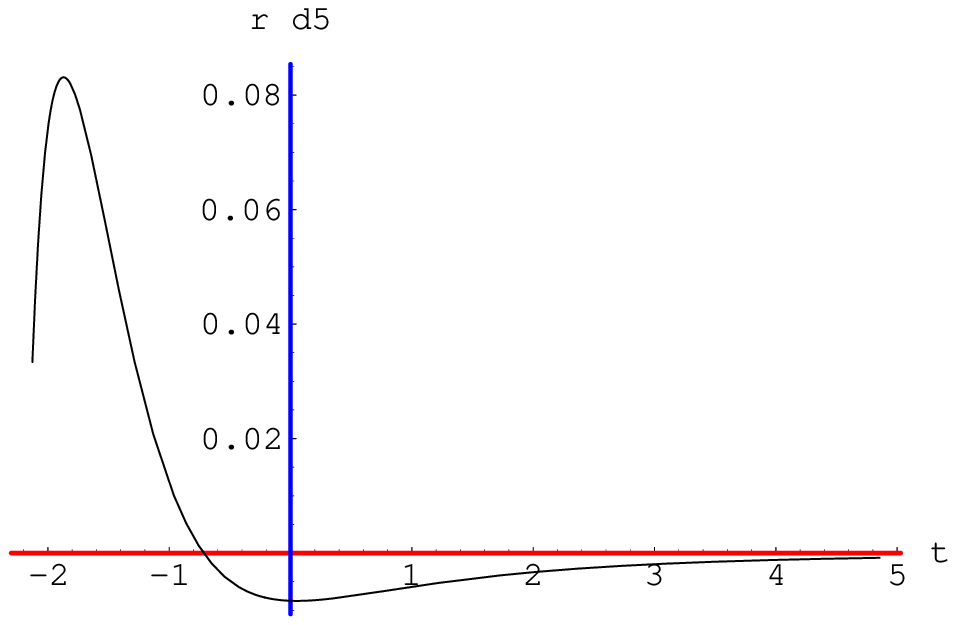}
  \caption{Plots of the energy densities as functions of the cosmic
    time $t=\tau(T)$ with the parameter choice $\omega =1$,
    $\kappa=0.8$ and for the $A_2$ solution with all the roots
    switched on. The first picture plots the behaviour of the total
    energy density. The second plots the ratio of the $D3$--brane
    contribution to the energy density with respect to the total
    density. The third plots the ratio of the $D3$--brane contribution
    with respect to the contribution of the dilaton $D5$--brane
    system. The fourth and the fifth picture plot the energy density
    of the $D3$--brane and of the \textit{dilaton}--$D$--string
    systems, respectively.}
  \label{energ2o1k08}
\end{figure}

The phenomenon is best appreciated by considering the plots of the
pressure eigenvalues (see figs. (\ref{press2o1k08})) and of the energy
density (see figs.  (\ref{energ2o1k08})). In the pressure plots we see
that there is a positive bump both in the directions 34 and 89, yet
the bump in 89 is anticipated at earlier times and it is bigger than
the bump in 34, the reason being the cooperation between the
$D3$--brane and $D$-string contributions. Even more instructive is the
plot of the energy densities.  

In fig.(\ref{energ2o1k08}) we see that the energy density of the
$D3$--brane has the usual positive bump, while the energy density of
the \textit{dilaton}--$D$--string system has a positive bump followed
by a smaller negative one, so that it passes through zero.

At the critical value $\kappa=\ft 32 \omega$ something very
interesting occurs in the behaviour of the scale factors.

\begin{figure}[!t]
  \centering
  \includegraphics[width=5cm]{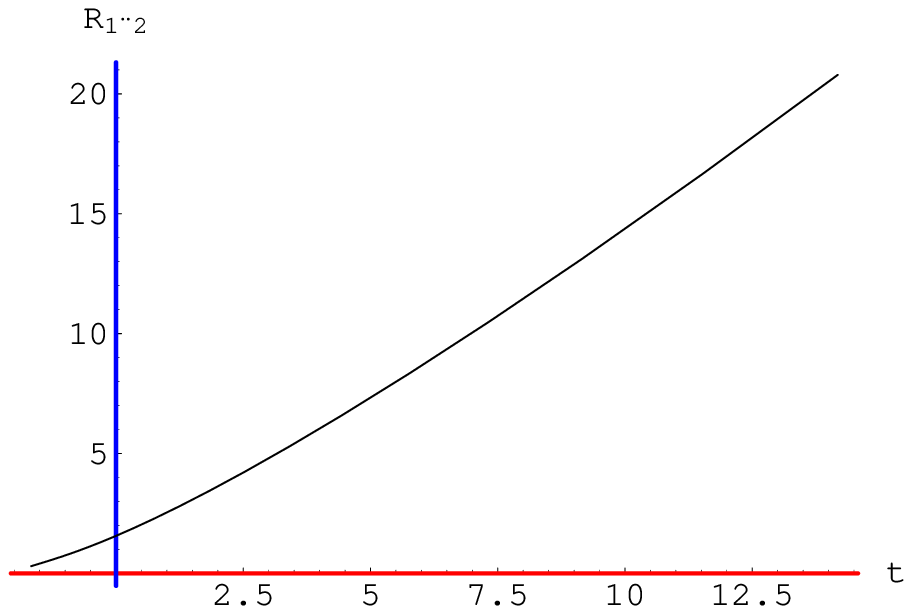}
  \includegraphics[width=5cm]{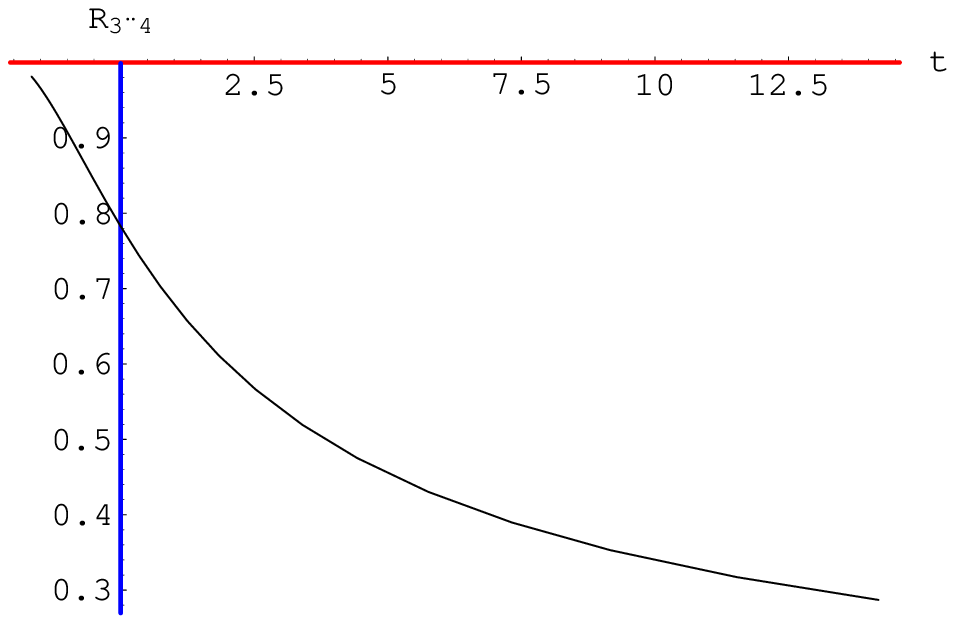}
  \includegraphics[width=5cm]{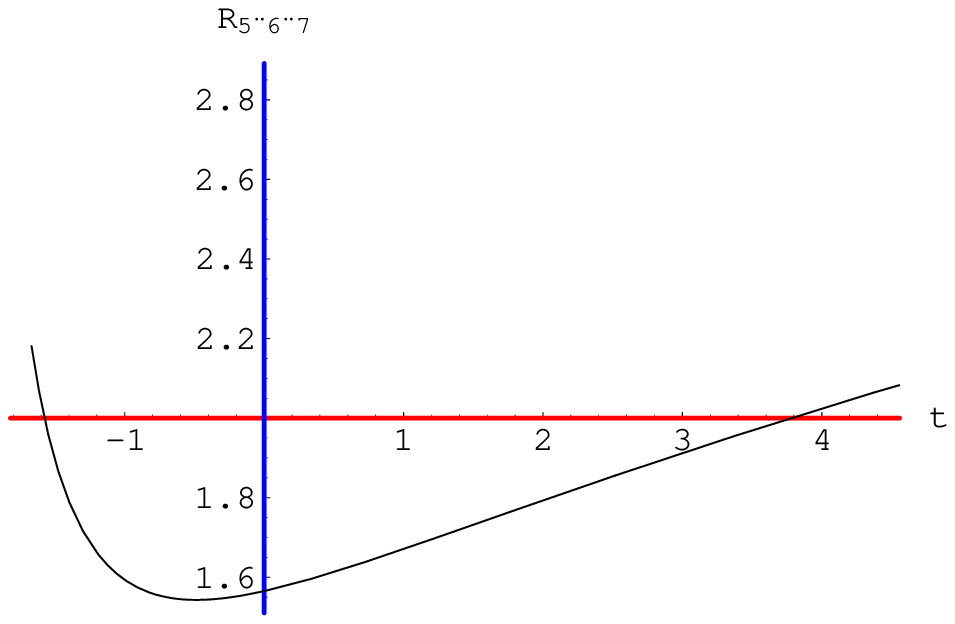}
  \includegraphics[width=5cm]{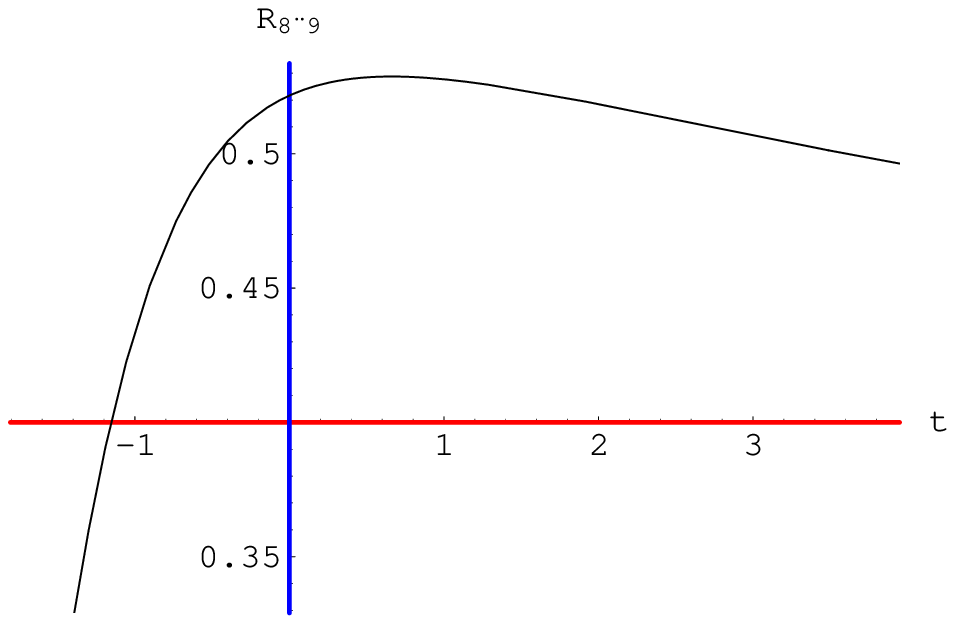}
  \caption{Plots of the scale factors $\overline{r}^2_{[\alpha]}$, 
    $\alpha= 1|2 \, ,\,3|4 \, , \, 5|6|7 \, , \, 8|9$ as functions of
    the cosmic time $t=\tau(T)$ with the critical choice of parameters
    $\omega =1$, $\kappa=3/2$ and for the $A_2$ solution with all the
    roots switched on.}
  \label{radiio1k15}
\end{figure}

As we see from fig.(\ref{radiio1k15}), the scale factor in the
direction 34, rather than starting from zero as in all other cases
starts from a finite value and then always decreases without
suffering a billiard bump. The bump is only in the scale factor 89.
Essentially this means that the positive energy of the $D3$ brane and
the negative one of the $D$-string exactly compensate at the origin
of time for these critical value of the parameters.

\subsection{Summarizing the above discussion and the cosmological billiard}

Summarizing what we have learned from the numerical analysis of the
type IIB cosmological backgrounds obtained by a specific oxidation of
the $A_2$ sigma model solutions we can say what follows.

The expansion or contraction of the cosmological scale factors in the
diagonal metric is driven by the presence of euclidean $D$--branes
which behave like instantons (S-branes). Their energy--density and
charge are localized functions of time. Alternatively we see that
these branes contribute rather sharp bumps in the eigenvalues of the
spatial part of the stress--energy tensor which we have named
\textit{pressures}.  Typically there are \textit{maxima} of these
pressures in the space directions parallell to the euclidean brane
world--volume and \textit{minima} of the same in the directions
transverse to the brane. These maxima and minima in the pressures
correspond to maxima and minima of the scale factors in the same
directions. Such inversions in the rate of expansion/contraction of
the scale factors is the \textit{cosmological billiard} phenomenon
originally envisaged by Damour et al. In the toy $A_2$ model we have
presented, we observe just one scattering, but this is due to the
insufficient number of branes (roots in the Lie algebra language) that
we have excited.  Indeed it is like we had only one wall of a Weyl
chamber. In subsequent publications we plan to study the phenomenon in
more complex situations with more algebraic roots switched on. What is
relevant in our opinion is that we were able to see the postulated
\textit{bumping phenomenon} in the context of exact smooth solutions
rather than in asymptotic limiting regimes.

\section{Conclusions and perspectives}

In \cite{Fre:2003ep} we developed a convenient mathematical framework
within three dimensional (ungauged) maximal supergravity where to
study homogeneous cosmological solutions of type II A or II B
theories. Our approach exploited the correspondence between
homogeneous time--dependent solutions in ten and three dimensions.
This mapping was realized through \emph{toroidal dimensional
  reduction} from $D=10$ to $D=3$ or through \emph{oxidation} from
$D=3$ to $D=10$.  The starting point of our study was the ${\rm
  E}_{8(8)}$ orbit described by three dimensional homogeneous
solutions and we defined the precise method for constructing a generic
representative of the orbit from the generating solution which is
defined only by the radii of the internal seven--torus and the
dilaton. Exploiting the solvable Lie algebra (or Iwasawa)
representation of the scalar manifold in the three dimensional theory
it was possible to control the ten dimensional interpretation of the
various bosonic fields. This allows for instance to construct a ten
dimensional solution characterized by certain (off--diagonal)
components of the metric or of the tensor fields by oxiding a three
dimensional solution in which the scalar fields associated with the
corresponding ${\rm E}_{8(8)}$ roots are switched on. As an example we
worked out in three dimensions the general homogeneous time--dependent
solution of an $A_2$ model in which the scalar fields span a ${\rm
  SL}(3,\mathbb{R})/{\rm SO}(3)$ submanifold of ${\rm E}_{8(8)}/{\rm
  SO}(16)$. It was shown that, depending on the embedding of ${\rm
  SL}(3,\mathbb{R})$ within ${\rm E}_{8(8)}$ the ten dimensional
solution obtained upon oxidation of the three dimensional one can have
radically different physical interpretations in terms of ten
dimensional fields. We then focused on one particular embedding for
which the axionic fields are interpreted as the components $B_{34}$,
$C_{89}$ and $C_{3489}$ of the type II B tensor fields, and
accomplished the oxidation of the solution to ten dimensions. Its
behavior, which has been described in detail in the previous section,
is characterized by an exchange of energy between the tensor fields
and the gravitational field which results in consecutive phases of
expansion and contraction of the cosmological scale factors along the
directions defined by the non vanishing components of the tensor
fields .  This background could be interpreted microscopically in
terms of a system of space--like or \emph{S--branes} (or SD--branes)
\cite{Gutperle:2002ai} along the directions $89$ and $3489$, coupled
to the Kalb--Ramond field. It is interesting to make contact with
the \emph{cosmological billiard} phenomenon describing the behavior of
solutions to Einstein equations near space--like singularities. In
this limit the evolution of the scale--factors/dilatons is described
by a null trajectory in a hyperbolic space which is reflected by
\emph{walls} or hyper--surfaces where the energy density of the
axionic fields diverges. Although ours is a different kind of analysis
which aims at the construction of exact smooth cosmological solutions,
we may retrieve a similar \emph{qualitative} description of the
evolution of the scale--factors/dilaton in relation to the evolution
of the axionic fields. In our formalism the logarithm $\sigma_i$ of
the scale factors associated with the internal directions together
with the ten dimensional dilaton $\phi$ are described by the vector
$h(t)$ in the Euclidean eight--dimensional space of the ${\rm
  E}_{8(8)}$ Cartan subalgebra. The kinetic term of an axion $\chi $
associated with the positive ${\rm E}_{8(8)}$ root $\alpha$ contains
the characteristic exponential factor $\exp{(-2\, \alpha\cdot h)}$.
The corresponding wall in the space of $h$ is defined by the equation
$ \alpha\cdot h= 0$ and the \emph{billiard} region by $\alpha\cdot
h\ge 0$. As it can be inferred from our solution, the evolution of
$h(t)$ is such that, if we denote by $h^\|$ the component of $h$ along
$\alpha$ and by $h^\perp$ its projection on the hyperplane
perpendicular to $\alpha$ (the wall), as the energy density of $\chi$
reaches its maximum (this temporal region corresponds to the
\emph{thickness} of the S--branes in the $A_2$ solution) $h$ undergoes
a reflection. This is most easily illustrated for example in the $A_2$
solution with just one root switched on (namely $\alpha[80]$). In this
case the component of $\dot{h}$ parallel to $\alpha[80]$ undergoes a
continuous sign inversion from negative values to positive ones while
$|\dot{h}^\perp|$ is constant and proportional to the time derivative
of the dilaton :
\begin{eqnarray}
\dot{h}^\|\rightarrow -\dot{h}^\| \,\,\,;\,\,\,|\dot{h}^\perp|\,=\,
\frac{k}{2\sqrt{6}}\,\propto\,\dot{\phi}\,=\,\mbox{const.}
\end{eqnarray}
 The roots in the $A_2$ system are not enough to define
a finite volume billiard which would result in an oscillatory
behavior of the solution. Indeed, using for $h$ the
parametrization in terms of the variables $x,\,y$, namely
$h=\{x,x,y,y,y,-y,-y,y\}$, the billiard region is defined by the
dominant walls $\alpha[69]\cdot h=0,\,\alpha[15]\cdot h=0$:
\begin{eqnarray}
x&\ge& 0\,;\,\,\,y\le -\frac{x}{3}
\end{eqnarray}
which is an open region. This explains the Kasner--like
(non--oscillatory) behavior of our solution for $t\rightarrow
+\infty$.\par 

More general ten dimensional homogeneous solutions deriving from the
coupling of gravity with purely metric 10 dimensional backgrounds were
analized by some of the authors of \cite{Fre:2003ep} in
\cite{Fre:2003tg}, in which they addressed the related question to define how
many independent orbits of cosmological solutions there are under the
action of the $U$-duality group.

In particular, exploiting first the classification of all the possible
regular embeddings $\mathbf{G}_r \hookrightarrow E_8, r \leq 8$
subalgebras, then the classification of the Weyl orbits of the
$\mathbf{G}_r$ root system within the $E_8$ root system, the authors
proved that all regular embeddings of each $A_r$ subalgebras inside
$E_{8(8)}$ fall into a single Weyl orbit, with the exception of $A_7$,
which falls into two distinct one.

They chose a purely
metric canonical representative embedding of the $A_2$ sigma model,
then oxided the three-dimensional solutions found \cite{Fre:2003ep}
and eventually studied their geometrical and physical
properties.  The simplest solution, which has only the highest of the
three roots associated to nihilpotents fields switched on and the
fundamental parameter $\kappa$ set to zero, provides an exact example
of Bianchi type $2A$ metric in four dimensions. Moreover, the solution
with $\kappa\neq0$ leads to a nontrivial evolution of the scale factor
in the fifth dimension which, after a Kaluza Klein reduction can be
reinterpreted as a Bianchi type $2A$ metric with scalar matter content.
The same holds for the solutions with all the roots switched on: they
produce further examples of type $2A$ Bianchi cosmologies with scalar
and vector matter contents.

All these solutions are homogeneous but not isotropic, and they show
the peculiar cosmological billiard feature.

A concrete step towards the understanding of billiard dynamics has been
done in \cite{Fre:2005si}. The main investigation in this sense would
involve derivation of exact solutions directly in a $D=2$ or $D=1$
context where Ka\v c--Moody symmetries become manifest. Although the
appearance of Ka\v c--Moody extensions is algebraically well
established, their exploitation in deriving solutions is not as clear
as the exploitation of ordinary symmetries.  In \cite{Fre:2005si}, the
authors try to clarify the field theoretical realization of the Ka\v
c--Moody extensions, this being the prerequisite for their utilization
in deriving billiard dynamics.  In particular they have shown that
there is a general mechanism underlying the affine Ka\v c--Moody
extension of the D=3 algebra $\mathbb{U}_{D=3}$ when stepping down to
$D=2$ and that this mechanism follows a general algebraic pattern for
all supergravity theories, independently of the number of supercharges
$N_Q$. This mechanism relies on the existence of two different
reduction schemes from $D=4$ to $D=2$, respectively named the Ehlers
reduction and the Matzner--Missner reduction, which are non locally
related to each other. Nicolai observed this phenomenon time ago in
the case of pure gravity (or better of N=1 pure
supergravity)\cite{NicolaiEHMM} and showed that one obtains two
identical lagrangians, each displaying an $\mathrm{SL(2,\mathbb{R})}$
symmetry. The fields appearing in one lagrangian have a non local
relation to those of the other lagrangian and one can put together
both $\mathrm{SL(2,\mathbb{R})}$ algebras. One algebra generates local
transformations on one set of fields the other algebra generates non
local ones. Together the six generators of the two
$\mathrm{SL(2,\mathbb{R})}$ provide a Chevalley basis for the Ka\v
c--Moody extension $\mathrm{SL(2,\mathbb{R})}^\wedge$ namely for
$A_1^\wedge$. The analysis in \cite{Fre:2005si} is an extension of the
argument by Nicolai. For a generic supergravity theory, the two
reduction schemes Ehlers and Matzner--Missner lead to two different
lagrangians with different local symmetries. The first is a normal
\textit{$\sigma$--model} the second is a \textit{twisted
  $\sigma$--model}. The authors discuss in detail the symmetries of
both theories, thus writing down a precise field theoretic realization
of the affine symmetries setting the basis to exploit them in billiard
dynamics.

%% file: lang_proj.tex
\chapter{Projection of generalized Langland's boundary states}
\label{sec:lang_proj}

This appendix contains a careful computation of the projection of
generalized Langlands boundary states \eqref{eq:lbs-} and
\eqref{eq:lbs+} on the constraints in equation
\eqref{eq:cont_bs}. Dealing with the compactified boson, the only
chiral fields involved besides Virasoro fields are the holomorphic and
antiholomorphic currents $J(\zeta)$ $\bar{J}(\bgz)$ generating the
Heisenberg algebra \eqref{eq:ha}. Being the current algebra abelian,
the possible gluing maps reduces to:
\begin{subequations}
\begin{gather}
 J(\zeta) \,=\,\bar{J}(\bgz)\\
 J(\zeta) \,=\, -\, \bar{J}(\bgz)
\end{gather}
\end{subequations}
which, trought radial quantization, are respectively mapped into
\begin{subequations}
  \label{conda}
  \begin{gather}
    \label{cpiua}
    \left( \mathfrak{a}_n \,+\, \overline{\mathfrak{a}}_{-n} \right) \,
    \Vert B \Rangle \,=\, 0, \\
    \label{cmenoa}
    \left( \mathfrak{a}_n \,-\, \overline{\mathfrak{a}}_{-n} \right) \,
    \Vert B \Rangle \,=\, 0.
  \end{gather}
\end{subequations}

The Sugawara construction \eqref{virasoro} ensures that this
conditions are sufficients to enforce conformal invariance encoded in
\eqref{eq:no_momflow}.

To work out the projection of \eqref{conda} over Generalized Langlands
boundary states, It is convenient to
introduce, for each given $n\geq 1$, the auxiliary states 
\begin{equation}
  \left| 
    \mathfrak{r}_{(\mu ,\nu )}^{\alpha }
    (S_{\varepsilon(k)}^{(-)})
  \right\rangle _{n} 
  \,\doteq\,
  \sum_{m_1, m_2} 
  \mathbb{A}_{m_1,m_2}^n (a_n^\alpha ,a_{-n}^\alpha )
  \frac{
    \left( \mathfrak{a}_{-n}^\alpha \right)^{m_1}
    \left( \overline{\mathfrak{a}}_{-n}^\alpha\right)^{m_2}}
  {\sqrt{n^{m_1 + m_2} m_1! m_2!}} 
  \left|
    (\mu^\alpha, \nu^\alpha)
  \right\rangle ,
\end{equation}
(one defines similarly the states 
$\left| \mathfrak{l}_{(\mu ,\nu)}^{\alpha} (S_{\varepsilon(k)}^{(+)})
 \right\rangle _{n}$
associated with \linebreak$\left| \mathfrak{r}_{(\mu ,\nu )}^{\alpha }
  (S_{\varepsilon(k)}^{(+)})\right\rangle $).

From condition \eqref{cmenoa} (for $n\,=\,0$) we get 
\begin{equation}
  \label{DirCond0}
  \lambda _{(\mu ,\nu )}^{\alpha }
  \,-\,
  \overline{\lambda }_{(\mu ,\nu)}^\alpha \,=\,0.
\end{equation}
which selects, at a generic value of the
ratio $\frac{L(k)}{R^\alpha(k)}$, the highest weight state
\begin{equation}
  \left| 
    (\mu^\alpha ,0)
  \right\rangle 
  \,=\,
  \left| 
    \lambda _{(\mu,0)}^{\alpha}
  \right\rangle 
  \,\otimes\, 
  \left| \overline{\lambda}_{(\mu,0)}^{\alpha}
  \right\rangle 
\end{equation}
\ie{} the winding number does not contribute to the left and right
momenta associated with this Verma module
\begin{equation}
  \begin{array}{lr}
    \mathfrak{a}_{0}^\alpha 
    \left| (\mu^\alpha, 0)\right\rangle 
    \,=\,
    \mu^\alpha \frac{L}{R^\alpha} 
    \left|(\mu^\alpha ,0)\right\rangle &
    \\
    & \mu^\alpha \,\in\, \mathbb{Z} \\
    \overline{\mathfrak{a}}_{0}^\alpha 
    \left| (\mu^\alpha, 0)\right\rangle 
    \,=\,
    \mu^\alpha \frac{L}{R^\alpha} 
    \left|(\mu^\alpha ,0)\right\rangle &
  \end{array}.
\end{equation}

Similarly, the constraint \eqref{cpiua} provides the condition
\begin{equation}
  \label{NeuCond0}
  \lambda _{(\mu ,\nu )}^{\alpha }
  \,+\,
  \overline{\lambda }_{(\mu ,\nu)}^\alpha \,=\,0.
\end{equation}
which selects, at generic value of the ratio 
$\frac{L(k)}{R^\alpha(k)}$, the highest weight state:
\begin{equation}
  \left| (0,\nu ^{\alpha })\right\rangle =\left| \lambda _{(0,\nu )}^{\alpha
    }\right\rangle \otimes \left| \overline{\lambda }_{(0,\nu )}^{\alpha
    }\right\rangle 
\end{equation}
such that 
\begin{equation}
  \begin{array}{lr}
    \mathfrak{a}_0^\alpha\,
    \left| (0, \nu^\alpha )\right\rangle  
    \,=\,
    \frac{1}{2} \nu^\alpha 
    \frac{R^\alpha}{L}
    \left| (0, \nu^\alpha )\right\rangle & \\
    & \nu^\alpha \,\in\, \mathbb{Z} \\
    \overline{\mathfrak{a}}_0^\alpha\,
    \left| (0, \nu^\alpha )\right\rangle  
    \,=\,
    - \frac{1}{2} \nu^\alpha 
    \frac{R^\alpha}{L}
    \left| (0, \nu^\alpha )\right\rangle &
  \end{array}.
\end{equation}

Before considering the conditions \eqref{conda} for
$n \,\neq\, 0$, let us impose the condition projecting the
$\left| \mathfrak{r}_{(\mu ,\nu )}^{\alpha }(S_{\varepsilon
    (k)}^{(-)})\right\rangle$ into an actual physical state:
\begin{equation}
  \left( \mathbb{L}_{0}^{\alpha }-\overline{\mathbb{L}}_{0}^{\alpha }\right)
  \left| \mathfrak{r}_{(\mu ,\nu )}^{\alpha }(S_{\varepsilon
      (k)}^{(-)})\right\rangle =0.
\end{equation}

From the commutation relation 
\begin{equation}
  \left[ \mathfrak{a}_{j}^{\alpha },(\mathfrak{a}_{-n}^{\alpha
    })^{m_{1}}\right] 
  = m_{1}j\delta _{j-n,0}(\mathfrak{a}_{-n}^{\alpha })^{m_{1}-1},
\end{equation}
one computes
\begin{gather}
  \left( \mathbb{L}_{0}^{\alpha }-\overline{\mathbb{L}}_{0}^{\alpha }\right)
  \left| \mathfrak{r}_{(\mu ,\nu )}^{\alpha }(S_{\varepsilon
      (k)}^{(-)})\right\rangle =\mu ^{\alpha }\nu ^{\alpha }\left|
  \mathfrak{r}_{(\mu 
      ,\nu )}^{\alpha }(S_{\varepsilon (k)}^{(-)})\right\rangle +e^{\sqrt{-1}%
    t_{+}^{\alpha }(\lambda _{(\mu ,\nu )}^{\alpha }+\overline{\lambda }_{(\mu
      ,\nu )}^{\alpha })}\cdot   \notag \\
  \\
  \cdot \prod_{n=1}^{\infty
  }\sum_{m_{1},m_{2}}\mathbb{A}_{m_{1},m_{2}}^{n}n(m_{1}-m_{2})\frac{\left(
      \mathfrak{a}% 
      _{-n}^{\alpha }\right) ^{m_{1}}\left( \mathfrak{a}_{-n}^{\alpha
      }\right) ^{m_{2}}% 
  }{\sqrt{n^{m_{1}+m_{2}}m_{1}!m_{2}!}}\left| (\mu ^{\alpha },\nu ^{\alpha
    })\right\rangle ,  \notag
\end{gather}
which, together with the  $U(1)$ constraints, (implying that
$\mu ^{\alpha }\nu ^{\alpha }=0$), shows that the physical states are
characterized by the further requirement
\begin{equation*}
  m_{1}=m_{2}.
\end{equation*}

For $j\neq 0$, the action of the operators $\mathfrak{a}_{j}^{\alpha
}$ on the boundary states $\left| \mathfrak{r}_{(\mu ,\nu )}^{\alpha
  }(S_{\varepsilon (k)}^{(\pm )})\right\rangle $ is non-trivial when
$j>0$, and it can be worked out by considering the auxiliary states
$\left| \mathfrak{l}_{(\mu ,\nu )}^{\alpha }(S_{\varepsilon (k)}^{(\pm
    )})\right\rangle
_{n}$ for each value of $n\geq 1$. In detail, let us consider the state $%
\left| \mathfrak{l}_{(\mu ,\nu )}^{\alpha }(S_{\varepsilon
    (k)}^{(-)})\right\rangle _{n}$, (with $m_{1}=m_{2}=m\geq{1}$; the
    action of the annihilator $\mathfrak{a}_{j}^{\alpha}$ being
    trivial on $m=0$),   

\begin{gather}
\mathfrak{a}_{j}^{\alpha }\left| \mathfrak{l}_{(\mu ,\nu )}^{\alpha
  }(S_{\varepsilon 
    (k)}^{(-)})\right\rangle _{n}=\sum_{m=1}^{\infty
}\mathbb{A}_{m}^{j}\frac{(mj)\left( \mathfrak{a}% 
    _{-j}^{\alpha }\right) ^{m-1}\left( \overline{\mathfrak{a}}_{-j}^{\alpha
    }\right) ^{m}}{j^{m}m!}\left| (\mu ^{\alpha },\nu ^{\alpha
  })\right\rangle   
\label{aAct} \\
=\overline{\mathfrak{a}}_{-j}^{\alpha
}\sum_{m=1}^{\infty}\mathbb{A}_{m}^{j}\;\frac{\left( \mathfrak{a}% 
    _{-j}^{\alpha }\right) ^{m_{1}-1}\left(
    \overline{\mathfrak{a}}_{-j}^{\alpha 
    }\right) ^{m_{2}-1}}{j^{m-1}(m-1)!}\left| (\mu ^{\alpha },\nu ^{\alpha
  })\right\rangle .  \notag
\end{gather}

From the definition of $\mathbb{A}_{m}^{j}$, and the following recursion
relation for Laguerre polynomials

\begin{equation}
  L_{m}^{-1}(x)=L_{m}(x)-L_{m-1}(x),
\end{equation}

we get
\begin{equation}
  \mathbb{A}_{m}^{j}=\;e^{2\pi s\sqrt{-1}}
  \mathbb{A}_{m-1}^{j}+e^{-2j|a_{j}^{\alpha
    }|^{2}}e^{2\pi sm\sqrt{-1}}L_{m}^{-1}(4j|a_{j}^{\alpha }|^{2}).
\end{equation}

Inserting this latter into (\ref{aAct}) provides

\begin{gather}
  \mathfrak{a}_{j}^{\alpha }
  \left| \mathfrak{l}_{(\mu ,\nu )}^{\alpha }
    (S_{\varepsilon(k)}^{(-)})\right\rangle_{n}
  \,=\,
  e^{2\pi s\sqrt{-1}}\overline{\mathfrak{a}}_{-j}^{\alpha }
  \left| \mathfrak{l}_{(\mu ,\nu )}^{\alpha }
    (S_{\varepsilon (k)}^{(-)})
  \right\rangle_{n}+ 
  \notag \\
  +e^{-2j|a_{j}^{\alpha }|^{2}}
  e^{2\pi sm\sqrt{-1}}
  \overline{\mathfrak{a}}_{-j}^{\alpha}\sum_{m}L_{m}^{-1}
  (4j|a_{j}^{\alpha }|^{2})
  \frac{\left( \mathfrak{a}_{-j}^{\alpha}\right)^{m-1}
    \left( 
      \overline{\mathfrak{a}}_{-j}^{\alpha}
    \right)^{m-1}}{j^{m-1}(m-1)!}
  \left| 
    (\mu ^{\alpha },\nu ^{\alpha})
  \right\rangle,  \notag
\end{gather}

from which it immediately follows that the Neumann condition  

\begin{equation}
  (\mathfrak{a}_{j}^{\alpha }
  \,-\,
  \overline{\mathfrak{a}}_{-j}^{\alpha })
  \left| 
    \mathfrak{l}_{(\mu ,\nu )}^{\alpha }
    (S_{\varepsilon (k)}^{(-)})
  \right\rangle_{n}^{(N)}
  \,=\,0
\end{equation}

requires $s\in \mathbb{Z}$\ \ and $L_{m}^{-1}(4j|a_{j}^{\alpha
}|^{2})=0$, \emph{viz.}\ $a_{j}^{\alpha }\equiv 0\;\forall j$, whereas
for the Dirichlet condition

\begin{equation}
  (\mathfrak{a}_{j}^{\alpha}
  \,+\,
  \overline{\mathfrak{a}}_{-j}^{\alpha })
  \left| 
    \mathfrak{l}_{(\mu ,\nu )}^{\alpha}
    (S_{\varepsilon (k)}^{(-)})
  \right\rangle_{n}^{(D)}
  \,=\,0
\end{equation}

we need $s\in \frac{1}{2}\mathbb{Z}$\ \ and $a_{j}^{\alpha }\equiv
0\;\forall j$. Since $\mathbb{A}_{m}^{n}(a_{n}^{\alpha }=0)=e^{2\pi
  ms\sqrt{-1}}$, and

\begin{equation}
  \sum_{m}\mathbb{A}_{m}^{n}
\frac{\left( \mathfrak{a}_{-n}^{\alpha }\right) ^{m}\left( 
      \overline{\mathfrak{a}}_{-n}^{\alpha }\right) ^{m}}{n^{m}m!}=\left\{ 
    \begin{tabular}{ll}
      $\exp \left( \frac{1}{n}\left( \mathfrak{a}_{-n}^{\alpha }\right) \left( 
          \overline{\mathfrak{a}}_{-n}^{\alpha }\right) \right) ,$ & $s\in
      \mathbb{Z}$ 
      \\   &  \\ 
      $\exp \left( 
        -\frac{1}{n}\left( \mathfrak{a}_{-n}^{\alpha }
        \right) 
        \left( 
          \overline{\mathfrak{a}}_{-n}^{\alpha }
        \right) 
      \right) ,$ 
      & $s\in \frac{1}{2}%
      \mathbb{Z}$%
    \end{tabular}
  \right. ,
\end{equation}
the corresponding boundary states are provided by 
\begin{equation}
  \left| \mathfrak{r}_\mu^{\alpha }(S_{\varepsilon
      (k)}^{(-)})\right\rangle ^{(D)}=e^{\sqrt{-1}t_{+}^{\alpha }\mu ^{\alpha }
    \frac{L}{R^{\alpha }}}\exp \left( \sum_{n=1}^{\infty }\frac{1}{n}\left( 
      \mathfrak{a}_{-n}^{\alpha }\right) 
    \left( \overline{\mathfrak{a}}_{-n}^{\alpha}
    \right) 
  \right) 
  \left| (\mu ^{\alpha },0)\right\rangle ,  \label{IshiDira}
\end{equation}
and 
\begin{equation}
  \left| \mathfrak{r}^{\alpha }(S_{\varepsilon
      (k)}^{(-)})\right\rangle ^{(N)}=\exp \left( \sum_{n=1}^{\infty }
    -\frac{1}{n}
    \left( \mathfrak{a}_{-n}^{\alpha }\right) \left( \overline{\mathfrak{a}}%
      _{-n}^{\alpha }\right) \right) \left| (0,\nu ^{\alpha })\right\rangle .
  \label{IshiNeua}
\end{equation}
Note that (up to the weighting factor in (\ref{IshiNeua})) these states are
nothing but the Dirichlet and Neumann Ishibashi states associated with the
free bosonic field $\left. X^{\alpha }(k)\right| _{-}$ on the circle $%
S_{\varepsilon (k)}^{(-)}$.

%% file: biobcft.tex
\chapter{Conformal properties of Boundary Insertion Operators}
\label{sec:biobcft}

Boundary Insertion Operators live on the ribbon graph, thus their
interactions are guided by the trivalent structure of $\Gamma$.  In
this appendix we report an exhaustive analysis which shows that this structure allows to define
all fundamental coefficients weighting the self-interaction of
boundary insertion operators. Two points functions are well defined on
the edges of the graph, while OPE coefficients naturally defines the
fusion of different boundary insertion operators interacting in
$N_2(T)$ tri-valent vertexes of $\Gamma$. Moreover, we can define a
set of sewing constraint these coefficients must satisfy. Remarkably,
these constraints are perfectly analogues to the sewing constraint
introduced in \cite{Lewellen:1991tb} for boundary conditions changing
operator, thus enforcing the analogy between these latters and BIOs.

Due to their CVO analogue structure, we are able to specify the
conformal properties of these operators. BIOs are primary operators of
the boundary chiral algebra, thus they have a well-defined conformal
dimension which, in this case, coincide with the highest weight of the
$V_\lambda(p,q)$ module of the $U(1)$ (Virasoro) algebra:
\begin{equation}
  \label{eq:BIOcd}
  H(p,q) \,=\,  \frac{1}{2} \lambda(p,q)^2 
\end{equation}

\begin{figure}[!t]
  \centering
  \includegraphics[width=.9\textwidth]{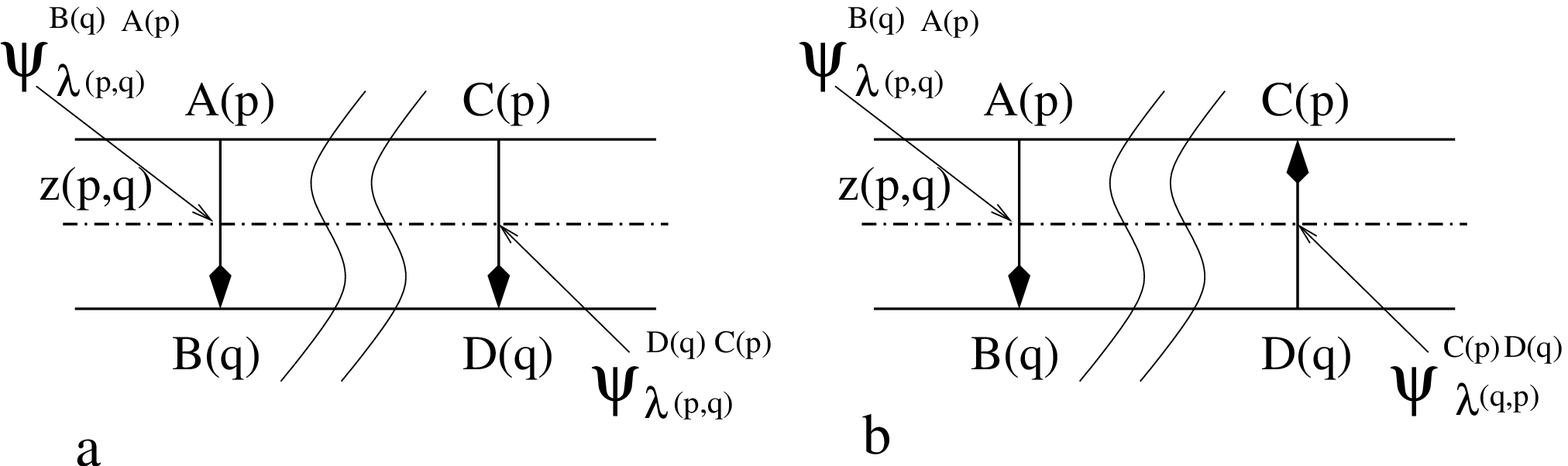}
  \caption{Two-point functions}
  \label{fig:BIO_2points}
\end{figure}

As usual, the conformal invariance fixes to zero the one point
function:
\begin{equation}
  \label{eq:BIO_1point}
  \langle \psi_{\lambda(p,q)}^{B(q) A(p)} \rangle\,=\,0
\end{equation} 
while, for the two-points function, it is a priori possible to
construct two types of correlators between BIOs which mediate the
changing in boundary conditions between the two adjacent cylinder
\cyl{p} and \cyl{q}:a first one will correlate two operators which
both mediate the changing in the ``$p$-to-$q$'' direction (or,
equivalently, which mediate both the changing in the ``$q$-to-$p$''
direction). The second type will correlate two BIOs which mediate one
in the ``$p$-to-$q$'' direction and the other in the ``$q$-to-$p$''
one (see figure \ref{fig:BIO_2points}).

The former situation, depicted in figure \ref{fig:BIO_2points}a, leads to:

\begin{equation}
\label{eq:2points}
  \langle 
  \psi_{\lambda(p,q)}^{B(q) A(p)} (x_1(p,q)) 
  \psi_{\lambda'(q,p)}^{C(p) D(q)} (x_2(q,p))
  \rangle \,=\, 
  \frac{b_{\lambda(p,q)}^{B(q) A(p)}
    \delta_{\lambda(p,q) \lambda'(p,q)}
    \delta^{A(p) C(p)}
    \delta^{B(q) D(q)}}
  {|x_1(p,q)\,-\,x_2(p,q)|^{2H(p,q)}}
\end{equation}

Note that the Kronecker delta function is used to let the two
boundary conditions on the same oriented edge of the strip to be
compatible. In the usual BCFT boundary fields correlators, the two
fields are taken to perform two serial switch in boundary condition,
and this leads to a different equalities between boundary conditions
indexes. Here, however, the two operators act in a parallel way, and
this justifies our choice for delta functions. 

Obviously, the second situation, depicted in figure
\ref{fig:BIO_2points}b, leads to the same result because of
transformation law \eqref{eq:conformal_bio}:
\begin{equation}
  \langle 
\label{eq:2points2}
  \psi_{\lambda(p,q)}^{B(q) A(p)} (x_1(p,q)) 
  \psi_{\lambda'(q,p)}^{C(p) D(q)} (x_2(q,p))
  \rangle \,=\, 
  \frac{b_{\lambda(p,q)}^{B(q) A(p)}
    \delta_{\lambda(p,q) \lambda'(p,q)}
    \delta^{A(p) C(p)}
    \delta^{B(q) D(q)}}
  {|x_1(p,q)\,-\,x_2(p,q)|^{2H(p,q)}}
\end{equation}

\begin{figure}[!t]
  \centering
  \includegraphics[width=\textwidth]{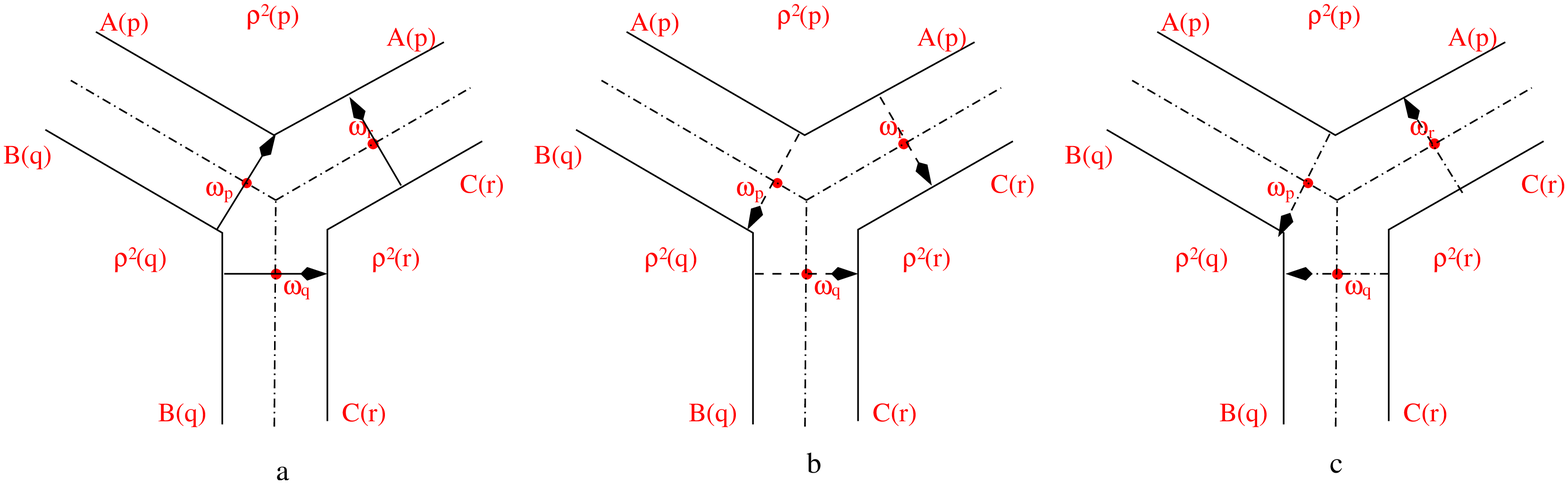}
  \caption{Two-point functions}
  \label{fig:BIO_ope}
\end{figure}

To analyze the conformal properties of BIOs and define completely the
equivalence between them and usual boundary operators in CFT, let
us consider what happens in each vertex of the underling triangulation
(the situation is the one depicted in figure \ref{fig:BIO_ope}).  Let
us consider the generic vertex of the triangulation $\rho^0(p,q,r)$
and its uniformizing neighborhood unit disk $\left(U_{\rho^0(p,q,r)},
  \omega(p,q,r)\right)$. The BIOs are well defined on the disk because
of the transition functions defined in \eqref{eq:edge_vertex_trans}.  The
vertexes of the triangulation are the natural point in which to define
the fusion between BIOs coming from different edges via their OPE.

Let us take three points in an
$\epsilon$-neighborhood of the vertex $\omega=0$, denoting with
\begin{align}
  \omega_r & \,=\, \frac{\epsilon}{3} \,e^\frac{\pi}{6} 
  \,\in\,U_{\rho^0(p,q,r)}\,\cap\,U_{\rho^1(r,p)} \\
  \omega_p & \,=\, \frac{\epsilon}{2} \,e^\frac{5\,\pi}{6}
  \,\in\,U_{\rho^0(p,q,r)}\,\cap\,U_{\rho^1(p,q)}\\  
  \omega_q & \,=\, \epsilon \,e^\frac{3\,\pi}{2}
  \,\in\,U_{\rho^0(p,q,r)}\,\cap\,U_{\rho^1(q,r)}  
\end{align}
their coordinates, and let us focus our attention the three fields
$\psi_{\lambda(r,p)}^{A(p) C(r)} (\omega_r)$,
$\psi_{\lambda(p,q)}^{B(q) A(p)} (\omega_p)$ and
$\psi_{\lambda(q,r)}^{C(r) B(q)} (\omega_q)$ which mediate pairwise
the boundary conditions between the boundaries of $\rho^2(r)$ and
$\rho^2(p)$, $\rho^2(p)$ and $\rho^2(q)$ and $\rho^2(q)$ and
$\rho^2(r)$ respectively (as usual, the direction of the action of
BIOs is given by the notation).

In the limit $\epsilon \rightarrow 0$ the product of the two fields
$\psi_{\lambda(r,p)}^{A(p) C(r)}$ and $\psi_{\lambda(q,r)}^{C(r)
  B(q)}$ will mediate the changing in boundary conditions from $B(q)$
to $A(p)$, thus the OPE of these two fields must be expressed as a
function of a $\psi_{\lambda(q,p)}^{A(p) B(q)}$-type field:
\begin{subequations}
  \label{eq:ope}
\begin{multline}
  \label{eq:ope1}
  \psi_{\lambda(r,p)}^{A(p) C(r)}(\omega_r) 
  \psi_{\lambda'(q,r)}^{C(r) B(q)}(\omega_q)
  \,\sim\, \\
  \sum_{\lambda''(q,p) \in \mathcal{Y}}
  \mathcal{C}^{A(p) C(r) B(q)}_{\lambda(r,p) \lambda'(q,r) \lambda''(q,p)}
  |\omega_r\,-\,\omega_q|^{H(q,p)\,-\,H(r,p)\,-\,H(q,r)}
  \psi_{\lambda''(q,p)}^{A(p) B(q)} (\omega_q)
\end{multline}
This situation is described by the continuous arrows in fig. \ref{fig:BIO_ope} 

In the same way we can write the other two naturally defined OPES:
\begin{multline}
  \label{eq:ope2}
  \psi_{\lambda(q,r)}^{C(r) B(q)}(\omega_q) 
  \psi_{\lambda'(p,q)}^{B(q) A(p)}(\omega_p)
  \,\sim\, \\
  \sum_{\lambda''(p,r) \in \mathcal{Y}}
  \mathcal{C}^{C(r) B(q) A(p)}_{\lambda(q,r) \lambda'(p,q) \lambda''(p,r)}
  |\omega_r\,-\,\omega_q|^{H(q,p)\,-\,H(r,p)\,-\,H(q,r)}
  \psi_{\lambda''(p,r)}^{C(r) A(p)} (\omega_p)
\end{multline}
described by the dashed arrows in fig. \ref{fig:BIO_ope}, and
\begin{multline}
  \label{eq:ope3}
  \psi_{\lambda(p,q)}^{B(q) A(p)}(\omega_p) 
  \psi_{\lambda'(r,p)}^{A(p) C(r)}(\omega_r)
  \,\sim\, \\
  \sum_{\lambda''(q,p) \in \mathcal{Y}}
  \mathcal{C}^{B(q) A(p) C(r)}_{\lambda(p,q) \lambda'(r,p) \lambda''(r,q)}
  |\omega_r\,-\,\omega_q|^{H(q,p)\,-\,H(r,p)\,-\,H(q,r)}
  \psi_{\lambda''(r,q)}^{B(q) C(r)} (\omega_r)
\end{multline}
\end{subequations}
described by the dotted arrows in fig. \ref{fig:BIO_ope}

The complete description of the interaction of the $N_0$ BCFTs passes
trough the determination of the OPE coefficients of the BIOs
$\mathcal{C}^{A(\cdot) B(\cdot) C(\cdot)}
_{\lambda(\cdot,\cdot) \lambda'(\cdot,\cdot) \lambda''(\cdot,\cdot)}$,
defined for $A,\,B,\,C\,\in\,\mathcal{A}$ and 
$\lambda,\,\lambda',\,\lambda''\,\in\,\mathcal{Y}$.

To obtain this result, it is first of all necessary to prove that our
description is totally equivalent with the usual BCFT formulation.
This can be achieved via demonstrating that the OPE coefficients and
the normalization factor entering in \eqref{eq:2points} satisfy both
the cyclic symmetry and the sewing constraint which derive from the
three points and four points functions in the usual formulation of a
Boundary Conformal Field Theory. Let us start
from considering the three points function which arise naturally in
the vertex $\rho^0(p,q,r)$, which must be cyclically invariant. Thus: 
\begin{multline}
  \label{eq:3points}
  \langle 
  \psi_{\lambda(r,p)}^{A(p) C(r)}(\omega_r)\,
  \psi_{\lambda'(q,r)}^{C(r) B(q)}(\omega_q)\,
  \psi_{\lambda''(p,q)}^{B(q) A(p)}(\omega_p)\,
  \rangle
  \,=\, \\
  \langle 
  \psi_{\lambda''(p,q)}^{B(q) A(p)}(\omega_p)\,
  \psi_{\lambda(r,p)}^{A(p) C(r)}(\omega_r)\,
  \psi_{\lambda'(q,r)}^{C(r) B(q)}(\omega_q)\,
  \rangle
  \,=\, \\
  \langle 
  \psi_{\lambda'(q,r)}^{C(r) B(q)}(\omega_q)\,
  \psi_{\lambda''(p,q)}^{B(q) A(p)}(\omega_p)\,
  \psi_{\lambda(r,p)}^{A(p) C(r)}(\omega_r)\,
  \rangle
\end{multline}

If we exploit the OPEs defined in \eqref{eq:ope}, from each identity
in the previous equation we obtain the following relations between the
OPEs coefficients:  
\begin{gather}
  \label{eq:cycl}
  \mathcal{C}^{A(p) C(r) B(q)}_{\lambda(r,p) \lambda'(q,r) \lambda''(q,p)} \,
  d^{A(p) B(q)}_{\lambda''(q,p)}
  \,=\,
  \mathcal{C}^{B(q) A(p) C(r)}_{\lambda''(p,q) \lambda(r,p) \lambda'(r,q)} \,
  d^{B(q) C(r)}_{\lambda'(r,q)}
  \\
  \mathcal{C}^{A(p) C(r) B(q)}_{\lambda(r,p) \lambda'(q,r) \lambda''(q,p)} \,
  d^{A(p) B(q)}_{\lambda''(q,p)}
  \,=\,
  \mathcal{C}^{C(r) B(q) A(p)}_{ \lambda'(q,r) \lambda''(p,q) \lambda(p,r)} \,
  d^{C(r) A(p)}_{\lambda(p,r)}
  \\
  \mathcal{C}^{B(q) A(p) C(r)}_{\lambda''(p,q) \lambda(r,p) \lambda'(r,q)} \,
  d^{B(q) C(r)}_{\lambda'(r,q)}
  \,=\,
  \mathcal{C}^{C(r) B(q) A(p)}_{ \lambda'(q,r) \lambda''(p,q) \lambda(p,r)} \,
  d^{C(r) A(p)}_{\lambda(p,r)}
\end{gather}

where the $d^{A(p) B(q)}_{\lambda''(q,p)}$ are the normalization
coefficients of the two point functions in the
$\left(U_{\rho^0(p,q,r)}, \omega(p,q,r)\right)$ frame:
\begin{equation}
  \label{eq:diff_frames}
  d^{A(p) B(q)}_{\lambda(q,p)} \,=\, 
  \left| \frac{d \omega}{d z(q,p)} \right|^{-H(q,p)}_{\omega=\omega_q}\,
  \left| \frac{d \omega}{d z(q,p)} \right|^{-H(q,p)}_{\omega=\omega_p}\,
  b^{A(p) B(q)}_{\lambda(q,p)},
\end{equation}

which satisfies the obvious property 
$d^{A(p) B(q)}_{\lambda(q,p)}\,=\, d^{B(q) A(p)}_{\lambda(p,q)}$.

Moreover, if we take the two different OPEs in the first term of
\eqref{eq:3points}, we obtain the usual sewing constraint: 
\begin{equation}
  \label{eq:3points_sewing}
  \mathcal{C}^{A(p) C(r) B(q)}_{\lambda(r,p) \lambda'(q,r) \lambda''(q,p)}\,
  \mathcal{C}^{A(p) B(q) A(p)}_{\lambda''(q,p) \lambda''(p,q) 0}
  \,=\,
  \mathcal{C}^{C(r) B(q) A(p)}_{ \lambda'(q,r) \lambda''(p,q) \lambda(p,r)} \,
  \mathcal{C}^{A(p) C(r) A(p)}_{\lambda(r,p) \lambda(p,r) 0}
\end{equation}
where $0$ is the singlet label.
To complete this analysis, we would need to check sewing constraints
arising from the four-points function behavior. We leave this further
investigation to the next chapter, in which will see
that the behavior of the four points function is fundamental to define
the action of BIOs on the full surface. 
This analysis, together with properties outlined in section proves the complete
equivalence between the usual action of boundary changing conditions
operators on the upper half plane (UHP) boundary, and our particular
model, in which BIOs merge dynamically the $N_0$ copies of an
UHP-defined BCFT.

%% file: bcft_deform.tex
\chapter{An introduction to truly marginal deformations of boundary
conformal field theories}
\label{ch:bcft_deform}

The coupling of a String theory with a generic background field
corresponds in a microscopic CFT description to a \emph{deformation}
of the model. This deformation is formally obtained modifying the
action with a perturbative term:
\begin{equation}
  \label{eq:action_deformation}
  S 
  \,\rightarrow\,
  S' \,=\,
  S \,+\, g \int d^2z \mathcal{O}(z,\overline{z})
\end{equation}

For the theory to remain conformally invariant, the operator must be a
\emph{marginal operator}, \ie{} $\mathcal{O}(z,\overline{z})$ must
at least have conformal dimension (1,1) in order to let the integrated
quantity be dimensionless.

Let us assume one is given a certain Lagrangian $\mathcal{L}$ defining
an exactly solvable Conformal Field Theory on a two dimensional manifold
$\Sigma$ (thus we know the central charge of the Virasoro
algebra, the spectrum of the model and the associated Vertex
algebra). A deformation of this model is obtained by adding a
perturbative term to the action:
\begin{equation}
  \label{eq:pert}
  \mathcal{L}\,\rightarrow \mathcal{L}'\,=\,\mathcal{L}\,\,+\,\sum_i
  g_i f_i(z,\overline{z})
\end{equation}
where $f_i(z,\overline{z})$ are some operators in the spectrum of the
theory defined by $L$, while $g_i$ are some coupling constants. In
this connection, a Conformal Field Theory can be defined as a fixed
point in the infinite dimensional space of theories parametrized by
the coefficients of possible operators such as in equation
\eqref{eq:pert}.

Let us consider the simpler case in which we add to $\mathcal{L}$ a
single operator $f(z,\bz)$ with coefficient $g$. In a Renormalization
Group (RG) approach, once the operator is added with a certain value
of the coefficient, the theory (\ie{} the coefficient value) flows
along a RG trajectory as the energy scale $\mu$ is changed. The
behavior of the coupling is characterized by the $\beta$-function 
\begin{equation}
  \label{eq:beta}
  \beta(g)\,\equiv\,
\mu \frac{\partial g}{\partial\mu}
\end{equation}

\begin{figure}[!ht]
  \centering
  \includegraphics[width=.75\textwidth]{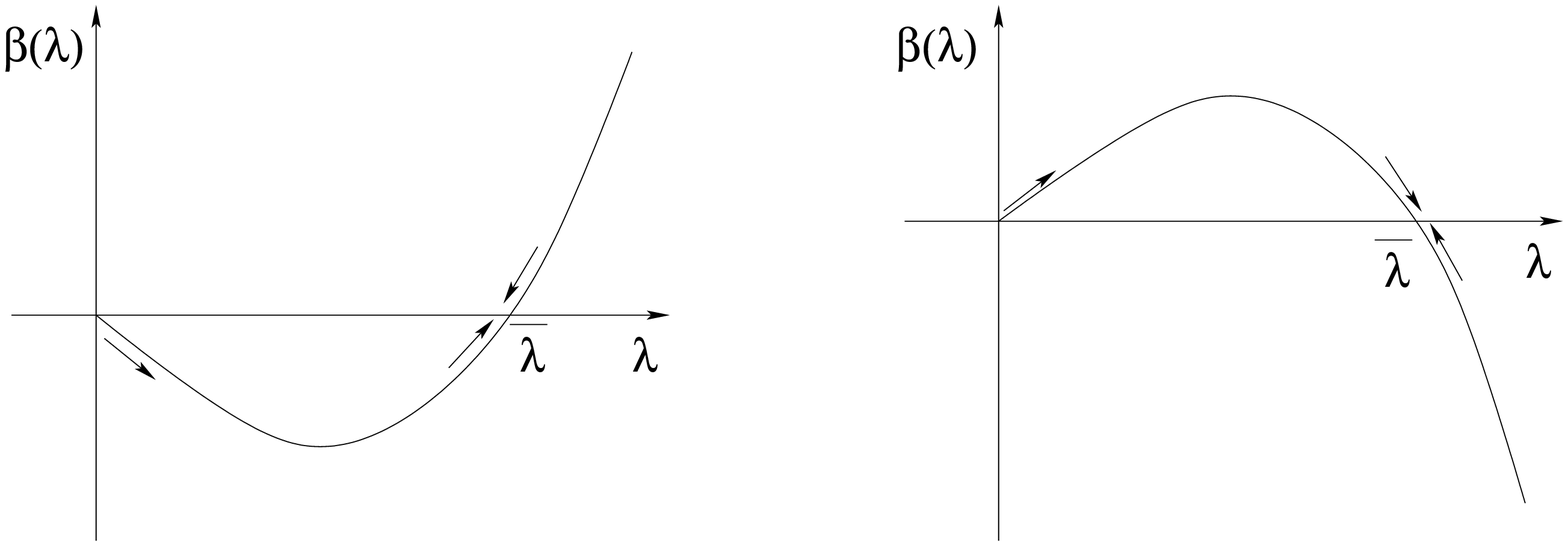}
  \caption{Flows}
  \label{fig:flows}
\end{figure}

One can imagine the existence of a fixed point of such flows when
$\beta(g)\,=\,0$ thus $g(\mu)\,=\,\overline{g}$ and
the coupling tend to that specific value, as shown in fig. \ref{fig:flows}.

On the left, $\overline{g}$ is an IR fixed point, since the
coupling tends to this value for decreasing $\mu$. On the contrary, on
the right, the theory is an UV fixed point, since the coupling
constant tends to the fixed point value for increasing energy scale.

The spectrum of operators associated with
$\mathcal{L}$ can be divided into the following categories:
\begin{itemize}
\item Operators whose conformal dimension is larger than (1,1), which
  are called \emph{irrelevant operators}. The correspondent coupling
  constant has a negative mass dimension. If one adds an irrelevant
  operator to a CFT exact Lagrangian, one finds that the coupling
  constant in the new Lagrangian decreases to zero in the IR limit,
  thus the name itself.
\item Operators whose conformal dimension is smaller than (1,1), which
  are called \emph{relevant operators}. The corresponding coupling
  constant has a positive mass dimension and, adding such an
  operator to $\mathcal{L}$ one finds that in the new model, $g$
  flows to a larger value in the IR. Moreover, if the theory is
  unitary, it will flow to a smaller $c$ value one. 
\item Operators whose dimension is exactly (1,1). They are called
  \emph{marginal operators}, and their associated coupling constants
  are dimensionless. They do not break the scale invariance of
  the model explicitly, thus they do not take away the CFT from the
  fixed point. However, the coupling constants can change under
  renormalization. This leads the marginal operators to subdivide into
  three classes:
  \begin{itemize}
  \item Marginal operators whose couplings in the modified Lagrangian
    are IR free on the worldsheet. Thus these operators turn out to be
    irrelevant ones. They differ from operators irrelevant at the
    classical level because they decrease logarithmically and not
    power-like towards the infrared. 
  \item  Marginal operators whose couplings $g_i$ under the
  modified Lagrangian are asymptotically free, thus increase
  logarithmically towards the infrared. They are then relevant
  operators.
  \item Marginal operators whose addiction to the Lagrangian
  $\mathcal{L}$ maintains the couplings $g_i$ dimensionless. This
  promotes $\mathcal{L}$ to a family of CFTs. These operators are
  called \emph{truly marginal} and form a basis for a family for
  neighborhood of CFTs. 
  \end{itemize}
\end{itemize}

\section{Deformations of a boundary conformal field theory}

We want to describe the effects of perturbations of a boundary
condition generated by marginal operators, in particular marginal
boundary fields. We will show that a certain class of perturbations,
which in \cite{Recknagel:1998ih} are called analytic deformations, are
truly marginal at all orders in the perturbation expansion. These
perturbations are induced by \emph{self-local} boundary operators of
dimension one.

\begin{figure}[t]
  \centering
  \includegraphics[width=.65\textwidth]{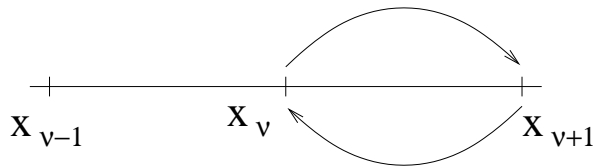}
  \caption{Curves along which correlators of bulk and boundary
    operators can be analytically continued}
  \label{fig:cont}
\end{figure}

\begin{defin}
\label{def:mut_loc}
  \textbf{Mutually local boundary fields}\\
  Let $\{\varphi_\lambda(z,\bar{z}) = \Phi
  (|\varphi_\lambda\rangle;\,z,\,\bz)\}$ and $\{\psi_\nu(x) = \Phi
  (|\psi_\nu\rangle;\,x)\}$ be respectively the collection of bulk and
  boundary fields of a given BCFT. Let us consider the correlator
  \begin{equation}
    \label{eq:c1}
    \langle 
  \psi_1(x_1) \cdots \psi_n(x_n)
  \psi_1(z_1,\bz_1) \cdots \psi_N(z_N,\bz_N)
  \rangle_\alpha, \qquad  x_\nu \,<\,x_{\nu+1}, 
  \end{equation}
where $\alpha$ denotes the given boundary condition on the real
axis. These functions are analytic in the variables
$z_i,\,i=1,\,\ldots,\,N$  throughout the whole UHP $\Im{z} > 0$. For
the variables $x_\nu,\,\nu=1,\,\ldots,\,N$, the analicity domain is
restricted to the interval $x \in ] x_{\nu - 1}, x_{\nu + !} [$ on the
boundary. In most cases, there are no unique analytic continuation of
the correlation functions to exchange the position of two two
neighboring boundary fields, since the result may depend on the
orientation of the path we analytically continue the correlator (see
fig. \ref{fig:cont}). 

In this connection, two boundary fields $\psi_{\nu_1}(x_1) = \Phi
(|\psi_{\nu_1}\rangle;\,x_1)$ and $\psi_{\nu_2}(x_2) = \Phi
(|\psi_{\nu_2}\rangle;\,x_2)$ are said to be \textsc{mutually local}
if 
\begin{equation}
  \label{eq:local}
  \Phi(|\psi_{\nu_1}\rangle;\,x_1)\,
\Phi(|\psi_{\nu_2}\rangle;\,x_2)
\,=\,
  \Phi(|\psi_{\nu_2}\rangle;\,x_2)\,
\Phi(|\psi_{\nu_1}\rangle;\,x_1), \qquad x_1 \,<\, x_2  
\end{equation}
where the last relation holds if included into an arbitrary
correlator function of bulk and boundary fields. Obviously, to make
sense it requires that an unique analytic continuation from  $x_1
\,<\, x_2$  to $x_2 \,<\, x_1$ does exist. 
\end{defin}

\begin{defin}
\label{def:selflocal}
\textbf{Self-local boundary fields}\\
A boundary field $\psi_{\nu}(x)$ is called \textsc{self-local} or
\textsc{analytic} if it is mutually local with respect to itself. 
\end{defin}

%%%%%%%%%%%%%%%%%%%%%%%%%%%%%%%%%%%%%%%%%%%%%%%%%%%%%%%%%%%%%% 
%%%%%%%%%%%%%%%%%%%%%%%%%%%%%%%%%%%%%%%%%%%%%%%%%%%%%%%%%%%%%% 

Let us notice that definition \ref{def:selflocal} constraints the OPE
of a self-local boundary field to contain only pole singularities. In
particular, the OPE of a self-local boundary field $\psi(x)$ with
$h_\psi=1$, is:
\begin{equation}
  \label{eq:opeh1}
  \psi(x_1)\,\psi(x_2) \,\sim\, \frac{K}{(x_1 - x_2)^2} 
\end{equation}
where $K$ is a real constant.

\subsection{The general prescription}

To describe truly marginal perturbations of a CFT in a general
contest, let us consider a Boundary Conformal Field Theory defined on
the Upper Half plane $\Im z \geq 0$.  Its state space will be labelled
by a couple of parameters $(\Omega,\alpha)$ where $\Omega$ identifies
the gluing map along the real line, while $\alpha$ is the given
boundary condition. We can use boundary operators
$\psi(x)\in\Psi(\mathcal{H})$ to define a new perturbed theory.
According to the general formula \eqref{eq:pert}, the deformed model
is obtained by adding to the action the operator:
\begin{equation}
\label{eq:pt}
  S'\,=\,g \int_{-\infty}^{+\infty} \frac{d x}{2\pi} \psi(x)
\end{equation}
Thus, perturbed correlators are obtained by the unperturbed ones by the
formal prescription:
\begin{multline}
\label{eq:def_pre}
  \langle 
  \varphi_1(z_1,\bz_1) \cdots \varphi_N(z_N,\bz_N) 
  \rangle_{\alpha; \lambda\varphi} \,=\,
  Z^{-1} \cdot   \langle 
  I_{\lambda\varphi} \varphi_1(z_1,\bz_1) \cdots \varphi_n(z_n,\bz_n) 
  \rangle_\alpha \\
  := Z^{-1}\sum_ng^n\idotsint_{x_i<x_{i+1}}\displaylimits
  \frac{d x_1}{2\pi} \cdots \frac{d x_n}{2\pi} 
  \langle 
  \psi(x_1) \cdots \psi(x_n)
  \varphi_1(z_1,\bz_1) \cdots \varphi_n(z_n,\bz_n)
  \rangle_\alpha\\
  =Z^{-1}\sum_n\frac{g^n}{n!}
  \sum_{\sigma \in S_n}
  \idotsint_{x_{\sigma(i)} <x_{\sigma(i+1)}}
  \frac{d x_1}{2\pi} \cdots \frac{d x_n}{2\pi} \\
\times
  \langle 
  \psi(x_1) \cdots \psi(x_n)
  \varphi_1(z_1,\bz_1) \cdots \varphi_n(z_n,\bz_n)
\rangle_\alpha
\end{multline}
where $g$ is the real coupling constant and the
second sum runs over all the elements of the permutation group $S_n$.
The operator $I_{g\varphi}$ has to be understood as the
path-ordered exponential of the perturbing term \eqref{eq:pt},
$I_{g\varphi} = \text{P} \exp{S'}$, and the normalization is
defined as the expectation value $Z = (A_0^\alpha)^{-1} \langle
I_{g\varphi} \rangle$. If other boundary fields are included in
the unperturbed correlator, prescription \eqref{eq:def_pre} has to be
modified in order to include all boundary fields in the path ordering.

To make sense of \eqref{eq:def_pre} beyond the formal level, integrals
need to be regularized, introducing suitable IR and UV cutoffs, while
coupling and field have to be renormalized. In the following, we will
not encounter IR-divergencies (which are usually cured by putting the
system in a finite box, \ie{} by considering finite-temperature
correlators). On the other hand, we will deal with UV-divergent
integrals. Thus, let us regularize them introducing a UV-cutoff
$\epsilon$, to let integrals domains be restricted to regions $|x_i -
x_j| > \epsilon$ and to let integrals became UV-finite before taking
the limit $\epsilon \rightarrow 0$.

In the following, we will analyze deformations generated by truly
marginal operators, \ie{} operators with $h = 1$. At RG fix-points,
al CFT deformations share the fact that local properties of the bulk
theory are not affected by the boundary condensate. Thus, boundary
deformations actually will only involve changes in boundary
conditions.

Let us start discussing the change induced by a boundary field
$\psi(x)$ on the two point function of the boundary field itself
$\langle \psi(x) \psi(x) \rangle_\alpha$. First order contributions
will involve the following integrals:
\begin{multline}
  \label{eq:bbint}
  \int_{- \infty}^{x_1 - \epsilon} 
  d x \,
  \langle \psi(x) \psi(x_1) \psi(x_2) \rangle_\alpha
  \,+\, \\
  \int_{x_1 + \epsilon}^{x_2 - \epsilon} 
  d x \,
  \langle \psi(x_1) \psi(x) \psi(x_2) \rangle_\alpha
  \,+\, 
  \int^{\infty}_{x_2 + \epsilon} 
  d x \,
  \langle \psi(x_1) \psi(x_2) \psi(x) \rangle_\alpha
\end{multline}

From the general form for a  three points function we get 
\begin{equation}
  \label{eq:3pgen}
  \langle \psi(x_1) \psi(x_2) \psi(x_3) \rangle_\alpha
  \,=\,
  \frac{
    C^\alpha_{\psi \psi \psi}
  }{
    (x_1 - x_2)\,(x_1 - x_3)\,(x_2 - x_3)}
\end{equation}
we easily notice that the first order contribution to the perturbed two
points function logarithmically diverges unless the structure
constants $C_{\psi \psi \psi}^\alpha$ would vanish. This divergence
would let the conformal weight $h_\psi$ flow away from $h_\psi = 1$,
thus a boundary field $\psi$ is not truly marginal unless $C_{\psi
  \psi \psi}^\alpha$. Moreover, if there are more boundary fields in a
BCFT, then a marginal field $\psi$ is truly marginal only if $C_{\psi
  \psi \psi'}^\alpha = 0$ for all marginal boundary fields $\{\psi'\}$
in the theory.  Equation \eqref{eq:opeh1}, together with
\eqref{eq:bf_1p}, shows that self-local operators satisfy this
(necessary but not sufficient - as a matter of fact true marginality
is not guaranteed at higher orders of the perturbative expansion) first
order condition.

\subsection{Truly marginal operators: the case of chiral fields}
The discussion can be pursued further: as a matter of fact, it is
possible to prove that \textsc{every self-local marginal boundary
  operator is indeed truly marginal to each order and it generates a
  deformation of a BCFT}. To this end, let us consider a perturbing
self-local marginal field $\psi(x)$ and let us rewrite the regularized
deformed correlators of bulk fields as:
\begin{multline}
  \label{eq:defc2}
  \langle 
  \varphi_1(z_1,\bz_1) \cdots \varphi_N(z_N,\bz_N) 
  \rangle_{\alpha; g\varphi}^\epsilon \,=\,\\
  Z^{-1}\sum_n\frac{g^n}{n!}
  \int_{-\infty}^{+\infty}
  \cdots
  \int_{-\infty}^{+\infty}
  \frac{d x_1}{2\pi} \cdots \frac{d x_n}{2\pi} 
  \langle 
  \psi(x_1) \cdots \psi(x_n)
  \varphi_1 \cdots \varphi_N
  \rangle_\alpha
\end{multline}
where the integrations are performed on the real axes with the UV-cutoff
described above. Including \eqref{eq:opeh1} in \eqref{eq:defc2}, we
can show that logarithmic divergences cancel out and the $\epsilon
\rightarrow 0$ limit is well defined. Due to self-locality, the final
result of the limit procedure is:
\begin{multline}
  \label{eq:defc3}
  \langle 
  \varphi_1(z_1,\bz_1) \cdots \varphi_N(z_N,\bz_N) 
  \rangle_{\alpha; g\varphi}\,=\,
  \lim_{\epsilon \rightarrow 0}
  \langle 
  \varphi_1(z_1,\bz_1) \cdots \varphi_N(z_N,\bz_N) 
  \rangle_{\alpha; g\varphi}^\epsilon \,=\,\\
  \sum_n\frac{g^n}{n!}
  \int_{\gamma_1}
  \cdots
  \int_{\gamma_n}
  \frac{d x_1}{2\pi} \cdots \frac{d x_n}{2\pi} 
  \langle 
  \psi(x_1) \cdots \psi(x_n)
  \varphi_1 \cdots \varphi_N
  \rangle_\alpha,
\end{multline}
where the various $\gamma_i$ are straight lines parallel to the real
axis parametrized as $\Im{z} = i \frac{\epsilon}{i}$. The rhs of
equation \eqref{eq:defc3} is finite and independent from $\epsilon$ as
long as $\epsilon < \Im{z_i}\,\forall i = 1, \ldots, N$, where
$\{z_i\}$ are the bulk fields insertion points. Thus, formula
\eqref{eq:defc3} allows to construct the perturbed bulk fields
correlators to all orders in perturbation theory.

The extension of the above formula to mixed correlator, containing
both bulk and boundary fields, is not straightforward. As a matter of
fact \eqref{eq:defc3} can be extended to:
\begin{multline}
  \label{eq:defbb}
  \langle 
  \psi_1 (u_1) \cdots \psi_M (u_M)\, 
  \varphi_1(z_1,\bz_1) \cdots \varphi_N(z_N,\bz_N) 
  \rangle_{\alpha; g\varphi}\,=\,\\
  \lim_{\epsilon \rightarrow 0}
  \langle 
  \varphi_1(z_1,\bz_1) \cdots \varphi_N(z_N,\bz_N) 
  \rangle_{\alpha; g\varphi}^\epsilon \,=\,\\
  Z^{-1}\sum_n\frac{g^n}{n!}
  \int_{-\infty}^{+\infty}
  \cdots
  \int_{-\infty}^{+\infty}
  \frac{d x_1}{2\pi} \cdots \frac{d x_n}{2\pi} 
  \langle  
  \psi(x_1) \cdots \psi(x_n)
  \psi_1 \cdots \psi_M\, 
  \varphi_1 \cdots \varphi_N
  \rangle_\alpha,
\end{multline}
if and only if the boundary fields are local with respect to the
perturbing field. Even in those cases, integrals on the rhs of
\eqref{eq:defbb} diverges when $\epsilon \rightarrow 0$ whenever the
iterated OPE of the perturbing field with one of the boundary fields
contains an even pole. Renormalization of such divergencies is
worked out introducing renormalized boundary operators:
\begin{equation}
  \label{eq:renbo}
  \tilde{\psi_i}
  \,=\,
  \left[
    e^{\frac{1}{2} g \psi}\,
    \psi_i
  \right](u_i)
  \,\doteq\,
  \sum_{n=0}^\infty
  \frac{g^n}{2^n n!} 
  \oint_{C_1}
  \frac{d x_1}{2 \pi}
  \oint_{C_n}
  \frac{d x_n}{2 \pi}\,
  \psi_i(u_i)\,
\psi(x_1) \cdots \psi(x_n)
\end{equation}
and contour integration (see \cite{Recknagel:1998ih} and references
therein), so that the perturbed correlator becomes:
\begin{multline}
  \label{eq:defbbren}
  \langle 
  \psi_1 (u_1) \cdots \psi_M (u_M)\, 
  \varphi_1(z_1,\bz_1) \cdots \varphi_N(z_N,\bz_N) 
  \rangle_{\alpha; g\varphi}\,=\,\\
  \lim_{\epsilon \rightarrow 0}
  \langle 
  \varphi_1(z_1,\bz_1) \cdots \varphi_N(z_N,\bz_N) 
  \rangle_{\alpha; g\varphi}^\epsilon \,=\,\\
  \sum_n\frac{g^n}{n!}
  \int_{\gamma_1}
  \cdots
  \int_{\gamma_n}
  \frac{d x_1}{2\pi} \cdots \frac{d x_n}{2\pi} 
  \langle  
  \psi(x_1) \cdots \psi(x_n)
  \tilde{\psi}_1 \cdots \tilde{\psi}_M\, 
  \varphi_1 \cdots \varphi_N
  \rangle_\alpha,
\end{multline}
where the $C_\mu$ are small circles surrounding the $\psi_i$
insertion point. 

Looking at \eqref{eq:renbo} the functional form of boundary fields is
left unchanged by the renormalization procedure, since integrals on
the rhs of \eqref{eq:renbo} pick up only simple poles. Thus, the
fields $\psi$ and $\tilde{\psi}$ have the same conformal dimension,
and $\tilde{\psi}$ can be regarded as the \textsc{rotation of $\psi_i$
  generated by the perturbing field $\psi$}.

Let us analyze deformations of $n$-point functions of the perturbing
field itself. Integrals on the rhs of \eqref{eq:renbo} vanish, since
OPE \eqref{eq:opeh1} does not contain simple poles. Thus,
$\tilde{\psi}$ and $\psi$ coincide, and equation \eqref{eq:opeh1}
forces all contour integrals in \eqref{eq:defbbren} to vanish if there
are no bulk fields inserted on the UHP. Hence, any perturbative
correction to the $n$-point function of $\psi$ vanish, \ie{} each
marginal boundary operator is truly marginal to all orders of the
perturbed theory.

Let us conclude this introductory section restricting our analysis to
boundary condensates made up by elements of the chiral a algebra, \ie{}
fields associated to elements of ${\mathcal{H}^{u(1)}}^0_1$ (see
comments after formula \eqref{eq:gen_vertex}). These marginal fields are local
with respect all bulk and boundary fields, thus formula
\eqref{eq:defbb} can be directly applied to compute deformations of arbitrary
bulk and boundary fields correlators. 

As remarked above, truly marginal perturbations actually generates
only changes in boundary conditions. Thus, let us analyze effects of
switching on a chiral boundary condensate on the gluing map $\Omega$
associate to a given BCFT. To this end, let us insert $W(z + 2 i
\delta) - \Omega \bar{W} (\bar{z} - 2 i \delta)$, with $z = \bar{z}$,
in an arbitrary unperturbed correlator of bulk fields. Then taking the
the $\delta \rightarrow 0^+$ limit, the chiral currents move towards
the boundary and the correlator vanishes in the limiting case. To
study deformations of the gluing map, let us perturb the above
relation introducing operator $I_{g,J(x)} \,=\, \text{P}[e^{-
  S_J'}]$:
\begin{multline}
  \label{eq:pertgl}
  0 \,=\,
  lim_{\delta \rightarrow 0^+}
  \text{P}[e^{- S_J'}] W(z + 2 i \delta) 
  \,-\, 
  \Omega \bar{W} (\bar{z} - 2 i \delta) 
  \,=\,\\
  lim_{\delta \rightarrow 0^+}
  \sum_n\frac{g^n}{n!}
  \int_{\gamma_1}
  \cdots
  \int_{\gamma_n}
  \frac{d x_1}{2\pi} \cdots \frac{d x_n}{2\pi} 
  J(x_1) \cdots J(x_n)
  W(z + 2 i \delta) 
  \,-\, 
  \Omega \bar{W} (\bar{z} - 2 i \delta) 
\end{multline}

Describing fields $W$ and $\Omega W$ in terms of the corresponding
states $\vert w \rangle,\,\vert \Omega w \rangle \in \mathcal
{H}^{(H)}_0$, we get from the above integrals:
\begin{align}
  \label{eq:pertgl2}
  0  & \,=\, 
  lim_{\delta \rightarrow 0^+}
  \sum_n\frac{g^n}{n!}
  \Phi( J_0^n \vert w \rangle \otimes \vert 0 \rangle;\, 
  z_\delta,\,\bar{z}_\delta)
  \,-\, 
  \Phi( \vert 0 \rangle \otimes \vert \Omega w \rangle ;\, 
  z_\delta,\,\bar{z}_\delta) \notag \\
  & \,=\,
  e^{i g J_0}\,W(z)\,e^{- i g J^0} \,-\, \Omega \bar{W}(\bar(z))
\end{align}
where $J_0$ is the zero mode of the UHP chiral current (more details
about this computation can be found in \cite{Recknagel:1998ih}). Thus,
conjugation with $e^{i g J_0}$ induces an inner automorphism on
the chiral algebra $\mathcal{W}$ defined by:
\begin{equation}
  \label{eq:aut}
  \gamma_J(W) \,\doteq\, e^{- i g J_0}\,W \,e^{i g J_0}
  \qquad 
  \forall\,W\,\in\,\mathcal{W} 
\end{equation}
Replacing \eqref{eq:aut} in the second line of \eqref{eq:pertgl2} we
get the perturbed gluing condition under the action of a chiral
marginal perturbation:
\begin{equation}
  \label{eq:pertglfin}
  W(z) \,=\,
\Omega \circ \gamma_{\bar{J}} (\bar{W})(\bar{z}) 
\end{equation}

Since $\gamma_J$ acts trivially on the Virasoro field, the gluing
condition $T = \bar{T}$ and those of all other generators which
commute with $J_0$ remain unchanged under the chiral
perturbation. Thus, these fields generate the same Ward identities ad
in the unperturbed theory.

Applying the construction outlined after formula \eqref{eq:cont_cond},
we can rephrase the result in equation \eqref{eq:pertglfin} in the
boundary state formalism. The chiral perturbed version of formula
\eqref{eq:cont_bs} turns out to be:
\begin{equation}
  \label{eq:def_bs}
  W_n \,-\,(-1)^{\bar{h}_{\overline{W}}} \Omega \circ \gamma_J
  (\overline{W}_{-n}) \Vert \Omega,\,\alpha\Rangle_{g,J},
\end{equation}
where the perturbed boundary state is related to the unperturbed one
via a simple rotation:
\begin{equation}
  \label{eq:bsrot}
  \Vert \Omega,\,\alpha\Rangle_{g,J} 
  \,=\,
  e^{i g J_0} \Vert \Omega,\,\alpha\Rangle
\end{equation}

%% file: formulae.tex
\chapter{Useful formulae}

This section contains a collection of useful equations and formulae.

\section{Simplicial String Duality}
\begin{itemize}
\item {\bf Direct products}\\
  If a group G is the direct product of groups, $G = G_1 \times G_2$,
  then, given any two elements $g_1 \in G_1$ and $g_2 \in G_2$ it
  holds:
  \begin{equation}
\label{eq:dir_prod}
    D^{j_1 \times j_2}_{m_1 n_1; m_2 n_2} (g_1 g_2) \,=\,
    D^{(1)j_1}_{m_1 n_1}(g_1)\,
    D^{(2)j_2}_{m_2 n_2}(g_2)
  \end{equation}
\item {\bf Clebsh-Gordan expansion}\\
  The product of two Wigner functions with the same argument can be
  expanded in the Clebsh-Gordan series:
  \begin{multline}
\label{eq:CG_series}
    D^{j_1}_{m_1\,n_1}(\Gamma) \,  D^{j_2}_{m_2\,n_2}(\Gamma) \,=\, \\
    \sum_{J = |j_1 - j_2|}^{j_1 + j_2} \sum_{|M|,|N| \leq J}
    C^{J\,M}_{j_1\, m_1\, j_2\, m_2\,}\,
    D^{J}_{M\,N}(\Gamma)\,
    C^{J\,N}_{j_1\, n_1\, j_2\, n_2\,}\,
  \end{multline}
\item {\bf Unitarity relations of Clebsh-Gordan coefficients}\\
  \begin{subequations}
    \label{eq:CG_unit}
    \begin{gather}
      \label{eq:CG_unit1}
      \sum_{m_1\,m_2} 
      C^{j\,m}_{j_1\,m_1\,j_2\,m_2}
      C^{j'\,m'}_{j_1\,m_1\,j_2\,m_2}
      \,=\,
      \delta_{j\,j'}\delta_{m\,m'} \\
      \label{eq:CG_unit2}
      \sum_{j\,m} 
      C^{j\,m}_{j_1\,m_1\,j_2\,m_2}
      C^{j\,m}_{j_1\,m_1'\,j_2\,m_2'}
      \,=\,
      \delta_{m_1\,m_1'}\delta_{m_2\,m_2'} 
    \end{gather}
  \end{subequations}
\end{itemize}
\section{Cosmological solution of supergravity}

\subsubsection{Bosonic Field Equations of type IIB supergravity}

The bosonic part of the equations can be formally obtained through
variation of the following action \footnote{Note that our $R$ is
  equal to $- \ft 1 2 R^{old}$, $R^{old}$ being the normalization of
  the scalar curvature usually adopted in General Relativity
  textbooks. The difference arises because in the traditional
  literature the Riemann tensor is not defined as the components of
  the curvature $2$-form $R^{ab}$ rather as $-2$ times such
  components.}:
\begin{multline}
  \label{bulkaction}
  S_{IIB} \,=\, \frac{1}{2 \kappa^2}
  \int d^{10} x
  \left[
    -2\,\sqrt{-\det g}\,R 
  \right] \\
  -\frac{1}{4 \kappa^2}
  \int
  d\varphi\,\wedge\,\star d \varphi \,+\,
  e^{- \varphi} F_{[3]}^{NS}  \wedge \star F_{[3]}^{NS} \,+\,
  e^{2 \varphi}\, F_{[1]}^{RR} \wedge \star F_{[1]}^{RR} \\\,+\,
  e^\varphi \,{F}_{[3]}^{RR} \wedge \star {F}_{[3]}^{RR} 
  \,+\, \frac{1}{2}\, {F}_{[5]}^{RR} \wedge \star {F}_{[5]}^{RR} \,-\,
  C_{[4]} \wedge F_{[3]}^{NS} \wedge F_{[3]}^{RR}
\end{multline}
where:
\begin{subequations}
  \label{bosecurve}
  \begin{align}
    F^{RR}_{[1]}& \,=\, dC_{[0]} \\
    F^{NS}_{[3]}& \,=\, dB_{[2]} \\
    F^{RR}_{[3]}& \,=\, dC_{[2]} -  \, C_{[0]} \,
    dB_{[2]} \\
    F^{RR}_{[5]}& \,=\, dC_{[4]}-  \ft 12 \,\left( B_{[2]} \wedge d C_{[2]}
      -  C_{[2]} \wedge d B_{[2]}\right) 
  \end{align}
\end{subequations}

It is important to stress though that the action \eqref{bulkaction} is
to be considered only a book keeping device since the $4$-form
$C_{[4]}$ is not free, its field strength $F_{[5]}^{RR}$ being subject
to the on-shell self-duality constraint:
\begin{equation}
  \label{selfonshell}
  F_{[5]}^{RR} = \star F_{[5]}^{RR}
\end{equation}
From the above action the corresponding equations of motion can be
obtained:
\begin{subequations}
  \begin{gather}
    d \star d \varphi 
    \,-\, 
    e^{2\varphi} \, 
    F^{RR}_{[1]} 
    \,\wedge\, 
    \star F^{RR}_{[1]}
    \,=\, -\frac{1}{2}\, 
    \left(
      e^{-\varphi} 
      F^{NS}_{[3]}\,\wedge\,\star F^{NS}_{[3]}
      \,-\,
      e^\varphi F^{RR}_{[3]}\,\wedge\,\star F^{RR}_{[3]}
    \right) 
    \label{NSscalapr} \\
    d \left( 
      e^{-\varphi}\,\star F_{[3]}^{NS}
    \right) 
    \,+\, 
    e^\varphi\,
    F^{RR}_{[1]}\,\wedge\,\star F^{RR}_{[3]}
    \,=\,
    -\,F_{[3]}^{RR}\,\wedge\,F^{RR}_{[5]}
    \label{3formNS}\\
    d
    \left( 
      e^{2\varphi} \star F^{RR}_{[1]}
    \right)
    \,=\, 
    - e^\varphi\,
    F^{NS}_{[3]}\,\wedge\,\star F^{RR}_{[3]}
    \label{RRscalapr} \\
    d
    \left( 
      e^\varphi 
      \star F_{[3]}^{RR} 
    \right) 
    \,=\,
    -\,F_{[5]}^{RR}\,\wedge\,F_{[3]}^{NS}
    \label{3formRR}\\
    d
    \star F^{RR}_{[5]} \,=\, 
-\,F^{NS}_{[3]} \,\wedge\,F^{RR}_{[3]}
    \label{f5RR}\\
-\,2 \,R_{{MN}} \,=\, 
\frac{1}{2}\partial_{{M}}\varphi
\partial_{{N}}\varphi+\frac{e^{2\varphi}}{2}
\partial_{{M}} C_{[0]} \partial_{{N}}
C_{[0]}+150
 {F}_{[5]{M}\cdot\cdot\cdot\cdot}
{F}_{[5]{N}}^{\phantom{{M}}\cdot\cdot\cdot\cdot}
\notag\\
+ 9 \left( e^{-\varphi}F_{[3]{M}\cdot\cdot}^{NS}\,
F_{[3]{N}}^{{NS}\phantom{{M}}\cdot\cdot} +e^{\varphi}{
F}_{[3]{M}\cdot\cdot}^{RR}
{ F}_{[3]{N}}^{RR\phantom{{M}}\cdot\cdot}\right) 
\notag\\
-\frac{3}{4}\,
g_{{MN}}\,\left(e^{-\varphi}F_{[3]\cdot\cdot\cdot}^{NS}
F_{[3]}^{NS\cdot\cdot\cdot}+e^{\varphi}{{F}}_{[3]\cdot\cdot\cdot}^{RR}{
F}^{RR\cdot\cdot\cdot}_{[3]}\right) \label{einsteinequa}
\end{gather}
\end{subequations}

It is not difficult to show, upon suitable identification of the
massless superstring fields, that this is the correct set of
equations which can be consistently obtained from the manifestly
$\mathrm{SU(1,1)}$ covariant formulation of type IIB supergravity
\cite{Schwarz:1983qr,Howe:1983sr,Castellani:1993ye}.

\subsubsection{A useful integral}
\begin{multline}
  \int \, e^{[ a\,x ]} \, \left ( \cosh[b \, x] \right) ^{1/4} \,=\, \\
  \frac{2^{7/4}}{4a-b}\, e^{[(a-b)x]} \, \,  _2 F_1\left( -\ft 18 + \ft
  {a}{2b} , - \ft 14, \ft 78 + \ft {a}{2b} \,; \, - e^{2bx} \right)
\label{star}
\end{multline}

%% file: roots.tex
\chapter{Listing of the $E_8$ positive roots}

\section{Listing according to height}

In this listing we present the roots of the $E_8$ Lie
algebra, giving their definition both in terms of the simple roots and
in the eucledian basis. The notation $a_{i|j}$ is introduced to denote
the height of the root (i).  The number (j) is introduced to
distinguish the roots of the same height

\begin{longtable}{|cp{2mm}cp{2mm}cp{2mm}p{41mm}|}
  \caption{Listing of positive $E_8$ rotts according to height}
  \label{tab:root_height}\\
  \hline\hline
  \text{label} & & \text{root} & & \text{Dynkin labels}  & &
  \text{Euclidian basis labels}\\
  \hline
  \endfirsthead
  
  \multicolumn{7}{c}%
  {{\bfseries \tablename\ \thetable{} -- continued from previous page}} \\
  \hline
  \text{label} & & \text{root} & & \text{Dynkin labels}  & &
  \text{Euclidian basis labels}\\
  \hline
  \endhead  
  
  \hline \multicolumn{7}{|r|}{{Continued on next page}} \\ \hline
  \endfoot
  
  \hline \hline
  \endlastfoot
  
  ${a_{1|1}}$ & = &${\alpha [} 1 ]$ & = & \{1,0,0,0,0,0,0,0\} & =
  &\{0,1,-1,0,0,0,0,0\} \\ ${a_{1|2}}$ & = &${\alpha [} 2 ]$& = &
  \{0,1,0,0,0,0,0,0\} & = &\{0,0,1,-1,0,0,0,0\} \\ ${a_{1|3}}$ & =
  &${\alpha [} 3 ]$& = & \{0,0,1,0,0,0,0,0\} & = &\{
  0,0,0,1,-1,0,0,0\} \\ ${a_{1|4}}$ & = &${\alpha [} 4 ]$& = &
  \{0,0,0,1,0,0,0,0\} & = &\{ 0,0,0,0,1,-1,0,0\} \\ ${a_{1|5}}$ & =
  &${\alpha [} 5 ]$& = & \{0,0,0,0,1,0,0,0\} & = &\{0,0,0,0,0,1,-1,0\}
  \\ ${a_{1|6}}$ & = & ${\alpha [} 6 ]$& = & \{0,0,0,0,0,1,0,0\} & =
  &\{ 0,0,0,0,0,1,1,0\} \\ ${a_{1|7}}$ & = &${\alpha [} 7 ]$& = &
  \{0,0,0,0,0,0,1,0\} & = & \{{$\scriptstyle
    -\frac{1}{2},-\frac{1}{2},-\frac{1}{2},-\frac{1}{2},
    -\frac{1}{2},-\frac{1}{2},-\frac{1}{2},-\frac{1}{2}$}\} \\
  ${a_{1|8}}$ & = &${\alpha [} 8 ]$& = & \{ 0,0,0,0,0,0,0,1 \} & =
  &\big\{ 1,-1,0,0,0,0,0,0\big\}\ \\
  ${a_{2|1}}$ & = &${\alpha [} 9 ]$& = & \{1,1,0,0,0,0,0,0\} & = &\{
  0,1,0,-1,0,0,0,0\} \\ ${a_{2|2}}$ & = &${\alpha [} 10 ]$& = &
  \{1,0,0,0,0,0,0,1\} & = &\big\{1,0,-1,0,0,0,0,0\big\} \\ ${a_{2|3}}$
  & = &${\alpha [} 11 ]$& = & \{0,1,1,0,0,0,0,0\} & =
  &\{0,0,1,0,-1,0,0,0\} \\ ${a_{2|4}}$ & = &${\alpha [} 12 ]$& = &
  \{0,0,1,1,0,0,0,0\} & = &\{ 0,0,0,1,0,-1,0,0\} \\ ${a_{2|5}}$ & =
  &${\alpha [} 13 ]$& = & \{0,0,0,1,1,0,0,0\} & =
  &\{0,0,0,0,1,0,-1,0\} \\ ${a_{2|6}}$ & = &${\alpha [} 14 ]$& = &
  \{0,0,0,1,0,1,0,0\} & = &\{ 0,0,0,0,1,0,1,0\} \\ ${a_{2|7}}$ & =
  &${\alpha [} 15 ]$& = & \{0,0,0,0,0,1,1,0\} & =
  &\{{$\scriptstyle-\frac{1}{2} , -\frac{1}{2} , -\frac{1}{2} ,
    -\frac{1}{2} , -\frac{1}{2} ,\frac{1}{2},\frac{1}{2},
    -\frac{1}{2}$}\} \  \\
  ${a_{3|1}}$ & = &${\alpha [} 16 ]$& = & \{1,1,1,0,0,0,0,0\} & =
  &\{0,1,0,0,-1,0,0,0\} \\ ${a_{3|2}}$ & = &${\alpha [} 17 ]$& = &
  \{1,1,0,0,0,0,0,1\} & = &\big\{1,0,0,-1,0,0,0,0\big\} \\ ${a_{3|3}}$
  & = &${\alpha [} 18 ]$& = & \{0,1,1,1,0,0,0,0\} & = &\{
  0,0,1,0,0,-1,0,0\} \\ ${a_{3|4}}$ & = &${\alpha [} 19 ]$& = &
  \{0,0,1,1,1,0,0,0\} & = &\{ 0,0,0,1,0,0,-1,0\} \\ ${a_{3|5}}$ & =
  &${\alpha [} 20 ]$& = & \{0,0,1,1,0,1,0,0\} & = &\{
  0,0,0,1,0,0,1,0\} \\ ${a_{3|6}}$ & = &${\alpha [} 21 ]$& = &
  \{0,0,0,1,0,1,1,0\} & = &\{{$\scriptstyle -\frac{1}{2} ,
    -\frac{1}{2} , -\frac{1}{2} , -\frac{1}{2} ,\frac{1}{2},
    -\frac{1}{2} ,\frac{1}{2}, -\frac{1}{2}$}\} \\ ${a_{3|7}}$ & =
  &${\alpha [} 22 ]$& = &
  \{0,0,0,1,1,1,0,0\} & = &\{0,0,0,0,1,1,0,0\} \\
 
  ${a_{4|1}}$ & = &${\alpha [} 23 ]$& = & \{1,1,1,1,0,0,0,0\} & =
  &\{0,1,0,0,0,-1,0,0\} \\ ${a_{4|2}}$ & = &${\alpha [} 24 ]$& = &
  \{0,1,1,1,1,0,0,0\} & = &\{0,0,1,0,0,0,-1,0\} \\ ${a_{4|3}}$ & =
  &${\alpha [} 25 ]$& = & \{0,1,1,1,0,1,0,0\} & = &\{
  0,0,1,0,0,0,1,0\} \\ ${a_{4|4}}$ & = &${\alpha [} 26 ]$& = &
  \{1,1,1,0,0,0,0,1\} & = &\{{1,0,0,0,-1,0,0,0}\}\\
  ${a_{4|5}}$ & = &${\alpha [} 27 ]$& = & \{0,0,1,1,0,1,1,0\} & =
  &\{{$\scriptstyle-\frac{1}{2},-\frac{1}{2},-\frac{1}{2}
    ,\frac{1}{2},-\frac{1}{2},-\frac{1}{2},\frac{1}{2},-\frac{1}{2}$}\}\\
  ${a_{4|6}}$ & = &${\alpha [} 28 ]$& = &
  \{0,0,1,1,1,1,0,0\} & = &\{0,0,0,1,0,1,0,0\}\\
  ${a_{4|7}}$ & = &${\alpha [} 29 ]$& = & \{0,0,0,1,1,1,1,0\} & =
  &\{{$\scriptstyle-\frac{1}{2} , -\frac{1}{2} , -\frac{1}{2} ,
    -\frac{1}{2} ,\frac{1}{2},\frac{1}{2}, -\frac{1}{2} ,
    -\frac{1}{2}$}\}
  \\
  ${a_{5|1}}$ & = &${\alpha [} 30 ]$& = & \{1,1,1,1,1,0,0,0\} & = &\{
  0,1,0,0,0,0,-1,0\} \\ ${a_{5|2}}$ & = &${\alpha [} 31 ]$& = &
  \{1,1,1,1,0,1,0,0\} & = &\{0,1,0,0,0,0,1,0\} \\ ${a_{5|3}}$ & =
  &${\alpha [} 32 ]$& = & \{0,1,1,1,0,1,1,0\} & =
  &\{{$\scriptstyle-\frac{1}{2} , -\frac{1}{2} ,\frac{1}{2},
    -\frac{1}{2} , -\frac{1}{2} , -\frac{1}{2} ,\frac{1}{2},
    -\frac{1}{2}$ }\} \\ ${a_{5|4}}$ & = &${\alpha [} 33 ]$& = &
  \{0,1,1,1,1,1,0,0\} & = &\{ 0,0,1,0,0,1,0,0\}
  \\
  ${a_{5|5}}$ & = &${\alpha [} 34 ]$& = & \{0,0,1,1,1,1,1,0\} & =
  &\{{$\scriptstyle-\frac{1}{2} , -\frac{1}{2} , -\frac{1}{2}
    ,\frac{1}{2}, -\frac{1}{2} ,\frac{1}{2}, -\frac{1}{2} ,
    -\frac{1}{2}$}\} \\ ${a_{5|6}}$ & = &${\alpha [} 35 ]$& = &
  \{1,1,1,1,0,0,0,1\} & = &\big\{1,0,0,0,0,-1,0,0\big\} \\ ${a_{5|7}}$
  & = &${\alpha [} 36 ]$& = &
  \{0,0,1,2,1,1,0,0\} & = &\{0,0,0,1,1,0,0,0\}  \\
  ${a_{6|1}}$ & = &${\alpha [} 37 ]$& = & \{1,1,1,1,0,1,1,0\} & =
  &\{{$\scriptstyle-\frac{1}{2} ,\frac{1}{2}, -\frac{1}{2} ,
    -\frac{1}{2} , -\frac{1}{2} , -\frac{1}{2} ,\frac{1}{2},
    -\frac{1}{2}$}\} \\ ${a_{6|2}}$ & = &${\alpha [} 38 ]$& = &
  \{1,1,1,1,1,1,0,0\} & = &\{0,1,0,0,0,1,0,0\} \\ ${a_{6|3}}$ & =
  &${\alpha [} 39 ]$& = & \{0,1,1,1,1,1,1,0\} & =
  &\{{$\scriptstyle-\frac{1}{2},-\frac{1}{2},\frac{1}{2},
    -\frac{1}{2},-\frac{1}{2},\frac{1}{2},-\frac{1}{2},-\frac{1}{2}$}\}\\ 
  ${a_{6|4}}$ & = &${\alpha [} 40 ]$& = &
  \{0,1,1,2,1,1,0,0\} & = &\{ 0,0,1,0,1,0,0,0\} \\ ${a_{6|5}}$ & =
  &${\alpha [} 41 ]$& = & \{0,0,1,2,1,1,1,0\} & =
  &\{{$\scriptstyle-\frac{1}{2} , -\frac{1}{2} , -\frac{1}{2}
    ,\frac{1}{2},\frac{1}{2}, -\frac{1}{2} , -\frac{1}{2} ,
    -\frac{1}{2}$}\} \\ ${a_{6|6}}$ & = &${\alpha [} 42 ]$& = &
  \{1,1,1,1,1,0,0,1\} & = &\big\{1,0,0,0,0,0,-1,0\big\} \\ ${a_{6|7}}$
  & = &${\alpha [} 43 ]$& = & \{1,1,1,1,0,1,0,1\}
  & = &\{{1,0,0,0,0,0,1,0}\} \\
  ${a_{7|1}}$ & = &${\alpha [} 44 ]$& = & \{1,1,1,1,1,1,1,0\} & =
  &\{{$\scriptstyle-\frac{1}{2} ,\frac{1}{2}, -\frac{1}{2} ,
    -\frac{1}{2} , -\frac{1}{2} ,\frac{1}{2}, -\frac{1}{2} ,
    -\frac{1}{2}$}\} \\ ${a_{7|2}}$ & = &${\alpha [} 45 ]$& = &
  \{1,1,1,2,1,1,0,0\} & = &\{ 0,1,0,0,1,0,0,0\} \\ ${a_{7|3}}$ & =
  &${\alpha [} 46 ]$& = & \{0,1,1,2,1,1,1,0\} & = &\{{$\scriptstyle
    -\frac{1}{2} , -\frac{1}{2} ,\frac{1}{2}, -\frac{1}{2}
    ,\frac{1}{2}, -\frac{1}{2} , -\frac{1}{2} , -\frac{1}{2}$}\} \\
  ${a_{7|4}}$ & = &${\alpha [} 47 ]$& = & \{0,1,2,2,1,1,0,0\} & =
  &\{0,0,1,1,0,0,0,0\} \\ ${a_{7|5}}$ & = &${\alpha [} 48 ]$& = &
  \{1,1,1,1,1,1,0,1\} & = &\big\{1,0,0,0,0,1,0,0\big\} \\ ${a_{7|6}}$
  & = &${\alpha [} 49 ]$& = & \{0,0,1,2,1,2,1,0\} & =
  &\{{$\scriptstyle-\frac{1}{2} , -\frac{1}{2} , -\frac{1}{2} ,
    \frac{1}{2},\frac{1}{2}, \frac{1}{2},\frac{1}{2}, -\frac{1}{2}$}\}
  \\ ${a_{7|7}}$ & = &${\alpha [} 50 ]$& = & \{1,1,1,1,0,1,1,1\} & =
  &\{{$\scriptstyle\frac{1}{2},-\frac{1}{2} , -\frac{1}{2} ,
    -\frac{1}{2} , -\frac{1}{2} , -\frac{1}{2} ,\frac{1}{2},
    -\frac{1}{2}$}\}
  \\
  ${a_{8|1}}$ & = &${\alpha [} 51 ]$& = & \{1,1,1,2,1,1,1,0\} & =
  &\{{$\scriptstyle-\frac{1}{2} ,\frac{1}{2}, -\frac{1}{2} ,
    -\frac{1}{2} ,\frac{1}{2}, -\frac{1}{2} , -\frac{1}{2} ,
    -\frac{1}{2}$}\} \\ ${a_{8|2}}$ & = &${\alpha [} 52 ]$& = &
  \{1,1,2,2,1,1,0,0\} & = &\{ 0,1,0,1,0,0,0,0\} \\ ${a_{8|3}}$ & =
  &${\alpha [} 53 ]$& = & \{0,1,1,2,1,2,1,0\} & =
  &\{{$\scriptstyle-\frac{1}{2} , -\frac{1}{2} ,\frac{1}{2},
    -\frac{1}{2} ,\frac{1}{2},\frac{1}{2}, \frac{1}{2},-\frac{1}{2}$
  }\} \\ ${a_{8|4}}$ & = &${\alpha [} 54 ]$& = & \{0,1,2,2,1,1,1,0\} &
  = &\{{$\scriptstyle-\frac{1}{2} , -\frac{1}{2}
    ,\frac{1}{2},\frac{1}{2}, -\frac{1}{2} , -\frac{1}{2} ,
    -\frac{1}{2} , -\frac{1}{2}$}\} \\ ${a_{8|5}}$ & = &${\alpha [} 55
  ]$& = & \{1,1,1,2,1,1,0,1\} & = &\big\{1,0,0,0,1,0,0,0\big\} \\
  ${a_{8|6}}$ & = &${\alpha [} 56 ]$& = & \{1,1,1,1,1,1,1,1\} & =
  &\{{$\scriptstyle\frac{1}{2},-\frac{1}{2} , -\frac{1}{2} ,
    -\frac{1}{2} , -\frac{1}{2} ,\frac{1}{2}, -\frac{1}{2} ,
    -\frac{1}{2}$}\}
  \\
  ${a_{9|1}}$ & = &${\alpha [} 57 ]$& = & \{1,1,1,2,1,2,1,0\} & =
  &\{{$\scriptstyle-\frac{1}{2} ,\frac{1}{2}, -\frac{1}{2} ,
    -\frac{1}{2} ,\frac{1}{2},\frac{1}{2},
    \frac{1}{2},-\frac{1}{2}$}\} \\ ${a_{9|2}}$ & = &${\alpha [} 58
  ]$& = & \{1,1,2,2,1,1,1,0\} & = &\{{$\scriptstyle-\frac{1}{2}
    ,\frac{1}{2}, -\frac{1}{2} ,\frac{1}{2}, -\frac{1}{2} ,
    -\frac{1}{2} , -\frac{1}{2} , -\frac{1}{2}$}\} \\ ${a_{9|3}}$ & =
  &${\alpha [} 59 ]$& = & \{1,2,2,2,1,1,0,0\} & = &\{0,1,1,0,0,0,0,0\}
  \\ ${a_{9|4}}$ & = &${\alpha [} 60 ]$& = & \{0,1,2,2,1,2,1,0\} & =
  &\{{$\scriptstyle-\frac{1}{2} , -\frac{1}{2}
    ,\frac{1}{2},\frac{1}{2}, -\frac{1}{2} ,\frac{1}{2},\frac{1}{2},
    -\frac{1}{2}$}\} \\ ${a_{9|5}}$ & = &${\alpha [} 61 ]$& = &
  \{1,1,2,2,1,1,0,1\} & = &\{{1,0,0,1,0,0,0,0}\} \\ ${a_{9|6}}$ & =
  &${\alpha [} 62 ]$& = & \{1,1,1,2,1,1,1,1\} & =
  &\{{$\scriptstyle\frac{1}{2},-\frac{1}{2} , -\frac{1}{2} ,
    -\frac{1}{2} ,\frac{1}{2}, -\frac{1}{2} , -\frac{1}{2} ,
    -\frac{1}{2}$}\} \\
  ${a_{10|1}}$ & = &${\alpha [} 63 ]$& = & \{1,1,2,2,1,2,1,0\} & =
  &\{{$\scriptstyle -\frac{1}{2} ,\frac{1}{2}, -\frac{1}{2}
    ,\frac{1}{2}, -\frac{1}{2} ,\frac{1}{2},\frac{1}{2},
    -\frac{1}{2}$}\} \\ ${a_{10|2}}$ & = &${\alpha [} 64 ]$& = &
  \{1,2,2,2,1,1,1,0\} & = &\{{$\scriptstyle-\frac{1}{2}
    ,\frac{1}{2},\frac{1}{2}, -\frac{1}{2} , -\frac{1}{2} ,
    -\frac{1}{2} , -\frac{1}{2} , -\frac{1}{2}$ }\} \\ ${a_{10|3}}$ &
  = &${\alpha [} 65 ]$& = & \{1,2,2,2,1,1,0,1\} & =
  &\big\{1,0,1,0,0,0,0,0\big\} \\ ${a_{10|4}}$ & = &${\alpha [} 66 ]$&
  = & \{1,1,2,2,1,1,1,1\} & =
  &\{{$\scriptstyle\frac{1}{2},-\frac{1}{2} , -\frac{1}{2}
    ,\frac{1}{2}, -\frac{1}{2} , -\frac{1}{2} , -\frac{1}{2} ,
    -\frac{1}{2}$}\} \\ ${a_{10|5}}$ & = &${\alpha [} 67 ]$& = &
  \{0,1,2,3,1,2,1,0\} & = &\{{$\scriptstyle-\frac{1}{2} , -\frac{1}{2}
    ,\frac{1}{2},\frac{1}{2}, \frac{1}{2},-\frac{1}{2} ,\frac{1}{2},
    -\frac{1}{2}$}\} \\ ${a_{10|6}}$ & = &${\alpha [} 68 ]$& = &
  \{1,1,1,2,1,2,1,1\} & = &\{{$\scriptstyle\frac{1}{2},-\frac{1}{2} ,
    -\frac{1}{2} , -\frac{1}{2} ,\frac{1}{2},\frac{1}{2},
    \frac{1}{2},-\frac{1}{2}$ }\}
  \\
  ${a_{11|1}}$ & = &${\alpha [} 69 ]$& = & \{2,2,2,2,1,1,0,1\} & =
  &\{{1,1,0,0,0,0,0,0}\} \\ ${a_{11|2}}$ & = &${\alpha [} 70 ]$& = &
  \{1,1,2,3,1,2,1,0\} & = &\{{$\scriptstyle-\frac{1}{2} ,\frac{1}{2},
    -\frac{1}{2} ,\frac{1}{2},\frac{1}{2}, -\frac{1}{2} ,\frac{1}{2},
    -\frac{1}{2}$}\} \\ ${a_{11|3}}$ & = &${\alpha [} 71 ]$& = &
  \{1,2,2,2,1,2,1,0\} & = &\{{$\scriptstyle-\frac{1}{2}
    ,\frac{1}{2},\frac{1}{2}, -\frac{1}{2} , -\frac{1}{2}
    ,\frac{1}{2},\frac{1}{2}, -\frac{1}{2}$}\} \\ ${a_{11|4}}$ & =
  &${\alpha [} 72 ]$& = & \{1,2,2,2,1,1,1,1\} & =
  &\{{$\scriptstyle\frac{1}{2},-\frac{1}{2} ,\frac{1}{2}, -\frac{1}{2}
    , -\frac{1}{2} , -\frac{1}{2} , -\frac{1}{2} , -\frac{1}{2}$}\} \\
  ${a_{11|5}}$ & = &${\alpha [} 73 ]$& = & \{1,1,2,2,1,2,1,1\} & =
  &\{{$\scriptstyle\frac{1}{2},-\frac{1}{2} , -\frac{1}{2}
    ,\frac{1}{2}, -\frac{1}{2} ,\frac{1}{2},\frac{1}{2},
    -\frac{1}{2}$}\} \\ ${a_{11|6}}$ & = &${\alpha [} 74 ]$& = &
  \{0,1,2,3,2,2,1,0\} & = &\{{$\scriptstyle-\frac{1}{2} , -\frac{1}{2}
    ,\frac{1}{2},\frac{1}{2}, \frac{1}{2},\frac{1}{2},-\frac{1}{2} ,
    -\frac{1}{2}$}\}\\
  ${a_{12|1}}$ & = &${\alpha [} 75 ]$& = & \{2,2,2,2,1,1,1,1\} & =
  &\{{$\scriptstyle\frac{1}{2},\frac{1}{2},-\frac{1}{2},-\frac{1}{2},
-\frac{1}{2},-\frac{1}{2},-\frac{1}{2},-\frac{1}{2}$}\}
  \\ ${a_{12|2}}$ & = &${\alpha [} 76 ]$& = & \{1,1,2,3,2,2,1,0\} & =
  &\{{$\scriptstyle-\frac{1}{2} ,\frac{1}{2}, -\frac{1}{2}
    ,\frac{1}{2},\frac{1}{2}, \frac{1}{2},-\frac{1}{2} ,
    -\frac{1}{2}$}\} \\ ${a_{12|3}}$ & = &${\alpha [} 77 ]$& = &
  \{1,2,2,3,1,2,1,0\} & = &\{{$\scriptstyle-\frac{1}{2}
    ,\frac{1}{2},\frac{1}{2}, -\frac{1}{2} ,\frac{1}{2}, -\frac{1}{2}
    ,\frac{1}{2}, -\frac{1}{2}$ }\} \\ ${a_{12|4}}$ & = &${\alpha [}
  78 ]$& = & \{1,2,2,2,1,2,1,1\} & =
  &\{{$\scriptstyle\frac{1}{2},-\frac{1}{2} ,\frac{1}{2}, -\frac{1}{2}
    , -\frac{1}{2} ,\frac{1}{2},\frac{1}{2}, -\frac{1}{2}$}\} \\
  ${a_{12|5}}$ & = &${\alpha [} 79 ]$& = & \{1,1,2,3,1,2,1,1\} & =
  &\{{$\scriptstyle\frac{1}{2},-\frac{1}{2} , -\frac{1}{2}
    ,\frac{1}{2},\frac{1}{2}, -\frac{1}{2} ,\frac{1}{2},
    -\frac{1}{2}$}\}\\
  ${a_{13|1}}$ & = &${\alpha [} 80 ]$& = & \{2,2,2,2,1,2,1,1\} & =
  &\{{$\scriptstyle\frac{1}{2},\frac{1}{2},-\frac{1}{2} , -\frac{1}{2}
    , -\frac{1}{2} ,\frac{1}{2},\frac{1}{2}, -\frac{1}{2}$ }\} \\
  ${a_{13|2}}$ & = &${\alpha [} 81 ]$& = & \{1,2,2,3,2,2,1,0\} & =
  &\{{$\scriptstyle-\frac{1}{2} ,\frac{1}{2},\frac{1}{2}, -\frac{1}{2}
    ,\frac{1}{2},\frac{1}{2}, -\frac{1}{2} , -\frac{1}{2}$}\} \\
  ${a_{13|3}}$ & = &${\alpha [} 82 ]$& = & \{1,2,2,3,1,2,1,1\} & =
  &\{{$\scriptstyle\frac{1}{2},-\frac{1}{2} ,\frac{1}{2}, -\frac{1}{2}
    ,\frac{1}{2}, -\frac{1}{2} ,\frac{1}{2}, -\frac{1}{2}$}\} \\
  ${a_{13|4}}$ & = &${\alpha [} 83 ]$& = & \{1,2,3,3,1,2,1,0\} & =
  &\{{$\scriptstyle-\frac{1}{2} ,\frac{1}{2},\frac{1}{2},
    \frac{1}{2},-\frac{1}{2} , -\frac{1}{2} ,\frac{1}{2},
    -\frac{1}{2}$}\} \\ ${a_{13|5}}$ & = &${\alpha [} 84 ]$& = &
  \{1,1,2,3,2,2,1,1\} & = &\{{$\scriptstyle\frac{1}{2},-\frac{1}{2} ,
    -\frac{1}{2} ,\frac{1}{2},\frac{1}{2}, \frac{1}{2},-\frac{1}{2} ,
    -\frac{1}{2}$}\}\\
  ${a_{14|1}}$ & = &${\alpha [} 85 ]$& = & \{2,2,2,3,1,2,1,1\} & =
  &\{{$\scriptstyle\frac{1}{2},\frac{1}{2},-\frac{1}{2} , -\frac{1}{2}
    ,\frac{1}{2}, -\frac{1}{2} ,\frac{1}{2}, -\frac{1}{2}$}\} \\
  ${a_{14|2}}$ & = &${\alpha [} 86 ]$& = & \{1,2,2,3,2,2,1,1\} & =
  &\{{$\scriptstyle\frac{1}{2},-\frac{1}{2} ,\frac{1}{2}, -\frac{1}{2}
    ,\frac{1}{2},\frac{1}{2}, -\frac{1}{2} , -\frac{1}{2}$}\} \\
  ${a_{14|3}}$ & = &${\alpha [} 87 ]$& = & \{1,2,3,3,2,2,1,0\} & =
  &\{{$\scriptstyle-\frac{1}{2} ,\frac{1}{2},\frac{1}{2},
    \frac{1}{2},-\frac{1}{2} ,\frac{1}{2}, -\frac{1}{2} ,
    -\frac{1}{2}$}\} \\ ${a_{14|4}}$ & = &${\alpha [} 88 ]$& = &
  \{1,2,3,3,1,2,1,1\} & = &\{{$\scriptstyle \frac{1}{2},-\frac{1}{2}
    ,\frac{1}{2}, \frac{1}{2},-\frac{1}{2} , -\frac{1}{2}
    ,\frac{1}{2}, -\frac{1}{2}$ }\}\\
  ${a_{15|1}}$ & = &${\alpha [} 89 ]$& = & \{2,2,2,3,2,2,1,1\} & =
  &\{{$\scriptstyle\frac{1}{2},\frac{1}{2},-\frac{1}{2} , -\frac{1}{2}
    ,\frac{1}{2},\frac{1}{2}, -\frac{1}{2} , -\frac{1}{2}$}\} \\
  ${a_{15|2}}$ & = &${\alpha [} 90 ]$& = & \{2,2,3,3,1,2,1,1\} & =
  &\{{$\scriptstyle\frac{1}{2},\frac{1}{2},-\frac{1}{2} ,
    \frac{1}{2},-\frac{1}{2} , -\frac{1}{2} ,\frac{1}{2},
    -\frac{1}{2}$}\} \\ ${a_{15|3}}$ & = &${\alpha [} 91 ]$& = &
  \{1,2,3,3,2,2,1,1\} & = &\{{$\scriptstyle\frac{1}{2},-\frac{1}{2}
    ,\frac{1}{2}, \frac{1}{2},-\frac{1}{2} ,\frac{1}{2}, -\frac{1}{2}
    , -\frac{1}{2}$}\} \\ ${a_{15|4}}$ & = &${\alpha [} 92 ]$& = &
  \{1,2,3,4,2,2,1,0\} & = &\{{$\scriptstyle-\frac{1}{2}
    ,\frac{1}{2},\frac{1}{2}, \frac{1}{2},\frac{1}{2},-\frac{1}{2} ,
    -\frac{1}{2} , -\frac{1}{2}$}\}  \\
  ${a_{16|1}}$ & = &${\alpha [} 93 ]$& = & \{2,2,3,3,2,2,1,1\} & =
  &\{{$\scriptstyle\frac{1}{2},\frac{1}{2},-\frac{1}{2} ,
    \frac{1}{2},-\frac{1}{2} ,\frac{1}{2}, -\frac{1}{2} ,
    -\frac{1}{2}$ }\} \\ ${a_{16|2}}$ & = &${\alpha [} 94 ]$& = &
  \{2,3,3,3,1,2,1,1\} & =
  &\{{$\scriptstyle\frac{1}{2},\frac{1}{2},\frac{1}{2}, -\frac{1}{2} ,
    -\frac{1}{2} , -\frac{1}{2},\frac{1}{2}, -\frac{1}{2}$}\} \\
  ${a_{16|3}}$ & = &${\alpha [} 95 ]$& = & \{1,2,3,4,2,2,1,1\} & =
  &\{{$\scriptstyle\frac{1}{2},-\frac{1}{2} ,\frac{1}{2},
    \frac{1}{2},\frac{1}{2},-\frac{1}{2} , -\frac{1}{2} ,
    -\frac{1}{2}$ }\} \\ ${a_{16|4}}$ & = &${\alpha [} 96 ]$& = &
  \{1,2,3,4,2,3,1,0\} & = &\{{$\scriptstyle-\frac{1}{2}
    ,\frac{1}{2},\frac{1}{2},
    \frac{1}{2},\frac{1}{2},\frac{1}{2},\frac{1}{2}, -\frac{1}{2}$}\}
  \\
  ${a_{17|1}}$ & = &${\alpha [} 97 ]$& = & \{2,2,3,4,2,2,1,1\} & =
  &\{{$\scriptstyle\frac{1}{2},\frac{1}{2},-\frac{1}{2} ,
    \frac{1}{2},\frac{1}{2},-\frac{1}{2} , -\frac{1}{2} ,
    -\frac{1}{2}$ }\} \\ ${a_{17|2}}$ & = &${\alpha [} 98 ]$& = &
  \{2,3,3,3,2,2,1,1\} & =
  &\{{$\scriptstyle\frac{1}{2},\frac{1}{2},\frac{1}{2}, -\frac{1}{2} ,
    -\frac{1}{2} ,\frac{1}{2}, -\frac{1}{2} , -\frac{1}{2}$}\} \\
  ${a_{17|3}}$ & = &${\alpha [} 99 ]$& = & \{1,2,3,4,2,3,1,1\} & =
  &\{{$\scriptstyle\frac{1}{2},-\frac{1}{2} ,\frac{1}{2},
    \frac{1}{2},\frac{1}{2},\frac{1}{2},\frac{1}{2}, -\frac{1}{2}$}\}
  \\ ${a_{17|4}}$ & = &${\alpha [} 100 ]$& = & \{1,2,3,4,2,3,2,0\} & =
  &\big\{-1,0,0,0,0,0,0,-1\big\}
  \\
  ${a_{18|1}}$ & = &${\alpha [} 101 ]$& = & \{2,2,3,4,2,3,1,1\} & =
  &\{{$\scriptstyle \frac{1}{2},\frac{1}{2},-\frac{1}{2} ,
    \frac{1}{2},\frac{1}{2},\frac{1}{2},\frac{1}{2}, -\frac{1}{2}$}\}
  \\ ${a_{18|2}}$ & = &${\alpha [} 102 ]$& = & \{2,3,3,4,2,2,1,1\} & =
  &\{{$\scriptstyle\frac{1}{2},\frac{1}{2},\frac{1}{2}, -\frac{1}{2}
    ,\frac{1}{2}, -\frac{1}{2} , -\frac{1}{2} , -\frac{1}{2}$}\} \\
  ${a_{18|3}}$ & = &${\alpha [} 103 ]$& = & \{1,2,3,4,2,3,2,1\} & =
  &\big\{0,-1,0,0,0,0,0,-1\big\}
  \\
  ${a_{19|1}}$ & = &${\alpha [} 104 ]$& = & \{2,2,3,4,2,3,2,1\} & =
  &\big\{0,0,-1,0,0,0,0,-1\big\} \\ ${a_{19|2}}$ & = &${\alpha [} 105
  ]$& = & \{2,3,3,4,2,3,1,1\} & =
  &\{{$\scriptstyle\frac{1}{2},\frac{1}{2},\frac{1}{2}, -\frac{1}{2}
    ,\frac{1}{2},\frac{1}{2}, \frac{1}{2},-\frac{1}{2}$}\} \\
  ${a_{19|3}}$ & = &${\alpha [} 106 ]$& = & \{2,3,4,4,2,2,1,1\} & =
  &\{{$\scriptstyle\frac{1}{2},\frac{1}{2},\frac{1}{2},\frac{1}{2},
    -\frac{1}{2} , -\frac{1}{2} , -\frac{1}{2} ,
    -\frac{1}{2}$}\}\\
  ${a_{20|1}}$ & = &${\alpha [} 107 ]$& = & \{2,3,3,4,2,3,2,1\} & =
  &\big\{0,0,0,-1,0,0,0,-1\big\} \\ ${a_{20|2}}$ & = &${\alpha [} 108
  ]$& = & \{2,3,4,4,2,3,1,1\} & =
  &\{{$\scriptstyle\frac{1}{2},\frac{1}{2},\frac{1}{2},\frac{1}{2},
    -\frac{1}{2} ,\frac{1}{2},\frac{1}{2}, -\frac{1}{2}$}\ 
  \\
  ${a_{21|1}}$ & = &${\alpha [} 109 ]$& = & \{2,3,4,4,2,3,2,1\} & =
  &\big\{ 0,0,0,0,-1,0,0,-1\big\} \\ ${a_{21|2}}$ & = &${\alpha [} 110
  ]$& = & \{2,3,4,5,2,3,1,1\} & =
  &\{{$\scriptstyle\frac{1}{2},\frac{1}{2},\frac{1}{2},\frac{1}{2},
    \frac{1}{2},-\frac{1}{2} ,\frac{1}{2}, -\frac{1}{2}$ }\}
  \\
  ${a_{22|1}}$ & = &${\alpha [} 111 ]$& = & \{2,3,4,5,2,3,2,1\} & =
  &\big\{0,0,0,0,0,-1,0,-1\big\} \\ ${a_{22|2}}$ & = &${\alpha [} 112
  ]$& = & \{2,3,4,5,3,3,1,1\} & = &\{{$\scriptstyle
    \frac{1}{2},\frac{1}{2},\frac{1}{2},\frac{1}{2},
    \frac{1}{2},\frac{1}{2},-\frac{1}{2} , -\frac{1}{2}$}\}\\
  ${a_{23|1}}$ & = &${\alpha [} 113 ]$& = & \{2,3,4,5,3,3,2,1\} & =
  &\big\{0,0,0,0,0,0,-1,-1\} \\ ${a_{23|2}}$ & = &${\alpha [} 114 ]$&
  = & \{2,3,4,5,2,4,2,1\} & = &\big\{0,0,0,0,0,0,1,-1\big\}\\
  ${a_{24|1}}$ & = &${\alpha [} 115 ]$& = & \{2,3,4,5,3,4,2,1\} & =
  &\big\{0,0,0,0,0,1,0,-1\big\}
  \\
  ${a_{25|1}}$ & = &${\alpha [} 116 ]$& = & \{2,3,4,6,3,4,2,1\} & =
  &\big\{0,0,0,0,1,0,0,-1\big\}
  \\
  ${a_{26|1}}$ & = &${\alpha [} 117 ]$& = & \{2,3,5,6,3,4,2,1\} & =
  &\big\{0,0,0,1,0,0,0,-1\big\}
  \\
  ${a_{27|1}}$ & = &${\alpha [} 118 ]$& = & \{2,4,5,6,3,4,2,1\} & =
  &\big\{ 0,0,1,0,0,0,0,-1\big\}
  \\
  ${a_{28|1}}$ & = &${\alpha [} 119 ]$& = & \{3,4,5,6,3,4,2,1\} & =
  &\big\{0,1,0,0,0,0,0,-1\big\}
  \\
  ${a_{29|1}}$ & = &${\alpha [} 120 ]$& = & \{3,4,5,6,3,4,2,2\} & =
  &\big\{1,0,0,0,0,0,0,-1\big\}
\end{longtable}

\section{Listing the roots according to the dimensional filtration}

\subsubsection{Roots in D[1], D[2], D[3], D[4], D[5] and D[6]}

\begin{longtable}{|ccccc|cc|}
  \caption{Listing of positive $E_8$ rots according to dimensional
  filtration: roots in D[1], D[2], D[3], D[4], D[5] and D[6]}
  \label{tab:root_filtr1-6}\\
  
  \hline\hline
  \text{label} & & \text{root $\#$} & & \text{Dynkin labels} &
  \text{Type IIB} & \text{type IIA}\\
  \hline
  \endfirsthead
  
  \multicolumn{7}{c}%
  {{\bfseries \tablename\ \thetable{} -- continued from previous page}} \\
  \hline
  \text{label} & & \text{root $\#$} & & \text{Dynkin labels} &
  \text{Type IIB} & \text{type IIA} \\
  \hline
  \endhead  
  
  \hline \multicolumn{7}{|r|}{{Continued on next page}} \\ \hline
  \endfoot
  
  \hline \hline
  \endlastfoot

{$d_{1|1}$} & = &  $\alpha [ 7 ]$& = &
 \{0,0,0,0,0,0,1,0\}   & $\rho$ & $C_{9}$ \\

{$d_{2|1}$} & = &  ${\alpha [} 6 ]$& = &
 \{0,0,0,0,0,1,0,0\}   & $B_{8\,9}$ & $\gamma_{8}^{\phantom{8}\,9}$\\
{$d_{2|2}$} & = &  ${\alpha [} 5 ]$& = &
 \{0,0,0,0,1,0,0,0\}   & $\gamma_{8}{}^{9}$ & $\gamma_{8}{}^{9}$ \\
{$d_{2|3}$} & = &  ${\alpha [} 15 ]$& = &
 \{0,0,0,0,0,1,1,0\}& $C_{8\,9}$ & $C_{8}$   \\

{$d_{3|1}$} & = &  ${\alpha [} 22 ]$& = &
 \{0,0,0,1,1,1,0,0\}   & $B_{7\,8}$ & $\gamma_{7}^{\phantom{7}\, 8}$ \\
{$d_{3|2}$} & = &  ${\alpha [} 4 ]$& = &
 \{0,0,0,1,0,0,0,0\}   &  $\gamma_{7}{}^{8}$ &$\gamma_{7}{}^{8}$\\
{$d_{3|3}$} & = &  ${\alpha [} 14 ]$& = &
 \{0,0,0,1,0,1,0,0\}   & $B_{7\,9}$ & $\gamma_{7}{}^{9}$\\
{$d_{3|4}$} & = &  ${\alpha [} 13 ]$& = &
 \{0,0,0,1,1,0,0,0\}   &  $\gamma_{7}{}^{9}$ & $B_{7\,9}$ \\
{$d_{3|5}$} & = &  ${\alpha [} 29 ]$& = &
 \{0,0,0,1,1,1,1,0\}  & $C_{7\,8}$ & $C_{7\,8\,9}$ \\
{$d_{3|6}$} & = &  ${\alpha [} 21 ]$& = &
 \{0,0,0,1,0,1,1,0\}   & $C_{7\,9}$ & $C_{7}$ \\

{$d_{4|1}$} & = &  ${\alpha [} 36 ]$& = &
 \{0,0,1,2,1,1,0,0\}   &$B_{6\,7}$ & $B_{6\,7}$ \\
{$d_{4|2}$} & = &  ${\alpha [} 3 ]$& = &
 \{0,0,1,0,0,0,0,0\}   &$\gamma_{6}{}^{7}$  & $\gamma_{6}{}^{7}$ \\
{$d_{4|3}$} & = &  ${\alpha [} 28 ]$& = &
 \{0,0,1,1,1,1,0,0\}   &$B_{6\,8}$ & $B_{6\,8}$\\
{$d_{4|4}$} & = &  ${\alpha [} 12 ]$& = &
 \{0,0,1,1,0,0,0,0\}   &$\gamma_{6}{}^{8}$  &$\gamma_{6}{}^{8}$ \\
{$d_{4|5}$} & = &  ${\alpha [} 20 ]$& = &
 \{0,0,1,1,0,1,0,0\}   &$B_{6\,9}$ &$\gamma_{6}{}^{9}$ \\
{$d_{4|6}$} & = &  ${\alpha [} 19 ]$& = &
 \{0,0,1,1,1,0,0,0\}   &$\gamma_{6}{}^{9}$ &$B_{6\,9}$ \\
{$d_{4|7}$} & = &  ${\alpha [} 41 ]$& = &
 \{0,0,1,2,1,1,1,0\}   &$C_{6\,7}$ & $C_{6\,7\,9}$ \\
{$d_{4|8}$} & = &  ${\alpha [} 34 ]$& = &
 \{0,0,1,1,1,1,1,0\}   &$C_{6\,8}$ & $C_{6\,8\,9}$\\
{$d_{4|9}$} & = &  ${\alpha [} 27 ]$& = &
 \{0,0,1,1,0,1,1,0\}   &$C_{6\,9}$ & $C_{6}$\\
{$d_{4|10}$} & = &  ${\alpha [} 49 ]$& = &
 \{0,0,1,2,1,2,1,0\}  &$C_{6\,7\,8\,9}$ & $C_{6\,7\,8}$ \\

{$d_{5|1}$} & = &  ${\alpha [} 47 ]$& = &
 \{0,1,2,2,1,1,0,0\}    &$B_{5\,6}$ & $B_{5\,6}$\\
{$d_{5|2}$} & = &  ${\alpha [} 2 ]$& = &
 \{0,1,0,0,0,0,0,0\}   &$\gamma_{5}{}^{6}$ & $\gamma_{5}{}^{6}$ \\
{$d_{5|3}$} & = &  ${\alpha [} 40 ]$& = &
 \{0,1,1,2,1,1,0,0\}   &$B_{5\,7}$ & $B_{5\,7}$ \\
{$d_{5|4}$} & = &  ${\alpha [} 11 ]$& = &
 \{0,1,1,0,0,0,0,0\}  &$\gamma_{5}{}^{7}$ &$\gamma_{5}{}^{7}$ \\
{$d_{5|5}$} & = &  ${\alpha [} 33 ]$& = &
 \{0,1,1,1,1,1,0,0\}   &$B_{5\,8}$ & $B_{5\,8}$\\
{$d_{5|6}$} & = &  ${\alpha [} 18 ]$& = &
 \{0,1,1,1,0,0,0,0\}   &$\gamma_{5}{}^{8}$  &$\gamma_{5}{}^{8}$  \\
{$d_{5|7}$} & = &  ${\alpha [} 25 ]$& = &
 \{0,1,1,1,0,1,0,0\}   &$B_{5\,9}$ &$\gamma_{5}{}^{9}$\\
{$d_{5|8}$} & = &  ${\alpha [} 24 ]$& = &
 \{0,1,1,1,1,0,0,0\}   &$\gamma_{5}{}^{9}$ &$B_{5\,9}$\\
{$d_{5|9}$} & = &  ${\alpha [} 54 ]$& = &
 \{0,1,2,2,1,1,1,0\}   &$C_{5\,6}$ & $C_{5\,6\,9}$ \\
{$d_{5|10}$} & = &  ${\alpha [} 46 ]$& = &
 \{0,1,1,2,1,1,1,0\}   &$C_{5\,7}$ & $C_{5\,7\,9}$\\
{$d_{5|11}$} & = &  ${\alpha [} 39 ]$& = &
 \{0,1,1,1,1,1,1,0\}  &$C_{5\,8}$ & $C_{5\,8\,9}$ \\
{$d_{5|12}$} & = &  ${\alpha [} 32 ]$& = &
 \{0,1,1,1,0,1,1,0\}   &$C_{5\,9}$ & $C_{5}$ \\
{$d_{5|13}$} & = &  ${\alpha [} 53 ]$& = &
 \{0,1,1,2,1,2,1,0\}   &$C_{5\,7\,8\,9}$ & $C_{5\,7\,8}$\\
{$d_{5|14}$} & = &  ${\alpha [} 60 ]$& = &
 \{0,1,2,2,1,2,1,0\}   &$C_{5\,6\,8\,9}$ & $C_{5\,6\,8}$\\
{$d_{5|15}$} & = &  ${\alpha [} 67 ]$& = &
 \{0,1,2,3,1,2,1,0\}   &$C_{5\,6\,7\,9}$ & $C_{5\,6\,7}$\\
{$d_{5|16}$} & = &  ${\alpha [} 74 ]$& = &
 \{0,1,2,3,2,2,1,0\}   &$C_{5\,6\,7\,8}$ & $C_{3\, 4\,\mu}$\\
  
{$d_{6|1}$} & = &  ${\alpha [} 59 ]$& = &
 \{1,2,2,2,1,1,0,0\}   &$B_{4\,5}$ & $B_{4\,5}$ \\
{$d_{6|2}$} & = &  ${\alpha [} 1 ]$& = &
 \{1,0,0,0,0,0,0,0\}   &$\gamma_{4}{}^{5}$ &$\gamma_{4}{}^{5}$\\
{$d_{6|3}$} & = &  ${\alpha [} 52 ]$& = &
 \{1,1,2,2,1,1,0,0\}   &$B_{4\,6}$ & $B_{4\,6}$\\
{$d_{6|4}$} & = &  ${\alpha [} 9 ]$& = &
 \{1,1,0,0,0,0,0,0\}   &$\gamma_{4}{}^{6}$&$\gamma_{4}{}^{6}$\\
{$d_{6|5}$} & = &  ${\alpha [} 45 ]$& = &
 \{1,1,1,2,1,1,0,0\}   &$B_{4\,7}$ & $B_{4\,7}$\\
{$d_{6|6}$} & = &  ${\alpha [} 16 ]$& = &
 \{1,1,1,0,0,0,0,0\}   &$\gamma_{4}{}^{7}$ &$\gamma_{4}{}^{7}$\\
{$d_{6|7}$} & = &  ${\alpha [} 38 ]$& = &
 \{1,1,1,1,1,1,0,0\}   &$B_{4\,8}$ & $B_{4\,8}$\\
{$d_{6|8}$} & = &  ${\alpha [} 23 ]$& = &
 \{1,1,1,1,0,0,0,0\}   &$\gamma_{4}{}^{8}$&$\gamma_{4}{}^{8}$\\
{$d_{6|9}$} & = &  ${\alpha [} 31 ]$& = &
 \{1,1,1,1,0,1,0,0\}   &$B_{4\,9}$ & $\gamma_{4}{}^{9}$\\
{$d_{6|10}$} & = &  ${\alpha [} 30 ]$& = &
 \{1,1,1,1,1,0,0,0\}   &$\gamma_{4}{}^{9}$ &$B_{4\,9}$\\
{$d_{6|11}$} & = &  ${\alpha [} 100 ]$& = &
 \{1,2,3,4,2,3,2,0\}    &$B_{3\,\mu}$ & $B_{3\,\mu}$\\
{$d_{6|12}$} & = &  ${\alpha [} 96 ]$& = &
 \{1,2,3,4,2,3,1,0\}   &$C_{3\,\mu}$ & $C_{3\,9\,\mu}$\\
{$d_{6|13}$} & = &  ${\alpha [} 64 ]$& = &
 \{1,2,2,2,1,1,1,0\}  &$C_{4\,5}$ & $C_{4\,5\,9}$ \\
{$d_{6|14}$} & = &  ${\alpha [} 58 ]$& = &
 \{1,1,2,2,1,1,1,0\}  &$C_{4\,6}$ & $C_{4\,6\,9}$ \\
{$d_{6|15}$} & = &  ${\alpha [} 51 ]$& = &
 \{1,1,1,2,1,1,1,0\}  &$C_{4\,7}$ & $C_{4\,7\,9}$ \\
{$d_{6|16}$} & = &  ${\alpha [} 44 ]$& = &
 \{1,1,1,1,1,1,1,0\}  &$C_{4\,8}$ & $C_{4\,8\,9}$ \\
{$d_{6|17}$} & = &  ${\alpha [} 37 ]$& = &
 \{1,1,1,1,0,1,1,0\}   &$C_{4\,9}$ & $C_{4}$\\
{$d_{6|18}$} & = &  ${\alpha [} 57 ]$& = &
 \{1,1,1,2,1,2,1,0\}  &$C_{4\,7\,8\,9}$ & $C_{4\,7\,8}$  \\
{$d_{6|19}$} & = &  ${\alpha [} 63 ]$& = &
 \{1,1,2,2,1,2,1,0\}  &$C_{4\,6\,8\,9}$ & $C_{4\,6\,8}$ \\
{$d_{6|20}$} & = &  ${\alpha [} 70 ]$& = &
 \{1,1,2,3,1,2,1,0\}   &$C_{4\,6\,7\,9}$ & $C_{4\,6\,7}$\\
{$d_{6|21}$} & = &  ${\alpha [} 76 ]$& = &
 \{1,1,2,3,2,2,1,0\}   &$C_{4\,6\,7\,8}$ & $C_{3\,5\,\mu}$\\
{$d_{6|22}$} & = &  ${\alpha [} 71 ]$& = &
 \{1,2,2,2,1,2,1,0\}   &$C_{4\,5\,8\,9}$ & $C_{4\,5\,8}$\\
{$d_{6|23}$} & = &  ${\alpha [} 77 ]$& = &
 \{1,2,2,3,1,2,1,0\}   &$C_{4\,5\,7\,9}$ & $C_{4\,5\,7}$\\
{$d_{6|24}$} & = &  ${\alpha [} 81 ]$& = &
 \{1,2,2,3,2,2,1,0\}   &$C_{4\,5\,7\,8}$ & $C_{3\,6\,\mu}$\\
{$d_{6|25}$} & = &  ${\alpha [} 83 ]$& = &
 \{1,2,3,3,1,2,1,0\}   &$C_{4\,5\,6\,9}$ & $C_{4\,5\,6}$\\
{$d_{6|26}$} & = &  ${\alpha [} 87 ]$& = &
 \{1,2,3,3,2,2,1,0\}   &$C_{4\,5\,6\,8}$ & $C_{3\,7\,\mu}$\\
{$d_{6|27}$} & = &  ${\alpha [} 92 ]$& = &
 \{1,2,3,4,2,2,1,0\}   &$C_{4\,5\,6\,7}$ & $C_{3\,8\,\mu}$\\
\end{longtable}

\subsubsection{Roots in D[7]}

\begin{itemize}
\item Electric with respect to the electric subgroup $\mathrm{SL(8)} \subset
\mathrm{E_{7(7)}} \subset\mathrm{ E_{8(8)}}$
\begin{longtable}{|ccccccc|cc|}
  \caption{Roots in D[7] electric with respect to 
    $\mathrm{SL(8)} \subset \mathrm{E_{7(7)}} \subset\mathrm{E_{8(8)}}$}
  \label{tab:root_filtr7el}\\
  
  \hline\hline
  \text{label} & & \text{root $\#$} & & \text{Dynkin labels} & &
  q-vector &  \text{Type IIB} & \text{type IIA}\\
  \hline
  \endfirsthead
  
  \multicolumn{9}{c}%
  {{\bfseries \tablename\ \thetable{} -- continued from previous page}} \\
  \hline
  \text{label} & & \text{root $\#$} & & \text{Dynkin labels} & &
  q-vector &  \text{Type IIB} & \text{type IIA}\\
  \hline
  \endhead  
  
  \hline \multicolumn{9}{|r|}{{Continued on next page}} \\ \hline
  \endfoot
  
  \hline \hline
  \endlastfoot

{$d_{7|1}$} & = &  ${\alpha [} 50 ]$& = &
 \{1,1,1,1,0,1,1,1\} & $\Rightarrow$ &  \{1,1,1,1,0,1,1\} &
 $C_{3\,9}$&$C_{3}$ \\ 
{$d_{7|2}$} & = &  ${\alpha [} 99 ]$& = &
 \{1,2,3,4,2,3,1,1\} & $\Rightarrow$ &  \{1,2,3,4,2,3,1\} &
 $C_{4\,\mu}$&$C_{4\,9\,\mu}$ \\
{$d_{7|3}$} & = &  ${\alpha [} 101 ]$& = &
 \{2,2,3,4,2,3,1,1\} & $\Rightarrow$ &  \{2,2,3,4,2,3,1\} &
 $C_{5\,\mu}$&$C_{5\,9\,\mu}$ \\
{$d_{7|4}$} & = &  ${\alpha [} 105 ]$& = &
 \{2,3,3,4,2,3,1,1\} & $\Rightarrow$ &  \{2,3,3,4,2,3,1\} &
 $C_{6\,\mu}$&$C_{6\,9\,\mu}$ \\
{$d_{7|5}$} & = &  ${\alpha [} 108 ]$& = &
 \{2,3,4,4,2,3,1,1\} & $\Rightarrow$ &  \{2,3,4,4,2,3,1\} &
 $C_{7\,\mu}$&$C_{7\,9\,\mu}$ \\
{$d_{7|6}$} & = &  ${\alpha [} 110 ]$& = &
 \{2,3,4,5,2,3,1,1\} & $\Rightarrow$ &  \{2,3,4,5,2,3,1\} &
 $C_{8\,\mu}$&$C_{8\,9\,\mu}$  \\
{$d_{7|7}$} & = &  ${\alpha [} 84 ]$& = &
 \{1,1,2,3,2,2,1,1\} & $\Rightarrow$ &  \{1,1,2,3,2,2,1\} &  $C_{3\,6\,7\,8}$&
$C_{4\,5\,\mu}$  \\
{$d_{7|8}$} & = &  ${\alpha [} 86 ]$& = &
 \{1,2,2,3,2,2,1,1\} & $\Rightarrow$ &  \{1,2,2,3,2,2,1\} & $C_{3\,5\,7\,8}$&
$C_{4\,6\,\mu}$  \\
{$d_{7|9}$} & = &  ${\alpha [} 91 ]$& = &
 \{1,2,3,3,2,2,1,1\} & $\Rightarrow$ &  \{1,2,3,3,2,2,1\} & $C_{3\,5\,6\,8}$&
$C_{4\,7\,\mu}$ \\
{$d_{7|10}$} & = &  ${\alpha [} 95 ]$& = &
 \{1,2,3,4,2,2,1,1\} & $\Rightarrow$ &  \{1,2,3,4,2,2,1\} &  $C_{3\,5\,6\,7}$&
$C_{4\,8\,\mu}$  \\
{$d_{7|11}$} & = &  ${\alpha [} 97 ]$& = &
 \{2,2,3,4,2,2,1,1\} & $\Rightarrow$ &  \{2,2,3,4,2,2,1\} & $C_{3\,4\,6\,7}$&
$C_{5\,8\,\mu}$ \\
{$d_{7|12}$} & = &  ${\alpha [} 102 ]$& = &
 \{2,3,3,4,2,2,1,1\} & $\Rightarrow$ &  \{2,3,3,4,2,2,1\} & $C_{3\,4\,5\,7}$&
$C_{6\,8\,\mu}$  \\
{$d_{7|13}$} & = &  ${\alpha [} 106 ]$& = &
 \{2,3,4,4,2,2,1,1\} & $\Rightarrow$ &  \{2,3,4,4,2,2,1\} & $C_{3\,4\,5\,6}$&
$C_{7\,8\,\mu}$   \\
{$d_{7|14}$} & = &  ${\alpha [} 93 ]$& = &
 \{2,2,3,3,2,2,1,1\} & $\Rightarrow$ &  \{2,2,3,3,2,2,1\} & $C_{3\,4\,6\,8}$&
$C_{5\,7\,\mu}$   \\
{$d_{7|15}$} & = &  ${\alpha [} 89 ]$& = &
 \{2,2,2,3,2,2,1,1\} & $\Rightarrow$ &  \{2,2,2,3,2,2,1\} & $C_{3\,4\,7\,8}$&
$C_{5\,6\,\mu}$ \\
{$d_{7|16}$} & = &  ${\alpha [} 98 ]$& = &
 \{2,3,3,3,2,2,1,1\} & $\Rightarrow$ &  \{2,3,3,3,2,2,1\} & $C_{3\,4\,5\,8}$&
$C_{6\,7\,\mu}$ \\
{$d_{7|17}$} & = &  ${\alpha [} 103 ]$& = &
 \{1,2,3,4,2,3,2,1\} & $\Rightarrow$ &  \{1,2,3,4,2,3,2\} &
 $B_{4\,\mu}$& $B_{4\,\mu}$ \\
{$d_{7|18}$} & = &  ${\alpha [} 104 ]$& = &
 \{2,2,3,4,2,3,2,1\} & $\Rightarrow$ &  \{2,2,3,4,2,3,2\} & $B_{5\,\mu}$&
$B_{5\,\mu}$ \\
{$d_{7|19}$} & = &  ${\alpha [} 107 ]$& = &
 \{2,3,3,4,2,3,2,1\} & $\Rightarrow$ &  \{2,3,3,4,2,3,2\} & $B_{6\,\mu}$&
$B_{6\,\mu}$ \\
{$d_{7|20}$} & = &  ${\alpha [} 109 ]$& = &
 \{2,3,4,4,2,3,2,1\} & $\Rightarrow$ &  \{2,3,4,4,2,3,2\} &
 $B_{7\,\mu}$& $B_{7\,\mu}$ \\
{$d_{7|21}$} & = &  ${\alpha [} 111 ]$& = &
 \{2,3,4,5,2,3,2,1\} & $\Rightarrow$ &  \{2,3,4,5,2,3,2\} & $B_{8\,\mu}$&
 $B_{8\,\mu}$ \\
{$d_{7|22}$} & = &  ${\alpha [} 114 ]$& = &
 \{2,3,4,5,2,4,2,1\} & $\Rightarrow$ &  \{2,3,4,5,2,4,2\} &
 $\gamma_\mu{}^{9}$ & $B_{9\mu}$  \\
{$d_{7|23}$} & = &  ${\alpha [} 8 ]$& = &
 \{0,0,0,0,0,0,0,1\} & $\Rightarrow$ &  \{0,0,0,0,0,0,0\} &
$\gamma_{3}{}^{4}$&$\gamma_{3}{}^{4}$   \\
{$d_{7|24}$} & = &  ${\alpha [} 10 ]$& = &
 \{1,0,0,0,0,0,0,1\} & $\Rightarrow$ &  \{1,0,0,0,0,0,0\} & $\gamma_{3}{}^{5}$&
$\gamma_{3}{}^{5}$  \\
{$d_{7|25}$} & = &  ${\alpha [} 17 ]$& = &
 \{1,1,0,0,0,0,0,1\} & $\Rightarrow$ &  \{1,1,0,0,0,0,0\}
&$\gamma_{3}{}^{6}$ &  $\gamma_{3}{}^{6}$   \\
{$d_{7|26}$} & = &  ${\alpha [} 26 ]$& = &
 \{1,1,1,0,0,0,0,1\} & $\Rightarrow$ &  \{1,1,1,0,0,0,0\} &
 $\gamma_{3}{}^{7}$ & $\gamma_{3}{}^{7}$  \\
{$d_{7|27}$} & = &  ${\alpha [} 35 ]$& = &
 \{1,1,1,1,0,0,0,1\} & $\Rightarrow$ &  \{1,1,1,1,0,0,0\} &
 $\gamma_{3}{}^{8}$ & $\gamma_{3}{}^{8}$  \\
{$d_{7|28}$} & = &  ${\alpha [} 43 ]$& = &
 \{1,1,1,1,0,1,0,1\} & $\Rightarrow$ &  \{1,1,1,1,0,1,0\} & $B_{3\,9}$&
$\gamma_{3}{}^9$ \\
\end{longtable}

\item Magnetic with respect to the electric subgroup $\mathrm{SL(8)} \subset
\mathrm{E_{7(7)}} \subset\mathrm{ E_{8(8)}}$
\begin{longtable}{|ccccccc|cc|}
  \caption{Roots in D[7] magnetic with respect to 
    $\mathrm{SL(8)} \subset \mathrm{E_{7(7)}} \subset\mathrm{E_{8(8)}}$}
  \label{tab:root_filtr7ma}\\
  
  \hline\hline
  \text{label} & & \text{root $\#$} & & \text{Dynkin labels} & &
  q-vector &  \text{Type IIB} & \text{type IIA}\\
  \hline
  \endfirsthead
  
  \multicolumn{9}{c}%
  {{\bfseries \tablename\ \thetable{} -- continued from previous page}} \\
  \hline
  \text{label} & & \text{root $\#$} & & \text{Dynkin labels} & &
  q-vector &  \text{Type IIB} & \text{type IIA}\\
  \hline
  \endhead  
  
  \hline \multicolumn{9}{|r|}{{Continued on next page}} \\ \hline
  \endfoot
  
  \hline \hline
  \endlastfoot

{$d_{7|29}$} & = &  ${\alpha [} 112 ]$& = &
 \{2,3,4,5,3,3,1,1\} & $\Rightarrow$ &  \{2,3,4,5,3,3,1\} & $C_{9\mu}$
 & $C_\mu$ \\ 
{$d_{7|30}$} & = &  ${\alpha [} 75 ]$& = &
 \{2,2,2,2,1,1,1,1\} & $\Rightarrow$ &  \{2,2,2,2,1,1,1\} & $C_{3\,4}$
 & $C_{3\,4\,9}$ \\
{$d_{7|31}$} & = &  ${\alpha [} 72 ]$& = &
 \{1,2,2,2,1,1,1,1\} & $\Rightarrow$ &  \{1,2,2,2,1,1,1\} &  $C_{3\,5}$ &
$C_{3\,5\,9}$ \\
{$d_{7|32}$} & = &  ${\alpha [} 66 ]$& = &
 \{1,1,2,2,1,1,1,1\} & $\Rightarrow$ &  \{1,1,2,2,1,1,1\} &  $C_{3\,6}$ &
$C_{3\,6\,9}$ \\
{$d_{7|33}$} & = &  ${\alpha [} 62 ]$& = &
 \{1,1,1,2,1,1,1,1\} & $\Rightarrow$ &  \{1,1,1,2,1,1,1\} & $C_{3\,7}$ &
$C_{3\,7\,9}$ \\
{$d_{7|34}$} & = &  ${\alpha [} 56 ]$& = &
 \{1,1,1,1,1,1,1,1\} & $\Rightarrow$ &  \{1,1,1,1,1,1,1\} & $C_{3\,8}$ &
$C_{3\,8\,9}$ \\
{$d_{7|35}$} & = &  ${\alpha [} 94 ]$& = &
 \{2,3,3,3,1,2,1,1\} & $\Rightarrow$ &  \{2,3,3,3,1,2,1\} & $C_{3\,4\,5\,9}$ &
$C_{3\,4\,5}$ \\
{$d_{7|36}$} & = &  ${\alpha [} 90 ]$& = &
 \{2,2,3,3,1,2,1,1\} & $\Rightarrow$ &  \{2,2,3,3,1,2,1\} & $C_{3\,4\,6\,9}$ &
$C_{3\,4\,6}$ \\
{$d_{7|37}$} & = &  ${\alpha [} 85 ]$& = &
 \{2,2,2,3,1,2,1,1\} & $\Rightarrow$ &  \{2,2,2,3,1,2,1\} & $C_{3\,4\,7\,9}$ &
$C_{3\,4\,7}$ \\
{$d_{7|38}$} & = &  ${\alpha [} 80 ]$& = &
 \{2,2,2,2,1,2,1,1\} & $\Rightarrow$ &  \{2,2,2,2,1,2,1\} & $C_{3\,4\,8\,9}$ &
$C_{3\,4\,8}$ \\
{$d_{7|39}$} & = &  ${\alpha [} 78 ]$& = &
 \{1,2,2,2,1,2,1,1\} & $\Rightarrow$ &  \{1,2,2,2,1,2,1\} &  $C_{3\,5\,8\,9}$ &
$C_{3\,5\,8}$ \\
{$d_{7|40}$} & = &  ${\alpha [} 73 ]$& = &
 \{1,1,2,2,1,2,1,1\} & $\Rightarrow$ &  \{1,1,2,2,1,2,1\} & $C_{3\,6\,8\,9}$ &
$C_{3\,6\,8}$ \\
{$d_{7|41}$} & = &  ${\alpha [} 68 ]$& = &
 \{1,1,1,2,1,2,1,1\} & $\Rightarrow$ &  \{1,1,1,2,1,2,1\} & $C_{3\,7\,8\,9}$ &
$C_{3\,7\,8}$ \\
{$d_{7|42}$} & = &  ${\alpha [} 82 ]$& = &
 \{1,2,2,3,1,2,1,1\} & $\Rightarrow$ &  \{1,2,2,3,1,2,1\} & $C_{3\,5\,7\,9}$ &
$C_{3\,5\,7}$ \\
{$d_{7|43}$} & = &  ${\alpha [} 88 ]$& = &
 \{1,2,3,3,1,2,1,1\} & $\Rightarrow$ &  \{1,2,3,3,1,2,1\} & $C_{3\,5\,6\,9}$ &
$C_{3\,5\,6}$ \\
{$d_{7|44}$} & = &  ${\alpha [} 79 ]$& = &
 \{1,1,2,3,1,2,1,1\} & $\Rightarrow$ &  \{1,1,2,3,1,2,1\} & $C_{3\,6\,7\,9}$ &
$C_{3\,6\,7}$ \\
{$d_{7|45}$} & = &  ${\alpha [} 69 ]$& = &
 \{2,2,2,2,1,1,0,1\} & $\Rightarrow$ &  \{2,2,2,2,1,1,0\} & $B_{3\,4}$ &
$B_{3\,4}$ \\
{$d_{7|46}$} & = &  ${\alpha [} 65 ]$& = &
 \{1,2,2,2,1,1,0,1\} & $\Rightarrow$ &  \{1,2,2,2,1,1,0\} & $B_{3\,5}$ &
$B_{3\,5}$ \\
{$d_{7|47}$} & = &  ${\alpha [} 61 ]$& = &
 \{1,1,2,2,1,1,0,1\} & $\Rightarrow$ &  \{1,1,2,2,1,1,0\} & $B_{3\,6}$ &
$B_{3\,6}$ \\
{$d_{7|48}$} & = &  ${\alpha [} 55 ]$& = &
 \{1,1,1,2,1,1,0,1\} & $\Rightarrow$ &  \{1,1,1,2,1,1,0\} & $B_{3\,7}$ &
$B_{3\,7}$ \\
{$d_{7|49}$} & = &  ${\alpha [} 48 ]$& = &
 \{1,1,1,1,1,1,0,1\} & $\Rightarrow$ &  \{1,1,1,1,1,1,0\} & $B_{3\,8}$
 &$B_{3\,8}$ \\ 
{$d_{7|50}$} & = &  ${\alpha [} 42 ]$& = &
 \{1,1,1,1,1,0,0,1\} & $\Rightarrow$ &  \{1,1,1,1,1,0,0\} &
 $\gamma_{3}{}^{9}$ & $B_{39}$ \\
{$d_{7|51}$} & = &  ${\alpha [} 119 ]$& = &
 \{3,4,5,6,3,4,2,1\} & $\Rightarrow$ &  \{3,4,5,6,3,4,2\} &
 $\gamma_{\mu}{}^{4}$ & $\gamma_{\mu}{}^{4}$ \\
{$d_{7|52}$} & = &  ${\alpha [} 118 ]$& = &
 \{2,4,5,6,3,4,2,1\} & $\Rightarrow$ &  \{2,4,5,6,3,4,2\} &
 $\gamma_{\mu}{}^{5}$  & $\gamma_{\mu}{}^{5}$ \\
{$d_{7|53}$} & = &  ${\alpha [} 117 ]$& = &
 \{2,3,5,6,3,4,2,1\} & $\Rightarrow$ &  \{2,3,5,6,3,4,2\} &
 $\gamma_{\mu}{}^{6}$ & $\gamma_{\mu}{}^{6}$ \\
{$d_{7|54}$} & = &  ${\alpha [} 116 ]$& = &
 \{2,3,4,6,3,4,2,1\} & $\Rightarrow$ &  \{2,3,4,6,3,4,2\} &
 $\gamma_{\mu}{}^{7}$ & $\gamma_{\mu}{}^{7}$ \\
{$d_{7|55}$} & = &  ${\alpha [} 115 ]$& = &
 \{2,3,4,5,3,4,2,1\} & $\Rightarrow$ &  \{2,3,4,5,3,4,2\} &
 $\gamma_{\mu}{}^{8}$ & $\gamma_{\mu}{}^{8}$ \\
{$d_{7|56}$} & = &  ${\alpha [} 113 ]$& = &
 \{2,3,4,5,3,3,2,1\} & $\Rightarrow$ &  \{2,3,4,5,3,3,2\} &
 $B_{9\,\mu}$ & $\gamma_{\mu 9}$ \\
\end{longtable}
\end{itemize}

\subsubsection{Roots in D[8]}
\begin{longtable}{|ccccc|cc|}
  \caption{Listing of positive $E_8$ rots according to dimensional
    filtration: roots in D[8]}
  \label{tab:root_filtr8}\\
  \hline\hline
  \text{label} & & \text{root $\#$} & & \text{Dynkin labels} &
  \text{Type IIB} & \text{type IIA}\\
  \hline
  ${d_{8|1}}$ & = &  ${\alpha [} 120 ]$& = &
  \{3,4,5,6,3,4,2,2\}&  $\gamma_\mu{}^4$& $\gamma_\mu{}^4$\\
  \hline\hline
\end{longtable}

%% file: ackn.tex
\chapter*{Acknowledgments}
First of all I would like to thank my supervisor, Mauro Carfora, and
Annalisa Marzuoli for their continuous support, help, encouragement
and teachings during these three years.  \vspace{.3cm}

I also thank my officemate, collaborator and friend, Claudio
Dappiaggi, for his help, his support and for all the fruitful discussions
we had these years\ldots and yes, obviously also for his patience the
many times I interrupted him with ``Scusa, Claudio, ho una domanda da
farti\ldots''.
\vspace{.3cm}

I am very grateful to Pietro Fr\'e, Floriana Gargiulo, Ksenya Rulik,
Alexander Sorin and Mario Trigiante for allowing me to work with them
during my first year of Ph.D.: you taught me a lot.  
\vspace{.3cm}

I also would like to thank the CERN Theoretical Division for its kind
hospitality during the past year, and all people I met there: I learnt
a lot from all of you.
\vspace{.3cm}

Especially, I thank Rodolfo Russo, for accepting to supervise my work
while I was at CERN, for the fruitful collaboration started there and
for the huge amount of things he taught and keep on teaching me. 
\vspace{.3cm}

I also would like to thank all people I met here in Pavia, and
especially my friend Umberto, who truly shared with me this three-year
experience.  
\vspace{.3cm}

Last but not least I thank Marco, obviously not only for his patience,
his continuous encouragement, for the many nights he spent at my side
while I was writing this thesis and for the help he gave me. Without
you I would not be what I am: you are the best part of me and you will
always be.
\newpage
\thispagestyle{empty}